\documentclass[aps, prx, 10pt, twocolumn, a4paper,  notitlepage]{revtex4-1}

\usepackage{amsmath}
\usepackage{amssymb}
\usepackage{textcomp}
\usepackage{upgreek}
\usepackage{units}

\usepackage{hyperref}
\hypersetup{colorlinks=true, citecolor=blue, urlcolor=blue, linkcolor=blue}

\usepackage[pdftex]{graphicx}

\usepackage{microtype}

\usepackage{txfonts}

\renewcommand{\vec}[1]{\ensuremath{\boldsymbol{#1}}}

\newcommand{\change}[1]{\textcolor{black}{#1}}

\AtBeginDocument{%
    \newwrite\bibnotes
    \def\bibnotesext{Notes.bib}
    \immediate\openout\bibnotes=\jobname\bibnotesext
    \immediate\write\bibnotes{@CONTROL{REVTEX41Control}}
    \immediate\write\bibnotes{@CONTROL{%
    apsrev41Control,author="08",editor="1",pages="1",title="0",year="1"}}
     \if@filesw
     \immediate\write\@auxout{\string\citation{apsrev41Control}}%
    \fi
}%

\begin{document}

\title{
\change{
Topological Hybrids of Magnons and Magnon Bound Pairs
}
}

\author{Alexander Mook}
\affiliation{Department of Physics, University of Basel, Klingelbergstrasse 82, CH-4056 Basel, Switzerland}

\author{Rhea Hoyer}
\affiliation{Department of Physics, University of Basel, Klingelbergstrasse 82, CH-4056 Basel, Switzerland}

\author{Jelena Klinovaja}
\affiliation{Department of Physics, University of Basel, Klingelbergstrasse 82, CH-4056 Basel, Switzerland}

\author{Daniel Loss}
\affiliation{Department of Physics, University of Basel, Klingelbergstrasse 82, CH-4056 Basel, Switzerland}

\begin{abstract}
	\change{
	We consider quantum condensed matter systems without particle-number conservation. Since the particle number is not a good quantum number, states belonging to different particle-number sectors can hybridize, which causes topological anticrossings in the spectrum. 
	The resulting spectral gaps support chiral edge excitations whose wavefunction is a superposition of states in the two hybridized sectors. 
	This situation is realized in fully saturated spin-anisotropic quantum magnets without spin conservation, in which single magnons hybridize with magnon bound pairs, i.e., two-magnon bound states. The resulting chiral edge excitations are exotic composites that carry mixed spin-multipolar character, inheriting spin-dipolar and spin-quadrupolar character from their single-particleness and two-particleness, respectively.
	In contrast to established topological magnons, the topological effects discussed here are of genuine quantum mechanical origin and vanish in the classical limit. We discuss implications for both intrinsic anomalous Hall-type transport and beyond-spintronics computation paradigms. 
	We conclude that fully polarized quantum magnets are a promising platform for topology caused by hybridizations \textit{between} particle-number sectors, complementing the field of ultracold atoms working with a conserved number of particles.
	}
\end{abstract}

\maketitle

%
%
\section{Introduction}
Topological band structure theory is a preeminent theme of research on metamaterials \cite{Xin2020}, the solid state \cite{Sato2017}, and ultracold atoms \cite{Cooper2019}. Besides its fundamental appeal, it promises applications in next-generation technologies relying on topological, robust properties of matter \cite{Klitzing1985, Nayak2008, Tian2017, Smejkal2018}. Typically, interacting quantum matter comes with particle-number conserving many-body interactions of the form $a^\dagger a^\dagger a a$ in a second-quantized language, with particle creators $a^\dagger$ and annihilators $a$. Such interactions are described, e.g., by (Bose-)Hubbard models \cite{Lewenstein2007, Cazalilla2011} and recent years saw tremendous progress on single and many-body topology \cite{Rachel2018}.
While particle-number conservation is natural to ultracold atoms in optical lattices and electrons in the solid state, other ubiquitous excitations in quantum condensed matter systems do not share this trait. For example, collective excitations of the lattice \change{(phonons}) or magnetic degrees of freedom \change{(e.g., magnons)} are nonconserved bosonic quasiparticles. As these excitations do not carry charge, they do not suffer from Joule heating and are envisioned to find application in low-energy computation paradigms \cite{Li2012phononics, Pirro2021} and quantum-hybrid systems \cite{Nakamura2019,Lachance_Quirion_2019}, requiring a fundamental understanding of particle-number nonconserving many-body interactions. It is well established that nonconserved particles can be topologically nontrivial on the single-particle level \cite{Liu2019phonons, Li2021topophonons, Malki2020review, McClarty2021review} and exhibit chiral edge states that might \change{be used} as unidirectional information channels \cite{Shindou13, Mook2015waveguide, Wang2018, Chumak2019, Aguilera2020, Pirro2021, Mook2020hinge}.
Unfortunately, a nonconserved number allows a particle to spontaneously decay into several particles via many-body processes, for example, of the form $a^\dagger a^\dagger a$ \cite{Masuda2006, Zhitomirsky2013, Hong2017, Verresen2019}.
Thus, even if these particles come with a nontrivial single-particle topology, they might be so strongly damped and lifetime-broadened that the notion of chiral edge states is questionable \cite{Chernyshev2016}.
Beyond this discouraging roadblock of damping, only a few facts are known about nonconserving many-body interactions in the context of quasiparticle topology: They can
(i) be suppressed in certain cases, reinstating the validity of the noninteracting theory \cite{McClarty2018, Mook2020QuantumDamping},
(ii) play a role in establishing non-Hermitian single-particle topology \cite{McClarty2019}, and
(iii) break symmetries of the noninteracting theory, qualitatively changing single-particle topology \cite{Mook2020Interactions}. References \onlinecite{McClarty2018, Mook2020QuantumDamping,McClarty2019, Mook2020Interactions} have in common that they focused on the many-body corrections to single-particle states, as obtained \change{by treating many-particle states as an \emph{incoherent} bath} within perturbation theory. This said, further progress in revealing qualitative effects of nonconserving many-body interactions should come from \change{nonperturbative \emph{coherent} theories}.

\change{Herein, we study topological excitations caused by the coherent hybridization of states belonging to different particle-number sectors.} This hybridization is only possible if the particle number is \textit{not} a good quantum number and, hence, unique to nonconserved particles. We demonstrate that the hybridization results in spectral gaps that can be topologically nontrivial and support chiral excitations at the edges of the sample. Because of the particle-sector hybridization, these exotic excitations are \change{composites} of states belonging to different sectors; they are neither one, nor two, nor any other integer multiple of the original constituents. Although particle number nonconservation enables spontaneous decays, the lifetime of the discussed excitations can be, in principle, infinite, establishing them as well-defined quasiparticles that surmount the aforementioned damping roadblock.

\change{For two states to hybridize and anticross upon coherent coupling, they have to cross in energy in the first place.} Typically, single-particle states are lowest in energy, followed by two-particle states, then three-particle states, and so on. The two-particle sector contains a scattering continuum---built from pairs of single-particle energies---and two-particle bound states (BS) \cite{Bethe1931, Wortis1963, Winkler2006}. Two-particle BS effectively behave as quasiparticles and their spectrum might be topological as well \cite{DiLiberto2016, Gorlach2017, Salerno2018, Qin2017, Qin2018, Stepanenko2020, Salerno2020}.
Here, however, we assume that both the single-particle and the BS spectrum are topologically trivial \change{before coherent coupling}. If the binding energy is strong enough for the BS to appear well below the continuum, they may energetically overlap with the single-particle states. A hybridization is possible, provided that the two states couple, that is, the coupling must not be forbidden by a $U(1)$ symmetry. Finally, the resulting spectral gap can only become Chern insulating if time-reversal symmetry (TRS) is broken. \change{To reiterate}, the requirements for \change{topological effects between states belonging to different particle-number sectors} are (1) strong many-body interactions, (2) a nonconserved particle number, and (3) broken TRS.

\begin{figure}
	\centering
	\includegraphics[width = \columnwidth]{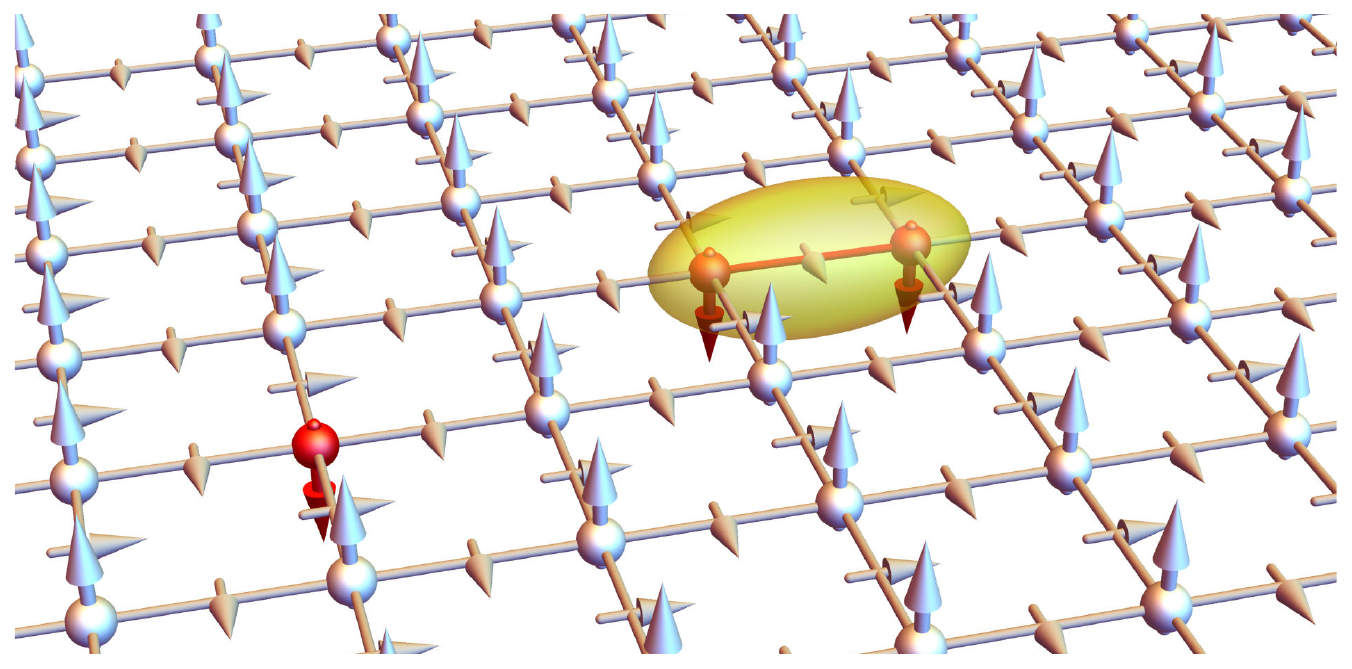}
	\caption{
		\change{Sketch of a quantum condensed matter system with topological hybridizations of states belonging to different particle-number sectors.} Single spin flips (left) and tightly bound double flips (right) are representatives of two different particle-number sectors in a fully polarized spin-$1/2$ quantum magnet. When particle-number conservation is broken by a spin-nonconserving magnetic interaction, illustrated here by Dzyaloshinskii-Moriya interaction indicated by little white arrows at the bond midpoints, single and two-particle sectors hybridize. As a result, there are topological spectral gaps that support chiral edge spin excitations. These inherit spin-dipolar character from the single-particle weight of their wave function and spin-quadrupolar character from the two-particle weight.
	}
	\label{fig:teaser}
\end{figure}

Particle-number sector hybridizations have already been observed for phonons in AlPO$_4$ \cite{Zawadowski1970, Ruvalds1970} and for magnetic excitations in quasi one-dimensional antiferromagnets (CD$_3$)$_4$NMnCl$_3$ \cite{Heilmann1981, Osano1982, Endoh1984} and CsVCl$_3$ \cite{Inami1997}. The most striking experimental evidence was presented only recently: inelastic neutron scattering data for the spin-$1$ antiferromagnet FeI$_2$ reveal a hybridization of single-magnons and single-ion BS \cite{Bai2021, Legros2020, Bai2021decay}. Motivated by this discovery and the burgeoning field of topological spin excitations, such as spinons \cite{Lee2015paramagnet, Kim2016Kane, Sonnenschein2017, Joshi2019}, triplons \cite{Romhanyi2015, mcclarty2017topological, Anisimov2019, Song2020spinon, Bhowmick2021Weyltriplon, Haldar2021}, and magnons \cite{Meier2003, Katsura2010, Hoogdalem2013, Zhang2013, Shindou13, Shindou13b, Mook2014edge, Shindou14, Mook2015interfaces, Owerre2016a, Mook2016Weyl, Xu2016, Nakata2017, Nakata2017AFM, Mook2017nodal, Li2018chiral, Mook2018duality, Diaz2019AFM, Kondo20192d, Kondo2019, Mook2019coplanar, Kim2019SkyrmionHallFerri, Malki2019, Diaz2020FM, Kondo2020NH, Hirosawa2020, Mook2020hinge, Corticelli2022}, we show that the above mentioned requirements for \change{topological effects between single particles and particle pairs} are met in fully saturated spin-anisotropic quantum magnets without spin conservation, as depicted in Fig.~\ref{fig:teaser}. In brevity, (1) the spin-space anisotropy binds single magnons together to form pairs whose energy is well below the two-magnon continuum and can overlap with single-magnon energies. (2) The spin-nonconserving magnetic interaction breaks the particle-number conservation for magnons, allowing hybridization of single-magnon states and two-magnon BS. Finally, (3) TRS is broken by the combination of magnetic ordering and spin nonconservation. 

To prove the general nature of the proposed physics, we detail \change{the topology of magnons and magnon pairs} in four selected models. They cover different spin lengths (spin-$1/2$ versus spin-$1$), different anisotropy mechanisms (Ising versus single-ion) and resulting types of BS (exchange BS versus single-ion BS), different lattice geometries (square versus triangular), and different mechanisms of $U(1)$-symmetry violation (antisymmetric versus symmetric transverse-longitudinal spin-spin interactions). In all cases, we find chiral edge excitations with a mixed single-\change{magnon} and two-\change{magnon} character that carry mixed spin-multipoles. Explicitly, these \change{hybrids} inherit spin-dipolar character from their one-magnon and spin-quadrupolar character from their two-magnon \change{weight}. As such, they are visible in experiments probing both the dynamical spin structure factor (e.g., inelastic neutron scattering \cite{Lovesey1977}) and higher-spin structure factors (e.g., resonant inelastic X-ray spectroscopy \cite{Nag2021}). With the chiral edge state lacking Joule heating and being immune to backscattering, our results \change{suggest} topological beyond-spintronics computation paradigms that utilize not only the spin dipole but also the spin quadrupole. 

Finally, we detail the differences between \change{the discussed} topology and ``conventional'' magnon topology. While the latter survives taking the classical limit of an infinite spin length, \change{the topological effects at hand} are a genuine nonperturbative quantum phenomenon that vanishes in this limit. 
Moreover, the \change{coherent mixing of particle-number sectors} leads to indirect experimental signatures in spin and heat transport because it gives rise to a Berry curvature and intrinsic anomalous Hall-type transport. We conclude that quantum magnets are an appealing playground to study \change{particle-number sector} mixing and its impact on topology and transport. Such studies will complement efforts in the field of ultracold atoms where topology within particle-number sectors is studied.

The remainder of the article is organized as follows. In Sec.~\ref{sec:PRE}, we set the stage by giving an introduction to many-magnon states in $U(1)$-symmetric, fully saturated ferromagnets (Sec.~\ref{sec:PREmanymag}), quickly reviewing spin-to-boson transformations (Sec.~\ref{sec:PREspintoboson}) and the origin of two-magnon BS (Sec.~\ref{sec:PREboundstates}), and discussing how to break $U(1)$-symmetry to enable \change{particle-number sector} hybridizations (Sec.~\ref{sec:PREbreakingU(1)}). We then specify the $U(1)$ symmetry breaking to antisymmetric exchange interaction in Sec.~\ref{sec:DMI} by detailing \change{spin excitation} topology both in spin-$1/2$ magnets with strong Ising anisotropy (Sec.~\ref{sec:spin12SQUARE}) and in spin-$1$ magnets with strong single-ion anisotropy (Sec.~\ref{sec:spin1SQUARE}). \change{Additionally, we} analyze bond-dependent symmetric off-diagonal exchange in Sec.~\ref{sec:SymmetricTransverse}; we show the \change{coherent hybridization} mechanism at work in spin-$1/2$ magnets (Sec.~\ref{sec:Spin12Triangular}) and spin-$1$ magnets (Sec.~\ref{sec:Spin1Triangular}). In Sec.~\ref{sec:Discussion}, we discuss the properties of chiral spin \change{excitations} (Sec.~\ref{sec:Properties}) and the differences to conventional magnon topology (Sec.~\ref{sec:harmonictopology}), comment on the implications of the \change{coherent particle-number sector} hybridization on intrinsic anomalous Hall-type transport phenomena (Sec.~\ref{sec:THE}), outline the role of \change{coherent} coupling beyond the single and two-particle spectrum (Sec.~\ref{sec:BeyondOneTwo}), and identify suitable material candidates for \change{experimental verification} (Sec.~\ref{sec:Materials}). We conclude in Sec.~\ref{sec:Conclusion}. Appendixes \ref{Appendix:reklis-construction}-\ref{App:Ising} provide additional details.

%
%
\section{Preliminaries}
\label{sec:PRE}

\subsection{Many-Magnon States in Saturated Magnets with Spin Conservation}
\label{sec:PREmanymag}
We consider collinear ferromagnets or field-polarized magnets with magnetization along the $z$-direction. If their spin Hamiltonian $H$ holds $U(1)$ symmetry, that is to say, if $H$ commutes with the $z$ component of the total spin, 
\begin{align}
  [\, H, S^z \,] = 0,
  \quad
  S^z = \sum_{l=1}^N S^z_{\vec{r}_l},
  \label{eq:Szconversation}
\end{align}
then $z$-spin is a good quantum number. Here, $N$ is the total number of spins and $S^\alpha_{\vec{r}_l}$ ($\alpha = x,y,z$) is a component of the spin operator $\vec{S}_{\vec{r}_l}$ at site $\vec{r}_l$. \change{Eigenstates} can be labelled by their $z$ spin \change{$\Delta S = 0,1,2,3,\ldots$ relative to the ground state, which is the fully polarized state, 
\begin{align}
    | 0 \rangle = \bigotimes_{l=1}^N | S^z_{\vec{r}_l} = S \rangle,
    \label{eq:GS}
\end{align}
($\Delta S = 0$). It is a tensor product of local $| S^z_{\vec{r}_l} = S \rangle$ states, which are eigenstates of the $z$-spin operator: $S^z_{\vec{r}_l} | S^z_{\vec{r}_l} = S \rangle = S | S^z_{\vec{r}_l} = S \rangle$, with spin quantum number $S = \frac{1}{2}, 1, \frac{3}{2}, 2, \ldots$; we set $\hbar = 1$ throughout.}

The $\Delta S = 1$ sector describes conventional dipolar single-magnon excitations. For periodic boundary conditions, their wave function reads
\begin{align}
	| \vec{k} \rangle = \frac{1}{\sqrt{N}} \sum_{l=1}^N \mathrm{e}^{\mathrm{i} \vec{k} \cdot \vec{r}_l } | \vec{r}_l \rangle
	\label{eq:onemagwave}
\end{align}
and is characterized by crystal momentum $\vec{k}$. It is built from superimposed local spin flips,
\begin{align}
    | \vec{r}_l \rangle = \frac{1}{\sqrt{2S}} S_{\vec{r}_l}^- |0\rangle,
\end{align}
caused by acting with the ladder operator 
$
	S_{\vec{r}_l}^- = S_{\vec{r}_l}^x - \mathrm{i} S_{\vec{r}_l}^y
$
on $| 0 \rangle$.

Quadrupolar two-magnon states are found in the $\Delta S = 2$ sector. Their basis can be written as \cite{Kecke2007}
\begin{align}
	| \vec{k}, \vec{r} \rangle 
	= 
	\frac{1}{\sqrt{N}} \sum_{l=1}^N \mathrm{e}^{\mathrm{i} \vec{k} \cdot \vec{r}_l } \mathrm{e}^{\mathrm{i} \vec{k} \cdot \vec{r} /2 } | \vec{r}_l, \vec{r}_{l} + \vec{r} \rangle,
	\label{eq:twomagwave}
\end{align} 
where states with two spin flips $| \vec{r}_l, \vec{r}_{l} + \vec{r} \rangle = \mathcal{N}^{-1}_{\vec{r}} S_{\vec{r}_l}^- S_{\vec{r}_l+\vec{r}}^- | 0 \rangle$ are superimposed. The spin flips' center-of-mass momentum is labeled by $\vec{k}$ and their relative distance vector by $\vec{r}$. The normalization factor reads $\mathcal{N}_{\vec{r}} = 2S$ for $\vec{r} \ne {0}$, and $\mathcal{N}_{\vec{0}} = \sqrt{4S(2S-1)}$ for $\vec{r}={0}$. In the latter case, two spin flips are located at a single site, requiring $S \ge 1$ \cite{Rastelli2011}.

Higher-spin excitations with $\Delta S > 2$ can be built in a similar fashion. They are associated with higher-order multipole moments, e.g., three-magnon states ($\Delta S = 3$) carry octupolar character, four-magnon states ($\Delta S = 4$) carry hexadecapolar character, and so on \cite{ChiuTsao1975, ChiuTsao1976, Momoi2006, Kecke2007, Sudan2009}.

We pay particular attention to the Hilbert subspace of one and two-magnon excitations because they are not only most frequently encountered as the relevant low-energy excitations but also are representatives of the simplest particle-number sectors. Both types of excitations are entirely transverse to the magnetization direction. Their multipolar character is revealed by studying the dynamics of transverse spin correlation functions of a single spin, $\langle S^{+}_{\vec{r}_l}(t) S^{-}_{\vec{r}_{l'}}(0) \rangle$, and a pair of spins at distance $\vec{r}$, $\langle S^{+}_{\vec{r}_l}(t) S^{+}_{\vec{r}_l+\vec{r}}(t) S^{-}_{\vec{r}_{l'}}(0)S^{-}_{\vec{r}_{l'}+\vec{r}}(0) \rangle$, respectively \cite{Toth2012} ($S_{\vec{r}_l}^+ = (S_{\vec{r}_l}^-)^\dagger$). The former probes spin-dipolar character because $S_{\vec{r}_l}^\pm$ is built from elements of the spin dipole operator $\vec{S}_{\vec{r}_l}$. Likewise, $Q^{-}_{\vec{r}_l,\vec{r}} \equiv S^{-}_{\vec{r}_l} S^{-}_{\vec{r}_l+\vec{r}}$ is built from elements of the spin-quadrupole operator $Q^{\alpha \beta}_{\vec{r}_l, \vec{r}_{l'}} = ( S^{\alpha}_{\vec{r}_l} S^{\beta}_{\vec{r}_{l'}} + S^{\beta}_{\vec{r}_l} S^{\alpha}_{\vec{r}_{l'}} ) / 2 - \delta_{\alpha, \beta} \vec{S}_{\vec{r}_l} \cdot \vec{S}_{\vec{r}_{l'}}/3$, explicitly \cite{Shannon2006}
\begin{align}
    Q^{-}_{\vec{r}_l,\vec{r}}
    =
    S^{-}_{\vec{r}_l} S^{-}_{\vec{r}_{l}+\vec{r}} 
    &= 
    S^{x}_{\vec{r}_l} S^{x}_{\vec{r}_{l}+\vec{r}} 
    - S^{y}_{\vec{r}_l} S^{y}_{\vec{r}_{l}+\vec{r}}
    - \mathrm{i} \left( S^{x}_{\vec{r}_l} S^{y}_{\vec{r}_{l}+\vec{r}} + S^{y}_{\vec{r}_l} S^{x}_{\vec{r}_{l}+\vec{r}} \right)
    \nonumber \\
    &=
    Q^{xx}_{\vec{r}_l, \vec{r}_{l}+\vec{r}} 
    -
    Q^{yy}_{\vec{r}_l, \vec{r}_{l}+\vec{r}} 
    -
    2 \mathrm{i} Q^{xy}_{\vec{r}_l, \vec{r}_{l}+\vec{r}} .
\end{align}

To directly see that the above correlation functions probe the one-magnon and two-magnon content of excitations, we consider the respective dynamical structure factors. The transverse dynamical spin structure factor reads
\begin{align}
	\mathcal{S}^{+-}(\vec{k},\omega)
	&=
	\frac{1}{2 \pi N} \sum_{l,l'=1}^N \mathrm{e}^{\mathrm{i} \vec{k} \cdot (\vec{r}_{l'}-\vec{r}_{l})}
	\int_{-\infty}^\infty \mathrm{e}^{\mathrm{i} \omega t} 
	\langle S^{+}_{\vec{r}_l}(t) S^{-}_{\vec{r}_{l'}}(0) \rangle
	\,\mathrm{d} t
	\nonumber \\
	&=
	2S\sum_{f} \left| \langle f | \vec{k} \rangle \right|^2 \delta( \omega - \varepsilon_f ),
	\label{eq:dynstrucdipole}
\end{align} 
from which one can read off that it measures the dipolar character of excitations, i.e., their ``single-magnonness'' [cf.~Eq.~\eqref{eq:onemagwave}]. We assume zero temperature and the sum runs over all excited states $|f\rangle$ with energy $\varepsilon_f$. Complementary, the quadrupolar character is captured by the dynamic spin-pair structure factor,
\begin{align}
	\mathcal{Q}^{+-}_{\vec{r}}(\vec{k},\omega) 
	&=
	\frac{1}{2 \pi N} \sum_{l,l'=1}^N \mathrm{e}^{\mathrm{i} \vec{k} \cdot (\vec{r}_{l'}-\vec{r}_{l})}
	\int_{-\infty}^\infty \mathrm{e}^{\mathrm{i} \omega t} 
	\langle Q^{+}_{\vec{r}_l,\vec{r}}(t) Q^{-}_{\vec{r}_{l'},\vec{r}}(0) \rangle
	\,\mathrm{d} t
	\nonumber \\
	&=
	\mathcal{N}^2_{\vec{r}}\sum_{f} \left| \langle f | \vec{k} , \vec{r} \rangle \right|^2 \delta( \omega - \varepsilon_f ),
	\label{eq:dynstrucquadrupole}
\end{align} 
that probes the two-magnon content of the eigenstates [cf.~Eq.~\eqref{eq:twomagwave}].

\subsection{Spin-to-boson transformations}
\label{sec:PREspintoboson}
Throughout the work, we use the spin language, facilitated by the ground state being the fully polarized state $|0\rangle$ in Eq.~\eqref{eq:GS}. However, we make contact with the particle language in our physical interpretations. The link between spin and particle language is provided by spin-to-boson transformations, such as the Holstein-Primakoff (HP) transformation \cite{Holstein1940} for general spin $S$,
\begin{subequations}
\begin{align}
	S^-_{\vec{r}_l} &= a^\dagger_{\vec{r}_l} \sqrt{2S - a^\dagger_{\vec{r}_l} a_{\vec{r}_l}}, \label{eq:HP1}
	\\
	S^+_{\vec{r}_l} &= \sqrt{2S - a^\dagger_{\vec{r}_l} a_{\vec{r}_l}} \, a_{\vec{r}_l}, 
	\\
	S^z_{\vec{r}_l} &= S - a^\dagger_{\vec{r}_l} a_{\vec{r}_l}.\label{eq:HP2}
\end{align}
\end{subequations}
The $a^{(\dagger)}_{\vec{r}_l}$'s denote bosonic annihilation (creation) operators \change{ that obey the commutation rule $[a_{\vec{r}_l}, a^\dagger_{\vec{r}_{l'}}] = \delta_{l,l'}$.
Using the transformation in Eq.~\eqref{eq:HP2}, one shows that spin conservation \eqref{eq:Szconversation} directly translates into particle-number conservation:
\begin{align}
	[ \, H, N \, ] = 0, \quad N = \sum_{l=1}^N a^\dagger_{\vec{r}_l} a_{\vec{r}_l}.
\end{align}
}
Plugging the transformation~\eqref{eq:HP1}-\eqref{eq:HP2} into a general spin Hamiltonian $H$ and expanding the square root leads to an expansion
\begin{align}
	H = H_0 + H_1 + H_2 + H_3 + H_4 + \ldots \label{eq:HPexpansion}
\end{align}
in $1/\sqrt{S}$ or, equivalently, the number of bosonic operators. 
The $r$-th sub-Hamiltonian contains $r$ bosonic operators and is $H_r = O( S^{2-r/2} )$. Thus, $H_0 = O(S^2)$ provides the classical ground state energy and $H_1$ vanishes if the magnetic order, about which the expansion is performed, is stable (or at least metastable). Single-particle physics is covered by the bilinear $H_2 \sim S a^\dagger a$. Particle-number conserving many-body interactions appear to lowest order in $H_4 \sim a^\dagger a^\dagger a a$. If attractive, they can lead to magnons binding together (see Sec.~\ref{sec:PREboundstates}). A coupling of particle-number sectors is provided by $H_3 \sim \sqrt{S} a^\dagger a a + \text{H.c.}$, which increases (decreases) the number of particles by one.

For spin-$1/2$ systems, one might also use the Matsuda-Matsubara transformation \cite{Matsubara1956}
\begin{subequations}
\begin{align}
	S^-_{\vec{r}_l} &= a_{\vec{r}_l}, 
	\\
	S^+_{\vec{r}_l} &= a^\dagger_{\vec{r}_l}, 
	\\
	S^z_{\vec{r}_l} &= a^\dagger_{\vec{r}_l} a_{\vec{r}_l} - \frac{1}{2},
\end{align}
\label{MMtrafo}
\end{subequations}
from spins to hard-core bosons. In contrast to the HP transformation, the Matsuda-Matsubara transformation does not lead to an infinite expansion but terminates at quartic order, assuming that only bilinear spin-spin interactions appear in the spin Hamiltonian.

\subsection{Two-Magnon Bound States}
\label{sec:PREboundstates}
The dynamic spin-pair structure factor $\mathcal{Q}^{+-}_{\vec{r}}(\vec{k},\omega)$ in Eq.~\eqref{eq:dynstrucquadrupole} comes with a subscript $\vec{r}$. It defines the distance vector between the two spins the paired dynamics of which is probed. We are particularly interested in the cases $|\vec{r}| = 0, 1$, covering the situations that a spin is paired with itself or its nearest neighbors, respectively. The respective two-magnon state $| \vec{k}, \vec{r} \rangle$ is associated with two tightly bound magnons, i.e., with two-magnon bound states (BS).

Bound states of magnons have been a fascinating research subject for almost a century \cite{Bethe1931}. They arise because of attractive magnon-magnon interactions that cause a binding energy \cite{Wortis1963}. That magnons in a ferromagnet interact attractively becomes apparent by comparing the energy associated with flipping two spins far apart from each other to that with flipping them right next to each other. In the first case, one has to pay an energy penalty given by the coordination number of the lattice for each of the two flips \change{(assuming only nearest-neighbor interactions)}. In the latter case, the effective ``boundary area'' of the flipped complex is reduced, which leads to a smaller energy penalty. \change{If this exchange energy gain overcomes the kinetic energy gain from two individual magnons, there appears an ``exchange BS'' whose energy is lower than the energy of two free single magnons \cite{Mattis2006, Rastelli2011}.} In the spectrum, BS appear below the two-magnon continuum of scattering states. In the limit of large magnon-magnon interactions, the BS split off from the two-magnon continuum for all quasimomenta \cite{Rastelli2011}; the two spin deviations are tightly bound together and may be considered a new effective quasiparticle.

With the two-magnon BS energies appearing well below the two-magnon continuum, they may overlap with the single-magnon energies. Since the two states belong to different spin sectors (or particle-number sectors), the crossing is protected by spin (or particle-number) conservation, i.e., by a $U(1)$ symmetry. Below, we explore the hybridization between single-particle states and two-magnon BS upon breaking the protecting $U(1)$ symmetry.

\subsection{Breaking $U(1)$ Symmetry}
\label{sec:PREbreakingU(1)}
In general, the bilinear interaction between two spins $\vec{S}_{\vec{r}_l}$ and $\vec{S}_{\vec{r}_{l'}}$ at different sites is given by $\vec{S}_{\vec{r}_l} \cdot \vec{I}_{\vec{r}_l, \vec{r}_{l'}} \vec{S}_{\vec{r}_{l'}}$, 
with the interaction matrix
\begin{align}
    \vec{I}_{\vec{r}_l, \vec{r}_{l'}}
    =
    \begin{pmatrix}
       J^{xx} & J^{xy} & J^{xz} \\
       J^{yx} & J^{yy} & J^{yz} \\
       J^{zx} & J^{zy} & J^{zz} 
    \end{pmatrix},
    \label{eq:matrixgeneral}
\end{align}
where we suppressed the site indexes to lighten notation. For a $U(1)$ symmetry, it must hold $J^{xx} = J^{yy}$, $J^{xy} = - J^{yx}$, and $J^{xz} = J^{yz} = J^{zx} = J^{zy} = 0$, i.e.,
\begin{align}
    \vec{I}_{U(1)}
    =
    \begin{pmatrix}
       J^{xx} & J^{xy} & 0 \\
       -J^{xy} & J^{xx} & 0 \\
       0 & 0 & J^{zz} 
    \end{pmatrix}.
    \label{eq:matrixU1}
\end{align}
This interaction matrix describes the XXZ model with additional antisymmetric off-diagonal exchange. To explicitly see that different $\Delta S$ spin sectors are not coupled, one writes
\begin{align}
    \vec{S}_{\vec{r}_l} \cdot \vec{I}_{U(1)}  \vec{S}_{\vec{r}_{l'}}
    =
    &\frac{J^{xx}}{2} \left( S^+_{\vec{r}_l} S^-_{\vec{r}_{l'}} + S^-_{\vec{r}_l} S^+_{\vec{r}_{l'}} \right)
    +
    J^{zz} S^z_{\vec{r}_l} S^z_{\vec{r}_{l'}}
    \nonumber \\
    &-
    \mathrm{i} \frac{J^{xy}}{2} \left( S^+_{\vec{r}_l} S^-_{\vec{r}_{l'}} - S^-_{\vec{r}_l} S^+_{\vec{r}_{l'}} \right)
\end{align}
and realizes that lowering and raising operators always appear together. They do not generate a net change of the $z$-spin quantum number or, equivalently, of the particle number because $S^+_{\vec{r}_l} S^-_{\vec{r}_{l'}} \sim a_{\vec{r}_l}^\dagger a_{\vec{r}_{l'}}, a_{\vec{r}_l}^\dagger a_{\vec{r}_l}^\dagger a_{\vec{r}_l} a_{\vec{r}_{l'}} , \ldots $ and $S^z_{\vec{r}_l} S^z_{\vec{r}_{l'}} \sim a^\dagger_{\vec{r}_l} a_{\vec{r}_{l'}}, a^\dagger_{\vec{r}_l} a_{\vec{r}_l} a^\dagger_{\vec{r}_{l'}} a_{\vec{r}_{l'}}$ after a HP transformation. Consequently, there are no matrix elements that couple two different particle number sectors and potential energetic crossings between sectors are protected.

Comparing the $U(1)$ symmetric interaction matrix in Eq.~\eqref{eq:matrixU1} with the general expression in Eq.~\eqref{eq:matrixgeneral}, one finds two different ways to break $U(1)$ symmetry and, consequently, couple different particle number sectors: 

(1) For $J^{xy} \ne -J^{yx}$ and/or $J^{xx} \ne J^{yy}$, there is anisotropic symmetric exchange that gives rise to terms $S^+_{\vec{r}_l} S^+_{\vec{r}_{l'}} \sim a_{\vec{r}_l} a_{\vec{r}_{l'}}$ and $S^-_{\vec{r}_l} S^-_{\vec{r}_{l'}} \sim a^\dagger_{\vec{r}_l} a^\dagger_{\vec{r}_{l'}}$ to lowest order. These terms couple particle sectors differing in spin by two. \change{In particular, they couple the ground state $\Delta S = 0$ sector to other even-$\Delta S$ sectors}. As a result, the fully polarized state is no longer the ground state. Instead, within spin-wave theory, the new ground state is approximately given by the Bogoliubov vacuum of magnons, \change{which is a squeezed state \cite{Kamra2020Niche}, whose quantum fluctuations reduce the magnitude of the ordered moment. We explicitly exclude this possibility of $U(1)$ symmetry breaking from our further considerations.}

(2) For $J^{xz}, J^{zx} \ne 0$ and/or $J^{yz}, J^{zy} \ne 0$, one obtains \emph{transverse-longitudinal} off-diagonal exchange. To lowest order, it gives rise to terms of the form $S^+_{\vec{r}_l} S^z_{\vec{r}_{l'}} \sim a_{\vec{r}_l} a^\dagger_{\vec{r}_{l'}} a_{\vec{r}_{l'}}$ and $S^-_{\vec{r}_l} S^z_{\vec{r}_{l'}} \sim a^\dagger_{\vec{r}_{l}} a^\dagger_{\vec{r}_{l'}} a_{\vec{r}_{l'}}$ that change the number of particles (or spin) by one. Importantly, below we show that in many cases the ground state ($\Delta S = 0$) is not coupled to the $\Delta S = 1$ sector. Thus, the fully polarized state remains the exact quantum mechanical ground state, facilitating the theoretical analysis of the coupling between single-magnon ($\Delta S = 1$) and two-magnon excitations ($\Delta S = 2$). From here on, we will study magnets with $U(1)$ symmetry breaking due to this transverse-longitudinal exchange interaction; it can come in an antisymmetric ($J^{xz} = -J^{zx}$ and $J^{yz} = -J^{zy}$, see Sec.~\ref{sec:DMI}) or symmetric form ($J^{xz} = J^{zx}$ and $J^{yz} = J^{zy}$,see Sec.~\ref{sec:SymmetricTransverse}).

\section{Coupling of Particle Number Sectors by Dzyaloshinskii-Moriya interaction}
\label{sec:DMI}
We begin our discussion by studying the influence of \emph{antisymmetric} exchange interaction, known as Dzyaloshinskii-Moriya interaction (DMI) \cite{Dzyaloshinsky58, Moriya60}. Explicitly, we consider the nearest-neighbor DMI spin Hamiltonian 
\begin{align}
    H_{\text{DMI}} 
    &= 
    \frac{1}{2} \sum_{\langle l,l' \rangle}  
    \vec{S}_{\vec{r}_l}
    \cdot
    \begin{pmatrix}
       0 & 0 & -D^y_{\vec{r}_l,\vec{r}_{l'}} \\
       0 & 0 & D^x_{\vec{r}_l,\vec{r}_{l'}} \\
       D^y_{\vec{r}_l,\vec{r}_{l'}} & -D^x_{\vec{r}_l,\vec{r}_{l'}} & 0
    \end{pmatrix} 
    \vec{S}_{\vec{r}_{l'}}
    \nonumber \\
    &=
    \frac{1}{2} \sum_{\langle l,l' \rangle} \vec{D}_{\vec{r}_l,\vec{r}_{l'}} \cdot \left( \vec{S}_{\vec{r}_l} \times \vec{S}_{\vec{r}_{l'}} \right),
    \label{eq:DMIham}
\end{align}
with DMI vectors 
\begin{align}
	\vec{D}_{\vec{r}_l,\vec{r}_{l'}} = D_{\vec{r}_l,\vec{r}_{l'}} \hat{\vec{z}} \times \hat{\vec{e}}_{\vec{r}_l,\vec{r}_{l'}}.
\end{align}
Here, $\hat{\vec{e}}_{\vec{r}_l,\vec{r}_{l'}}$ is the bond unit vector from site $\vec{r}_l$ to site $\vec{r}_{l'}$. We restrict ourselves to two-dimensional magnets, such that $\hat{\vec{e}}_{\vec{r}_l,\vec{r}_{l'}} \perp \hat{\vec{z}}$, i.e., the DMI vectors are orthogonal to the magnetization direction. Below, we will refer to such an interaction as ``transverse DMI.''
DMI is the leading-order spin-orbit correction to the magnetic interaction matrix and requires the absence of a center of inversion at the bond midpoint \cite{Moriya60}. Thus, DMI can be intrinsic to particular crystal structures or it can be controlled by electric fields \cite{Katsura2005} or by growing the spin system on a substrate \cite{Fert1980, Levy1981, Zakeri2010, Wang2020YIGDMI}. 

\begin{figure}
	\centering
	\includegraphics[scale=1]{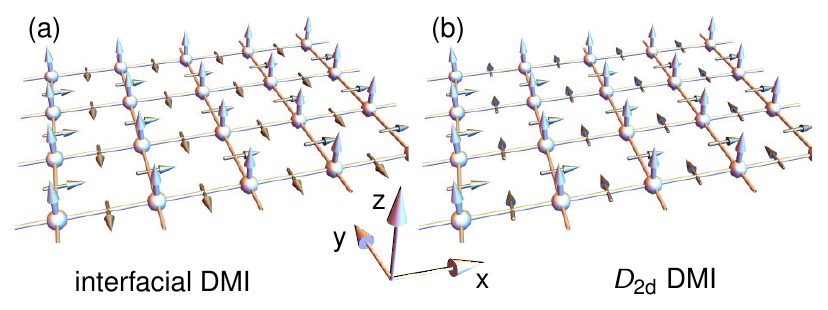}
    \caption{
        Transverse Dzyaloshinskii-Moriya interaction (DMI) in square-lattice ferromagnets. Large arrows at the vertices of the square lattice indicate spin moments, while small arrows at the bonds indicate DMI vectors. The DMI vector directions are given in the convention that the bonds are directed along the $x$ and $y$-directions, respectively. (a) Interfacial DMI as it occurs by mounting the square lattice onto a substrate and, hence, breaking the structural inversion symmetry. (b) DMI respecting $D_\text{2d}$ symmetry that is intrinsic to $D_\text{2d}$-symmetric materials. Note that in contrast to (a) the direction of DMI vectors is flipped for bonds along the $x$-direction.
    }
	\label{fig:dmi}
\end{figure}

Provided that periodic boundary conditions are assumed and the formation of spin spirals or skyrmions is prohibited either by easy-axis anisotropy or external fields, transverse DMI does not compromise the ferromagnetic product ground state in Eq.~\eqref{eq:GS}. This is due to the antisymmetry of DMI and discrete translational invariance. To show this explicitly, we consider a square lattice with $N_x$ ($N_y$) sites along the $x$ ($y$) direction ($N = N_x N_y$) and a lattice constant $a=1$. We write $(i,j)$ as shorthand for $\vec{r}_l$, with $i$ and $j$ indexing the $x$ and $y$ coordinate of a lattice site. The DMI Hamiltonian can then be written in the form
\begin{align}
    H_{\text{DMI}} 
    &= 
	\sum_{i=1}^{N_x} \sum_{j=1}^{N_y} 
	\left[
    D_{x} \hat{\vec{y}} \cdot \left( \vec{S}_{i,j} \times \vec{S}_{i+1,j} \right)
    +
    D_{y} \hat{\vec{x}} \cdot \left( \vec{S}_{i,j} \times \vec{S}_{i,j+1} \right) \right],
    \label{eq:DMIsquare}
\end{align}
where $D_x$ ($D_y$) is the amplitude of the DMI of bonds along the $x$ ($y$) direction.
For $D_x = -D_y$, one obtains interfacial DMI, as depicted in Fig.~\ref{fig:dmi}(a). In contrast, for $D_x = D_y$, $D_\text{{2d}}$-type DMI shown in Fig.~\ref{fig:dmi}(b) is realized. 
Using $S_{i,j}^z |0 \rangle = S |0 \rangle$, we find
\begin{align}
    H_\text{DMI}
    |0 \rangle
    &=
    S
    \sum_{i=1}^{N_x} \sum_{j=1}^{N_y} 
	\left[
    D_{x} \left( S^x_{i+1,j} - S^x_{i,j} \right)
    +
    D_{y} \left( S^y_{i,j} - S^y_{i,j+1} \right) \right]
     |0 \rangle
    \nonumber \\
    &= 0
\end{align}
because $\sum_{i,j} ( S^x_{i+1,j} - S^x_{i,j} ) = 0$ and $\sum_{i,j} ( S^y_{i,j} - S^y_{i,j+1} ) = 0$. This result applies to any other two-dimensional lattice. Without periodic boundary conditions, the cancellation is not perfect, with the spin operators at the edges of the material remaining. Then, DMI causes a boundary-bound chiral twist of the ground state polarization \cite{Wilson2013, Meynell2014}.

Although transverse DMI does not contribute to the ground state energy of periodic systems, it \textit{does} contribute to the spectrum by linking spin (particle) sectors differing in spin (particle number) by one. In particular, it causes a hybridization between dipolar single magnons and quadrupolar two-magnon excitations. To see this, we note that the terms of $H_\text{DMI}$ in Eq.~\eqref{eq:DMIsquare} are of the form $S^\pm_{\vec{r}_l} S^z_{\vec{r}_{l'}} \sim a^{(\dagger)}_{\vec{r}_l} a^{\dagger}_{\vec{r}_{l'}} a_{\vec{r}_{l'}}$, changing the spin quantum number of the excitation (or: the number of bosons) by $\mp 1$. Consequently, in general,
\begin{align}
    \langle \vec{k} | H_\text{DMI} | \vec{k}, \vec{r} \rangle 
    \sim \sqrt{S} \ne 0,
    \label{eq:DMIaction}
\end{align}
allowing \change{particle-number sector} mixing and avoided crossings. For DMI restricted to nearest neighbors, as in Eq.~\eqref{eq:DMIsquare}, $\langle \vec{k} | H_\text{DMI} | \vec{k}, \vec{r} \rangle \ne 0$ only for $|\vec{r}| = 1$ (in units of the lattice constant), i.e., spin flips can only be created or destroyed right next to one another.

With the general possibility of avoided crossings between states belonging to different particle-number sectors established, we now turn to the question whether the resulting spectral gaps can be topologically nontrivial. For topological band gaps to appear, (effective) time-reversal symmetry (TRS) $\mathcal{T}' = \mathcal{R} \mathcal{T}$ must be broken. It is built from actual TRS $\mathcal{T}$, which flips the ground state spin direction, and a rotation $\mathcal{R}$ by $180^\circ$ in spin space about an in-plane axis. Without DMI, $\mathcal{T}'$ is present because $\mathcal{R}$ leaves spin-conserving XXZ-type interactions of the form $\vec{I}_{\vec{r}_l,\vec{r}_{l'}} = \text{diag}(J^{xx}, J^{xx}, J^{zz})$ invariant. In contrast, $H_\text{DMI}$ is not invariant under $\mathcal{R}$, breaking $\mathcal{T}'$ and enabling nontrivial Chern numbers. Importantly, the two DMI configurations in Fig.~\ref{fig:dmi} are related by the $\mathcal{T}'$ operation: After $\mathcal{T}$ has flipped the magnetization direction, a $180^\circ$-rotation in spin space, say, about the $y$-axis, maps the texture back onto itself. DMI vectors along the $x$-direction flip sign, mapping one DMI configuration onto the other. Hence, as far as their spin excitations are concerned, the two ferromagnets in Fig.~\ref{fig:dmi} are effectively time-reversed partners and we expect opposite chirality of topological edge states.

Below, we present two examples of DMI-induced \change{coherent coupling of states belonging to different particle-number sectors} in quantum magnets: (i) in a spin-$1/2$ magnet with strong Ising exchange anisotropy (see Sec.~\ref{sec:spin12SQUARE}) and (ii) in a spin-$1$ magnet with strong single-ion \change{(or onsite)} anisotropy (see Sec.~\ref{sec:spin1SQUARE}).

%
%
\subsection{Spin-$1/2$ Square-Lattice Magnets with Ising Anisotropy}
\label{sec:spin12SQUARE}
To prove that anticrossings between different particle sectors hold topological information and the gapped phase can support chiral edge states, we specify our analysis to the $J_1$-$J_3$ spin-$1/2$ square lattice with Ising exchange anisotropy. The full spin Hamiltonian reads $H = H_\text{c} + H_\text{DMI}$, with $H_\text{DMI}$ as given in Eq.~\eqref{eq:DMIsquare} and the spin-conserving (c) term as
\begin{align}
	H_\text{c}
	= 
	\sum_{i=1}^{N_x} \sum_{j=1}^{N_y} 
	&\left[
		\vec{S}_{i,j} \cdot \vec{I}_1 \left( \vec{S}_{i+1,j}+\vec{S}_{i,j+1} \right)
	\right.
	\nonumber \\
	&\left.
		+\vec{S}_{i,j} \cdot \vec{I}_3 \left( \vec{S}_{i+2,j}+\vec{S}_{i,j+2} \right)
		- B S^z_{i,j}
	\right].
	\label{eq:H0square}
\end{align}
We have introduced a magnetic field $B$ along $z$-direction and XXZ-type exchange interaction for nearest, $\vec{I}_1 = \text{diag}(J_1,J_1,J_1^z)$, and third-nearest neighbors, $\vec{I}_3 = \text{diag}(J_3,J_3,J_3^z)$. We consider ferromagnetic coupling $J_1^{(z)} < 0$ and exclusively concentrate on easy-axis Ising anisotropy, i.e., $|J_1| \le |J_1^{z}|$. 

\subsubsection{One and Two-Magnon Spectrum}
In the following, we assume $N_x = N_y = n$ and evaluate the spectrum numerically. We restrict the Hilbert space to one- and two-magnon states and introduce the basis
\begin{align}
	\change{\vec{\Psi}(\vec{k}) = \left( \langle \vec{k}, \vec{r}_1 | , \langle \vec{k}, \vec{r}_2 |, \ldots, \langle \vec{k}, \vec{r}_R |, \langle\vec{k}| \right)^\text{T}.}
	\label{eq:PsiSquare}
\end{align}
Here, the last component is the single-magnon state and the remaining components are two-magnon states, where $\vec{r}_i$ with $i=1,\ldots, R$ labels all possible relative distances $\vec{r}$ between two spin flips on a square lattice with dimension $n \times n$ and periodic boundary conditions. For the correct counting of two-magnon states, we use the construction by Reklis~\cite{Reklis1974}, resulting in $R = (n^2-1)/2$ for odd $n$. \change{Further details are provided in Appendix \ref{Appendix:reklis-construction}. 
}
\change{
By expanding the full Hamiltonian in the basis \eqref{eq:PsiSquare}, we obtain 
$
	H_\text{c} + H_\text{DMI}
	\approx
	\sum_{\vec{k},\vec{k}'}
	\vec{\Psi}^\dagger (\vec{k}') M(\vec{k}', \vec{k})  \vec{\Psi}(\vec{k})
$
}
with the effective Hamilton matrix 
\change{
$
	M(\vec{k}',\vec{k}) = \delta_{\vec{k}'\vec{k}} M(\vec{k})
$,
where
}
\begin{align}
	M(\vec{k})  
	= 
	\begin{pmatrix}
		M_2(\vec{k}) & \vec{D}(\vec{k}) \\
		\vec{D}^\dagger(\vec{k}) & M_1(\vec{k})
	\end{pmatrix}.
	\label{eq:Msquare}
\end{align}
The momentum Kronecker symbol $\delta_{\vec{k}' \vec{k}}$ is a result of discrete translational invariance and momentum being a good quantum number.
In the one-magnon block, we find
\begin{subequations}
\begin{align}
	M_{1}(\vec{k}) 
	&= \langle \vec{k} | H_\text{c} | \vec{k}\rangle
	\\
	&= 
	B - 2 \left( J_{1}^{z} + J_{3}^{z} \right) + J_{1} \left( \cos k_{x} + \cos k_{y} \right) \nonumber \\
	& \quad + J_{3} \left[ \cos \left(2 k_{x} \right) + \cos \left( 2 k_{y}\right) \right],
\end{align}
\end{subequations}
which is the single-magnon energy. $M_{1}(\vec{k})$ is scalar because there is only a single one-magnon band on a Bravais lattice. The two-magnon block $M_2(\vec{k})$ has dimension $R \times R$ and its elements
$
	[M_2(\vec{k})]_{ij} = \langle \vec{k},\vec{r}_i| H_\text{c} |\vec{k},\vec{r}_j\rangle
$
\change{are given in Appendix~\ref{Appendix:reklis-construction}.}
\change{Finally, the particle sector coupling between the two blocks is encoded in the vector $\vec{D}(\vec{k})$ with elements 
$
	D_i(\vec{k})
	=
	\langle \vec{k}, \vec{r}_i | H_\text{DMI} |\vec{k} \rangle
$,
whose explicit expression is also given in Appendix~\ref{Appendix:reklis-construction}. Importantly, only two elements of $\vec{D}(\vec{k})$ are nonzero, which are those for which $| \vec{k}, \vec{r}_i \rangle$ contains two spin flips on nearest-neighbor sites. This is because the DMI in Eq.~\eqref{eq:DMIsquare} acts between nearest neighbors and, as such, can only add or annihilate a spin flip next to another.} 

\begin{figure*}
    \centering
    \includegraphics[width=\textwidth]{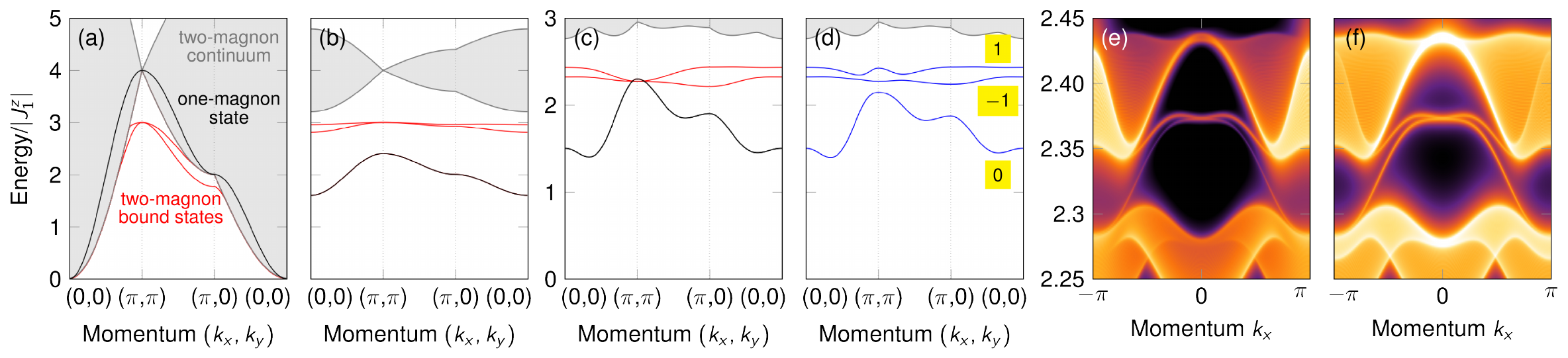}
    \caption{
    Topology of spin excitations in spin-$1/2$ quantum magnets on the square lattice.
    (a-d) Bulk one and two-magnon spectrum of spin-$1/2$ quantum magnets on the square lattice \change{with $N = 61^2$ lattice sites and periodic boundary conditions}. Black lines indicate the single-magnon dispersion in the absence of DMI. Gray areas delimit the two-magnon continuum. Two-magnon bound states are highlighted in red. 
    One- and two-magnon hybrids are colored in blue \change{and Chern numbers indicated by yellow labels}. Parameters taken are (a) $J_1/J_1^z = 1$, (b) $J_1/J_1^z = 0.2$, (c) $J_1/J_1^z = 0.2$, $J_3/J_1^z = -0.1$, $J^z_3/J_1^z = -0.3$, and $B/|J_1^z|=0.3$ and (d) $J_1/J_1^z = 0.2$, $J_3/J_1^z = -0.1$, $J^z_3/J_1^z = -0.3$, $B/|J_1^z|=0.3$, and $D_x/J_1^z = D_y/J_1^z = -0.1$. Parameters not explicitly given are set to zero. (e,f) Dynamical correlation functions of one and two spins, $\mathcal{S}^{+-}(k_x,\omega)$ and $\mathcal{Q}^{+-}_{\vec{x}}(k_x,\omega) + \mathcal{Q}^{+-}_{\vec{y}}(k_x,\omega)$, respectively encoding the dipolar and quadrupolar character of excitations. \change{A nanoribbon geometry is applied, with periodic (open) boundary conditions along the $x$ ($y$) direction. The width along the $y$ direction is $80$ unit cells. Chiral edge states are clearly visible in the topological band gap. (The plotted energy interval corresponds to the second and third blue band in panel (d) counted from bottom to top.)} Logarithmic color scale with orange/black denoting large/small values. Magnetic parameters as in (d) and a numerical Lorentzian linewidth broadening of $10^{-3} |J_1^z|$  were assumed.
    }
    \label{fig:squareS12}
\end{figure*}

Before turning to the numerical solution, we point out that some intuition on the influence of the different exchange parameters on the single- and two-magnon spectrum is obtained by the Matsuda-Matsubara transformation 
in Eq.~(\ref{MMtrafo}); for example, the expression $\vec{S}_{i,j} \cdot \vec{I}_1 \vec{S}_{i+1,j}$ is transformed into
\begin{align}
	\frac{J_1}{2} \left( a^\dagger_{i,j} a_{i+1,j} + a_{i,j} a^\dagger_{i+1,j} \right)
		+ J_1^z \left( a^\dagger_{i,j} a_{i,j} - \frac{1}{2} \right) \left( a^\dagger_{i+1,j} a_{i+1,j} - \frac{1}{2} \right).
\end{align}
It shows that $J_1$ takes over the role of a hopping and $J_1^z$ that of both an on-site potential and particle-number conserving interactions. Since $J_1^z < 0$, the interaction potential is attractive and binds magnons together. A similar transformation is found for the third-neighbor interaction. As it is chosen antiferromagnetically, the interaction potential associated with it is repulsive and does not lead to BS.

We now perform a numerical diagonalization of $M(\vec{k})$ to obtain the coupled spectra of the single and two-magnon sectors. For $J_1^z = J_1$ and other parameters zero, we obtain the isotropic nearest-neighbor model, for which two-magnon BS are well known \cite{Wortis1963, Rastelli2011}, see Fig.~\ref{fig:squareS12}(a). There are two BS, associated with the pair of spin flips being located along the $x$ or $y$-direction. The binding energy of the BS is largest at the Brillouin zone corner $\vec{k}=(\pi,\pi)$, where the BS are also degenerate. For $|J_1^z| > |J_1|$, the Goldstone mode is lifted and magnon-magnon interactions increase, causing a separation of the BS from the continuum [Fig.~\ref{fig:squareS12}(b)]. Including antiferromagnetic $J_3$ and $J_3^z$ changes the dispersion in a way that the BS overlap with the single-magnon energies [Fig.~\ref{fig:squareS12}(c)]. As expected, the antiferromagnetic third-nearest neighbor interaction does not lead to new two-magnon BS. Finally, as advertised, finite DMI causes a hybridization of single-magnon states and two-magnon BS, which results in spectral gaps [Fig.~\ref{fig:squareS12}(d)]. 

\begin{figure}
	\centering
	\includegraphics[width=1\columnwidth]{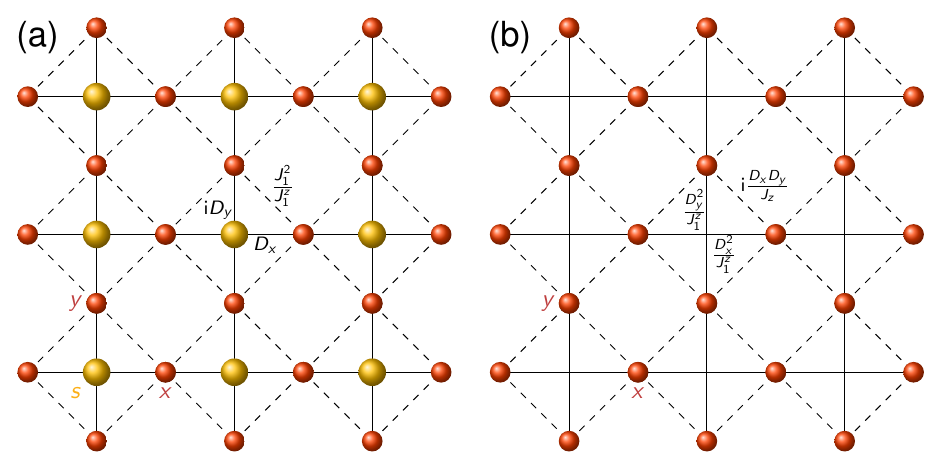}
    \caption{
        Real-space visualization of (a) the effective three-band model (Sec.~\ref{sec:effective3Spin12Square}) and (b) the effective two-band model (Sec.~\ref{sec:effective2Spin12Square}) for the hybridization of exchange bound states (BS) and single-magnon states in spin-$1/2$ square-lattice magnets with DMI and strong Ising anisotropy. (a) Lieb-lattice three-band model. Three types of particles are considered, one creating single spin flips ($s$) and the others two spin flips at nearest-neighbor sites along the $x$-direction ($x$) or the $y$-direction ($y$). The $s$ ($x,y$) particles live at the sites (bond) of the original square lattice. Solid and dashed bonds indicate effective hopping processes that mix particle species. Solid bonds are associated with processes that convert $s$ particles into $x,y$ particles. These processes derive from particle-number (or spin) nonconserving DMI in first-order perturbation theory. Dashed bonds convert $x$ into $y$ particles or vice versa and are associated with two spin hopping processes caused by the nearest-neighbor exchange interaction, rendering them a second-order perturbation. 
        (b) Checkerboard-lattice two-band model. If the energy of $s$ particles is well-separated from that of $x$ and $y$ particles, $s$ particles may be integrated out. This procedure gives rise to a checkerboard lattice with second-order virtual transitions that are quadratic in DMI and cause complex hoppings.
        In both (a) and (b), several further-neighbor hopping processes that do not mix particle species (derived from $J_1$ and $J_3$) are \emph{not} depicted for clarity. Full details are given in Appendix \ref{sec:AppSpin12}.    
        }
	\label{fig:Lieb}
\end{figure}

\subsubsection{Effective Three-Band Model}
\label{sec:effective3Spin12Square}
To describe the hybridization of the two BS with the single-magnon mode below the continuum, analytical progress can be made by deriving an effective three-band model within perturbation theory in the limit of large $J_1^z$, in which the dominating nearest-neighbor Ising anisotropy causes such strong attractive magnon-magnon interactions that the exchange BS split off from the continuum. There are at least two possibilities to derive such an effective model. In principle, one can directly perform a Schrieffer-Wolff transformation on $M(\vec{k})$ in Eq.~\eqref{eq:Msquare}. We find, however, that an equivalent method in real space provides more physical intuition. We restrict our attention to the relevant low-energy sector and introduce three flavors of particles:
\begin{subequations}
\begin{align}
	s^\dagger_{\vec{r}} | 0 \rangle &= |\vec{r} \rangle, \\
	x^\dagger_{\vec{r}} | 0 \rangle &= |\vec{r}, \vec{r} + \hat{\vec{x}} \rangle, \\
	y^\dagger_{\vec{r}} | 0 \rangle &= |\vec{r} , \vec{r} + \hat{\vec{y}} \rangle.
\end{align}
\end{subequations}
The $s$ flavor is a single spin flip at lattice site $\vec{r} = (i,j)$ and the others are two spin flips at nearest-neighbor sites connected either by an $x$ or $y$ bond. We derive an effective hopping model for these particles within second-order perturbation theory, as detailed in Appendix \ref{sec:AppSpin12}. Here, we note that the resulting hopping model is defined on the Lieb lattice, with the $s$ particles living on the sites and the $x$ and $y$ particles on the bonds of the original square lattice, as depicted in Fig.~\ref{fig:Lieb}(a). The change in the structural lattice is associated with the center of mass of double-spin-flip states not coinciding with a lattice site; such an effect was recently also discussed in the context of doublons \cite{Salerno2020}. Hopping within the $s$-sublattice or within the $x$ and $y$-sublattice derives from the number-conserving exchange interaction [see dashed lines in Fig.~\ref{fig:Lieb}(a)]. Hopping from the $s$-sublattice onto the $x$ or $y$-sublattice takes an interconversion from $s$ into $x$ or $y$ particles (or \textit{vice versa}), and is only possible by DMI [solid lines in Fig.~\ref{fig:Lieb}(a)].

\begin{figure}
	\centering
	\includegraphics[width=1\columnwidth]{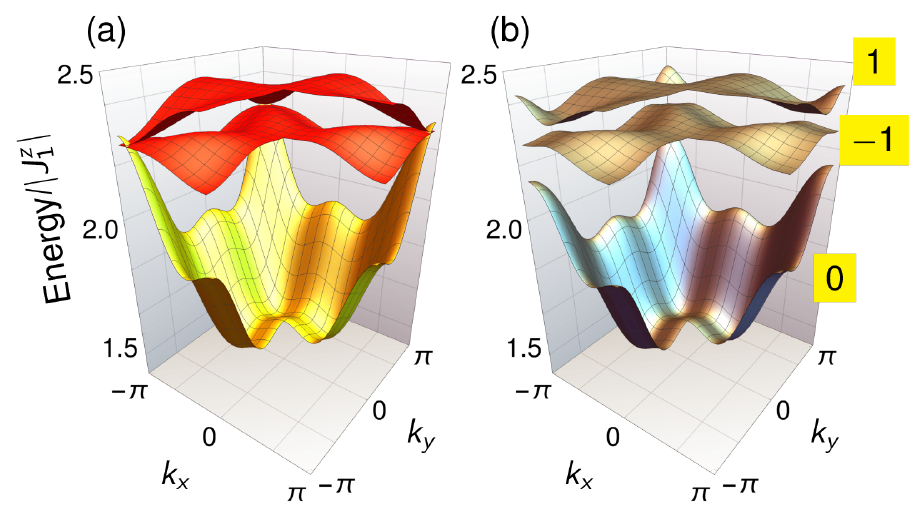}
    \caption{
        One-magnon states and two-magnon BS in spin-$1/2$ square-lattice ferromagnets with strong Ising ansisotropy as derived from the effective three-band model in Eq.~\eqref{eq:effective_ham_3band}. (a) Without DMI, i.e., $D_x = D_y = 0$, single-magnon excitations (yellow band) and two-magnon excitations (red bands) cross without hybridization because of particle-number conservation. (b) With DMI, $D_x/J_1^z = D_y/J_1^z = -0.1$, the particle-number is no longer conserved and the states hybridize, causing avoided crossings and topological band gaps. \change{Chern numbers are indicated by yellow labels.}
        Other parameters read $J_1/J_1^z = 0.2$, $J_3/J_1^z = -0.1$, $J_3^z/J_1^z = -0.3$, $B/|J_1^z|=0.3$. (Note that $J_1^z<0$.)
        }
	\label{fig:Spin12Bands}
\end{figure}

The final expression of the effective Hamiltonian in reciprocal space is given by
$
	H^\text{eff}
	=
	\sum_{\vec{k}} 
		\vec{C}^\dagger_{\vec{k}}  
		H(\vec{k})
		 \vec{C}_{\vec{k}}
$, 
where
$
	\vec{C}^\dagger_{\vec{k}} = (s^\dagger_{\vec{k}}, x^\dagger_{\vec{k}}, y^\dagger_{\vec{k}})  U
$
is a vector of Fourier transformed creation operators and $U$ a diagonal unitary matrix defined in Appendix \ref{sec:AppSpin12}. The effective Hamiltonian kernel
reads
\begin{align}
	H(\vec{k}) = \begin{pmatrix}
		E_1(\vec{k}) & -\mathrm{i} D_x \sin \frac{k_x}{2} & D_y\sin \frac{k_y}{2} \\
		\mathrm{i} D_x \sin \frac{k_x}{2} & E_2^{x,y}(\vec{k}) & \frac{2J^2_1}{J_1^z} \cos \frac{k_x}{2} \cos \frac{k_y}{2} \\
	D_y \sin \frac{k_y}{2} & \frac{2J^2_1}{J_1^z} \cos \frac{k_x}{2} \cos \frac{k_y}{2} & E_2^{y,x} (\vec{k}) 
	\end{pmatrix},
	\label{eq:effective_ham_3band}
\end{align}
with the single-magnon dispersion
\begin{align}
    E_1(\vec{k}) = B &- 2 (J_1^z + J_3^z ) + J_1 (\cos k_x + \cos k_y ) \nonumber \\
    &+ J_3 [ \cos (2 k_x) + \cos (2 k_y) ]
\end{align}
and
\begin{align}
    E_2^{\alpha,\beta}(\vec{k})
 	 &=
 	 2 B - 3 J_1^z - 4 J_3^z 
 	 + J_3 \cos k_\alpha
 	 + \frac{J_1^2}{J_1^z} \cos^2 \frac{k_\alpha}{2}
 	 \nonumber \\
 	 &\quad+ \frac{J_3^2}{J_1^z} \cos^2 k_\alpha
 	 + 2 \frac{J_1^2}{J_1^z} \cos^2 \frac{k_\beta}{2}
 	 + 2 \frac{J_3^2}{J_1^z} \cos^2 k_\beta.
 	 \label{eq:E2}
\end{align}
As expected, for zero DMI, the single-magnon ($s$ particles) and two-magnon sectors ($x$ and $y$ particles) are decoupled and $H(\vec{k})$ additionally obeys $H(-\vec{k})=H^\ast(\vec{k})$ due to $\mathcal{T}'$ symmetry. Finite DMI breaks both spin conservation and $\mathcal{T}'$ symmetry and couples the particle sectors already in \emph{first-order} perturbation theory. Figures \ref{fig:Spin12Bands}(a,b) depict example band structures without and with DMI, respectively. The emergence of avoided crossings upon $U(1)$ symmetry breaking due to DMI is clearly visible.

The topological phases of this effective three-band model are characterized by a triple of Chern numbers $(C_1, C_2, C_3)$, ordered from the energetically lowest to highest band. The Chern number of the $n$-th band is given by \cite{Kohmoto1985}
\begin{align}
	C_n = -\frac{1}{2 \pi} \int_{\text{BZ}} \Omega^z_n( \vec{k})\, \mathrm{d} \vec{k} 
	\label{eq:Chernnumber}
\end{align}
as an integral of the Berry curvature  
$
    \vec{\Omega}_n(\vec{k}) = \vec{\nabla}_{\vec{k}} \times \mathrm{i} \langle \vec{u}_n(\vec{k}) | \vec{\nabla}_{\vec{k}} | \vec{u}_n(\vec{k}) \rangle,
$
over the first Brillouin zone (BZ). Here, $| \vec{u}_n(\vec{k}) \rangle$ is the respective eigenvector of $H(\vec{k})$ in Eq.~\eqref{eq:effective_ham_3band}. We find that at least two out of the three bands are always topologically nontrivial for parameters consistent with perturbation theory.
For example, the three bands shown in Fig.~\ref{fig:Spin12Bands}(b) are characterized by the Chern number triple $(0,-1,1)$. The reason for two bands being topologically nontrivial is discussed below within an effective two-band model. 
We stress that the topological character of these bands is not an artifact of the effective perturbative model. We have obtained identical Chern numbers of the three bands from the eigenvectors of the full matrix $M(\vec{k})$ in Eq.~\eqref{eq:Msquare}, relying on the link variable formula derived in Ref.~\onlinecite{Fukui2005}.

According to the bulk-boundary correspondence \cite{hatsugaiChernNumberEdge1993, hatsugaiEdgeStatesInteger1993}, we expect one chiral edge state in the band gap that separates the topmost band from the lower two bands in Fig.~\ref{fig:Spin12Bands}(b). For a slab with open boundary conditions along the $y$-direction, we show the edge spectrum in Fig.~\ref{fig:squareS12}(e) that clearly exhibits this chiral edge state. For simplicity, we neglected the DMI-induced boundary twist of the ground state magnetization because it is expected to merely renormalize the boundary states without compromising the existence of the chiral edge states dictated by bulk topology. From the dynamical spin structure factor $\mathcal{S}^{+-}(k_x,\omega)$ and spin-pair structure factor $\mathcal{Q}^{+-}_{\vec{x}}(k_x,\omega)+\mathcal{Q}^{+-}_{\vec{y}}(k_x,\omega)$ in Fig.~\ref{fig:squareS12}(e) and Fig.~\ref{fig:squareS12}(f), respectively, we read off that the chiral mode carries both spin-dipolar and spin-quadrupolar character.

\subsubsection{Effective Two-Band Model: Bound States Only}
\label{sec:effective2Spin12Square}
Above, we mentioned that two out of three bands are always topologically nontrivial. Here, we derive an effective two-band model to explain why.
In the pure Ising limit, defined by $J_3 = J_3^z = J_1 = B = 0$, single-magnon states have an energy $-2J_1^z$ and two-magnon BS $-3J_1^z$. Hence, for small values of $J_3$, $J_3^z$, and $J_1$, single and two-magnon BS do \emph{not} overlap. Still, topological spin excitations can be found, albeit only in the two-magnon sector by integrating out single-magnon states in Hamiltonian \eqref{eq:effective_ham_3band}. The resulting effective two-band model is defined on the checkerboard lattice [cf.~Fig.~\ref{fig:Lieb}(b)] and the Hamiltonian reads
$
    H^\text{eff,2}
    = \sum_{\vec{k}} \vec{\mathcal{C}}^\dagger_{\vec{k}}
		h(\vec{k}) 
		\vec{\mathcal{C}}_{\vec{k}},
$
with 
$
	\vec{\mathcal{C}}^\dagger_{\vec{k}} = (x^\dagger_{\vec{k}}, y^\dagger_{\vec{k}}) \mathcal{U}
$ 
and 
$
	\mathcal{U}=\text{diag}(\mathrm{e}^{\mathrm{i} k_x/2}, \mathrm{e}^{\mathrm{i} k_y/2})
$. The two-band Hamilton kernel 
\begin{align}
    h(\vec{k})
    =
    \begin{pmatrix}
	\tilde{E}_2^{x,y}(\vec{k}) & \tilde{\Delta}(\vec{k})  \\
	\tilde{\Delta}^\ast(\vec{k}) & \tilde{E}_2^{y,x} (\vec{k}) 
	\end{pmatrix}
	\label{eq:effective_ham_2band}
\end{align}
contains
\begin{subequations}
\begin{align}
 	 \tilde{E}_2^{\alpha,\beta}(\vec{k})
 	 &=
 	 E_2^{\alpha,\beta}(\vec{k})
 	 + \frac{D_\alpha^2}{2\left( B-J_1^z \right) } \left( 1-\cos k_\alpha \right),
 	 \label{eq:E2Spin12}
 	 \\
 	 \tilde{\Delta}(\vec{k}) 
 	 &=
 	 \frac{2J^2_1}{J_1^z} \cos \frac{k_x}{2} \cos \frac{k_y}{2} 
 	 + \mathrm{i} \frac{D_x D_y}{B - J_1^z} \sin \frac{k_x}{2} \sin \frac{k_y}{2},
 	 \label{eq:DeltaSpin12}
\end{align}
\end{subequations}
where $E_2^{\alpha,\beta}(\vec{k})$ is given in Eq.~\eqref{eq:E2}. 

\begin{figure}
	\centering
	\includegraphics[width=1\columnwidth]{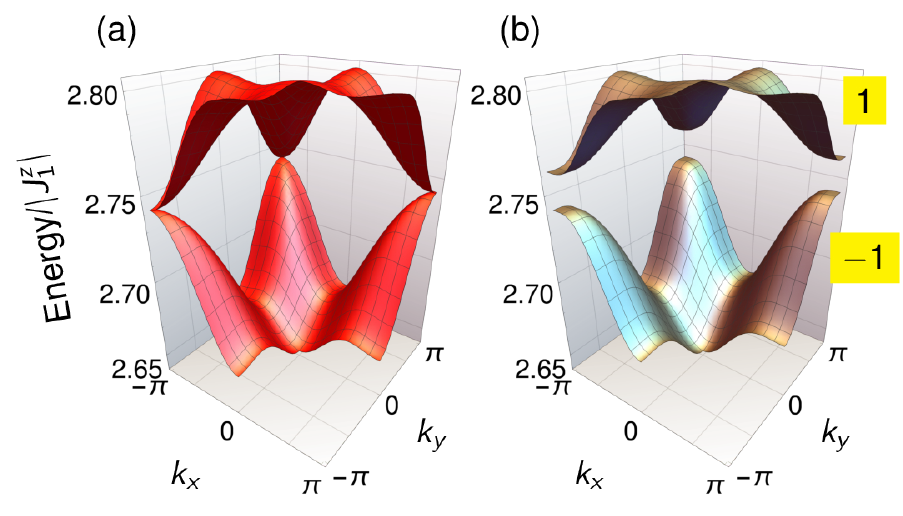}
    \caption{
        Two-magnon BS in spin-$1/2$ square-lattice ferromagnets with strong Ising anisotropy as derived from the effective two-band model in Eq.~\eqref{eq:effective_ham_2band}. (a) Without DMI, i.e., $D_x = D_y = 0$, two-magnon BS touch quadratically at the corners of the Brillouin zone. (b) With DMI, $D_x = D_y = 0.1$, virtual processes via single-magnon states break time-reversal symmetry, gapping out the quadratic touching point and causing a topological band gap. \change{Chern numbers are indicated by yellow labels.}
        Other parameters read $J_1/J_1^z = 0.2$, $J_3/J_1^z = -0.05$, $J_3^z/J_1^z = -0.05$, $B=0$. (Note that $J_1^z<0$.)
        }
	\label{fig:Spin12BandsBS}
\end{figure}

Without DMI, the BS energies touch quadratically at $(k_x,k_y) = (\pi,\pi)$, as depicted in Fig.~\ref{fig:Spin12BandsBS}(a) [also visible in Fig.~\ref{fig:Spin12Bands}(a)]. This touching point is already found in the spin-isotropic limit ($J_1 = J_1^z$) [see Fig.~\ref{fig:squareS12}(a)] \cite{Wortis1963}. It carries a Berry flux of $2\pi$ and is stabilized by TRS and $C_4$ symmetry, similar to what is known for noninteracting electronic gapless topological semimetals on the checkerboard lattice \cite{Sun2008, Chong2008, Sun2009}. However, with DMI, second-order virtual transitions via the single-magnon state gap out this touching point by breaking TRS [see Fig.~\ref{fig:Spin12BandsBS}(b)]. The gap is given by $(D_x^2+D_y^2)/(J_1^z-B)$; it scales with $1/B$ because the energy spacing between single-magnon states and two-magnon BS is linear in $B$.
Moreover, the gap is always topologically nontrivial. To show so, we invoke the general topological analysis of two-level systems and decompose the Hamilton kernel 
$
	h(\vec{k})
	=
	d_0(\vec{k}) \sigma_0
	+
	\vec{d}(\vec{k}) \cdot \vec{\sigma}
$
into Pauli matrices $\sigma_i$ ($i=0,\ldots,3$), $\vec{\sigma} = (\sigma_1,\sigma_2,\sigma_3)$. 
We define
$
    \vec{d}(\vec{k}) = (d_1(\vec{k}),d_2(\vec{k}),d_3(\vec{k}))
$
with
\begin{subequations}
\begin{align}
    d_0(\vec{k}) &= \frac{1}{2} [\tilde{E}^{x,y}_2(\vec{k})+\tilde{E}^{y,x}_2(\vec{k)}], \\
	d_1(\vec{k}) &= \frac{1}{2} [\tilde{\Delta}(\vec{k})+\tilde{\Delta}^\ast(\vec{k)}], \\
	d_2(\vec{k}) &=	\mathrm{i}\frac{1}{2} [\tilde{\Delta}(\vec{k})-\tilde{\Delta}^\ast(\vec{k)}], \\
	d_3(\vec{k}) &=	\frac{1}{2} [\tilde{E}^{x,y}_2(\vec{k})-\tilde{E}^{y,x}_2(\vec{k)}].
\end{align}
\end{subequations}

For a two-band model, the Chern number of the lower band can be written as \cite{Sticlet2012}
\begin{align}
    C = - \frac{1}{2} \sum_{\vec{k} \in \mathcal{D}} \mathrm{sgn}\left[ d_3(\vec{k}) \right]
    \mathrm{sgn}\left[ W(\vec{k}) \right],
    \label{eq:ChernSticlet}
\end{align}
where 
\begin{align}
    W(\vec{k}) 
    \equiv 
    \frac{\partial d_1(\vec{k})}{\partial k_x} \frac{\partial d_2(\vec{k})}{\partial k_y} - \frac{\partial d_1(\vec{k})}{\partial k_y} \frac{\partial d_2(\vec{k})}{\partial k_x} 
\end{align}
and $\mathcal{D}$ is the set of $k$-points where $d_1(\vec{k}) = d_2(\vec{k}) = 0$. From Eq.~\eqref{eq:DeltaSpin12} one reads off that this is the case at $\vec{k}_1^\mathcal{D} \equiv (0,\pi)$ and $\vec{k}_2^\mathcal{D} \equiv (\pi,0)$. At these points, $W(\vec{k}_1^\mathcal{D}) = - W(\vec{k}_2^\mathcal{D})$, with
\begin{align}
    W(\vec{k}_1^\mathcal{D}) 
    =
    D_x D_y \frac{J_1^2}{2 J_1^z \left( J_1^z - B\right)}, 
    \label{eq:Wtopology}
\end{align}
and
\begin{subequations}
\begin{align}
    d_3(\vec{k}_1^\mathcal{D}) 
    &=
    J_3
    -
    \frac{J_1^2}{2 J_1^z}
    + 
    \frac{D_y^2}{2( J_1^z  - B)} ,
    \\
    d_3(\vec{k}_2^\mathcal{D}) 
    &=
    -\left( J_3
    -
    \frac{J_1^2}{2 J_1^z}
    + 
    \frac{D_x^2}{2( J_1^z - B)}\right).
\end{align}
\end{subequations}
Since $\mathrm{sgn}[d_3(\vec{k}_1^\mathcal{D})] = -\mathrm{sgn}[ d_3(\vec{k}_2^\mathcal{D})]$, the Chern number reads
\begin{align}
    C
    &=
    -\mathrm{sgn} \left[ d_3(\vec{k}_1^\mathcal{D}) \right]
    \mathrm{sgn} \left[ W(\vec{k}_1^\mathcal{D}) \right].
    \label{eq:C1twoband}
\end{align}
We obtain $|C| = 1$, rendering the spectrum always topologically nontrivial for nonzero DMI. Note that DMI enters as $\mathrm{sgn} ( D_x D_y )$ in Eq.~\eqref{eq:C1twoband}, i.e., only the \emph{sign of the product} of $D_x$ and $D_y$ is relevant. This finding makes the time-reversal relation between interfacial and $D_\text{2d}$ DMI in Fig.~\ref{fig:dmi} explicit, in accordance with the discussion in Sec.~\ref{sec:DMI}.

Although the effective two-band model only accounts for two-magnon BS and the excitations carry predominantly spin-quadrupolar character, spin-dipolar character is not zero but perturbatively small. Let $\Delta E$ denote the energy difference between two-magnon BS and single-magnon excitations. Then, the dipolar character of the wave function is proportional to $D/\Delta E$, with $D$ being either $D_x$ or $D_y$. The amplitude of the single-spin dynamical structure factor $\mathcal{S}^{+-}(\vec{k},\omega)$ hence scales with $(D/\Delta E)^2 \ne 0$.
This is at variance with scenarios in which topological two-magnon BS appear due to spin-conserving interactions \cite{Qin2017, Qin2018}. There, two-magnon BS are fully spin-quadrupolar and invisible in the dynamical spin structure factor.

%
%
\subsection{Spin-$1$ Square-Lattice Magnets with Single-Ion Anisotropy}
\label{sec:spin1SQUARE}

\begin{figure}
	\centering
	\includegraphics[width=\columnwidth]{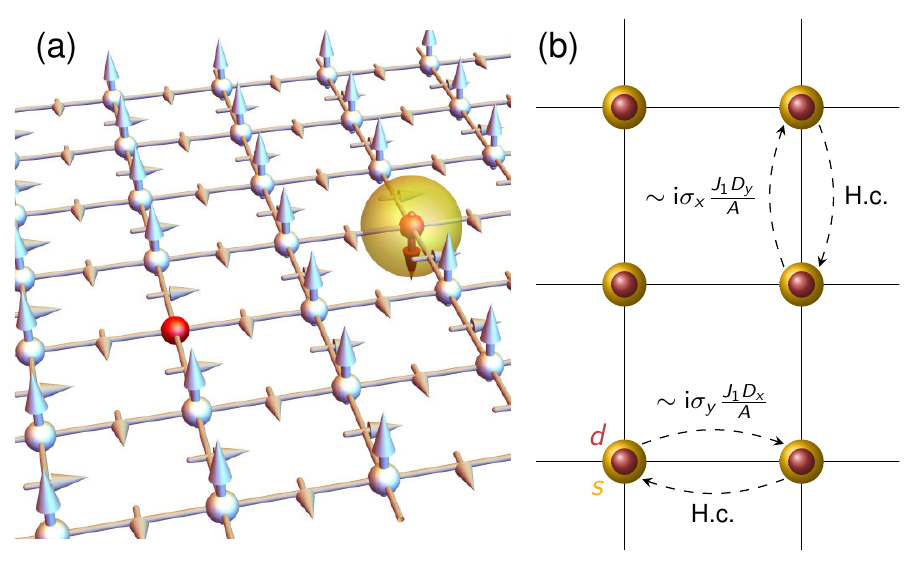}
    \caption{
    Spin-$1$ square lattice ferromagnet with strong single-ion anisotropy.
    (a) Sketch of the saturated ground state with two types of excitations (red). The single-magnon excitation is built from delocalized single spin deviations (left). Two-magnon excitations consist of two spin deviations, that is, a full spin flip (right).
    (b) Effective hopping model on the square lattice, with each site hosting two flavors of particles, $s$ and $d$, respectively denoting single spin deviations and double deviations. Upon hopping, a DMI-induced bond-dependent operation mixes the internal degrees of freedom, as indicated by Pauli matrices $\sigma_x$ and $\sigma_y$. This mixing causes a hybridization of single-magnon and two-magnon excitations and gives rise to topological spectral gaps.
    }
	\label{fig:spin1intro}
\end{figure}

So far, we discussed spin-$1/2$ quantum magnets with exchange BS. By enlarging the local Hilbert space to spin-$1$ states, a novel type of BS emerges, a single-ion BS (SIBS) built from two spin flips at the \textit{same} site \cite{Silberglitt1970, Tonegawa1970}, see Fig.~\ref{fig:spin1intro}(a). Instead of Ising anisotropy, it is susceptive to single-ion anisotropy $A$ and we consider the spin Hamiltonian $H = H_\text{c} + H_\text{DMI}$, where the spin-conserving piece reads
\begin{align}
	H_\text{c}
	= 
	\sum_{i=1}^{N_x} \sum_{j=1}^{N_y} 
	\left[
	    J_1 \vec{S}_{i,j} \cdot \left( \vec{S}_{i+1,j} + \vec{S}_{i,j+1} \right)
	    - A \left( S_{i,j}^z \right)^2
	    - B  S_{i,j}^z
	\right],
	\label{eq:HamiltonianSpin1}
\end{align}
with $J_1<0$, and $H_\text{DMI}$ as given in Eq.~\eqref{eq:DMIsquare}.
In the limit $A \gg |J_1|$, the SIBS separates from both the continuum and the two exchange BS and may overlap with the single-magnon energy. Again, we invoke perturbation theory and define two flavors of particles, 
\begin{subequations}
\begin{align}
	s^\dagger_{\vec{r}} |0\rangle &= |\vec{r}\rangle, \\
	d^\dagger_{\vec{r}} |0\rangle &= | \vec{r}, \vec{r}\rangle.
\end{align}
\end{subequations}
Here, $s$ particles create a single spin deviation and $d$ particles a double deviation, that is to say, a full flip of $S=1$. In contrast to the effective model of exchange BS in spin-$1/2$ magnets, the spin-$1$ effective model is defined on the square lattice, as derived in Appendix \ref{sec:AppSpin1}. Each site hosts two flavors of excitations or ``internal degrees of freedom'' and the hopping between sites is accompanied by a DMI-induced bond-dependent operation within this internal space [see Fig.~\ref{fig:spin1intro}(b)], which is conceptually similar to the Qi-Wu-Zhang model of Chern insulators \cite{Qi2006, Asboth2016}.
In Fourier space, the effective hopping model reads 
$
	H^\text{eff}
	=
	\sum_{\vec{k}} 
		\vec{c}^\dagger_{\vec{k}} 
		H(\vec{k})
		\vec{c}_{\vec{k}},
$
with $\vec{c}^\dagger_{\vec{k}} = (s^\dagger_{\vec{k}}, d^\dagger_{\vec{k}})$ and the Hamilton kernel
\begin{align}
    H(\vec{k}) = \begin{pmatrix}
        E_1(\vec{k}) & \Delta(\vec{k}) \\
        \Delta^\ast(\vec{k}) & E_2(\vec{k})
    \end{pmatrix},
    \label{eq:hamkernelSpin1}
\end{align}
where
\begin{subequations}
\begin{align}
	E_1(\vec{k}) &=	A + B - 4 J_1 + 2 J_1 \left( \cos k_x + \cos k_y \right)	\nonumber \\
	&\quad - \frac{2}{A+B} \left( D_x^2 \sin^2 \frac{k_x}{2} + D_y^2 \sin^2 \frac{k_y}{2}  \right), \label{eq:E1Spin1}
	\\
	E_2(\vec{k}) &=
		2B-8J_1 -\frac{2J_1^2}{A} \left(\cos^2 \frac{k_x}{2} + \cos^2 \frac{k_y}{2} \right), \label{eq:E2Spin1}
	\\
    \Delta(\vec{k}) &= \frac{J_1}{\sqrt{2}} \left( \frac{1}{2A} + \frac{1}{A+B} \right)
     \left( \mathrm{i} D_x \sin k_x - D_y \sin k_y \right)
	. \label{eq:Delta}
\end{align}
\end{subequations}
That the DMI-induced off-diagonal coupling $\Delta(\vec{k})$ in Eq.~\eqref{eq:Delta} between the two flavors of particles is a second-order process and, hence, proportional to $D_{x(y)} J_1 / A$ can be understood as follows. First-order DMI processes can only link single-magnon states with exchange BS but not with SIBS, that is to say, $\langle \vec{k} | H_\text{DMI} | \vec{k},\vec{0} \rangle = 0$. Instead, a second-order process is necessary that---starting from the SIBS---first splits the two spin deviations by letting one of them hop to a nearest-neighbor site and then DMI annihilates the remaining spin deviation. 

\begin{figure}
	\centering
	\includegraphics[width=1\columnwidth]{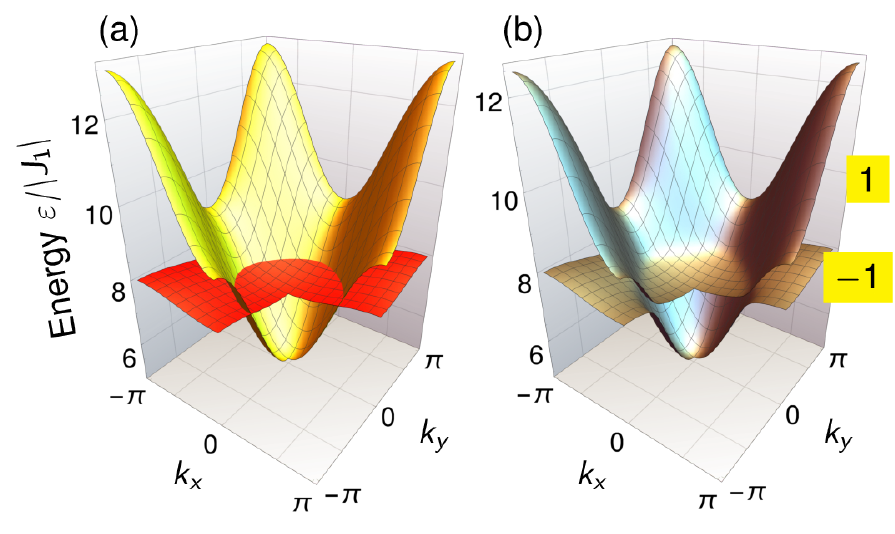}
    \caption{
        One-magnon states and two-magnon BS in spin-$1$ square-lattice ferromagnets with strong single-ion ansisotropy as derived from the effective two-band model in Eq.~\eqref{eq:hamkernelSpin1}. (a) Without DMI, i.e., $D_x = D_y = 0$, the single-magnon excitation (yellow band) crosses the single-ion bound state (red band) without hybridization because of particle-number conservation. (b) With DMI, $D_x/|J_1| = D_y/|J_1| = 0.8$, the particle-number is no longer conserved and the states hybridize, causing avoided crossings and a topological band gap. \change{Chern numbers are indicated by yellow labels. However, note that the band gap is not global.}
        Other parameters read $A/|J_1| = 5$ and $B=0$.}
	\label{fig:Spin1BandsBS}
\end{figure}

For zero DMI, as depicted in Fig.~\ref{fig:Spin1BandsBS}(a), the two bands touch along a closed line in reciprocal space if $B_\Gamma \le B \le B_M$, with 
\begin{subequations}
\begin{align}
    B_\Gamma &\equiv A + 8 J_1 + 4 \frac{J_1^2}{A} , \\
    B_M &\equiv A.
\end{align}
\end{subequations}
\change{These fields are defined by band touching conditions at the $\Gamma$ [$\vec{k} = (0,0)$] and $M$ point [$\vec{k} = (\pi,\pi)$], respectively.}
There is an intermediate field $B_\Gamma < B_X < B_M$, with
\begin{align}
    B_X \equiv A + 4J_1 + 2 \frac{J_1^2}{A},
\end{align}
such that, for $B < B_X$, the nodal line is centered about the BZ origin and, for $B>B_X$, about the BZ corner. \change{At $B=B_X$, the nodal line cuts through the $X$ points [$\vec{k} = (\pi,0), (0,\pi)$].}
Finite $D_x$ and $D_y$ splits this degeneracy and causes a topological gap [see Fig.~\ref{fig:Spin1BandsBS}(b)], which can be shown by an analysis similar to the one from Eq.~\eqref{eq:ChernSticlet} onwards. Although general results can be obtained in a closed form, they are rather unwieldy. Thus, we concentrate on the limit of small DMI such that terms proportional to $D^2_{x(y)}/A$ in Eq.~\eqref{eq:E1Spin1} can be ignored. We then obtain
\begin{align}
    C = &\text{sgn} \left( D_x D_y \right) \nonumber \\ 
    	&\times\left[ 
        \frac{\text{sgn}(B_M-B) + \text{sgn}\left( B_\Gamma-B \right)}{2}
        -
        \text{sgn} \left( B_X-B \right)
    \right]
    \label{eq:ChernNumberSpin1DMI}
\end{align}
and encounter again the factor $\text{sgn} ( D_x D_y )$ that explicitly emphasizes the time-reversal relation between interfacial and $D_\text{2d}$ DMI in Fig.~\ref{fig:dmi}; see also discussion in Sec.~\ref{sec:DMI}.
According to Eq.~\eqref{eq:ChernNumberSpin1DMI}, the following topological phase diagram is found:
\begin{align}
    C = 
    \text{sgn} \left( D_x D_y \right)
    \begin{cases}
        0, & B<B_\Gamma \\
        -1, & B_\Gamma<B<B_X \\
        1, & B_X<B<B_M \\
        0, & B_M<B
    \end{cases}
    .
\end{align}
While a chiral edge state is expected in the nontrivial phases, it turns out that the bands---although being separated by a gap at every point in reciprocal space---extend over overlapping energy intervals, such that upon projection onto the edge, there is no global gap. This property is a result of both $E_1(\vec{k})$ and $E_2(\vec{k})$ in Eqs.~\eqref{eq:E1Spin1} and \eqref{eq:E2Spin1}, respectively, having the same curvature. 

\begin{figure}
	\centering
	\includegraphics[width=1\columnwidth]{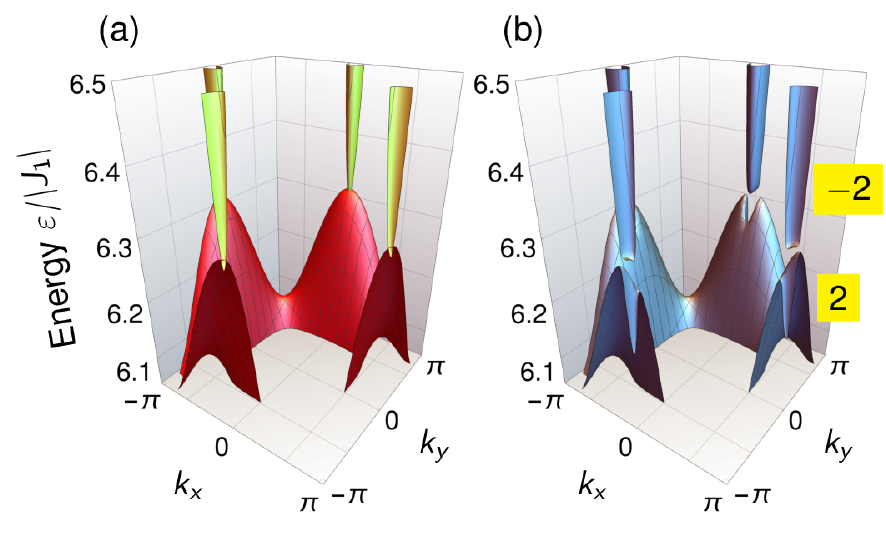}
    \caption{
        Same as Fig.~\ref{fig:Spin1BandsBS} but with additional antiferromagnetic second-nearest neighbor exchange interaction $J_2/|J_1| = 1$ and other parameters taken as $B/|J_1| = 3.3$ and $A/|J_1| = 7$. \change{(a) Without DMI, there are two nodal lines situated around $\vec{k}=(0,\pi)$ and $\vec{k}=(\pi,0)$. (b) With DMI, there is a global topological spectral gap. Chern numbers are indicated by yellow labels.}}
	\label{fig:Spin1BandsBSJ2}
\end{figure}

We can remedy this situation by including an antiferromagnetic second-nearest neighbor exchange interaction $J_2$ in the spin-conserving Hamiltonian $H_\text{c}$ in Eq.~\eqref{eq:HamiltonianSpin1}; thus, we replace 
$
    H_\text{c} \to H_\text{c} + J_2 \sum_{i=1}^{N_x} \sum_{j=1}^{N_y} 
	    \vec{S}_{i,j} \cdot \left( \vec{S}_{i+1,j+1} + \vec{S}_{i-1,j+1} \right)
$,
with $J_2 > 0$ but $J_2 < J_2^\text{max} \equiv ( B/2-J_1 )/2$ to keep the ferromagnetic ground state stable. The such amended effective model assumes the form of Eq.~\eqref{eq:hamkernelSpin1} with new entries (indicated by a tilde)
\begin{subequations}
\begin{align}
	\tilde{E}_1(\vec{k}) 
	&=
	E_1(\vec{k}) 
	- 4 J_2 + 2 J_2 \left[ \cos \left( k_x + k_y \right) + \cos \left( k_x - k_y \right) \right],
	\label{eq:E1Spin1J2}
	\\
	\tilde{E}_2(\vec{k}) 
	&=
	E_2(\vec{k}) - 8 J_2 - \frac{2J_2^2}{ A } \left[ 
		    \cos^2\left( \frac{k_x+k_y}{2} \right)
		    +
		    \cos^2\left( \frac{k_x-k_y}{2} \right)
		\right] , 
		\label{eq:E2Spin1J2}
	\\
    \tilde{\Delta}(\vec{k}) 
    &= 
    \Delta(\vec{k}) .
    \label{eq:DeltaSpin1J2}
\end{align}
\end{subequations}
Accordingly, we redefine the relevant magnetic fields as
\begin{align}
    \tilde{B}_\Gamma &\equiv A + 8 \left( J_1 + J_2 \right) + 4 \frac{
    J_1^2+J_2^2}{A},
    \\
    \tilde{B}_M &\equiv    
    A + 8 J_2 + 4 \frac{J_2^2}{A}.
\end{align}
Note that $\tilde{B}_X = B_X$ does not change upon inclusion of $J_2$. When neglecting the terms proportional to $D_{x(y)}^2/A$ in Eq.~\eqref{eq:E1Spin1J2}, the Chern number $\tilde{C}$ is still given by Eq.~\eqref{eq:ChernNumberSpin1DMI} with the appropriate replacement of the fields. The crucial difference to the case without $J_2$ is that $\tilde{B}_X < \tilde{B}_\Gamma$ for 
\begin{align}
    J_2 > J_2^* \equiv \sqrt{A^2 - A J_1 - \frac{J_1^2}{2}} - A  > 0.
\end{align}
This change of relative magnitude in the fields allows phases with higher Chern numbers:
\begin{align}
    \tilde{C} = 
    \text{sgn} \left( D_x D_y \right)
    \begin{cases}
        0, & B<\tilde{B}_X \\
        2, & \tilde{B}_X<B<\tilde{B}_\Gamma \\
        1, & \tilde{B}_\Gamma<B<\tilde{B}_M \\
        0, & \tilde{B}_M<B
    \end{cases}
    ,
    \quad
    J_2 > J_2^*.
\end{align}

\begin{figure}
	\centering
	\includegraphics[scale=1]{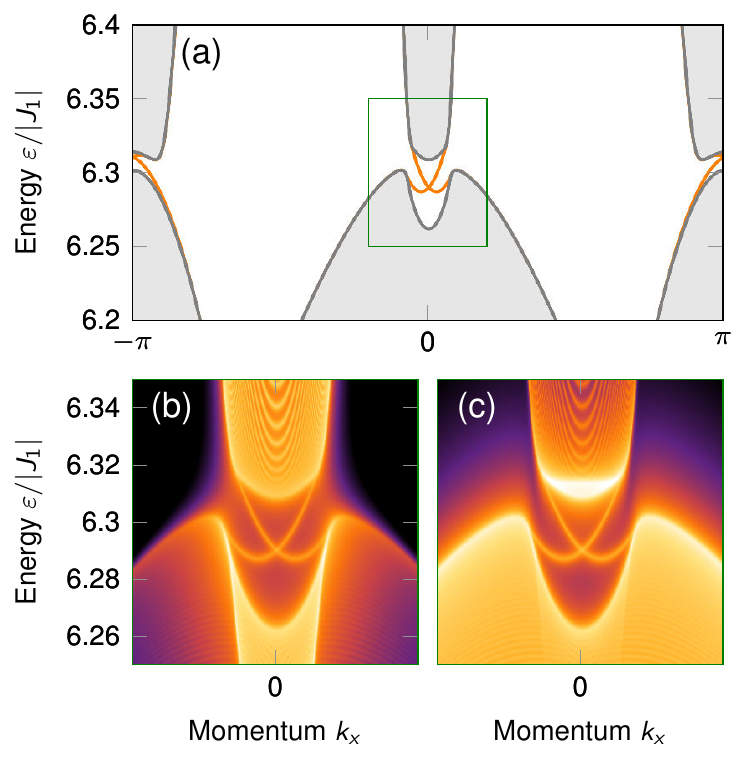}
    \caption{Topological chiral edge spin excitation built from single magnons and single-ion two-magnon BS in spin-$1$ quantum magnets with strong single-ion anisotropy. (a) Spin excitation spectrum of a slab with open (periodic) boundary conditions along the $y$ ($x$) direction. Gray continua depict the bulk states, orange lines indicate topological chiral edge states. \change{There are two chiral states per edge, in accordance with the Chern number. The green rectangle indicates the region zoomed into in panels (b) and (c).} (b,c) Dynamical correlation functions of one and two spins, $\mathcal{S}^{+-}(k_x,\omega)$ and $\mathcal{Q}^{+-}_{\vec{0}}(k_x,\omega)$, respectively encoding the dipolar and quadrupolar character of excitations. Logarithmic color scale with orange/black denoting large/small values. \change{The chiral edge state carries both spin-dipolar and spin-quadrupolar character.} Parameters read $J_2/|J_1| = 1$, $D_x/|J_1| = D_y/|J_1| = 0.8$, $B /|J_1|=3.3$, and $A/|J_1| = 7$ \change{and the slab width is given by $250$ unit cells.}}
	\label{fig:spin1}
\end{figure}

We find that a global band gap can be established within the topologically nontrivial phase with $\tilde{C} = - 2$. As an example, Figs.~\ref{fig:Spin1BandsBSJ2}(a) and (b) show the bulk single-magnon and SIBS dispersion for $J_2^* < J_2 < J_2^\text{max}$ without and with DMI, respectively. The hybridization between the two branches appears in the immediate vicinity of the $X$ points: $(k_x,k_y) = (0,\pi)$ and $(k_x,k_y) = (\pi,0)$. Fig.~\ref{fig:spin1}(a) shows the spectrum for periodic boundary conditions along the $x$-direction but open boundary conditions along the $y$-direction. In accordance with the Chern number, two chiral modes are found per edge. Their dipolar and quadrupolar character is revealed by the respective dynamical structure factors in Figs.~\ref{fig:spin1}(b) and (c), respectively.

\section{Coupling of Particle Number Sectors by Symmetric Off-Diagonal Interaction}
\label{sec:SymmetricTransverse}

\begin{figure}
	\centering
	\includegraphics[width=1\columnwidth]{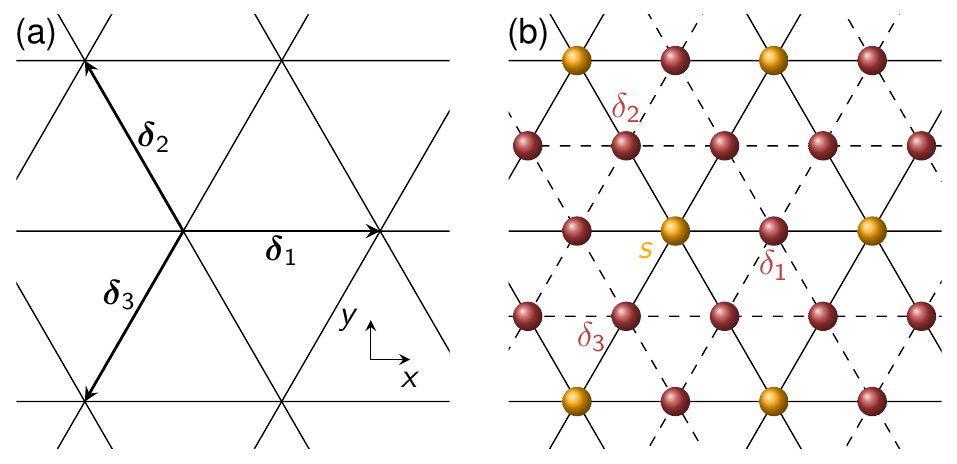}
    \caption{
        Spin-$1/2$ triangular-lattice ferromagnet with strong Ising anisotropy. (a) Structural lattice with nearest-neighbor vectors $\vec{\delta}_i$ ($i=1,2,3$). (b) Effective hopping model in the limit of strong anisotropy. Four types of particles are considered, one single spin-flip, $s$, and three double spin flips, $\delta_i$ ($i=1,2,3$). While the $s$ particles live on the original triangular lattice, the $\delta$ particles live on the bond midpoints, which form a kagome lattice. Solid bonds indicate $J^{z \pm}$-induced hoppings that convert $s$ particles into $\delta$ particle (and vice versa) because of broken particle-number (spin) conservation. Dashed bonds indicate particle-number conserving hoppings that mix $\delta$ particles among each other.}
	\label{fig:Spin12Triangular}
\end{figure}

We now turn to transverse-longitudinal \emph{symmetric} off-diagonal exchange interaction. Instead of square lattices, we focus on triangular lattices because they intrinsically allow this type of magnetic interactions \cite{Li2015Triangular, Li2016Triangular, Maksimov2019}. Let the exchange interaction between nearest-neighbor spins on the triangular lattice read $\vec{S}_{\vec{r}_l} \cdot \vec{I}_{\vec{\delta}_\alpha} \vec{S}_{\vec{r}_l+\vec{\delta}_\alpha}$. Here, $\vec{r}_l$ denotes a lattice site and $\vec{\delta}_\alpha$ with $\alpha = 1,2,3$ are three vectors to nearest neighbors [see Fig.~\ref{fig:Spin12Triangular}(a)]; explicitly:
\begin{subequations}
\begin{align}
    \vec{\delta}_1 &= \left( 1,0 \right)^\text{T}, \\
    \vec{\delta}_2 &= \frac{1}{2} \left( -1,\sqrt{3} \right)^\text{T}, \\
    \vec{\delta}_3 &= \frac{1}{2} \left( -1,-\sqrt{3} \right)^\text{T}.
\end{align}
\end{subequations}
Lattice symmetries restrict the exchange matrix to the following shape \cite{Li2015Triangular, Li2016Triangular, Maksimov2019}: 
\begin{align}
    \vec{I}_{\vec{\delta}_\alpha} = 
    \begin{pmatrix}
        J + 2 J^{\pm\pm} \cos{\phi_{\alpha}} & - 2 J^{\pm\pm} \sin{\phi_{\alpha}} & - J^{z\pm} \sin{\phi_{\alpha}}
        \\
        - 2 J^{\pm\pm} \sin{\phi_{\alpha}} & J - 2 J^{\pm\pm} \cos{\phi_{\alpha}} & J^{z\pm} \cos{\phi_{\alpha}} 
        \\
        - J^{z\pm} \sin{\phi_{\alpha}} & J^{z\pm} \cos{\phi_{\alpha}}  & J^z
    \end{pmatrix},
    \label{eq:matrixshapetriangular}
\end{align}
where $\phi_{\alpha}$ is the angle the bond vector $\vec{\delta}_\alpha$ makes with the $x$-axis; explicitly, $\phi_1 = 0$, $\phi_2 = 2 \pi/3$, and $\phi_3 = - 2 \pi/3$. As explained in Sec.~\ref{sec:PREbreakingU(1)}, our objective is to clarify the role of the transverse-longitudinal off-diagonal symmetric exchange $J^{z\pm}$. Hence, we set the transverse-transverse exchange $J^{\pm\pm}$ to zero and write the full spin Hamiltonian as $H = H_\text{c} + H_{z\pm}$, where $H_\text{c}$ contains spin-conserving interactions and
\begin{align}
    H_{z\pm}    
    =
    - \mathrm{i} \frac{J^{z \pm}}{2} \sum_{\vec{r}_l} \sum_{\alpha = 1}^3 
    &\left[
    \gamma^\ast_{\alpha}
    \left( S_{\vec{r}_l}^+ S_{\vec{r}_l+\vec{\delta}_\alpha}^z + S_{\vec{r}_l}^z S_{\vec{r}_l+\vec{\delta}_\alpha}^+ \right)
    \right.
    \nonumber \\
    -
    &\left. \gamma_{\alpha}
    \left( S_{\vec{r}_l}^- S_{\vec{r}_l+\vec{\delta}_\alpha}^z + S_{\vec{r}_l}^z S_{\vec{r}_l+\vec{\delta}_\alpha}^- \right)
    \right],
    \label{eq:Zpm}
\end{align}
with $\gamma_{\alpha} = \mathrm{e}^{\mathrm{i} \phi_{\alpha} }$. Similar to transverse DMI, $J^{z\pm}$ does not compromise the fully polarized ferromagnetic ground state, i.e.,
\begin{align}
    H_{z\pm} | 0 \rangle
    &=
    \mathrm{i} \frac{J^{z \pm}}{2} \sum_{\vec{r}_l} \sum_{\alpha = 1}^3 
    \gamma_{\alpha}
    \left( S_{\vec{r}_l}^- S_{\vec{r}_l+\vec{\delta}_\alpha}^z + S_{\vec{r}_l}^z S_{\vec{r}_l+\vec{\delta}_\alpha}^- \right)
    | 0 \rangle
    \nonumber \\
    &=
    \mathrm{i} S \frac{J^{z \pm}}{2} \sum_{\vec{r}_l} \sum_{\alpha = 1}^3 
    \gamma_{\alpha}
    \left( S_{\vec{r}_l}^- + S_{\vec{r}_l+\vec{\delta}_\alpha}^- \right)
    | 0 \rangle
    = 0
\end{align}
because $\sum_{\alpha = 1}^3 \gamma_\alpha = 0$ owing to the $C_3$ symmetry of the lattice. Periodic boundary conditions were assumed. Also similar to DMI, $J^{z \pm}$ breaks the effective TRS $\mathcal{T}'$.

Below, we present two examples of \change{particle sector} coupling induced by symmetric transverse-longitudinal bond-dependent off-diagonal exchange $J^{z \pm}$: (i) spin-$1/2$ magnets with strong Ising anisotropy (see Sec.~\ref{sec:Spin12Triangular}) and (ii) spin-$1$ magnets with strong single-ion anisotropy (see Sec.~\ref{sec:Spin1Triangular}).

\subsection{Spin-$1/2$ Triangular-Lattice Magnets with Ising Anisotropy}
\label{sec:Spin12Triangular}
First, we consider spin-$1/2$ triangular-lattice magnets with strong Ising anisotropy. The full spin Hamiltonian $H = H_\text{c} + H_{z\pm}$ has a spin-conserving part,
\begin{align}
    H_\text{c}
    = 
    \sum_{\vec{r}_l} \sum_{\alpha = 1}^3 
    \left(
    \vec{S}_{\vec{r}_l} \cdot \vec{I}_1 \vec{S}_{\vec{r}_l+\vec{\delta}_\alpha} 
    +
    \vec{S}_{\vec{r}_l} \cdot \vec{I}_3 \vec{S}_{\vec{r}_l+2\vec{\delta}_\alpha} 
    \right)
    - B \sum_l S_{\vec{r}_l}^z,
\end{align}
and $H_{z\pm}$ as given in Eq.~\eqref{eq:Zpm}.
We defined $\vec{I}_1 = \text{diag}(J_1,J_1,J_1^z)$ and $\vec{I}_3 = \text{diag}(J_3,J_3,J_3^z)$, with ferromagnetic nearest-neighbor exchange, $J_1, J_1^z <0$ and antiferromagnetic third-nearest-neighbor exchange $J_3, J_3^z>0$.

The two-magnon BS problem on the triangular lattice was considered in Ref.~\onlinecite{Wada1975} in the isotropic limit $J_1=J_1^z$ (and $J_3 = J_3^z = 0$) for general $S$. Two exchange BS were found below the two-magnon continuum in the vicinity of the corners of the Brillouin zone.
Conceptually, the spin-anisotropic limit is similar to that of the spin-$1/2$ magnet on the square lattice and general arguments can be carried over. In particular, ferromagnetic nearest-neighbor exchange, $J_1^z<0$, binds magnons together and BS energies separate from the two-magnon continuum in the limit $|J_1| \ll |J_1^z|$. However, we point out that there are three BS in the anisotropic limit, as expected for a lattice with coordination number six because spin flips can pair up along any of the three directions given by $\vec{\delta}_1$, $\vec{\delta}_2$, and $\vec{\delta}_3$. We will show below that the dispersion of the BS in the anisotropic limit mimicks that of free particles on the kagome lattice. The kagome lattice supports three bands, two of which are degenerate at the Brillouin zone corner (linear Dirac cone crossing) and the third is well-separated at much higher energies. Due to its higher energy, this state is buried by the two-magnon continuum in the isotropic limit, which explains the presence of only two BS in Ref.~\onlinecite{Wada1975}.

\subsubsection{Effective four-band model}
Again, we proceed with a perturbation analysis in the limit of strong nearest-neighbor Ising anisotropy $J_1^z$. We introduce four types of particles, $s$, $\delta_1$, $\delta_2$, and $\delta_3$, respectively associated with single spin flips and two neighboring spin flips connected by a $\vec{\delta}_\alpha$ bond, i.e.,
\begin{subequations}
\begin{align}
    s^\dagger_{\vec{r}} | 0 \rangle &= | \vec{r} \rangle, \\
    \delta^\dagger_{\alpha,\vec{r}} | 0 \rangle &= | \vec{r}, \vec{r} + \vec{\delta}_\alpha \rangle ,
\end{align}
\end{subequations}
with $\alpha = 1,2,3$.
Since the $\delta_\alpha$ particles live on the midpoints of the bonds of the original triangular lattice, they form a kagome lattice [see Fig.~\ref{fig:Spin12Triangular}(b)]. The $s$ particles live on the vertices of the triangular lattice that coincide with the midpoints of the hexagons of the kagome lattice. We assume that $J_3,J_3^z>0$ are antiferromagnetic, such that third-neighbor interaction does not cause magnon binding, ensuring that the relevant low-energy basis is given by the $s$ and $\delta$ particles.

As detailed in Appendix \ref{sec:AppSpin12Triangular}, a perturbation analysis in the limit of dominating nearest-neighbor Ising exchange $J_1^z$ is carried out to first order in perturbing hoppings, which is sufficient to both couple particle-number sectors and introduce finite hopping of two-magnon BS. In contrast to the square lattice, BS obtain finite hopping amplitudes already at first order in $J_1$ because of the coordination of the triangular lattice. In Fourier space, one obtains the effective model 
$
    H^\text{eff}
    = \sum_{\vec{k}} \vec{\mathcal{C}}^\dagger_{\vec{k}} 
		H(\vec{k}) 
		 \vec{\mathcal{C}}_{\vec{k}},
$
with 
$
    \vec{\mathcal{C}}^\dagger_{\vec{k}} = (s^\dagger_{\vec{k}}, \delta^\dagger_{1,\vec{k}}, \delta^\dagger_{2,\vec{k}}, \delta^\dagger_{3,\vec{k}}) U
$
and $U$ given in Appendix \ref{sec:AppSpin12Triangular}. 
The four-band Hamilton kernel reads
\begin{align}
    H(\vec{k}) = \begin{pmatrix}
        E_1(\vec{k}) & \Gamma_1 c_1  & \Gamma_2 c_2  & \Gamma_3 c_3 \\
        \Gamma^\ast_1 c_1 & E_2^{(1)}(\vec{k}) & J_1 c_3 & J_1 c_2 \\ 
        \Gamma^\ast_2 c_2 & J_1 c_3 & E_2^{(2)}(\vec{k}) & J_1 c_1 \\ 
        \Gamma^\ast_3 c_3 & J_1 c_2 & J_1 c_1 & E_2^{(3)}(\vec{k})
    \end{pmatrix},
    \label{eq:four-band-model}
\end{align}
with $c_\alpha = \cos \frac{k_\alpha}{2}$, and $\Gamma_\alpha = \mathrm{i} J^{z\pm} \gamma_\alpha^\ast$, and 
\begin{subequations}
\begin{align}
   & E_1(\vec{k}) = B - 3 \left( J_1^z + J_3^z \right) + \sum_{\alpha =1}^3 \left[ J_1 \cos k_\alpha + J_3 \cos ( 2 k_\alpha ) \right],
    \\
   & E_2^{(\alpha)}(\vec{k}) = 2B - 5J_1^z - 6J_3^z + J_3 \cos k_\alpha ,
\end{align}
\end{subequations}
where $k_\alpha = \vec{k} \cdot \vec{\delta}_\alpha$.

\begin{figure}
	\centering
	\includegraphics[width=1\columnwidth]{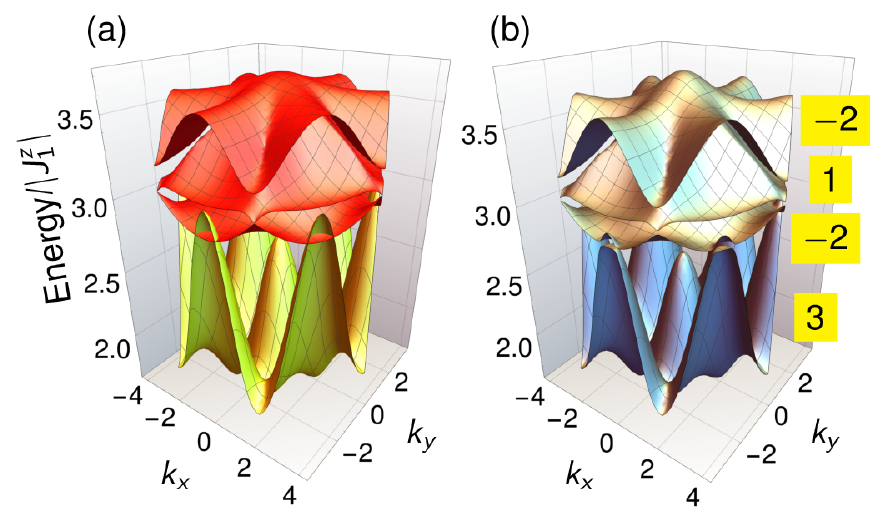}
    \caption{
        One-magnon states and two-magnon BS in spin-$1/2$ triangular-lattice ferromagnets with strong Ising ansisotropy as derived from the effective four-band model in Eq.~\eqref{eq:four-band-model}. (a) Without transverse-longitudinal off-diagonal exchange, i.e., $J^{z\pm} = 0$, single-magnon excitations (yellow band) and two-magnon excitations (red bands) cross without hybridization because of particle-number conservation. (b) With transverse-longitudinal off-diagonal exchange, $J^{z\pm}/|J_1^z| = 0.2$, the particle-number is no longer conserved and the states hybridize, causing avoided crossings and topological band gaps. \change{Chern numbers are indicated by yellow labels.}
        Other parameters read $J_1/|J_1^z| = -0.1$, $J_3/|J_1^z| = J_3^z/|J_1^z| = 0.28$, $B=0$. (Note that $J_1^z<0$.)
        }
	\label{fig:Spin12BandsTriangular}
\end{figure}

Figure \ref{fig:Spin12BandsTriangular} depicts the one-magnon and two-magnon BS spectrum, as obtained by diagonalizing Hamilton matrix \eqref{eq:four-band-model}. For zero $J^{z\pm}$, the particle number (spin) is conserved and states belonging to different particle sectors do not hybridize [see Fig.~\ref{fig:Spin12BandsTriangular}(a)]. Once particle number conservation is broken, i.e., $J^{z\pm} \ne 0$, the nodal-line crossings of one-magnon and two-magnon states get lifted [see Fig.~\ref{fig:Spin12BandsTriangular}(b)].  

\begin{figure}
	\centering
	\includegraphics[scale=1]{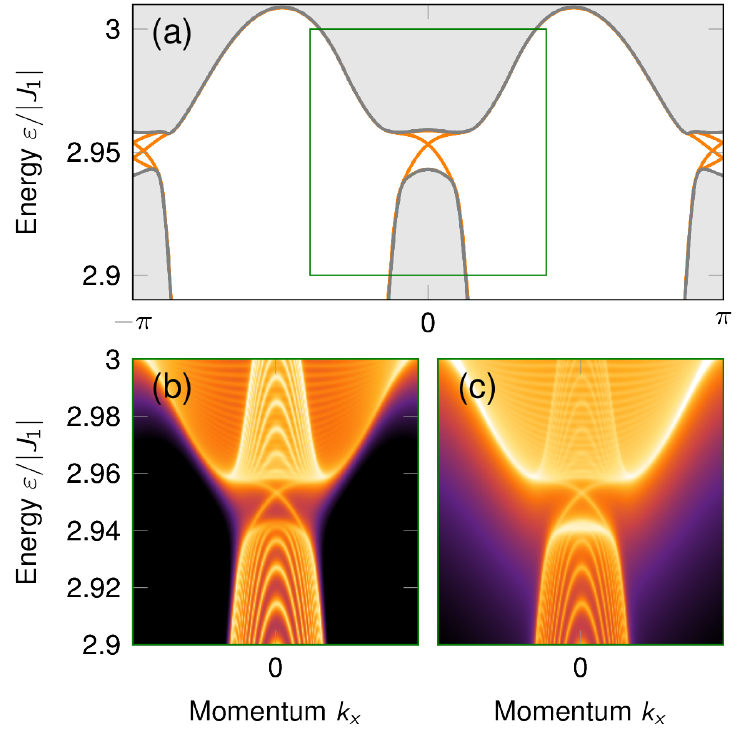}
    \caption{Topological chiral edge spin excitation built from single magnons and two-magnon BS in spin-$1/2$ triangular-lattice quantum magnets with strong Ising anisotropy. (a) Spin excitation spectrum of a slab with open (periodic) boundary conditions along the $y$ ($x$) direction. Gray continua depict the bulk states, orange lines indicate topological chiral edge states. \change{There are three chiral states per edge, in accordance with the Chern number. The green rectangle shows the region zoomed into in panels (b) and (c).} (b,c) Dynamical correlation functions of one and two spins, $\mathcal{S}^{+-}(k_x,\omega)$ and $\mathcal{Q}^{+-}_{\vec{\delta}_1}(k_x,\omega)+\mathcal{Q}^{+-}_{\vec{\delta}_2}(k_x,\omega)+\mathcal{Q}^{+-}_{\vec{\delta}_3}(k_x,\omega)$, respectively encoding the dipolar and quadrupolar character of excitations. \change{The chiral edge state carries both spin-dipolar and spin-quadrupolar character.} Logarithmic color scale with orange/black denoting large/small values. Parameters as in Fig.~\ref{fig:Spin12BandsTriangular}(b). \change{The slab width is given by $160$ unit cells.}}
	\label{fig:spin12triangularslab}
\end{figure}

Hamiltonian \eqref{eq:four-band-model} supports a rich topological phase diagram. For example, for the parameters in Fig.~\ref{fig:Spin12BandsTriangular}(b), the quadruple of Chern numbers reads $(3,-2,1,-2)$, ordered from the energetically lowest to the highest band. Thus, the lowest band gap between the single-magnon and two-magnon BS bands must support three chiral edge modes. A calculation with open boundary conditions verifies this prediction, as depicted in Fig.~\ref{fig:spin12triangularslab}(a). The edge modes carry both spin-dipolar [cf.~Fig.~\ref{fig:spin12triangularslab}(b)] and spin-quadrupolar character [Fig.~\ref{fig:spin12triangularslab}(c)].
In contrast, for $J_3 = J_3^z = 0$, the BS are well-separated from the (topologically trivial) single-magnon mode at lower energies and we find a Chern quadruple $(0,1,-2,1)$, with the BS bands always being topologically nontrivial, as explained within an effective three-band model below.

\subsubsection{Effective three-band model: Bound states only}
We derive an effective three-band model for the BS to understand their topological nature when they do \textit{not} overlap with the single-magnon states. Then, BS interact by virtual second-order hopping processes via single-magnon states. Such an effective kagome-lattice three-band model is given by
\begin{align}
    H'(\vec{k}) = \begin{pmatrix}
        E_2^{(1)}(\vec{k})  & J_1 c_3  & J_1 c_2  \\ 
        J_1 c_3 & E_2^{(2)}(\vec{k})  & J_1 c_1 \\ 
        J_1 c_2  & J_1 c_1  & E_2^{(3)}(\vec{k}),
    \end{pmatrix}
    +
    \Lambda(\vec{k}),
    \label{eq:three-band-model-triangular}
\end{align}
where $\Lambda(\vec{k})$ has the following elements:
\begin{align}
    \Lambda_{\alpha \beta}(\vec{k})
    =
    \frac{ \left( J^{z\pm} \right)^2 }{B-2J_1^z} \gamma_\alpha \gamma_\beta^\ast c_\alpha c_\beta
    .
\end{align}
We have obtained $H'(\vec{k})$ from Eq.~\eqref{eq:four-band-model} by means of a Schrieffer-Wolff transformation. Alternatively, it can be derived directly in real space. The elements of $\Lambda(\vec{k})$ correspond to effective first, second, and third-nearest neighbor hoppings on the kagome lattice, all of which come with the same amplitude. Importantly, the first two are complex ($\alpha \ne \beta$) and the BS acquire a TRS-breaking phase upon hopping. These additional hoppings render $H'(\vec{k})$ a generalized version of the kagome-lattice models studied, for example, in Refs.~\onlinecite{Ohgushi2000, Zhang2013, Mook2014, Mook2014edge}, for which topological phases are well established.

\begin{figure}
	\centering
	\includegraphics[width=1\columnwidth]{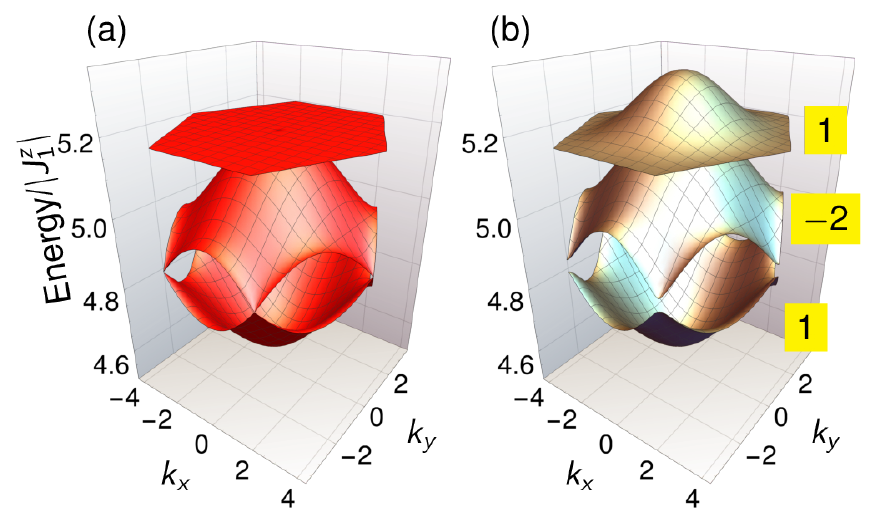}
    \caption{
    	Two-magnon BS in spin-$1/2$ triangular-lattice ferromagnets with strong Ising anisotropy as derived from the effective three-band model in Eq.~\eqref{eq:three-band-model-triangular}. (a) Without transverse-longitudinal off-diagonal exchange, i.e., $J^{z\pm} = 0$, two-magnon BS touch linearly at the K and K' points of the hexagonal Brillouin zone. (b) For $J^{z\pm}/|J_1^z| = 0.3$, virtual processes via single-magnon states break time-reversal symmetry, gapping out the Dirac cones and introducing topological band gaps. \change{Chern numbers are indicated by yellow labels.}
        Other parameters read $J_1/|J_1^z| = -0.2$, $J_3/|J_1^z| = J_3^z/|J_1^z| = 0.0$, $B=0$. (Note that $J_1^z<0$.)
        }
	\label{fig:Spin12BandsTriangularBSonly}
\end{figure}

In the following, we concentrate on the case $J_3 = J_3^z = 0$. If additionally $J^{z\pm} =0$, then $H'(\vec{k})$ assumes the form of a nearest-neighbor tight-binding model on the kagome lattice. As shown in Fig.~\ref{fig:Spin12BandsTriangularBSonly}(a), the dispersion of two-magnon BS exhibits the usual spectral characteristics of particles hopping on the kagome lattice: (i) linear Dirac cones at the Brillouin zone corners, (ii) a quadratic band touching at the Brillouin zone center, and (iii) a flat band. Once particle number conservation is broken, i.e., $J^{z\pm} \ne 0$, both the Dirac cones and the quadratic touching point get lifted, as depicted in Fig.~\ref{fig:Spin12BandsTriangularBSonly}(b). These liftings come with nonzero Berry curvature, leading to the Chern number triple $(1,-2,1)$.

\subsection{Spin-$1$ Triangular-Lattice Magnets with Single-Ion Anisotropy}
\label{sec:Spin1Triangular}
As the last model of topology originating from particle-number sector coupling in quantum magnets, we focus on spin-$1$ magnets on the triangular lattice. Similar to its cousin on the square lattice (see Sec.~\ref{sec:spin1SQUARE}), single-ion anisotropy establishes SIBS. We consider the spin Hamiltonian $H = H_\text{c} + H_{z\pm}$, with $H_{z\pm}$ given in Eq.~\eqref{eq:Zpm} and the spin-conserving piece being
\begin{align}
    H _\text{c} 
    &= 
    \sum_{\vec{r}_l} 
    \left[-A \left( S_{\vec{r}_l}^z \right)^2 - B S_{\vec{r}_l}^z
    \right. \nonumber \\
    &\left. \quad 
	+ \sum_{\alpha = 1}^3  
    \left( J_1 \vec{S}_{\vec{r}_l} \cdot \vec{S}_{\vec{r}_l+\vec{\delta}_\alpha} 
    +
    J_2 \vec{S}_{\vec{r}_l} \cdot \vec{S}_{\vec{r}_l+\vec{\delta}_\alpha-\vec{\delta}_{\alpha-1}}  \right)\right].
\end{align}
Here, $\alpha-1$ is to be understood cyclically such that $\vec{\delta}_\alpha - \vec{\delta}_{\alpha-1}$ connects second-nearest neighbors. 
We invoke perturbation theory once again and define two flavors of particles, 
$
	s^\dagger_{\vec{r}_l} |0\rangle = |\vec{r}_l\rangle
$,
$ 
	d^\dagger_{\vec{r}_l} |0\rangle = | \vec{r}_l, \vec{r}_l\rangle
$, for single spin deviations and double deviations, respectively. These particles both live on the vertices of the original triangular lattice and $J^{z\pm}$ causes a bond-dependent mixing between them. Details are given in Appendix \ref{sec:AppSpin1Triangular}. In reciprocal space, we arrive at $H^\text{eff} = \sum_{\vec{k}} \vec{c}^\dagger_{\vec{k}} H(\vec{k}) \vec{c}_{\vec{k}}$, with $\vec{c}^\dagger_{\vec{k}} = (s^\dagger_{\vec{k}}, d^\dagger_{\vec{k}})$ and the kernel
\begin{align}
    H(\vec{k}) = \begin{pmatrix}
        E_1(\vec{k}) 
        &
        \Delta(\vec{k}) 
        \\
        \Delta^\ast(\vec{k}) 
        &
        E_2(\vec{k}) 
    \end{pmatrix},
    \label{eq:effectivehamiltonianspin1triangular}
\end{align}
where
\begin{subequations}
\begin{align}
     E_1(\vec{k}) &= A + B- 6(J_1 + J_2)
     	- \frac{\left( J^{z\pm} \right)^2}{A+B} \sum_{\alpha = 1}^3 \left( 1+\cos k_\alpha \right)
     	\nonumber \\
     	&\quad + 2 \sum_{\alpha = 1}^3 \left[ J_1 \cos k_\alpha + J_2 \cos (k_\alpha-k_{\alpha-1}) \right],
     \\
     E_2(\vec{k}) &= 2B - 12 (J_1 + J_2) \nonumber \\
     	&\quad- \frac{1}{A}\sum_{\alpha = 1}^3 \left[ J_1^2 (1 + \cos k_\alpha) + J_2^2 \left(1 + \cos (k_\alpha-k_{\alpha-1}) \right) \right],
     \\
     \Delta(\vec{k}) &= -\frac{\mathrm{i} J^{z\pm} J_1}{\sqrt{2}} \left( \frac{1}{A+B} + \frac{1}{2A} \right)
    \sum_{\alpha = 1}^3 \gamma^\ast_{\alpha} \cos k_\alpha.
\end{align}
\end{subequations}
Similar to the square-lattice model, particle-sector coupling is a second-order process proportional to $J^{z\pm}J_1/A$. Starting from a $d$ particle, a $J_1$ hopping first seperates the two spin deviations and then $J^{z\pm}$ annihilates the remaining spin deviation.

\begin{figure}
	\centering
	\includegraphics[width=\columnwidth]{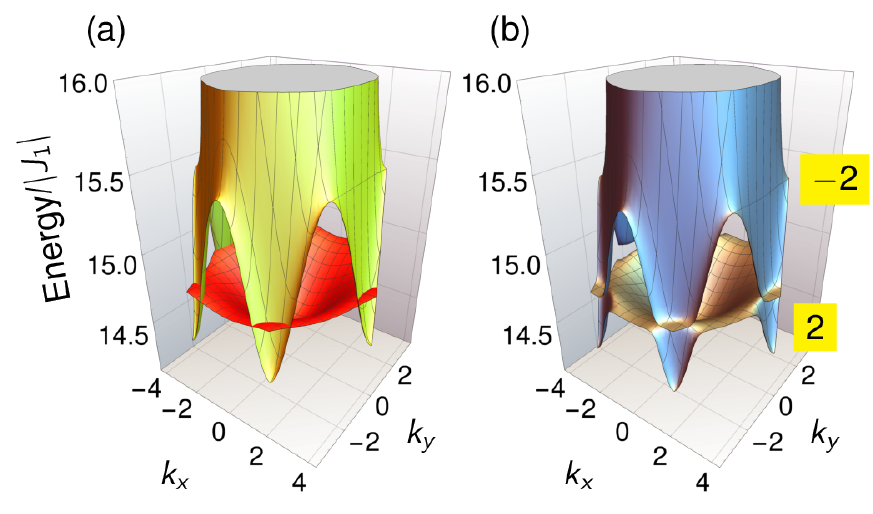}
    \caption{
    One-magnon states and two-magnon BS in fully polarized spin-$1$ triangular-lattice magnets with strong single-ion ansisotropy as derived from the effective two-band model in Eq.~\eqref{eq:effectivehamiltonianspin1triangular}. (a) Without $J^{z\pm}$, the single-magnon excitation (yellow band) crosses the single-ion bound state (red band) without hybridization because of particle-number conservation. (b) For $J^{z\pm}/|J_1| = 1$, the particle-number is no longer conserved and the states hybridize, causing avoided crossings and a topological band gap. \change{Chern numbers are indicated by yellow labels.}
        Other parameters read $J_2 = 0$, $A/|J_1| = 10$ and $B/|J_1|=13.5$. (Note that $J_1>0$.)}
	\label{fig:Spin1BandsTriangular}
\end{figure}

\begin{figure}
	\centering
	\includegraphics[scale=1]{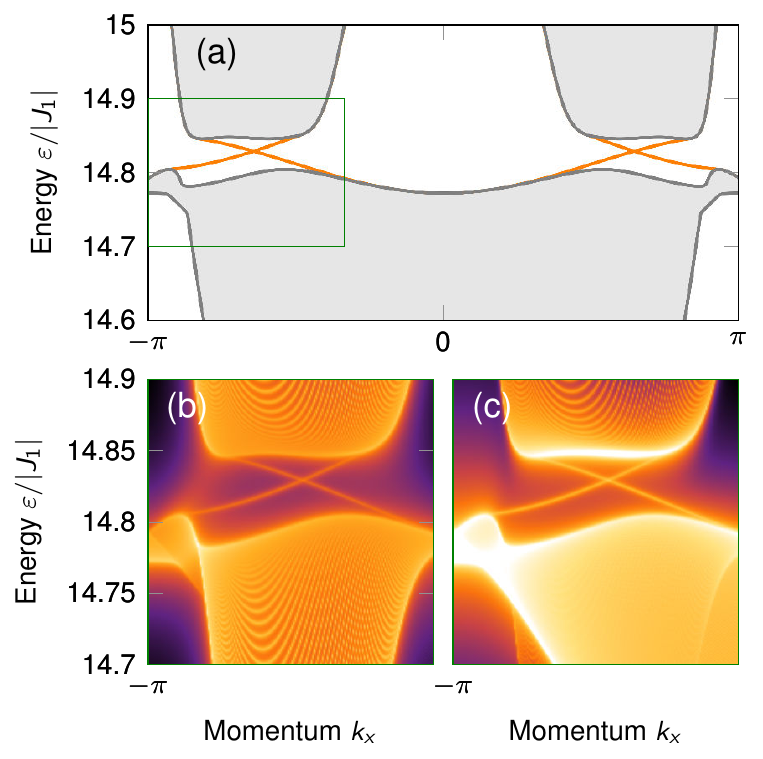}
    \caption{Topological chiral edge spin excitation built from single magnons and single-ion two-magnon BS in spin-$1$ triangular-lattice quantum magnets with strong single-ion anisotropy. (a) Spin excitation spectrum of a slab with open (periodic) boundary conditions along the $y$ ($x$) direction. Gray continua depict the bulk states, orange lines indicate topological chiral edge states. \change{There are two chiral states per edge, in accordance with the Chern number. The green rectangle indicates the region zoomed into in panels (b) and (c).} (b,c) Dynamical correlation functions of one and two spins, $\mathcal{S}^{+-}(k_x,\omega)$ and $\mathcal{Q}^{+-}_{\vec{0}}(k_x,\omega)$, respectively encoding the dipolar and quadrupolar character of excitations. Logarithmic color scale with orange/black denoting large/small values. Parameters as in Fig.~\ref{fig:Spin1BandsTriangular}(b). \change{The slab width is given by $200$ unit cells.}}
	\label{fig:spin1triangularslab}
\end{figure}

To go beyond the considerations so far, which assumed ferromagnetic nearest-neighbor exchange, here we assume \textit{antiferromagnetic} $J_1>0$ as an example and set $J_2 = 0$ for simplicity. However, such large a magnetic field $B$ is applied that the field-polarized phase becomes stable. If the two bands overlap for $J^{z\pm} = 0$, as is the case depicted in Fig.~\ref{fig:Spin1BandsTriangular}(a), a topological gap appears for $J^{z\pm} \ne 0$ [cf.~Fig.~\ref{fig:Spin1BandsTriangular}(b)]. The Chern numbers are $(2,-2)$ and two edge states are expected, as confirmed for a slab calculation [cf.~Fig.~\ref{fig:spin1triangularslab}(a)]. Once again, the chiral edge states carry both spin dipolar and spin quadrupolar character, as visible in the spin and spin-pair dynamical structure factors in Figs.~\ref{fig:spin1triangularslab}(b) and (c), respectively. These results show that topological gaps between states belonging to different particle-number sectors is also found in field-polarized antiferromagnets, extending the possible material spectrum.

\section{Discussion}
\label{sec:Discussion}
\change{Here, we discuss the spin-multipolar and lifetime properties of the discovered chiral edge states (Sec.~\ref{sec:Properties}), point out the difference of the magnon topology at hand to ``conventional'' magnon topology (Sec.~\ref{sec:harmonictopology}), propose experimental signatures of the discussed topological effects (Sec.~\ref{sec:THE}), mention implications of particle-sector coupling beyond the one and two-particle sectors (Sec.~\ref{sec:BeyondOneTwo}), and identify suitable material candidates for experimental verification (Sec.~\ref{sec:Materials}).}

\subsection{Properties of Chiral Spin Hybrids: Spin Multipoles and Lifetime}
\label{sec:Properties}
For all cases of \change{particle-number sector coupling considered above, we have shown that the edge states are hybrids of single-magnon and two-magnon states.} As a general trend, we find that the chiral modes are predominantly quadrupolar. This is to be expected, given that the bandwidth (the density of states) of two-magnon BS is considerably smaller (larger) than that of single magnons. So, a particle sector coupling mixes many two-magnon BS into a few single-magnon states, rendering the resulting chiral modes predominantly two-magnon-like. Nonetheless, \change{the finite single-magnon character renders the chiral edge mode visible in experiments that are subject to the dipole selection rule.} 

The chiral mode being immune to backscattering and carrying spin-dipolar as well as spin-quadrupolar character \change{suggests} topological \change{beyond-spintronics} computation paradigms \cite{Romhanyi2019}. A chiral spin hybrid can couple via its spin dipole to an electronic spin accumulation at an interface to a normal metal. The information then propagates unidirectionally, unimpeded by defects and disorder, and gets partially converted into spin quadrupolar information, establishing a spin quadrupolar current \cite{Hell2013}. It may be read out by coupling to the hybrid's spin quadrupole, potentially possible by quantum dots with spins larger than $1/2$ \cite{misiorny2013} or the magnetic quadrupole moment of Bloch electrons \cite{Gao2018, Shitade2018}. 

Both for the spectral detection as well as for any applications, it is important that the quasiparticles are long-lived. Naively, particle-number nonconservation is associated with spontaneous decays and, hence, a finite quasiparticle lifetime is expected even at zero temperature. However, as a result of energy conservation, spontaneous decays are strongly kinematically restricted \cite{Zhitomirsky2013}. A particle can only spontaneously decay into a continuum of states if there are two (or more) decay products at lower energies available. For the quantum magnets discussed here, this is never the case because the lowest-energy continuum is the two-magnon continuum that has much higher energy than the single-magnon and two-magnon BS in the limit of strong spin anisotropy. Consequently, the chiral edge states do not suffer from many-body decays and their lifetime is, in principle, infinite. Of course, this argument only applies to the pristine magnetic system without defects or coupling to phonons. \change{We expect the latter perturbation to be responsible for a finite edge state lifetime, which calls for a detailed analysis that takes the many-body environment of the solid state into account.}

\subsection{Comparison to ``Harmonic'' Magnon Topology}
\label{sec:harmonictopology}
Topological effects arising from a particle-number sector coupling go beyond the semiclassical ``harmonic'' magnon topology described by linear spin-wave theory 
(e.g., cf.~Refs.~\onlinecite{Katsura2010, Onose2010, Ideue2012, Zhang2013, Hoogdalem2013, Shindou13, Shindou13b, Mook2014, Mook2014edge, Shindou14, Mook2015interfaces, Mook2015waveguide, Owerre2016a, Li2016Weyl, Mook2016Weyl, Mook2017nodal, Nakata2017, Nakata2017AFM, Mook2018duality, Aguilera2020, Mook2020hinge, Neumann2022}) 
and frequently used to fit topological magnon spectra found in inelastic scattering data \cite{Chisnell2015, Chen2018CrI3, Cai2021CrBr3, zhu2021topological, Bao2018, yao2018topological, Wang2019Diracmagnon, elliot2021order}. To make this point explicit, we recall that single-magnon topology is found at the same order as the single-magnon energies, that is, at order $S$ in the bosonic bilinear $H_2$, as obtained from expansion \eqref{eq:HPexpansion} of the HP transformation. Consequently, the ratio of topological gaps $\Delta = O(S)$ and the magnon bandwidth $W = O(S)$ scales with $\Delta/W = O(S^0)$ and topology survives taking the classical limit $S \to \infty$. \change{In contrast, the effects discussed in the present work vanish in the classical limit: the topological gaps}, $\Delta'$, appear due to matrix elements of the form
\begin{align}
	\langle \text{two magnons} | H_3 | \text{one magnon} \rangle \propto \sqrt{S}, \label{eq:H3scaling}
\end{align}
such that $\Delta'/W = O(S^{-1/2})$ vanishes as $S \to \infty$. Similarly, the binding energy of two-magnon BS---being a result of attractive interactions in $H_4$---is of order $O(S^{-1})$ relative to the single-magnon energies. Taking $S \to \infty$, BS disappear by merging into the lower threshold of the two-magnon continuum, as discussed, for example, in Ref.~\onlinecite{Rastelli2011}.  
We conclude that the \change{present} topological effects are an example of ``quantum magnon topology,'' an umbrella term encompassing any topological effects that disappear in the classical limit.

\subsection{Experimental Signatures}
\label{sec:THE}
\subsubsection{Spectroscopy}
As pointed out above, the most drastic effects of particle-number sector coupling are anticrossings between single-particle and two-particle states. Such splittings were recently observed in the antiferromagnet FeI$_2$ \cite{Bai2021} by means of inelastic neutron scattering.
For the detection of chiral edge states, other, more surface-sensitive methods could be used. Possibilities include parametric amplification of chiral edge magnons \cite{Malz2019}, Raman scattering \cite{Perreault2016}, spin-polarized scanning tunneling microscopy \cite{Feldmeier2020}, nitrogen-vacancy center relaxometry \cite{Rustagi2020}, and spin-resolved inelastic electron spectroscopy \cite{DosSantos2018}.

Besides topological edge states, nontopological effects of particle-number sector coupling already lead to spectroscopic signatures. We recall that a state's spin moment $\Delta S_n(\vec{k})$ can be extracted from its energy $E_n(\vec{k})$ by
$
	\Delta S_n(\vec{k}) = \partial E_n(\vec{k}) / \partial B_z
$,
where $n$ is the state index; we merged the $g$-factor and Bohr magneton into the field. Without spin (or particle) sector coupling, $\Delta S_n(\vec{k}) = \text{const.} \in \mathbb{N}$ is the particle number. Such measurements were performed on FeI$_2$ by means of 
infrared absorption \cite{Fert1978, Petitgrand1980}, neutron scattering \cite{Petitgrand1979}, and electron spin resonance \cite{Katsumata2000}. However, with coupling, spin sectors mix and $\Delta S_n(\vec{k})$ can assume any value, as recently seen in FeI$_2$ by terahertz spectroscopy \cite{Legros2020}.

\subsubsection{Transport}
\change{The physics of particle-number sector coupling discussed here} contributes to the recently debated puzzle of anomalous thermal Hall effects (THE) in systems with only a \emph{single} single-magnon band as is the case for Bravais lattices. Since an isolated band cannot carry Berry curvature, it seems that an intrinsic anomalous THE is impossible. There are two proposals how an anomalous THE can appear nonetheless, one for $S=1/2$ \cite{Park2020} and another for $S = \infty$ \cite{Carnahan2021}. For $S=1/2$, Ref.~\onlinecite{Park2020} proposed a mean-field Schwinger boson description that predicts an excitation spectrum with Rashba-like split spinon bands the Berry curvature of which causes an intrinsic anomalous THE. For classical spins, $S = \infty$, Ref.~\onlinecite{Carnahan2021} identified temperature-induced chiral spin fluctuations as the source of a THE, relying on Landau-Lifshitz spin dynamics simulations for the evaluation of current-current correlation functions in the Kubo formula \cite{Mook2016spindynamics, Mook2017noncollinear}.

Our results complement the aforementioned efforts by proposing an alternative intrinsic origin of the anomalous THE for any $S<\infty$. Due to the coupling between particle-number sectors, two-magnon states endow the single-magnon band with a finite Berry curvature. Our effective models in the spin-anisotropic limit show this mechanism explicitly. However, we stress that the spin-anisotropic limit is \emph{not} necessary to obtain a finite Berry curvature of the single-magnon band. Even for isotropic Heisenberg exchange and a magnetic field large enough to prevent the formation of spin spirals or skyrmions, there is a finite Berry curvature of the single-magnon band. Importantly, and again in contrast to the ``harmonic'' magnon topology (cf.~Sec.~\ref{sec:harmonictopology}), the Berry curvature induced by particle-number sector coupling is $\Omega = O(S^{-1})$ [instead of $\Omega = O(S^{0})$]. Thus, while any ferromagnet with $S<\infty$ and particle-number sector coupling can, in principle, exhibit intrinsic anomalous Hall-type transport effects, the proposed mechanism is inactive in the classical limit; the anomalous intrinsic thermal Hall conductivity $\kappa_{xy} = O(S^{-1})$ vanishes as $S \to \infty$. It remains an open theoretical problem how to evaluate the intrinsic transverse Hall conductivities beyond the noninteracting boson approximation of Refs.~\onlinecite{Katsura2010, Matsumoto2011a, Matsumoto2011b, Matsumoto2014}.

\subsection{Beyond One And Two-Particle Sectors}
\label{sec:BeyondOneTwo}
Throughout the present work, we have restricted the Hilbert space to one and two-magnon excitations and neglected coupling beyond. The presented particle-number sector coupling straightforwardly applies to higher-magnon-number sectors and topological hybridizations between, e.g., spin-quadrupolar and spin-octupolar excitations can be expected. However, the higher up in energy, the more likely the situation that the topological bands are buried by a continuum of states. 

The effects of particle-number sector coupling change qualitatively, if there is an energy well below any continua for which there are BS  of \textit{all} particle-number sectors. Then, coupling appears between an infinite number of states and causes fractionalization, i.e., a local spin-flip operator $S^-$ creates not one but multiple quasiparticles. Such a situation appears in the one-dimensional spin-$1/2$ Ising model with transverse DMI, a situation we explore for completeness in Appendix \ref{App:Ising}.

\subsection{Material Candidates}
\label{sec:Materials}
Our results predict two broad material classes as suitable platforms for the detection of \change{topology originating from a coupling of particle-number sectors:} chiral magnets (see Sec.~\ref{sec:DMI}) and triangular-lattice magnets (see Sec.~\ref{sec:SymmetricTransverse}), both with strong spin-anisotropy either in the form of Ising or single-ion anisotropy. Although we have restricted our study of chiral magnets to the square lattice, the general arguments carry over to any other two-dimensional lattice. Large DMI requires a broken inversion symmetry and strong spin-orbit coupling. It has been predicted in Janus monolayers of, for example, manganese dichalcogenides \cite{Liang2020}, chromium dichalcogenides \cite{Cui2020}, and other van der Waals magnets \cite{Yuan2020Janus, Zhang2020Janus, Shen2021Janus}. Additionally, DMI can be engineered by interfacial symmetry breaking \cite{Fert1980, Levy1981, Zakeri2010, Wang2020YIGDMI}, electric fields \cite{Katsura2005}, and chemisorbed oxygen \cite{Chen2020chemi}. It may hence be feasible to engineer DMI in systems that naturally come with strong spin-space anisotropy, such as molecular magnets \cite{Koch2003}.
On the other hand, triangular-lattice magnets with strong spin-orbit coupling are very common in the large class of transition metal dihalides and trihalides \cite{McGuire2017}, an example being the antiferromagnet FeI$_2$ that was recently shown to exhibit a spectral gap caused by \change{coupling of single magnons and SIBS} \cite{Bai2021}. However, its magnetic point group $2/m1'$ \cite{Gelard1974, Gallego2019} is not compatible with ferromagnetism, rendering Chern insulating behavior impossible and we conclude that the spectral gap is topologically trivial. This situation may change upon application of a magnetic field and entering the spin flop phase. Yet another candidate material family are the rare-earth chalcogenides \cite{Li2015Triangular, Li2016Triangular, Maksimov2019}, for which field-polarization may overcome the tendency for spin-liquid behavior.

\section{Conclusion}
\label{sec:Conclusion}
We have introduced the notion of topology that arises from a hybridization between two different particle-number sectors. The wave function of the resulting chiral edge states was shown to carry weight in both particle-number sectors, rendering it a hybrid without definite particle number. We specified our considerations to topological hybrids of single-magnon and two-magnon states in quantum ferromagnets, \change{for which spin nonconservation induces particle-number nonconservation.} In contrast to semiclassical topological magnons, the present chiral edge hybrids carry a mixed spin-multipolar character, that is, a spin quadrupole beyond the spin dipole. This finding not only \change{suggests beyond-spintronics paradigms} but may also lay the foundation of ``quantum topological magnonics,'' comprising magnon topology and related Hall-type transport phenomena that vanish in the classical limit.

On a fundamental level, our results highlight that even fully saturated ferromagnets---arguably the ``least quantum'' of all quantum magnets because their classical and quantum ground states coincide---support exotic topological quasiparticles that escape the classical Landau-Lifshitz theory. This finding establishes quantum magnets as a rich platform for studying topological phenomena that are inaccessible in particle-number conserving systems, an example being ultracold atoms. As such, the topological effects discussed here go beyond the celebrated connection between atomic and magnetic systems that encompasses phenomena such as Bose-Einstein condensation \cite{giamarchi2008bose, Zapf2014} and the Efimov effect \cite{Nishida2013}. 
We hope that our discovery will inspire further research on topological magnetic quasiparticles. It will be interesting to \change{explore spin space group \cite{Corticelli2022} and topological quantum chemistry arguments beyond linear spin-wave theory \cite{Corticelli2022b} and to} investigate how particle-number sector hybridizations \change{rectify longitudinal and transverse heat and spin (multipole) transport \cite{ Zyuzin2016SNE, Cheng2016SNE, Zhang2018, Zyuzin2018, Mook2019SSE, Li2020SNE, Mook2020MSHE, Li2020Edelstein, Neumann2020}.} \change{Since our general considerations on particle-number sector coupling go far beyond magnetic excitations, they apply, in principle, to any nonconserved bosonic collective excitation in quantum condensed matter, such as phonons \cite{Cohen1969,Ziman2001}.}

\begin{acknowledgements}
	We thank Sebasti\'{a}n A.~D\'{\i}az for helpful discussions.
    This work was supported by the Georg H. Endress Foundation, the Swiss National Science Foundation, NCCR QSIT, and NCCR SPIN. This project received funding from the European Union's Horizon 2020 research and innovation program (ERC Starting Grant, Grant No.~757725).
\end{acknowledgements}

\appendix

\section{Two-magnon states on the square lattice}
\label{Appendix:reklis-construction}

\begin{figure}
    \centering
    \includegraphics[width=0.9\columnwidth]{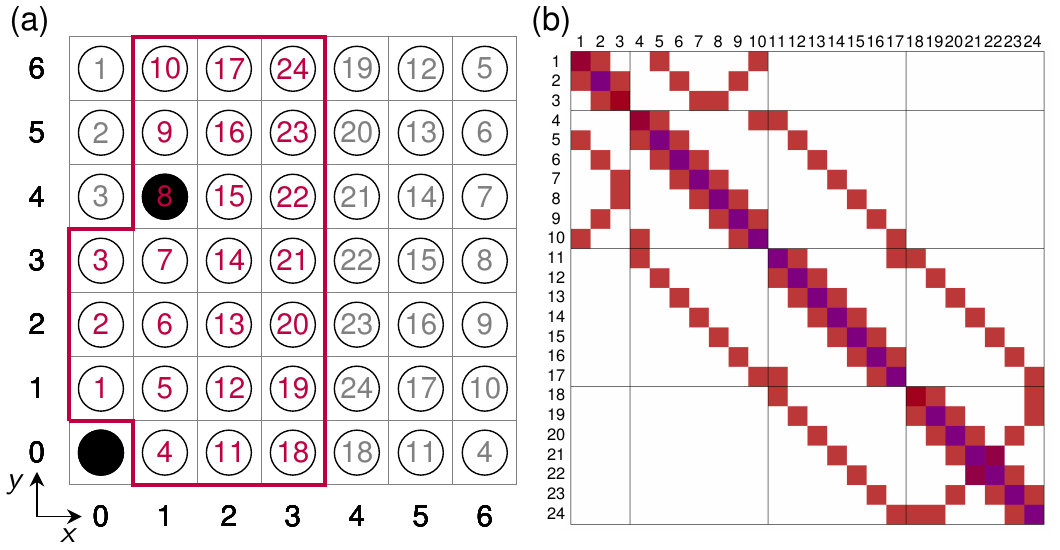}
    \caption{
		\change{Basis of the two-magnon Hilbert space for spin-$1/2$ square-lattice ferromagnets of dimension $7 \times 7$.    
    (a) Pictorial representation of the two-magnon basis state $| \vec{k}, \vec{r}_8 \rangle$, with $\vec{r}_8 = (x_8,y_8) = (1,4)$. By definition, one spin flip is situated at the origin $(0,0)$ and the other spin flip indicates the relative distance. States with spin flips within the area enclosed by the purple line provide the complete set of basis states for a fixed choice of the momentum quantum number $\vec{k}$. States with spin flips outside the purple area can be associated with states within the area because of periodic boundary conditions, as indicated by the gray numerical labels. This construction scheme follows Ref.~\onlinecite{Reklis1974}. (b) Graphical representation of the two-magnon Hamilton matrix $M_2(\vec{k})$, with $J_{3}$ and $J_{3}^{z}$ set to zero. Colored entries are nonzero, with different colors indicating different entries.}}
    \label{fig:s12-2D-Reklis}
\end{figure}

\change{
For the numerical treatment of the two-magnon states $|\vec{k},\vec{r}\rangle$ we follow the method of Ref.~\onlinecite{Reklis1974}, which we review in the following. For a square lattice with $N = n \times n$ lattice sites ($n$ odd) and periodic boundary conditions, there are $n$ possible values each for the momentum quantum numbers $k_x$ and $k_y$, and $R = (n^{2} -1)/2$ possible values for the relative distance $\vec{r}$ between spins. The dimension of the two-magnon Hilbert space is thus $n^2 (n^2-1)/2$. 
}

\change{
Let us take $n=7$ as an example and assume that $\vec{k}$ is fixed. There are $R = 24$ basis states associated with the variation of $\vec{r}$. We label these states by $\vec{r}_i = (x_i,y_i)$ with $1 \le i \le R$. The diagram in Fig.~\ref{fig:s12-2D-Reklis}(a) is a pictorial representation of the basis state $|\vec{k},\vec{r}_8 \rangle$ with $\vec{r}_8 = (1,4)$. It is to be read as follows: two spins are flipped with a relative distance vector $(1,4)$, where by definition one of the spin flips is situated at the orgin $(0, 0)$ and the second flipped spin resides inside the area enclosed by the purple line. For the ordering of these two-magnon states, we use the following scheme: starting from $x_{i} = 0$, we first vary the $y$ index of $\boldsymbol{r}_{i} = (x_{i}, y_{i})$ from smallest to largest possible values followed by varying the $x$ index similarly. Purple numerical labels in Fig.~\ref{fig:s12-2D-Reklis}(a) indicate the resulting ordering. 
}

\change{
Using the above notation, the elements of the two-magnon Hamilton matrix $M_2(\vec{k})$ read
\begin{align}
	[&M_2(\vec{k})]_{ij} 
	\nonumber \\
	&=
	\langle \vec{k},\vec{r}_i| H_\text{c} |\vec{k},\vec{r}_j\rangle
	\nonumber \\
	&= \delta_{x_i, x_j}  \delta_{y_i, y_j}
	\left[2 B + J_{1}^{z} \left( \delta_{x_i, 1} \delta_{y_i, 0} +  \delta_{x_i, 0} \delta_{y_i, 1} - 4 \right) 
	\right. \nonumber \\	
	&\qquad \left.	
	+ J_{3}^{z} \left( \delta_{x_i, 2} \delta_{y_i, 0} +  \delta_{x_i, 0} \delta_{y_i, 2} - 4 \right) \right] 
	\nonumber \\
	&\quad + J_{1}\left[ \left( \delta_{x_i, x_j+1} \delta_{y_i, y_j} +  \delta_{x_i, x_j-1} \delta_{y_i, y_j} \right) \cos \frac{k_{x}}{2} 
	\right. \nonumber \\	
	&\qquad \left.	
	+ \left( \delta_{x_i, x_j} \delta_{y_i, y_j+1} 
	+  \delta_{x_i, x_j} \delta_{y_i, y_j-1} \right) \cos \frac{k_{y}}{2}  \right] \nonumber \\
	&\quad +  J_{3} \left[ \left( \delta_{x_i, x_j+2} \delta_{y_i, y_j} +  \delta_{x_i, x_j-2} \delta_{y_i, y_j} + \delta_{x_i, 2-x_j} \delta_{y_i, - y_j} \right) \cos k_{x} \right. \nonumber \\
	& \qquad \left. + \left( \delta_{x_i, x_j} \delta_{y_i, y_j+2} +  \delta_{x_i, x_j} \delta_{y_i, y_j-2} + \delta_{x_i, -x_j} \delta_{y_i, 2 - y_j} \right) \cos k_{y} \right].
	\label{eq:M2details}
\end{align}
Instead of the states inside the purple area in Fig.~\ref{fig:s12-2D-Reklis}(a), we could have used those outside because periodic boundary conditions establish a one-to-one correspondence between the two sets, as made explicit by the gray numerical labels. Thus, for each index $i$, there are two equivalent coordinate pairs---for example, $i=8$ yields $(x_8, y_8) = (1,4) \equiv (6,3)$---which have to be accounted for when evaluating the Kronecker symbols in Eq.~\eqref{eq:M2details}. Explicitly, this means that $\delta_{x_i,x_j} = 1$ if $x_i = x_j$ holds for one of the two equivalent values for $x_i$ (and $x_j$). Moreover, periodic boundary conditions imply that $x_i = x_{i+n}$. The same set of rules applies for $y$. As an example, Figure \ref{fig:s12-2D-Reklis}(b) shows the two-magnon matrix $M_{2}(\boldsymbol{k})$ at a general $\vec{k}$ for $J_3 = J_3^z = 0$. Note in particular that row $8$ has finite entries in columns $3$, $7$, $9$, and $15$, which are associated with the spin flip at $(1,4)$ hopping to the west, south, north, and east, respectively. Hopping to the west means $(1,4) \to (0,4)$, which is outside the purple area in Fig.~\ref{fig:s12-2D-Reklis}(a). However, by virtue of the aforementioned one-to-one correspondence, the coordinate $(0,4)$ is identical to $(0,3)$, which carries the index $3$.
}

\change{
Finally, we note that the matrix elements coupling particle-number sectors of the full one and two-magnon Hamilton matrix $M(\vec{k})$ in Eq.~\eqref{eq:Msquare} are given by
\begin{align}
	D_i(\vec{k})
	&=
	\langle \vec{k}, \vec{r}_i | H_\text{DMI} |\vec{k} \rangle
	\nonumber 
	\\
	&=	
	 \delta_{{x}_i, 1}\delta_{y_i, 0}  \mathrm{i} D_{x} \sin \frac{k_{x}}{2} + \delta_{x_i, 0}\delta_{y_i, 1} D_{y} \sin \frac{k_{y}}{2}.
\end{align}
Only two elements of $D(\vec{k})$ are nonzero, which are those with $|\vec{r}_i| = \sqrt{x_i^2+y_i^2} = 1$ because the DMI in Eq.~\eqref{eq:DMIsquare} acts between nearest neighbors. According to Fig.~\ref{fig:s12-2D-Reklis}(a), this condition is met only for $\vec{r}_1 = (0,1)$ and $\vec{r}_4 = (1,0)$.
}

\section{Derivation Of Effective Hamiltonians}
\label{sec:Appendix}
In the limit of strong spin anisotropy (either Ising or single-ion type), BS separate from the continuum throughout the entire Brillouin zone. For the effective description of BS physics, the derivation of effective Hamiltonians has been proven useful \cite{DiLiberto2016, Gorlach2017, Salerno2018, Qin2017, Qin2018, Stepanenko2020, Salerno2020}. Here, we follow a similar approach and additionally take into account the single-particle state.

We write the full spin Hamiltonian as
\begin{align}
	H = H_0 + V,
	\label{eq:decompositionperturbation}
\end{align}
where $H_0$ is that part of the spin Hamiltonian responsible for the magnon binding, that is to say, the Ising or single-ion anisotropy. Other spin-spin interactions enter the perturbation $V$, assuming their energy scales to be much smaller than that of the anisotropy.
The matrix elements of an effective low-energy Hamiltonian $H^\text{eff}$ are then given by \cite{CohenTannoudji1998} 
\begin{align}
	H^\text{eff}_{\alpha\beta}
	&=	
	\langle \alpha | H_0 | \beta \rangle
	+
	\langle \alpha | V | \beta \rangle
	\nonumber \\
	&+
	\frac{1}{2} \sum_{v} \langle \alpha | V | v \rangle \langle v | V | \beta \rangle \left( \frac{1}{E_\alpha - E_v} + \frac{1}{E_\beta - E_v}  \right).
	\label{eq:perturbationtheory1}
\end{align} 
\change{Here, $| \alpha \rangle$ and $| \beta \rangle$ denote eigenstates of $H_0$ with energy $E_\alpha = \langle \alpha | H_0 |\alpha \rangle$ and $E_\beta = \langle \beta | H_0 |\beta \rangle$, respectively, spanning the sub-Hilbert space of interest, that is, single-magnon states and BS. Virtual states $|v\rangle$ with energy $E_v = \langle v | H_0 |v \rangle$ are drawn from the two-magnon continuum. It must hold $E_\alpha, E_\beta \ne E_v$.}
Below, we provide the details for all models considered in this work.

\subsection{Spin-$1/2$ magnets on the square lattice}
\label{sec:AppSpin12}
Here, we show how to arrive at Eq.~\eqref{eq:effective_ham_3band}. First, we separate the spin Hamiltonian according to Eq.~\eqref{eq:decompositionperturbation}:
\begin{subequations}
\begin{align}
    H_0 &= \sum_{i,j}^{N_x,N_y} \left( 
        J_1^z S_{i,j}^z S_{i+1,j}^z
        + J_1^z S_{i,j}^z S_{i,j+1}^z
        - B S_{i,j}^z
    \right),
    \\
    V &= \sum_{i,j}^{N_x,N_y} \left[
        J_1 \left( S_{i,j}^x S_{i+1,j}^x
        + S_{i,j}^x S_{i,j+1}^x
        + S_{i,j}^y S_{i+1,j}^y
        + S_{i,j}^y S_{i,j+1}^y \right)
        \right.
        \nonumber \\
        &\quad+ J_3 \vec{S}_{i,j} \cdot \vec{S}_{i+2,j} + J_3 \vec{S}_{i,j} \cdot \vec{S}_{i,j+2}
        \nonumber \\
        &\quad\left. + D_x \hat{\vec{y}} \cdot \left( \vec{S}_{i,j} \times \vec{S}_{i+1,j} \right)
		+ D_y \hat{\vec{x}} \cdot \left( \vec{S}_{i,j} \times \vec{S}_{i,j+1}  \right)
    \right].
\end{align}
\end{subequations}
The low-energy Hilbert space is spanned by single-spin-flip states $|l,m\rangle = S^-_{l,m} |0\rangle$ and two-spin-flip states, with spin flips being nearest neighbors, i.e., either $|l,m;l+1,m\rangle = S^-_{l,m} S^-_{l+1,m} |0\rangle$ or $|l,m;l,m+1\rangle = S^-_{l,m} S^-_{l,m+1} |0\rangle$. Virtual (or intermediate) states, labeled by $v$, contain two spin flips at sites that are not nearest neighbors, that is, $| \lambda, \mu; \lambda', \mu' \rangle = S^-_{\lambda,\mu} S^-_{\lambda',\mu'} |0\rangle$ with $\lambda$, $\lambda'$, $\mu$, and $\mu'$ chosen appropriately. These states make up the continuum.

In a second-quantized formulation, we introduce a vector $\vec{c}^\dagger_{l,m} = ( s^\dagger_{l,m}, x^\dagger_{l,m}, y^\dagger_{l,m} )$ composed from single-flip, $s^\dagger_{l,m}$, and double spin-flip particle creators, $x^\dagger_{l,m}$ and $y^\dagger_{l,m}$. They are respectively defined as $|l,m\rangle = s^\dagger_{l,m} |0\rangle$, $|l,m;l+1,m\rangle = x^\dagger_{l,m} |0\rangle$, and $|l,m;l,m+1\rangle = y^\dagger_{l,m} |0\rangle$. These particles live on the Lieb lattice, with $s$-type particles being defined on the sites and the $x$-type and $y$-type particles on the bonds of the original square lattice. After evaluating Eq.~\eqref{eq:perturbationtheory1} for all possible hopping processes, the effective hopping model is found to read
\begin{align}
	H^\text{eff}
	&=
	\sum_{l,m} \left[ 
		\vec{c}^\dagger_{l,m} \vec{\mu} \vec{c}_{l,m}
		+ 
		\left(
		\vec{c}^\dagger_{l+1,m} \vec{t}_x \vec{c}_{l,m}	
		+
		\text{H.\,c.}
		\right)
		\right.
		\nonumber \\
	&+		 
		\left(
		\vec{c}^\dagger_{l,m+1} \vec{t}_y \vec{c}_{l,m}	
		+
		\text{H.\,c.}
		\right)
	    +
	    \left(
		\vec{c}^\dagger_{l+2,m} \vec{t}_{2x} \vec{c}_{l,m}	
		+
		\text{H.\,c.}
		\right)
		\nonumber \\
	&+ 
		\left.
		\left(
		\vec{c}^\dagger_{l,m+2} \vec{t}_{2y} \vec{c}_{l,m}	
		+
		\text{H.\,c.}
		\right)
		+ 
		\left(
		\vec{c}^\dagger_{l+1,m-1} \vec{t}_{x\bar{y}} \vec{c}_{l,m}	
		+
		\text{H.\,c.}
		\right)
	\right],
\end{align}
with matrices
\begin{subequations}
\begin{align}
    \vec{\mu} 
    &=
    \begin{pmatrix}
        B-2\left(J_1^z+J_3^z \right) & - \frac{D_x}{2} & - \mathrm{i} \frac{D_y}{2} \\
        -\frac{D_x}{2} & \mu & \frac{J_1^2}{2J_1^z} \\
        \mathrm{i} \frac{D_y}{2} & \frac{J_1^2}{2J_1^z} & \mu
    \end{pmatrix},
    \\
    \vec{t}_x 
    &=
    \begin{pmatrix}
        \frac{J_1}{2} & \frac{D_x}{2} & 0 \\
        0 & \frac{J_3}{2} + \frac{J_1^2}{4J_1^z} & 0\\
        0 & \frac{J_1^2}{2J_1^z} & \frac{J_1^2}{2J_1^z}
    \end{pmatrix},
    \,
    \vec{t}_y 
    =
    \begin{pmatrix}
        \frac{J_1}{2} & 0 &  \mathrm{i} \frac{D_y}{2} \\
        0 & \frac{J_1^2}{2J_1^z} & \frac{J_1^2}{2J_1^z} \\
        0 & 0 & \frac{J_3}{2} + \frac{J_1^2}{4 J_1^z}
    \end{pmatrix},
    \\
    \vec{t}_{2x} 
    &=
    \begin{pmatrix}
        \frac{J_3}{2} & 0 & 0 \\
        0 & \frac{J_3^2}{4J_1^z} & 0 \\
        0 & 0 & \frac{J_3^2}{2J_1^z}
    \end{pmatrix},
    \,
    \vec{t}_{2y} 
    =
    \begin{pmatrix}
        \frac{J_3}{2} & 0 & 0 \\
        0 & \frac{J_3^2}{2J_1^z} & 0 \\
        0 & 0 & \frac{J_3^2}{4J_1^z}
    \end{pmatrix},
    \,
    \vec{t}_{x\bar{y}} 
    =
    \begin{pmatrix}
        0 & 0 &  0 \\
        0 & 0 & 0 \\
        0 & \frac{J_1^2}{2J_1^z} & 0
    \end{pmatrix},
\end{align}
\end{subequations}
where
\begin{align}
    \mu = 2B-3J_1^z -4 J_3^z + \frac{3 J_1^2}{2 J_1^z} + \frac{3 J_3^2}{2J_1^z}.
\end{align}
We perform a Fourier transformation,
\begin{subequations}
\begin{align}
	\vec{c}_{l,m} &= \frac{1}{\sqrt{N}} \sum_{\vec{k}} \mathrm{e}^{\mathrm{i}\vec{k}\cdot\vec{r}_{l,m}} \vec{c}_{\vec{k}}, \\
	\vec{c}^\dagger_{l,m} &= \frac{1}{\sqrt{N}} \sum_{\vec{k}} \mathrm{e}^{-\mathrm{i}\vec{k}\cdot\vec{r}_{l,m}} \vec{c}^\dagger_{\vec{k}},
\end{align}
\end{subequations}
which brings the effective model to the form
\begin{subequations}
\begin{align}
	H^\text{eff}
	&=
	\sum_{\vec{k}} 
		\vec{c}^\dagger_{\vec{k}}
		H'(\vec{k})
		\vec{c}_{\vec{k}},
\end{align}
\end{subequations}
with the Hamilton kernel
\begin{align}
    H'(\vec{k}) = \vec{\mu} 
			&+ \mathrm{e}^{-\mathrm{i} k_x}  \vec{t}_x 
			+ \mathrm{e}^{\mathrm{i} k_x} \vec{t}^\dagger_x
			+ \mathrm{e}^{-\mathrm{i} k_y}  \vec{t}_y 
			+ \mathrm{e}^{\mathrm{i} k_y} \vec{t}^\dagger_y
	\nonumber \\			
			&+ \mathrm{e}^{-2\mathrm{i} k_x}  \vec{t}_{2x} 
			+ \mathrm{e}^{2\mathrm{i} k_x} \vec{t}^\dagger_{2x}
			+ \mathrm{e}^{-2\mathrm{i} k_y}  \vec{t}_{2y} 
			+ \mathrm{e}^{2\mathrm{i} k_y} \vec{t}^\dagger_{2y}
	\nonumber \\
			&+ \mathrm{e}^{-\mathrm{i} (k_x - k_y)} \vec{t}_{x\bar{y}} 
			+ \mathrm{e}^{\mathrm{i} (k_x - k_y)} \vec{t}^\dagger_{x\bar{y}}.
\end{align}
For the sake of a wieldy mathematical expression, we perform a unitary transformation 
$
	H(\vec{k}) = U^\dagger H'(\vec{k})  U
$
and
$
	\vec{C}_{\vec{k}} = U^\dagger \vec{c}_{\vec{k}}
$
,
where 
$
	U=\mathrm{diag}(1, \mathrm{e}^{\mathrm{i} k_x/2}, \mathrm{e}^{\mathrm{i} k_y/2}).
$
Our final result for the Hamilton kernel is given in Eq.~\eqref{eq:effective_ham_3band}.

\subsection{Spin-$1$ magnets on the square lattice}
\label{sec:AppSpin1}
Here, we present the derivation of Eqs.~\eqref{eq:hamkernelSpin1} and Eqs.~\eqref{eq:E1Spin1J2}-\eqref{eq:DeltaSpin1J2}, including nearest and second-nearest-neighbor exchange interaction.
In the limit $A \gg |J_1|, |J_2|$, we may concentrate on the low-energy sector spanned by single spin-deviation states $|l,m\rangle \equiv \frac{1}{\sqrt{2}} S_{l,m}^- |0\rangle$ and the SIBS, i.e., states $|l,m; l,m \rangle \equiv \frac{1}{2} S_{l,m}^- S_{l,m}^- |0\rangle$ with two spin deviations localized at the same site (full flip of the $S=1$ spins).
We decompose the Hamiltonian to bring it into the form of Eq.~\eqref{eq:decompositionperturbation}:
\begin{subequations}
\begin{align}
	H_0 &= - \sum_{i,j}^{N_x,N_y} \left(  A S_{i,j}^z S_{i,j}^z + B S_{i,j}^z \right), \\
	V &= \sum_{i,j}^{N_x,N_y} \left[ 
		  J_1 \vec{S}_{i,j} \cdot \left( \vec{S}_{i+1,j} + \vec{S}_{i,j+1} \right)
		  \right. \nonumber \\		  
		  & \quad+
		  J_2 \vec{S}_{i,j} \cdot \left( \vec{S}_{i+1,j+1} + \vec{S}_{i-1,j+1} \right)
		\nonumber \\
		&\left. \quad+ D_x \hat{\vec{y}} \cdot \left( \vec{S}_{i,j} \times \vec{S}_{i+1,j}  \right)
		+ D_y \hat{\vec{x}} \cdot \left( \vec{S}_{i,j} \times \vec{S}_{i,j+1}  \right)
		\right].
\end{align}
\end{subequations}
In a second-quantized formulation, we introduce a vector $\vec{c}^\dagger_{l,m} = ( s^\dagger_{l,m}, d^\dagger_{l,m} )$ composed from single-deviation, $s^\dagger_{l,m}$, and double-deviation particle creators, $d^\dagger_{l,m}$. They are respectively defined as $|l,m\rangle = s^\dagger_{l,m} |0\rangle$ and $|l,m;l,m\rangle = d^\dagger_{l,m} |0\rangle$ and live on the sites of the original square lattice. Then, using Eq.~\eqref{eq:perturbationtheory1}, the effective hopping model reads
\begin{align}
	H^\text{eff}
	&=
	\sum_{l,m} \left[ 
		\vec{c}^\dagger_{l,m} \vec{\mu} \vec{c}_{l,m}
		+ 
		\left(
		\vec{c}^\dagger_{l+1,m} \vec{t}_x \vec{c}_{l,m}	
		+
		\text{H.\,c.}
		\right)
		\right.\nonumber \\		
	&\quad + 
		\left(
		\vec{c}^\dagger_{l,m+1} \vec{t}_y \vec{c}_{l,m}	
		+
		\text{H.\,c.}
		\right)
		+
		\left(
		\vec{c}^\dagger_{l+1,m+1} \vec{t}_{xy} \vec{c}_{l,m}	
		+
		\text{H.\,c.}
		\right)
		\nonumber \\
		&\left. \quad + 
		\left(
		\vec{c}^\dagger_{l-1,m+1} \vec{t}_{\overline{x}y} \vec{c}_{l,m}	
		+
		\text{H.\,c.}
		\right)
	\right],
\end{align}
with the on-site energy and hoppings given by
\begin{subequations}
\begin{align}
	\vec{\mu} &= \begin{pmatrix}
		\mu_1 & 0 \\
		0 & \mu_2
	\end{pmatrix},
	\\
	\vec{t}_x &= \begin{pmatrix}
		J_1 + \frac{D_x^2}{2(A+B)} & 
		-\frac{J_1 D_x}{2\sqrt{2}}\left( \frac{1}{2A} + \frac{1}{A+B} \right) \\
		\frac{J_1 D_x}{2\sqrt{2}}\left( \frac{1}{2A} + \frac{1}{A+B} \right) & -\frac{J_1^2}{2A}
	\end{pmatrix},
	\\
	\vec{t}_y &= \begin{pmatrix}
		J_1 + \frac{D_y^2}{2(A+B)} & -\mathrm{i}\frac{J_1 D_y}{ 2\sqrt{2}}\left( \frac{1}{2A} + \frac{1}{A+B} \right) \\
		-\mathrm{i}\frac{J_1 D_y}{2\sqrt{2}}\left( \frac{1}{2A} + \frac{1}{A+B} \right) & -\frac{J_1^2}{2A}
	\end{pmatrix},
	\\
	\vec{t}_{xy} &= \begin{pmatrix}
		J_2 & 0 \\
		0 & -\frac{J_2^2}{2A}
	\end{pmatrix},
	\quad
	\vec{t}_{\overline{x}y} = \begin{pmatrix}
		J_2 & 0 \\
		0 & -\frac{J_2^2}{2A}
	\end{pmatrix},
\end{align}
\end{subequations}
where
\begin{subequations}
\begin{align}
	\mu_1 &= A+B-4(J_1 + J_2) - \frac{ D_x^2+D_y^2}{A+B}, \\
	\mu_2 &= 2B-8(J_1+J_2) - \frac{2}{A} \left( J_1^2+J_2^2 \right).
\end{align}
\end{subequations}
After a Fourier transformation, we obtain Eq.~\eqref{eq:hamkernelSpin1} (for $J_2=0$) and Eqs.~\eqref{eq:E1Spin1J2}-\eqref{eq:DeltaSpin1J2} (for $J_2 \ne 0$).

\subsection{Spin-$1/2$ magnets on the triangular lattice}
\label{sec:AppSpin12Triangular}
The derivation of Eq.~\eqref{eq:four-band-model} is analogous to that of the effective spin-$1/2$ model on the square lattice in Appendix \ref{sec:AppSpin12}, the only difference being that we truncate the perturbation theory in Eq.~\eqref{eq:perturbationtheory1} already at first order. This is sufficient to capture the dispersion of BS on triangular lattices. 
The decomposition in Eq.~\eqref{eq:decompositionperturbation} is achieved by defining
\begin{subequations}
\begin{align}
    H_0 &= \sum_{\vec{r}_l} \left( - B S_{\vec{r}_l}^z  		 
    + J_1^z \sum_{\alpha = 1}^3 
     S^z_{\vec{r}_l} S^z_{\vec{r}_l+\vec{\delta}_\alpha}  
    \right)
    ,
    \\
    V &= \sum_{\vec{r}_l} \sum_{\alpha = 1}^3 
    \left\{
    \frac{J_1}{2} \left( 
        S^+_{\vec{r}_l} S^-_{\vec{r}_l+\vec{\delta}_\alpha} 
        +
        S^-_{\vec{r}_l} S^+_{\vec{r}_l+\vec{\delta}_\alpha} 
    \right)
    +
    J_3
        \vec{S}_{\vec{r}_l} \cdot \vec{S}_{\vec{r}_l+2\vec{\delta}_\alpha}
    \right.
    \nonumber \\
    &
 	\qquad - \frac{\mathrm{i} J^{z \pm}}{2} 
    \left[
    \gamma^\ast_{\alpha}
    \left( S_{\vec{r}_l}^+ S_{\vec{r}_l+\vec{\delta}_\alpha}^z + S_{\vec{r}_l}^z S_{\vec{r}_l+\vec{\delta}_\alpha}^+ \right)
    \right. \nonumber \\
    &\left.\left. \qquad \qquad 
    -     
    \gamma_{\alpha}
    \left( S_{\vec{r}_l}^- S_{\vec{r}_l+\vec{\delta}_\alpha}^z + S_{\vec{r}_l}^z S_{\vec{r}_l+\vec{\delta}_\alpha}^- \right)
    \right] \right\}.
\end{align}
\end{subequations}
For notational ease, we introduce $\vec{c}^\dagger_{\vec{r}_l} = ( s^\dagger_{\vec{r}_l}, \delta^\dagger_{1,\vec{r}_l}, \delta^\dagger_{2,\vec{r}_l}, \delta^\dagger_{3,\vec{r}_l} )$ composed from single-flip, $s^\dagger_{\vec{r}_l} |0\rangle = | \vec{r}_l \rangle$, and double spin-flip particle creators, $\delta^\dagger_{\alpha,\vec{r}_l} |0\rangle = |\vec{r}_l, \vec{r}_l + \vec{\delta}_\alpha \rangle$ ($\alpha =1,2,3$). With their center of mass being located at the midpoints of bonds, the $\delta$ particles live on a kagome lattice, whose hexagons host the vertices of the original triangular lattice where the $s$ particles reside.
The resulting hopping model reads
\begin{align}
	H^\text{eff} 
	&= 
	\sum_{l=1}^N 
	\left[ \vec{c}_{\vec{r}_l}^{\dagger} \vec{\mu} \vec{c}_{\vec{r}_l} 
	+ \sum_{\alpha = 1}^{3} \left( \vec{c}_{\vec{r}_l + \vec{\delta}_\alpha}^{\dagger} \vec{t}_{\vec{\delta}_\alpha} \vec{c}_{\vec{r}_l} +  \text{H.c.}\right)
	\right. \nonumber \\ 
	&\left. \quad + \sum_{\alpha = 1}^{3}  \left( \vec{c}_{\vec{r}_l + 2 \vec{\delta}_\alpha}^{\dagger} \vec{t}_{2\vec{\delta}_\alpha} \vec{c}_{\vec{r}_l} + \text{H.c.} \right) \right],
\end{align}
with the matrices
\begin{align}
	\vec{\mu} = 
	\begin{pmatrix}
		B - 3 \left( J_{1}^{z} + J_{3}^{z} \right) & \mathrm{i} \frac{J^{z\pm}}{2} \gamma^{*}_{1} & \mathrm{i} \frac{J^{z\pm}}{2} \gamma^{*}_{2} & \mathrm{i} \frac{J^{z\pm}}{2} \gamma^{*}_{3} \\
		-\mathrm{i} \frac{J^{z\pm}}{1} \gamma_{1} & \mu & 0 & 0 \\
		-\mathrm{i} \frac{J^{z\pm}}{2} \gamma_{2} & 0 & \mu & 0 \\
		-\mathrm{i} \frac{J^{z\pm}}{3} \gamma_{3} & 0 & 0 & \mu
	\end{pmatrix},
\end{align}
where
$
	\mu = 2B - 5 J_{1}^{z} - 6 J_{3}^{z},
$
and
\begin{subequations}
\begin{align}
	\vec{t}_{\vec{\delta}_1} &= 
	\begin{pmatrix}
		\frac{J_{1}}{2} & \mathrm{i} \frac{J^{z\pm}}{2} \gamma^{*}_{1} & 0 & 0 \\
		0 & \frac{J_{3}}{2} & 0 & 0 \\
		0 & \frac{J_{1}}{2} & 0 & 0 \\
		0 & \frac{J_{1}}{2} & 0 & 0 
		\end{pmatrix}, \;
	\vec{t}_{\vec{\delta}_2} = 
	\begin{pmatrix}
		\frac{J_{1}}{2} & 0 & \mathrm{i} \frac{J^{z\pm}}{2} \gamma^{*}_{2} & 0 \\
		0 & 0 & \frac{J_{1}}{2} & 0 \\
		0 & 0 & \frac{J_{3}}{2} & 0 \\
		0 & 0 & \frac{J_{1}}{2} & 0 
		\end{pmatrix},
	\\
	\vec{t}_{\vec{\delta}_3} &= 
	\begin{pmatrix}
		\frac{J_{1}}{2} & 0 & 0 & \mathrm{i} \frac{J^{z\pm}}{2} \gamma^{*}_{3} \\
		0 & 0 & 0 & \frac{J_{1}}{2} \\
		0 & 0 & 0 & \frac{J_{1}}{2} \\
		0 & 0 & 0 & \frac{J_{3}}{2}  
		\end{pmatrix},
	\;
	\vec{t}_{2\vec{\delta}_1} = \vec{t}_{2\vec{\delta}_2} = \vec{t}_{2\vec{\delta}_3} =
	\begin{pmatrix}
		\frac{J_{3}}{2} & 0 & 0 & 0 \\
		0 & 0 & 0 & 0 \\
		0 & 0 & 0 & 0 \\
		0 & 0 & 0 & 0  
		\end{pmatrix}.
\end{align}
\end{subequations}
A Fourier transformation leads to 
$
	H^\text{eff}
	=
	\sum_{\vec{k}} 
		\vec{c}^\dagger_{\vec{k}}  
		H'(\vec{k})
		 \vec{c}_{\vec{k}},
$
with
$
	\vec{c}^\dagger_{\vec{k}} = (s^\dagger_{\vec{k}}, \delta^\dagger_{1,\vec{k}}, \delta^\dagger_{2,\vec{k}}, \delta^\dagger_{3,\vec{k}})
$
and the Hamilton kernel
\begin{align}
    &H'(\vec{k}) \nonumber \\
    &= 
    \vec{\mu} + \sum_{\alpha = 1}^3 \left( \mathrm{e}^{- \mathrm{i} \vec{k} \cdot \vec{\delta}_{\alpha}} \vec{t}_{\vec{\delta}_\alpha} + \mathrm{e}^{\mathrm{i} \vec{k} \cdot \vec{\delta}_{\alpha}} \vec{t}_{\vec{\delta}_\alpha}^{\dagger} + \mathrm{e}^{- \mathrm{i} \vec{k} \cdot 2\vec{\delta}_{\alpha}} \vec{t}_{2\vec{\delta}_\alpha} + \mathrm{e}^{\mathrm{i} \vec{k} \cdot 2\vec{\delta}_{\alpha}} \vec{t}_{2\vec{\delta}_\alpha}^{\dagger} \right).
\end{align}
A unitary transformation 
$
	H(\vec{k}) = U^\dagger H'(\vec{k}) U
$,
with
$
	U = \mathrm{diag}(1, \mathrm{e}^{\mathrm{i} \vec{k} \cdot \vec{\delta}_{1} /2}, \mathrm{e}^{\mathrm{i} \vec{k} \cdot \vec{\delta}_{2} /2}, \mathrm{e}^{\mathrm{i} \vec{k} \cdot \vec{\delta}_{3} /2})
$
and
$\vec{C}_{\vec{k}} = U^\dagger \vec{c}_{\vec{k}}$,
results in Eq.~\eqref{eq:four-band-model}.

\subsection{Spin-$1$ magnets on the triangular lattice}
\label{sec:AppSpin1Triangular}
For the derivation of Eq.~\eqref{eq:effectivehamiltonianspin1triangular}, we focus on the limit of $A \gg |J_1|, |J_2|$ and the two-dimensional subspace spanned by single-magnon excitations and SIBS. The spin Hamiltonian decomposition according to Eq.~\eqref{eq:decompositionperturbation} reads
\begin{align}
    H 
    &= 
    -\sum_{\vec{r}_l} 
    \left[ A \left( S_{\vec{r}_l}^z \right)^2 + B S_{\vec{r}_l}^z \right],
    \\
    V 
    &= 
    \sum_{\vec{r}_l}  \sum_{\alpha = 1}^3  
    \left\{ 
    J_1 \vec{S}_{\vec{r}_l} \cdot \vec{S}_{\vec{r}_l+\vec{\delta}_\alpha} 
    +
    J_2 \vec{S}_{\vec{r}_l} \cdot \vec{S}_{\vec{r}_l+\vec{\delta}_\alpha-\vec{\delta}_{\alpha-1}} 
    \right. \nonumber \\
    &\qquad - \mathrm{i} \frac{J^{z \pm}}{2} 
    \left[
    \gamma^\ast_{\alpha}
    \left( S_{\vec{r}_l}^+ S_{\vec{r}_l+\vec{\delta}_\alpha}^z + S_{\vec{r}_l}^z S_{\vec{r}_l+\vec{\delta}_\alpha}^+ \right)
    \right.
    \nonumber \\
    &\qquad \quad \left. \left. -\gamma_{\alpha}
    \left( S_{\vec{r}_l}^- S_{\vec{r}_l+\vec{\delta}_\alpha}^z + S_{\vec{r}_l}^z S_{\vec{r}_l+\vec{\delta}_\alpha}^- \right)
    \right]
    \right\}.
\end{align}
After introducing $s$ and $d$ particles similar to the spin-$1$ model on the square lattice (see Appendix~\ref{sec:AppSpin1}, and comprising them in the vector $\vec{c}^\dagger_{\vec{r}_l} = (s^\dagger_{\vec{r}_l}, d^\dagger_{\vec{r}_l})$, we obtain the following effective hopping model:
\begin{align}
	H^\text{eff} 
	&= 
	\sum_{\vec{r}_{l}} \left[ \vec{c}_{\vec{r}_{l}}^{\dagger}  \vec{\mu} \vec{c}_{\vec{r}_{l}} + \sum_{\alpha = 1}^{3} \left( \vec{c}_{\vec{r}_{l} + \vec{\delta}_{\alpha}}^{\dagger} \vec{t}_{\alpha}  \vec{c}_{\vec{r}_{l}} + \text{H.c.}\right) 
	\right.\nonumber \\	
	&\left. \qquad + \sum_{\alpha = 1}^{3} \left( \vec{c}_{\vec{r}_{l} + \vec{\delta}_{\alpha} - \vec{\delta}_{\alpha-1}}^{\dagger}  \vec{t}_{\alpha \overline{\alpha-1}} \vec{c}_{\vec{r}_{l}} + \text{H.c.}\right) \right],
\end{align}
with matrices
\begin{subequations}
\begin{align}
	\vec{\mu} &= 
	\begin{pmatrix}
		\mu_1 & 0 \\
		0 & \mu_2 
	\end{pmatrix},
	\qquad
	\vec{t}_{\alpha\overline{\alpha-1}} 
	=	 
	\begin{pmatrix}
		J_{2} & 0\\
		0 & -\frac{J_{2}^{2}}{2A}
	\end{pmatrix},
	\\
	\vec{t}_{\alpha} &= 
	\begin{pmatrix}
		J_{1} - \frac{\left( J^{z\pm} \right)^{2}}{ 2 \left(A+B\right)} 
		& -\frac{ \mathrm{i}\gamma^{*}_{\alpha} J^{z\pm} J_{1}}{2 \sqrt{2}} \left( \frac{1}{A+B} + \frac{1}{2A} \right)  \\
		\frac{ \mathrm{i}\gamma_{\alpha} J^{z\pm} J_{1}}{2 \sqrt{2}} \left( \frac{1}{A+B} + \frac{1}{2A} \right) & -\frac{J_{1}^{2}}{2A}
	\end{pmatrix}, 
\end{align} 
\end{subequations}
where
\begin{subequations}
\begin{align}
	\mu_1 &= A+B - 6 \left( J_{1} + J_{2} \right) - 3 \frac{\left( J^{z\pm} \right)^{2}}{A+B}, 
	\\
	\mu_2 &= 2B - 12 \left( J_{1} + J_{2} \right) - 3 \frac{J_{1}^{2} + J_{2}^{2}}{A}.
\end{align}
\end{subequations}
After a Fourier transformation, we arrive at Eq.~\eqref{eq:effectivehamiltonianspin1triangular}.

\section{Coupling of particle-number sectors in spin-1/2 Ising chains with transverse DMI}
\label{App:Ising}
To explore particle-number sector coupling beyond the interactions between one and two-magnon states, we consider the spin-$1/2$ Ising chain with transverse DMI, which was already studied in Refs.~\onlinecite{Derzhko2006, Soltani2019}. Its spin Hamiltonian reads
\begin{align}
	H = \sum_{i=1}^N \left( -J S^z_i S^z_{i+1} 
	+ D \vec{x} \cdot \vec{S}_i \times \vec{S}_{i+1} \right).
\end{align}
For notational ease, we flipped the sign convention with respect to the main text; $J>0$ is assumed to stabilize the fully polarized ground state $|0\rangle$. Periodic boundary conditions are assumed, such that the DMI does not compromise $|0\rangle$ as long as $D<|J/2|$. A Jordan-Wigner transformation \cite{Jordan1928} maps $H$ onto a free-fermion problem $H = \sum_k \varepsilon_k (\gamma^\dagger_k \gamma_k-1/2)$, where $\gamma^{(\dagger)}_k$ annihilates (creates) a fermion with dispersion $\varepsilon_k = J/2 - D \sin k $ \cite{Derzhko2006, Soltani2019}. Here, however, we stay in the spin language to discuss particle-number sector coupling and to derive spin structure factors.

We introduce $m$-magnon domain states (or BS),
\begin{align}
    | k; m\rangle 
    =
    \frac{1}{\sqrt{N}} \sum_{l=1}^N \mathrm{e}^{\mathrm{i} k l } \mathrm{e}^{\mathrm{i} k (m-1)/2} | l, l+ 1, l+2, \ldots, l+ m-1 \rangle,
\end{align}
in which $m$ spin flips line up to form a single domain:
\begin{align}
	| l, l+ 1, l+2, \ldots, l+m-1 \rangle
    =
    S_l^-
    S_{l+1}^-
    S_{l+2}^-
    \ldots
    S_{l+m-1}^-
    |0\rangle.
\end{align}
One verifies that the $| k; m\rangle$ states are eigenstates of $H$ with eigenenergy $J$. The DMI enables the domains to grow and shrink by coupling sectors with incremental $m$; the corresponding matrix elements are
\begin{align}
   D(k)  
   \equiv
   \langle k; m | H_\text{DMI} | k; m+1 \rangle
   =
   D \sin\frac{k}{2}.
\end{align}
Thus, in the sub-Hilbert space of $m$-magnon domain states $|k;m\rangle$, the (infinite) Hamilton matrix reads
\begin{align} 
    \begin{pmatrix}
      J & D(k) \\
      D(k) & J & D(k)\\
       & D(k) & J & D(k)\\
       & & D(k) & J & D(k)\\
       & & & \ddots &\ddots &\ddots &
    \end{pmatrix}.
    \label{eq:HamApprox}
\end{align}
Since matrix \eqref{eq:HamApprox} is symmetric, tridiagonal, and Toeplitz, its eigenvalues $\epsilon_m(k)$ and eigenvectors $w_m(k)$ are known in closed form, if we restrict it to finite dimension $M$ \cite{Kulkarni1999, Noschese2012}. They read
\begin{subequations}
\begin{align}
    \epsilon_m(k) &= J + 2 D(k) \cos \left( \frac{m \pi}{M+1} \right),
    \label{eq:eigenenergies-Ising}
    \\
    w_m(k) &= \frac{1}{\zeta} \left[ \sin \left( \frac{m \pi}{M+1}\right), \sin\left(  \frac{2m \pi}{M+1}\right), \ldots, \sin\left(  \frac{M m \pi}{M+1}\right) \right]^\text{T}, 
    \label{eq:eigenvectors-Ising}
\end{align}
\end{subequations}
($m=1,\ldots, M$), where $\zeta = \sqrt{ (M+1)/2 }$ is a normalization factor. In the thermodynamic limit ($M \to \infty$), the dynamical spin-multipolar structure factors read
\begin{align}
    \mathcal{S}^f(\omega,k) 
    &=
    \lim_{M \to \infty} \sum_{m=1}^M | \langle w_m(k) | k; f \rangle |^2
    \delta\left( \omega-\epsilon_m(k) \right).
    \label{eq:multipolarSF}
\end{align}
They probe the dynamics of $f$ neighboring $S^-$ operators; $\mathcal{S}^1(\omega,k) = \mathcal{S}^{+-}(\omega,k)$ is the usual spin structure factor and $\mathcal{S}^2(\omega,k) = \mathcal{Q}^{+-}_1(\omega,k)$ the quadrupolar structure factor. By plugging Eqs.~\eqref{eq:eigenenergies-Ising} and \eqref{eq:eigenvectors-Ising} into Eq.~\eqref{eq:multipolarSF}, we arrive at
\begin{align}
    \mathcal{S}^f(\omega,k) 
    &=
    \frac{1}{\pi | D(k) |} \frac{1-T_f^2(\xi)}{\sqrt{1-T_1^2(\xi)}},
    \quad
    \xi \equiv \frac{\omega-J}{ 2 D(k) },
\end{align}
where
\begin{align}
    T_f(\xi) = \cos \left( f \arccos(\xi) \right), \quad |\xi| \le 1
\end{align}
is the $f$-th Chebyshev polynomial of first kind. In particular, for $f=1$, we obtain
\begin{align}
    \mathcal{S}^{1}(\omega,k) = \mathcal{S}^{+-}(\omega,k) 
    &=
    \frac{1}{\pi |D(k)|} 
    \sqrt{1-\left( \frac{\omega-J}{ 2 D(k) } \right)^2}.
\end{align}

\begin{figure}
    \centering
    \includegraphics[width=\columnwidth]{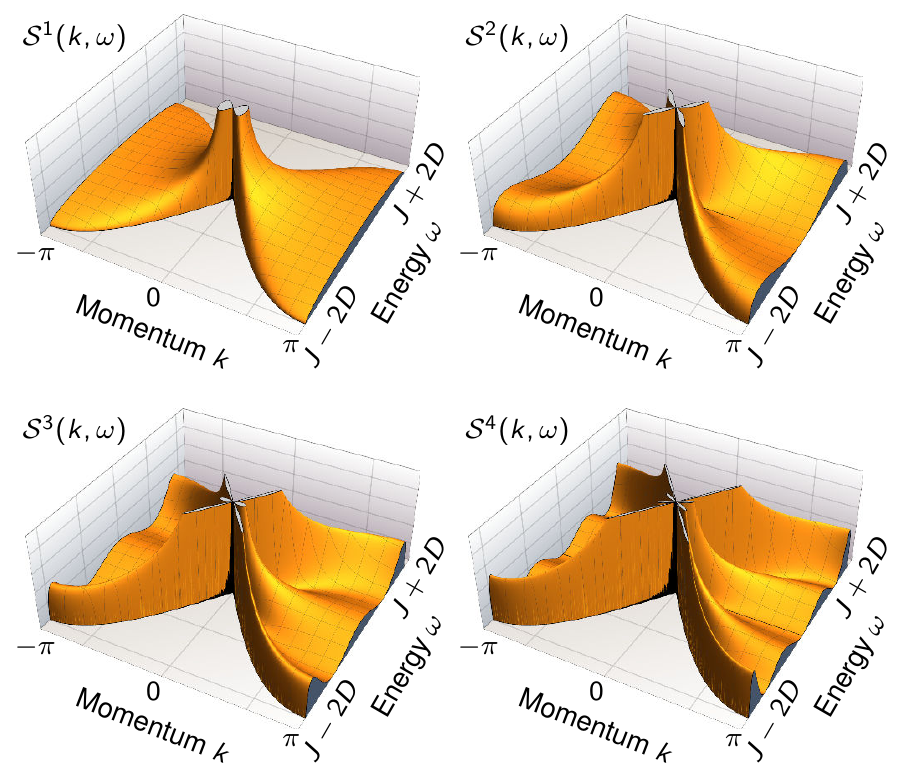}
    \caption{Dynamical multipole structure factors $\mathcal{S}^{f}(\omega,k)$ for $f=1,2,3,4$ of a spin-$1/2$ Ising chain with transverse DMI. \change{$\mathcal{S}^{f}(\omega,k)$ is nonzero if $| \omega - J | \le 2 |D(k)|$ and comes with a square-root edge singularity. Instead of exhibiting quasiparticle features, the structure factors are broad continua, owing to the fractionalization of excitations associated with locally acting with $S^-$ on the fully polarized ground state.}}
    \label{fig:ISING-dyn}
\end{figure}

The first four multipolar dynamical structure factors are depicted in Fig.~\ref{fig:ISING-dyn}. None of them has quasiparticle structure. Instead they are broad, structureless continua. This is because a single flip caused by $S^-$---or, in fact, any number of neighboring flipped spins---decays into two domain walls due to particle-number sector coupling. This fractionalization is similar to that in Ising spin chains in transverse field \cite{Suzuki2013}. We note that $\mathcal{S}^{f}(\omega,k)$ comes with square-root edge singularities for all $f$.

\newpage


\bibliography{bib}

\begin{thebibliography}{195}%
\makeatletter
\providecommand \@ifxundefined [1]{%
 \@ifx{#1\undefined}
}%
\providecommand \@ifnum [1]{%
 \ifnum #1\expandafter \@firstoftwo
 \else \expandafter \@secondoftwo
 \fi
}%
\providecommand \@ifx [1]{%
 \ifx #1\expandafter \@firstoftwo
 \else \expandafter \@secondoftwo
 \fi
}%
\providecommand \natexlab [1]{#1}%
\providecommand \enquote  [1]{``#1''}%
\providecommand \bibnamefont  [1]{#1}%
\providecommand \bibfnamefont [1]{#1}%
\providecommand \citenamefont [1]{#1}%
\providecommand \href@noop [0]{\@secondoftwo}%
\providecommand \href [0]{\begingroup \@sanitize@url \@href}%
\providecommand \@href[1]{\@@startlink{#1}\@@href}%
\providecommand \@@href[1]{\endgroup#1\@@endlink}%
\providecommand \@sanitize@url [0]{\catcode `\\12\catcode `\$12\catcode
  `\&12\catcode `\#12\catcode `\^12\catcode `\_12\catcode `\%12\relax}%
\providecommand \@@startlink[1]{}%
\providecommand \@@endlink[0]{}%
\providecommand \url  [0]{\begingroup\@sanitize@url \@url }%
\providecommand \@url [1]{\endgroup\@href {#1}{\urlprefix }}%
\providecommand \urlprefix  [0]{URL }%
\providecommand \Eprint [0]{\href }%
\providecommand \doibase [0]{http://dx.doi.org/}%
\providecommand \selectlanguage [0]{\@gobble}%
\providecommand \bibinfo  [0]{\@secondoftwo}%
\providecommand \bibfield  [0]{\@secondoftwo}%
\providecommand \translation [1]{[#1]}%
\providecommand \BibitemOpen [0]{}%
\providecommand \bibitemStop [0]{}%
\providecommand \bibitemNoStop [0]{.\EOS\space}%
\providecommand \EOS [0]{\spacefactor3000\relax}%
\providecommand \BibitemShut  [1]{\csname bibitem#1\endcsname}%
\let\auto@bib@innerbib\@empty
\bibitem [{\citenamefont {Xin}\ \emph {et~al.}(2020)\citenamefont {Xin},
  \citenamefont {Siyuan}, \citenamefont {Harry}, \citenamefont {Minghui},\ and\
  \citenamefont {Yanfeng}}]{Xin2020}%
  \BibitemOpen
  \bibfield  {author} {\bibinfo {author} {\bibfnamefont {L.}~\bibnamefont
  {Xin}}, \bibinfo {author} {\bibfnamefont {Y.}~\bibnamefont {Siyuan}},
  \bibinfo {author} {\bibfnamefont {L.}~\bibnamefont {Harry}}, \bibinfo
  {author} {\bibfnamefont {L.}~\bibnamefont {Minghui}}, \ and\ \bibinfo
  {author} {\bibfnamefont {C.}~\bibnamefont {Yanfeng}},\ }\bibfield  {title}
  {\enquote {\bibinfo {title} {Topological mechanical metamaterials: A brief
  review},}\ }\href {\doibase 10.1016/j.cossms.2020.100853} {\bibfield
  {journal} {\bibinfo  {journal} {Current Opinion in Solid State and Materials
  Science}\ }\textbf {\bibinfo {volume} {24}},\ \bibinfo {pages} {100853}
  (\bibinfo {year} {2020})}\BibitemShut {NoStop}%
\bibitem [{\citenamefont {Sato}\ and\ \citenamefont {Ando}(2017)}]{Sato2017}%
  \BibitemOpen
  \bibfield  {author} {\bibinfo {author} {\bibfnamefont {M.}~\bibnamefont
  {Sato}}\ and\ \bibinfo {author} {\bibfnamefont {Y.}~\bibnamefont {Ando}},\
  }\bibfield  {title} {\enquote {\bibinfo {title} {Topological superconductors:
  a review},}\ }\href {\doibase 10.1088/1361-6633/aa6ac7} {\bibfield  {journal}
  {\bibinfo  {journal} {Reports on Progress in Physics}\ }\textbf {\bibinfo
  {volume} {80}},\ \bibinfo {pages} {076501} (\bibinfo {year}
  {2017})}\BibitemShut {NoStop}%
\bibitem [{\citenamefont {Cooper}\ \emph {et~al.}(2019)\citenamefont {Cooper},
  \citenamefont {Dalibard},\ and\ \citenamefont {Spielman}}]{Cooper2019}%
  \BibitemOpen
  \bibfield  {author} {\bibinfo {author} {\bibfnamefont {N.~R.}\ \bibnamefont
  {Cooper}}, \bibinfo {author} {\bibfnamefont {J.}~\bibnamefont {Dalibard}}, \
  and\ \bibinfo {author} {\bibfnamefont {I.~B.}\ \bibnamefont {Spielman}},\
  }\bibfield  {title} {\enquote {\bibinfo {title} {Topological bands for
  ultracold atoms},}\ }\href {\doibase 10.1103/RevModPhys.91.015005} {\bibfield
   {journal} {\bibinfo  {journal} {Rev. Mod. Phys.}\ }\textbf {\bibinfo
  {volume} {91}},\ \bibinfo {pages} {015005} (\bibinfo {year}
  {2019})}\BibitemShut {NoStop}%
\bibitem [{\citenamefont {v.~Klitzing}\ and\ \citenamefont
  {Ebert}(1985)}]{Klitzing1985}%
  \BibitemOpen
  \bibfield  {author} {\bibinfo {author} {\bibfnamefont {K.}~\bibnamefont
  {v.~Klitzing}}\ and\ \bibinfo {author} {\bibfnamefont {G.}~\bibnamefont
  {Ebert}},\ }\bibfield  {title} {\enquote {\bibinfo {title} {Application of
  the quantum {H}all effect in metrology},}\ }\href {\doibase
  10.1088/0026-1394/21/1/004} {\bibfield  {journal} {\bibinfo  {journal}
  {Metrologia}\ }\textbf {\bibinfo {volume} {21}},\ \bibinfo {pages} {11--18}
  (\bibinfo {year} {1985})}\BibitemShut {NoStop}%
\bibitem [{\citenamefont {Nayak}\ \emph {et~al.}(2008)\citenamefont {Nayak},
  \citenamefont {Simon}, \citenamefont {Stern}, \citenamefont {Freedman},\ and\
  \citenamefont {Das~Sarma}}]{Nayak2008}%
  \BibitemOpen
  \bibfield  {author} {\bibinfo {author} {\bibfnamefont {C.}~\bibnamefont
  {Nayak}}, \bibinfo {author} {\bibfnamefont {S.~H.}\ \bibnamefont {Simon}},
  \bibinfo {author} {\bibfnamefont {A.}~\bibnamefont {Stern}}, \bibinfo
  {author} {\bibfnamefont {M.}~\bibnamefont {Freedman}}, \ and\ \bibinfo
  {author} {\bibfnamefont {S.}~\bibnamefont {Das~Sarma}},\ }\bibfield  {title}
  {\enquote {\bibinfo {title} {Non-abelian anyons and topological quantum
  computation},}\ }\href {\doibase 10.1103/RevModPhys.80.1083} {\bibfield
  {journal} {\bibinfo  {journal} {Rev. Mod. Phys.}\ }\textbf {\bibinfo {volume}
  {80}},\ \bibinfo {pages} {1083--1159} (\bibinfo {year} {2008})}\BibitemShut
  {NoStop}%
\bibitem [{\citenamefont {Tian}\ \emph {et~al.}(2017)\citenamefont {Tian},
  \citenamefont {Yu}, \citenamefont {Shi},\ and\ \citenamefont
  {Wang}}]{Tian2017}%
  \BibitemOpen
  \bibfield  {author} {\bibinfo {author} {\bibfnamefont {W.}~\bibnamefont
  {Tian}}, \bibinfo {author} {\bibfnamefont {W.}~\bibnamefont {Yu}}, \bibinfo
  {author} {\bibfnamefont {J.}~\bibnamefont {Shi}}, \ and\ \bibinfo {author}
  {\bibfnamefont {Y.}~\bibnamefont {Wang}},\ }\bibfield  {title} {\enquote
  {\bibinfo {title} {The property, preparation and application of topological
  insulators: A review},}\ }\href {\doibase 10.3390/ma10070814} {\bibfield
  {journal} {\bibinfo  {journal} {Materials}\ }\textbf {\bibinfo {volume}
  {10}},\ \bibinfo {pages} {814} (\bibinfo {year} {2017})}\BibitemShut
  {NoStop}%
\bibitem [{\citenamefont {{\v{S}}mejkal}\ \emph {et~al.}(2018)\citenamefont
  {{\v{S}}mejkal}, \citenamefont {Mokrousov}, \citenamefont {Yan},\ and\
  \citenamefont {MacDonald}}]{Smejkal2018}%
  \BibitemOpen
  \bibfield  {author} {\bibinfo {author} {\bibfnamefont {L.}~\bibnamefont
  {{\v{S}}mejkal}}, \bibinfo {author} {\bibfnamefont {Y.}~\bibnamefont
  {Mokrousov}}, \bibinfo {author} {\bibfnamefont {B.}~\bibnamefont {Yan}}, \
  and\ \bibinfo {author} {\bibfnamefont {A.~H.}\ \bibnamefont {MacDonald}},\
  }\bibfield  {title} {\enquote {\bibinfo {title} {Topological
  antiferromagnetic spintronics},}\ }\href {\doibase 10.1038/s41567-018-0064-5}
  {\bibfield  {journal} {\bibinfo  {journal} {Nature Physics}\ }\textbf
  {\bibinfo {volume} {14}},\ \bibinfo {pages} {242--251} (\bibinfo {year}
  {2018})}\BibitemShut {NoStop}%
\bibitem [{\citenamefont {Lewenstein}\ \emph {et~al.}(2007)\citenamefont
  {Lewenstein}, \citenamefont {Sanpera}, \citenamefont {Ahufinger},
  \citenamefont {Damski}, \citenamefont {Sen(De)},\ and\ \citenamefont
  {Sen}}]{Lewenstein2007}%
  \BibitemOpen
  \bibfield  {author} {\bibinfo {author} {\bibfnamefont {M.}~\bibnamefont
  {Lewenstein}}, \bibinfo {author} {\bibfnamefont {A.}~\bibnamefont {Sanpera}},
  \bibinfo {author} {\bibfnamefont {V.}~\bibnamefont {Ahufinger}}, \bibinfo
  {author} {\bibfnamefont {B.}~\bibnamefont {Damski}}, \bibinfo {author}
  {\bibfnamefont {A.}~\bibnamefont {Sen(De)}}, \ and\ \bibinfo {author}
  {\bibfnamefont {U.}~\bibnamefont {Sen}},\ }\bibfield  {title} {\enquote
  {\bibinfo {title} {Ultracold atomic gases in optical lattices: mimicking
  condensed matter physics and beyond},}\ }\href {\doibase
  10.1080/00018730701223200} {\ \textbf {\bibinfo {volume} {56}},\ \bibinfo
  {pages} {243--379} (\bibinfo {year} {2007})}\BibitemShut {NoStop}%
\bibitem [{\citenamefont {Cazalilla}\ \emph {et~al.}(2011)\citenamefont
  {Cazalilla}, \citenamefont {Citro}, \citenamefont {Giamarchi}, \citenamefont
  {Orignac},\ and\ \citenamefont {Rigol}}]{Cazalilla2011}%
  \BibitemOpen
  \bibfield  {author} {\bibinfo {author} {\bibfnamefont {M.~A.}\ \bibnamefont
  {Cazalilla}}, \bibinfo {author} {\bibfnamefont {R.}~\bibnamefont {Citro}},
  \bibinfo {author} {\bibfnamefont {T.}~\bibnamefont {Giamarchi}}, \bibinfo
  {author} {\bibfnamefont {E.}~\bibnamefont {Orignac}}, \ and\ \bibinfo
  {author} {\bibfnamefont {M.}~\bibnamefont {Rigol}},\ }\bibfield  {title}
  {\enquote {\bibinfo {title} {One dimensional bosons: From condensed matter
  systems to ultracold gases},}\ }\href {\doibase 10.1103/RevModPhys.83.1405}
  {\bibfield  {journal} {\bibinfo  {journal} {Rev. Mod. Phys.}\ }\textbf
  {\bibinfo {volume} {83}},\ \bibinfo {pages} {1405--1466} (\bibinfo {year}
  {2011})}\BibitemShut {NoStop}%
\bibitem [{\citenamefont {Rachel}(2018)}]{Rachel2018}%
  \BibitemOpen
  \bibfield  {author} {\bibinfo {author} {\bibfnamefont {S.}~\bibnamefont
  {Rachel}},\ }\bibfield  {title} {\enquote {\bibinfo {title} {Interacting
  topological insulators: a review},}\ }\href {\doibase
  10.1088/1361-6633/aad6a6} {\bibfield  {journal} {\bibinfo  {journal} {Reports
  on Progress in Physics}\ }\textbf {\bibinfo {volume} {81}},\ \bibinfo {pages}
  {116501} (\bibinfo {year} {2018})}\BibitemShut {NoStop}%
\bibitem [{\citenamefont {Li}\ \emph {et~al.}(2012)\citenamefont {Li},
  \citenamefont {Ren}, \citenamefont {Wang}, \citenamefont {Zhang},
  \citenamefont {H\"anggi},\ and\ \citenamefont {Li}}]{Li2012phononics}%
  \BibitemOpen
  \bibfield  {author} {\bibinfo {author} {\bibfnamefont {N.}~\bibnamefont
  {Li}}, \bibinfo {author} {\bibfnamefont {J.}~\bibnamefont {Ren}}, \bibinfo
  {author} {\bibfnamefont {L.}~\bibnamefont {Wang}}, \bibinfo {author}
  {\bibfnamefont {G.}~\bibnamefont {Zhang}}, \bibinfo {author} {\bibfnamefont
  {P.}~\bibnamefont {H\"anggi}}, \ and\ \bibinfo {author} {\bibfnamefont
  {B.}~\bibnamefont {Li}},\ }\bibfield  {title} {\enquote {\bibinfo {title}
  {Colloquium: Phononics: Manipulating heat flow with electronic analogs and
  beyond},}\ }\href {\doibase 10.1103/RevModPhys.84.1045} {\bibfield  {journal}
  {\bibinfo  {journal} {Rev. Mod. Phys.}\ }\textbf {\bibinfo {volume} {84}},\
  \bibinfo {pages} {1045--1066} (\bibinfo {year} {2012})}\BibitemShut {NoStop}%
\bibitem [{\citenamefont {Pirro}\ \emph {et~al.}(2021)\citenamefont {Pirro},
  \citenamefont {Vasyuchka}, \citenamefont {Serga},\ and\ \citenamefont
  {Hillebrands}}]{Pirro2021}%
  \BibitemOpen
  \bibfield  {author} {\bibinfo {author} {\bibfnamefont {P.}~\bibnamefont
  {Pirro}}, \bibinfo {author} {\bibfnamefont {V.~I.}\ \bibnamefont
  {Vasyuchka}}, \bibinfo {author} {\bibfnamefont {A.~A.}\ \bibnamefont
  {Serga}}, \ and\ \bibinfo {author} {\bibfnamefont {B.}~\bibnamefont
  {Hillebrands}},\ }\bibfield  {title} {\enquote {\bibinfo {title} {Advances in
  coherent magnonics},}\ }\href {\doibase 10.1038/s41578-021-00332-w}
  {\bibfield  {journal} {\bibinfo  {journal} {Nat. Rev. Mater.}\ }\textbf
  {\bibinfo {volume} {1-22}} (\bibinfo {year} {2021}),\
  10.1038/s41578-021-00332-w}\BibitemShut {NoStop}%
\bibitem [{\citenamefont {Nakamura}(2019)}]{Nakamura2019}%
  \BibitemOpen
  \bibfield  {author} {\bibinfo {author} {\bibfnamefont {K.}~\bibnamefont
  {Nakamura}},\ }\href {\doibase 10.1007/978-3-030-11924-9} {\emph {\bibinfo
  {title} {Quantum Phononics}}}\ (\bibinfo  {publisher} {Springer International
  Publishing},\ \bibinfo {year} {2019})\BibitemShut {NoStop}%
\bibitem [{\citenamefont {Lachance-Quirion}\ \emph {et~al.}(2019)\citenamefont
  {Lachance-Quirion}, \citenamefont {Tabuchi}, \citenamefont {Gloppe},
  \citenamefont {Usami},\ and\ \citenamefont
  {Nakamura}}]{Lachance_Quirion_2019}%
  \BibitemOpen
  \bibfield  {author} {\bibinfo {author} {\bibfnamefont {D.}~\bibnamefont
  {Lachance-Quirion}}, \bibinfo {author} {\bibfnamefont {Y.}~\bibnamefont
  {Tabuchi}}, \bibinfo {author} {\bibfnamefont {A.}~\bibnamefont {Gloppe}},
  \bibinfo {author} {\bibfnamefont {K.}~\bibnamefont {Usami}}, \ and\ \bibinfo
  {author} {\bibfnamefont {Y.}~\bibnamefont {Nakamura}},\ }\bibfield  {title}
  {\enquote {\bibinfo {title} {Hybrid quantum systems based on magnonics},}\
  }\href {\doibase 10.7567/1882-0786/ab248d} {\bibfield  {journal} {\bibinfo
  {journal} {Applied Physics Express}\ }\textbf {\bibinfo {volume} {12}},\
  \bibinfo {pages} {070101} (\bibinfo {year} {2019})}\BibitemShut {NoStop}%
\bibitem [{\citenamefont {Liu}\ \emph {et~al.}(2019)\citenamefont {Liu},
  \citenamefont {Chen},\ and\ \citenamefont {Xu}}]{Liu2019phonons}%
  \BibitemOpen
  \bibfield  {author} {\bibinfo {author} {\bibfnamefont {Y.}~\bibnamefont
  {Liu}}, \bibinfo {author} {\bibfnamefont {X.}~\bibnamefont {Chen}}, \ and\
  \bibinfo {author} {\bibfnamefont {Y.}~\bibnamefont {Xu}},\ }\bibfield
  {title} {\enquote {\bibinfo {title} {Topological phononics: From fundamental
  models to real materials},}\ }\href {\doibase 10.1002/adfm.201904784}
  {\bibfield  {journal} {\bibinfo  {journal} {Advanced Functional Materials}\
  }\textbf {\bibinfo {volume} {30}},\ \bibinfo {pages} {1904784} (\bibinfo
  {year} {2019})}\BibitemShut {NoStop}%
\bibitem [{\citenamefont {Li}\ \emph {et~al.}(2021)\citenamefont {Li},
  \citenamefont {Liu}, \citenamefont {Baronett}, \citenamefont {Liu},
  \citenamefont {Wang}, \citenamefont {Li}, \citenamefont {Chen}, \citenamefont
  {Li}, \citenamefont {Zhu},\ and\ \citenamefont {Chen}}]{Li2021topophonons}%
  \BibitemOpen
  \bibfield  {author} {\bibinfo {author} {\bibfnamefont {J.}~\bibnamefont
  {Li}}, \bibinfo {author} {\bibfnamefont {J.}~\bibnamefont {Liu}}, \bibinfo
  {author} {\bibfnamefont {S.~A.}\ \bibnamefont {Baronett}}, \bibinfo {author}
  {\bibfnamefont {M.}~\bibnamefont {Liu}}, \bibinfo {author} {\bibfnamefont
  {L.}~\bibnamefont {Wang}}, \bibinfo {author} {\bibfnamefont {R.}~\bibnamefont
  {Li}}, \bibinfo {author} {\bibfnamefont {Y.}~\bibnamefont {Chen}}, \bibinfo
  {author} {\bibfnamefont {D.}~\bibnamefont {Li}}, \bibinfo {author}
  {\bibfnamefont {Q.}~\bibnamefont {Zhu}}, \ and\ \bibinfo {author}
  {\bibfnamefont {X.-Q.}\ \bibnamefont {Chen}},\ }\bibfield  {title} {\enquote
  {\bibinfo {title} {Computation and data driven discovery of topological
  phononic materials},}\ }\href {\doibase 10.1038/s41467-021-21293-2}
  {\bibfield  {journal} {\bibinfo  {journal} {Nature Communications}\ }\textbf
  {\bibinfo {volume} {12}},\ \bibinfo {pages} {1204} (\bibinfo {year}
  {2021})}\BibitemShut {NoStop}%
\bibitem [{\citenamefont {Malki}\ and\ \citenamefont
  {Uhrig}(2020)}]{Malki2020review}%
  \BibitemOpen
  \bibfield  {author} {\bibinfo {author} {\bibfnamefont {M.}~\bibnamefont
  {Malki}}\ and\ \bibinfo {author} {\bibfnamefont {G.~S.}\ \bibnamefont
  {Uhrig}},\ }\bibfield  {title} {\enquote {\bibinfo {title} {Topological
  magnetic excitations},}\ }\href {\doibase 10.1209/0295-5075/132/20003}
  {\bibfield  {journal} {\bibinfo  {journal} {{EPL} (Europhysics Letters)}\
  }\textbf {\bibinfo {volume} {132}},\ \bibinfo {pages} {20003} (\bibinfo
  {year} {2020})}\BibitemShut {NoStop}%
\bibitem [{\citenamefont {McClarty}(2021)}]{McClarty2021review}%
  \BibitemOpen
  \bibfield  {author} {\bibinfo {author} {\bibfnamefont {P.~A.}\ \bibnamefont
  {McClarty}},\ }\bibfield  {title} {\enquote {\bibinfo {title} {Topological
  magnons: A review},}\ }\href {\doibase
  10.1146/annurev-conmatphys-031620-104715} {\bibfield  {journal} {\bibinfo
  {journal} {Annual Review of Condensed Matter Physics}\ }\textbf {\bibinfo
  {volume} {13}} (\bibinfo {year} {2021}),\
  10.1146/annurev-conmatphys-031620-104715}\BibitemShut {NoStop}%
\bibitem [{\citenamefont {Shindou}\ \emph
  {et~al.}(2013{\natexlab{a}})\citenamefont {Shindou}, \citenamefont
  {Matsumoto}, \citenamefont {Murakami},\ and\ \citenamefont
  {Ohe}}]{Shindou13}%
  \BibitemOpen
  \bibfield  {author} {\bibinfo {author} {\bibfnamefont {R.}~\bibnamefont
  {Shindou}}, \bibinfo {author} {\bibfnamefont {R.}~\bibnamefont {Matsumoto}},
  \bibinfo {author} {\bibfnamefont {S.}~\bibnamefont {Murakami}}, \ and\
  \bibinfo {author} {\bibfnamefont {J.-i.}\ \bibnamefont {Ohe}},\ }\bibfield
  {title} {\enquote {\bibinfo {title} {Topological chiral magnonic edge mode in
  a magnonic crystal},}\ }\href {\doibase 10.1103/PhysRevB.87.174427}
  {\bibfield  {journal} {\bibinfo  {journal} {Phys. Rev. B}\ }\textbf {\bibinfo
  {volume} {87}},\ \bibinfo {pages} {174427} (\bibinfo {year}
  {2013}{\natexlab{a}})}\BibitemShut {NoStop}%
\bibitem [{\citenamefont {Mook}\ \emph
  {et~al.}(2015{\natexlab{a}})\citenamefont {Mook}, \citenamefont {Henk},\ and\
  \citenamefont {Mertig}}]{Mook2015waveguide}%
  \BibitemOpen
  \bibfield  {author} {\bibinfo {author} {\bibfnamefont {A.}~\bibnamefont
  {Mook}}, \bibinfo {author} {\bibfnamefont {J.}~\bibnamefont {Henk}}, \ and\
  \bibinfo {author} {\bibfnamefont {I.}~\bibnamefont {Mertig}},\ }\bibfield
  {title} {\enquote {\bibinfo {title} {Magnon waveguide with nanoscale
  confinement constructed from topological magnon insulators},}\ }\href
  {\doibase 10.1103/PhysRevB.91.174409} {\bibfield  {journal} {\bibinfo
  {journal} {Phys. Rev. B}\ }\textbf {\bibinfo {volume} {91}},\ \bibinfo
  {pages} {174409} (\bibinfo {year} {2015}{\natexlab{a}})}\BibitemShut
  {NoStop}%
\bibitem [{\citenamefont {Wang}\ \emph {et~al.}(2018)\citenamefont {Wang},
  \citenamefont {Zhang},\ and\ \citenamefont {Wang}}]{Wang2018}%
  \BibitemOpen
  \bibfield  {author} {\bibinfo {author} {\bibfnamefont {X.~S.}\ \bibnamefont
  {Wang}}, \bibinfo {author} {\bibfnamefont {H.~W.}\ \bibnamefont {Zhang}}, \
  and\ \bibinfo {author} {\bibfnamefont {X.~R.}\ \bibnamefont {Wang}},\
  }\bibfield  {title} {\enquote {\bibinfo {title} {Topological magnonics: A
  paradigm for spin-wave manipulation and device design},}\ }\href {\doibase
  10.1103/PhysRevApplied.9.024029} {\bibfield  {journal} {\bibinfo  {journal}
  {Phys. Rev. Applied}\ }\textbf {\bibinfo {volume} {9}},\ \bibinfo {pages}
  {024029} (\bibinfo {year} {2018})}\BibitemShut {NoStop}%
\bibitem [{\citenamefont {Chumak}(2019)}]{Chumak2019}%
  \BibitemOpen
  \bibfield  {author} {\bibinfo {author} {\bibfnamefont {A.~V.}\ \bibnamefont
  {Chumak}},\ }\bibfield  {title} {\enquote {\bibinfo {title} {Magnon
  spintronics},}\ }in\ \href {\doibase 10.1201/9780429423079-6} {\emph
  {\bibinfo {booktitle} {Spintronics Handbook: Spin Transport and Magnetism,
  Second Edition}}}\ (\bibinfo  {publisher} {{CRC} Press},\ \bibinfo {year}
  {2019})\ pp.\ \bibinfo {pages} {247--302}\BibitemShut {NoStop}%
\bibitem [{\citenamefont {Aguilera}\ \emph {et~al.}(2020)\citenamefont
  {Aguilera}, \citenamefont {Jaeschke-Ubiergo}, \citenamefont {Vidal-Silva},
  \citenamefont {Torres},\ and\ \citenamefont {Nunez}}]{Aguilera2020}%
  \BibitemOpen
  \bibfield  {author} {\bibinfo {author} {\bibfnamefont {E.}~\bibnamefont
  {Aguilera}}, \bibinfo {author} {\bibfnamefont {R.}~\bibnamefont
  {Jaeschke-Ubiergo}}, \bibinfo {author} {\bibfnamefont {N.}~\bibnamefont
  {Vidal-Silva}}, \bibinfo {author} {\bibfnamefont {L.~E. F.~F.}\ \bibnamefont
  {Torres}}, \ and\ \bibinfo {author} {\bibfnamefont {A.~S.}\ \bibnamefont
  {Nunez}},\ }\bibfield  {title} {\enquote {\bibinfo {title} {Topological
  magnonics in the two-dimensional van der waals magnet {CrI}$_3$},}\ }\href
  {\doibase 10.1103/PhysRevB.102.024409} {\bibfield  {journal} {\bibinfo
  {journal} {Phys. Rev. B}\ }\textbf {\bibinfo {volume} {102}},\ \bibinfo
  {pages} {024409} (\bibinfo {year} {2020})}\BibitemShut {NoStop}%
\bibitem [{\citenamefont {Mook}\ \emph
  {et~al.}(2021{\natexlab{a}})\citenamefont {Mook}, \citenamefont {D\'{\i}az},
  \citenamefont {Klinovaja},\ and\ \citenamefont {Loss}}]{Mook2020hinge}%
  \BibitemOpen
  \bibfield  {author} {\bibinfo {author} {\bibfnamefont {A.}~\bibnamefont
  {Mook}}, \bibinfo {author} {\bibfnamefont {S.~A.}\ \bibnamefont {D\'{\i}az}},
  \bibinfo {author} {\bibfnamefont {J.}~\bibnamefont {Klinovaja}}, \ and\
  \bibinfo {author} {\bibfnamefont {D.}~\bibnamefont {Loss}},\ }\bibfield
  {title} {\enquote {\bibinfo {title} {Chiral hinge magnons in second-order
  topological magnon insulators},}\ }\href {\doibase
  10.1103/PhysRevB.104.024406} {\bibfield  {journal} {\bibinfo  {journal}
  {Phys. Rev. B}\ }\textbf {\bibinfo {volume} {104}},\ \bibinfo {pages}
  {024406} (\bibinfo {year} {2021}{\natexlab{a}})}\BibitemShut {NoStop}%
\bibitem [{\citenamefont {Masuda}\ \emph {et~al.}(2006)\citenamefont {Masuda},
  \citenamefont {Zheludev}, \citenamefont {Manaka}, \citenamefont {Regnault},
  \citenamefont {Chung},\ and\ \citenamefont {Qiu}}]{Masuda2006}%
  \BibitemOpen
  \bibfield  {author} {\bibinfo {author} {\bibfnamefont {T.}~\bibnamefont
  {Masuda}}, \bibinfo {author} {\bibfnamefont {A.}~\bibnamefont {Zheludev}},
  \bibinfo {author} {\bibfnamefont {H.}~\bibnamefont {Manaka}}, \bibinfo
  {author} {\bibfnamefont {L.-P.}\ \bibnamefont {Regnault}}, \bibinfo {author}
  {\bibfnamefont {J.-H.}\ \bibnamefont {Chung}}, \ and\ \bibinfo {author}
  {\bibfnamefont {Y.}~\bibnamefont {Qiu}},\ }\bibfield  {title} {\enquote
  {\bibinfo {title} {Dynamics of composite {H}aldane spin chains in
  {IPA-CuCl}$_{3}$},}\ }\href {\doibase 10.1103/PhysRevLett.96.047210}
  {\bibfield  {journal} {\bibinfo  {journal} {Phys. Rev. Lett.}\ }\textbf
  {\bibinfo {volume} {96}},\ \bibinfo {pages} {047210} (\bibinfo {year}
  {2006})}\BibitemShut {NoStop}%
\bibitem [{\citenamefont {Zhitomirsky}\ and\ \citenamefont
  {Chernyshev}(2013)}]{Zhitomirsky2013}%
  \BibitemOpen
  \bibfield  {author} {\bibinfo {author} {\bibfnamefont {M.~E.}\ \bibnamefont
  {Zhitomirsky}}\ and\ \bibinfo {author} {\bibfnamefont {A.~L.}\ \bibnamefont
  {Chernyshev}},\ }\bibfield  {title} {\enquote {\bibinfo {title} {Colloquium:
  Spontaneous magnon decays},}\ }\href {\doibase 10.1103/RevModPhys.85.219}
  {\bibfield  {journal} {\bibinfo  {journal} {Rev. Mod. Phys.}\ }\textbf
  {\bibinfo {volume} {85}},\ \bibinfo {pages} {219--242} (\bibinfo {year}
  {2013})}\BibitemShut {NoStop}%
\bibitem [{\citenamefont {Hong}\ \emph {et~al.}(2017)\citenamefont {Hong},
  \citenamefont {Qiu}, \citenamefont {Matsumoto}, \citenamefont {Tennant},
  \citenamefont {Coester}, \citenamefont {Schmidt}, \citenamefont {Awwadi},
  \citenamefont {Turnbull}, \citenamefont {Agrawal},\ and\ \citenamefont
  {Chernyshev}}]{Hong2017}%
  \BibitemOpen
  \bibfield  {author} {\bibinfo {author} {\bibfnamefont {T.}~\bibnamefont
  {Hong}}, \bibinfo {author} {\bibfnamefont {Y.}~\bibnamefont {Qiu}}, \bibinfo
  {author} {\bibfnamefont {M.}~\bibnamefont {Matsumoto}}, \bibinfo {author}
  {\bibfnamefont {D.~A.}\ \bibnamefont {Tennant}}, \bibinfo {author}
  {\bibfnamefont {K.}~\bibnamefont {Coester}}, \bibinfo {author} {\bibfnamefont
  {K.~P.}\ \bibnamefont {Schmidt}}, \bibinfo {author} {\bibfnamefont {F.~F.}\
  \bibnamefont {Awwadi}}, \bibinfo {author} {\bibfnamefont {M.~M.}\
  \bibnamefont {Turnbull}}, \bibinfo {author} {\bibfnamefont {H.}~\bibnamefont
  {Agrawal}}, \ and\ \bibinfo {author} {\bibfnamefont {A.~L.}\ \bibnamefont
  {Chernyshev}},\ }\bibfield  {title} {\enquote {\bibinfo {title} {Field
  induced spontaneous quasiparticle decay and renormalization of quasiparticle
  dispersion in a quantum antiferromagnet},}\ }\href {\doibase
  10.1038/ncomms15148} {\bibfield  {journal} {\bibinfo  {journal} {Nature
  Communications}\ }\textbf {\bibinfo {volume} {8}},\ \bibinfo {pages} {15148}
  (\bibinfo {year} {2017})}\BibitemShut {NoStop}%
\bibitem [{\citenamefont {Verresen}\ \emph {et~al.}(2019)\citenamefont
  {Verresen}, \citenamefont {Moessner},\ and\ \citenamefont
  {Pollmann}}]{Verresen2019}%
  \BibitemOpen
  \bibfield  {author} {\bibinfo {author} {\bibfnamefont {R.}~\bibnamefont
  {Verresen}}, \bibinfo {author} {\bibfnamefont {R.}~\bibnamefont {Moessner}},
  \ and\ \bibinfo {author} {\bibfnamefont {F.}~\bibnamefont {Pollmann}},\
  }\bibfield  {title} {\enquote {\bibinfo {title} {Avoided quasiparticle decay
  from strong quantum interactions},}\ }\href {\doibase
  10.1038/s41567-019-0535-3} {\bibfield  {journal} {\bibinfo  {journal} {Nature
  Physics}\ }\textbf {\bibinfo {volume} {15}},\ \bibinfo {pages} {750--753}
  (\bibinfo {year} {2019})}\BibitemShut {NoStop}%
\bibitem [{\citenamefont {Chernyshev}\ and\ \citenamefont
  {Maksimov}(2016)}]{Chernyshev2016}%
  \BibitemOpen
  \bibfield  {author} {\bibinfo {author} {\bibfnamefont {A.~L.}\ \bibnamefont
  {Chernyshev}}\ and\ \bibinfo {author} {\bibfnamefont {P.~A.}\ \bibnamefont
  {Maksimov}},\ }\bibfield  {title} {\enquote {\bibinfo {title} {Damped
  topological magnons in the kagome-lattice ferromagnets},}\ }\href {\doibase
  10.1103/PhysRevLett.117.187203} {\bibfield  {journal} {\bibinfo  {journal}
  {Phys. Rev. Lett.}\ }\textbf {\bibinfo {volume} {117}},\ \bibinfo {pages}
  {187203} (\bibinfo {year} {2016})}\BibitemShut {NoStop}%
\bibitem [{\citenamefont {McClarty}\ \emph {et~al.}(2018)\citenamefont
  {McClarty}, \citenamefont {Dong}, \citenamefont {Gohlke}, \citenamefont
  {Rau}, \citenamefont {Pollmann}, \citenamefont {Moessner},\ and\
  \citenamefont {Penc}}]{McClarty2018}%
  \BibitemOpen
  \bibfield  {author} {\bibinfo {author} {\bibfnamefont {P.~A.}\ \bibnamefont
  {McClarty}}, \bibinfo {author} {\bibfnamefont {X.-Y.}\ \bibnamefont {Dong}},
  \bibinfo {author} {\bibfnamefont {M.}~\bibnamefont {Gohlke}}, \bibinfo
  {author} {\bibfnamefont {J.~G.}\ \bibnamefont {Rau}}, \bibinfo {author}
  {\bibfnamefont {F.}~\bibnamefont {Pollmann}}, \bibinfo {author}
  {\bibfnamefont {R.}~\bibnamefont {Moessner}}, \ and\ \bibinfo {author}
  {\bibfnamefont {K.}~\bibnamefont {Penc}},\ }\bibfield  {title} {\enquote
  {\bibinfo {title} {Topological magnons in {K}itaev magnets at high fields},}\
  }\href {\doibase 10.1103/PhysRevB.98.060404} {\bibfield  {journal} {\bibinfo
  {journal} {Phys. Rev. B}\ }\textbf {\bibinfo {volume} {98}},\ \bibinfo
  {pages} {060404} (\bibinfo {year} {2018})}\BibitemShut {NoStop}%
\bibitem [{\citenamefont {Mook}\ \emph
  {et~al.}(2020{\natexlab{a}})\citenamefont {Mook}, \citenamefont {Klinovaja},\
  and\ \citenamefont {Loss}}]{Mook2020QuantumDamping}%
  \BibitemOpen
  \bibfield  {author} {\bibinfo {author} {\bibfnamefont {A.}~\bibnamefont
  {Mook}}, \bibinfo {author} {\bibfnamefont {J.}~\bibnamefont {Klinovaja}}, \
  and\ \bibinfo {author} {\bibfnamefont {D.}~\bibnamefont {Loss}},\ }\bibfield
  {title} {\enquote {\bibinfo {title} {Quantum damping of skyrmion crystal
  eigenmodes due to spontaneous quasiparticle decay},}\ }\href {\doibase
  10.1103/PhysRevResearch.2.033491} {\bibfield  {journal} {\bibinfo  {journal}
  {Phys. Rev. Research}\ }\textbf {\bibinfo {volume} {2}},\ \bibinfo {pages}
  {033491} (\bibinfo {year} {2020}{\natexlab{a}})}\BibitemShut {NoStop}%
\bibitem [{\citenamefont {McClarty}\ and\ \citenamefont
  {Rau}(2019)}]{McClarty2019}%
  \BibitemOpen
  \bibfield  {author} {\bibinfo {author} {\bibfnamefont {P.~A.}\ \bibnamefont
  {McClarty}}\ and\ \bibinfo {author} {\bibfnamefont {J.~G.}\ \bibnamefont
  {Rau}},\ }\bibfield  {title} {\enquote {\bibinfo {title} {Non-{H}ermitian
  topology of spontaneous magnon decay},}\ }\href {\doibase
  10.1103/PhysRevB.100.100405} {\bibfield  {journal} {\bibinfo  {journal}
  {Phys. Rev. B}\ }\textbf {\bibinfo {volume} {100}},\ \bibinfo {pages}
  {100405} (\bibinfo {year} {2019})}\BibitemShut {NoStop}%
\bibitem [{\citenamefont {Mook}\ \emph
  {et~al.}(2021{\natexlab{b}})\citenamefont {Mook}, \citenamefont {Plekhanov},
  \citenamefont {Klinovaja},\ and\ \citenamefont
  {Loss}}]{Mook2020Interactions}%
  \BibitemOpen
  \bibfield  {author} {\bibinfo {author} {\bibfnamefont {A.}~\bibnamefont
  {Mook}}, \bibinfo {author} {\bibfnamefont {K.}~\bibnamefont {Plekhanov}},
  \bibinfo {author} {\bibfnamefont {J.}~\bibnamefont {Klinovaja}}, \ and\
  \bibinfo {author} {\bibfnamefont {D.}~\bibnamefont {Loss}},\ }\bibfield
  {title} {\enquote {\bibinfo {title} {Interaction-stabilized topological
  magnon insulator in ferromagnets},}\ }\href {\doibase
  10.1103/PhysRevX.11.021061} {\bibfield  {journal} {\bibinfo  {journal} {Phys.
  Rev. X}\ }\textbf {\bibinfo {volume} {11}},\ \bibinfo {pages} {021061}
  (\bibinfo {year} {2021}{\natexlab{b}})}\BibitemShut {NoStop}%
\bibitem [{\citenamefont {Bethe}(1931)}]{Bethe1931}%
  \BibitemOpen
  \bibfield  {author} {\bibinfo {author} {\bibfnamefont {H.}~\bibnamefont
  {Bethe}},\ }\bibfield  {title} {\enquote {\bibinfo {title} {Zur {T}heorie der
  {M}etalle},}\ }\href {\doibase 10.1007/bf01341708} {\bibfield  {journal}
  {\bibinfo  {journal} {Zeitschrift f\"{u}r Physik}\ }\textbf {\bibinfo
  {volume} {71}},\ \bibinfo {pages} {205--226} (\bibinfo {year}
  {1931})}\BibitemShut {NoStop}%
\bibitem [{\citenamefont {Wortis}(1963)}]{Wortis1963}%
  \BibitemOpen
  \bibfield  {author} {\bibinfo {author} {\bibfnamefont {M.}~\bibnamefont
  {Wortis}},\ }\bibfield  {title} {\enquote {\bibinfo {title} {Bound states of
  two spin waves in the {H}eisenberg ferromagnet},}\ }\href {\doibase
  10.1103/PhysRev.132.85} {\bibfield  {journal} {\bibinfo  {journal} {Phys.
  Rev.}\ }\textbf {\bibinfo {volume} {132}},\ \bibinfo {pages} {85--97}
  (\bibinfo {year} {1963})}\BibitemShut {NoStop}%
\bibitem [{\citenamefont {Winkler}\ \emph {et~al.}(2006)\citenamefont
  {Winkler}, \citenamefont {Thalhammer}, \citenamefont {Lang}, \citenamefont
  {Grimm}, \citenamefont {Denschlag}, \citenamefont {Daley}, \citenamefont
  {Kantian}, \citenamefont {B\"{u}chler},\ and\ \citenamefont
  {Zoller}}]{Winkler2006}%
  \BibitemOpen
  \bibfield  {author} {\bibinfo {author} {\bibfnamefont {K.}~\bibnamefont
  {Winkler}}, \bibinfo {author} {\bibfnamefont {G.}~\bibnamefont {Thalhammer}},
  \bibinfo {author} {\bibfnamefont {F.}~\bibnamefont {Lang}}, \bibinfo {author}
  {\bibfnamefont {R.}~\bibnamefont {Grimm}}, \bibinfo {author} {\bibfnamefont
  {J.~H.}\ \bibnamefont {Denschlag}}, \bibinfo {author} {\bibfnamefont {A.~J.}\
  \bibnamefont {Daley}}, \bibinfo {author} {\bibfnamefont {A.}~\bibnamefont
  {Kantian}}, \bibinfo {author} {\bibfnamefont {H.~P.}\ \bibnamefont
  {B\"{u}chler}}, \ and\ \bibinfo {author} {\bibfnamefont {P.}~\bibnamefont
  {Zoller}},\ }\bibfield  {title} {\enquote {\bibinfo {title} {Repulsively
  bound atom pairs in an optical lattice},}\ }\href {\doibase
  10.1038/nature04918} {\bibfield  {journal} {\bibinfo  {journal} {Nature}\
  }\textbf {\bibinfo {volume} {441}},\ \bibinfo {pages} {853--856} (\bibinfo
  {year} {2006})}\BibitemShut {NoStop}%
\bibitem [{\citenamefont {Di~Liberto}\ \emph {et~al.}(2016)\citenamefont
  {Di~Liberto}, \citenamefont {Recati}, \citenamefont {Carusotto},\ and\
  \citenamefont {Menotti}}]{DiLiberto2016}%
  \BibitemOpen
  \bibfield  {author} {\bibinfo {author} {\bibfnamefont {M.}~\bibnamefont
  {Di~Liberto}}, \bibinfo {author} {\bibfnamefont {A.}~\bibnamefont {Recati}},
  \bibinfo {author} {\bibfnamefont {I.}~\bibnamefont {Carusotto}}, \ and\
  \bibinfo {author} {\bibfnamefont {C.}~\bibnamefont {Menotti}},\ }\bibfield
  {title} {\enquote {\bibinfo {title} {Two-body physics in the
  {S}u-{S}chrieffer-{H}eeger model},}\ }\href {\doibase
  10.1103/PhysRevA.94.062704} {\bibfield  {journal} {\bibinfo  {journal} {Phys.
  Rev. A}\ }\textbf {\bibinfo {volume} {94}},\ \bibinfo {pages} {062704}
  (\bibinfo {year} {2016})}\BibitemShut {NoStop}%
\bibitem [{\citenamefont {Gorlach}\ and\ \citenamefont
  {Poddubny}(2017)}]{Gorlach2017}%
  \BibitemOpen
  \bibfield  {author} {\bibinfo {author} {\bibfnamefont {M.~A.}\ \bibnamefont
  {Gorlach}}\ and\ \bibinfo {author} {\bibfnamefont {A.~N.}\ \bibnamefont
  {Poddubny}},\ }\bibfield  {title} {\enquote {\bibinfo {title}
  {Interaction-induced two-photon edge states in an extended {H}ubbard model
  realized in a cavity array},}\ }\href {\doibase 10.1103/PhysRevA.95.033831}
  {\bibfield  {journal} {\bibinfo  {journal} {Phys. Rev. A}\ }\textbf {\bibinfo
  {volume} {95}},\ \bibinfo {pages} {033831} (\bibinfo {year}
  {2017})}\BibitemShut {NoStop}%
\bibitem [{\citenamefont {Salerno}\ \emph {et~al.}(2018)\citenamefont
  {Salerno}, \citenamefont {Di~Liberto}, \citenamefont {Menotti},\ and\
  \citenamefont {Carusotto}}]{Salerno2018}%
  \BibitemOpen
  \bibfield  {author} {\bibinfo {author} {\bibfnamefont {G.}~\bibnamefont
  {Salerno}}, \bibinfo {author} {\bibfnamefont {M.}~\bibnamefont {Di~Liberto}},
  \bibinfo {author} {\bibfnamefont {C.}~\bibnamefont {Menotti}}, \ and\
  \bibinfo {author} {\bibfnamefont {I.}~\bibnamefont {Carusotto}},\ }\bibfield
  {title} {\enquote {\bibinfo {title} {Topological two-body bound states in the
  interacting {H}aldane model},}\ }\href {\doibase 10.1103/PhysRevA.97.013637}
  {\bibfield  {journal} {\bibinfo  {journal} {Phys. Rev. A}\ }\textbf {\bibinfo
  {volume} {97}},\ \bibinfo {pages} {013637} (\bibinfo {year}
  {2018})}\BibitemShut {NoStop}%
\bibitem [{\citenamefont {Qin}\ \emph {et~al.}(2017)\citenamefont {Qin},
  \citenamefont {Mei}, \citenamefont {Ke}, \citenamefont {Zhang},\ and\
  \citenamefont {Lee}}]{Qin2017}%
  \BibitemOpen
  \bibfield  {author} {\bibinfo {author} {\bibfnamefont {X.}~\bibnamefont
  {Qin}}, \bibinfo {author} {\bibfnamefont {F.}~\bibnamefont {Mei}}, \bibinfo
  {author} {\bibfnamefont {Y.}~\bibnamefont {Ke}}, \bibinfo {author}
  {\bibfnamefont {L.}~\bibnamefont {Zhang}}, \ and\ \bibinfo {author}
  {\bibfnamefont {C.}~\bibnamefont {Lee}},\ }\bibfield  {title} {\enquote
  {\bibinfo {title} {Topological magnon bound states in periodically modulated
  heisenberg xxz chains},}\ }\href {\doibase 10.1103/PhysRevB.96.195134}
  {\bibfield  {journal} {\bibinfo  {journal} {Phys. Rev. B}\ }\textbf {\bibinfo
  {volume} {96}},\ \bibinfo {pages} {195134} (\bibinfo {year}
  {2017})}\BibitemShut {NoStop}%
\bibitem [{\citenamefont {Qin}\ \emph {et~al.}(2018)\citenamefont {Qin},
  \citenamefont {Mei}, \citenamefont {Ke}, \citenamefont {Zhang},\ and\
  \citenamefont {Lee}}]{Qin2018}%
  \BibitemOpen
  \bibfield  {author} {\bibinfo {author} {\bibfnamefont {X.}~\bibnamefont
  {Qin}}, \bibinfo {author} {\bibfnamefont {F.}~\bibnamefont {Mei}}, \bibinfo
  {author} {\bibfnamefont {Y.}~\bibnamefont {Ke}}, \bibinfo {author}
  {\bibfnamefont {L.}~\bibnamefont {Zhang}}, \ and\ \bibinfo {author}
  {\bibfnamefont {C.}~\bibnamefont {Lee}},\ }\bibfield  {title} {\enquote
  {\bibinfo {title} {Topological invariant and cotranslational symmetry in
  strongly interacting multi-magnon systems},}\ }\href {\doibase
  10.1088/1367-2630/aa9556} {\bibfield  {journal} {\bibinfo  {journal} {New
  Journal of Physics}\ }\textbf {\bibinfo {volume} {20}},\ \bibinfo {pages}
  {013003} (\bibinfo {year} {2018})}\BibitemShut {NoStop}%
\bibitem [{\citenamefont {Stepanenko}\ and\ \citenamefont
  {Gorlach}(2020)}]{Stepanenko2020}%
  \BibitemOpen
  \bibfield  {author} {\bibinfo {author} {\bibfnamefont {A.~A.}\ \bibnamefont
  {Stepanenko}}\ and\ \bibinfo {author} {\bibfnamefont {M.~A.}\ \bibnamefont
  {Gorlach}},\ }\bibfield  {title} {\enquote {\bibinfo {title}
  {Interaction-induced topological states of photon pairs},}\ }\href {\doibase
  10.1103/PhysRevA.102.013510} {\bibfield  {journal} {\bibinfo  {journal}
  {Phys. Rev. A}\ }\textbf {\bibinfo {volume} {102}},\ \bibinfo {pages}
  {013510} (\bibinfo {year} {2020})}\BibitemShut {NoStop}%
\bibitem [{\citenamefont {Salerno}\ \emph {et~al.}(2020)\citenamefont
  {Salerno}, \citenamefont {Palumbo}, \citenamefont {Goldman},\ and\
  \citenamefont {Di~Liberto}}]{Salerno2020}%
  \BibitemOpen
  \bibfield  {author} {\bibinfo {author} {\bibfnamefont {G.}~\bibnamefont
  {Salerno}}, \bibinfo {author} {\bibfnamefont {G.}~\bibnamefont {Palumbo}},
  \bibinfo {author} {\bibfnamefont {N.}~\bibnamefont {Goldman}}, \ and\
  \bibinfo {author} {\bibfnamefont {M.}~\bibnamefont {Di~Liberto}},\ }\bibfield
   {title} {\enquote {\bibinfo {title} {Interaction-induced lattices for bound
  states: Designing flat bands, quantized pumps, and higher-order topological
  insulators for doublons},}\ }\href {\doibase
  10.1103/PhysRevResearch.2.013348} {\bibfield  {journal} {\bibinfo  {journal}
  {Phys. Rev. Research}\ }\textbf {\bibinfo {volume} {2}},\ \bibinfo {pages}
  {013348} (\bibinfo {year} {2020})}\BibitemShut {NoStop}%
\bibitem [{\citenamefont {Zawadowski}\ and\ \citenamefont
  {Ruvalds}(1970)}]{Zawadowski1970}%
  \BibitemOpen
  \bibfield  {author} {\bibinfo {author} {\bibfnamefont {A.}~\bibnamefont
  {Zawadowski}}\ and\ \bibinfo {author} {\bibfnamefont {J.}~\bibnamefont
  {Ruvalds}},\ }\bibfield  {title} {\enquote {\bibinfo {title} {Indirect
  coupling and antiresonance of two optic phonons},}\ }\href {\doibase
  10.1103/PhysRevLett.24.1111} {\bibfield  {journal} {\bibinfo  {journal}
  {Phys. Rev. Lett.}\ }\textbf {\bibinfo {volume} {24}},\ \bibinfo {pages}
  {1111--1114} (\bibinfo {year} {1970})}\BibitemShut {NoStop}%
\bibitem [{\citenamefont {Ruvalds}\ and\ \citenamefont
  {Zawadowski}(1970)}]{Ruvalds1970}%
  \BibitemOpen
  \bibfield  {author} {\bibinfo {author} {\bibfnamefont {J.}~\bibnamefont
  {Ruvalds}}\ and\ \bibinfo {author} {\bibfnamefont {A.}~\bibnamefont
  {Zawadowski}},\ }\bibfield  {title} {\enquote {\bibinfo {title} {Two-phonon
  resonances and hybridization of the resonance with single-phonon states},}\
  }\href {\doibase 10.1103/PhysRevB.2.1172} {\bibfield  {journal} {\bibinfo
  {journal} {Phys. Rev. B}\ }\textbf {\bibinfo {volume} {2}},\ \bibinfo {pages}
  {1172--1175} (\bibinfo {year} {1970})}\BibitemShut {NoStop}%
\bibitem [{\citenamefont {Heilmann}\ \emph {et~al.}(1981)\citenamefont
  {Heilmann}, \citenamefont {Kjems}, \citenamefont {Endoh}, \citenamefont
  {Reiter}, \citenamefont {Shirane},\ and\ \citenamefont
  {Birgeneau}}]{Heilmann1981}%
  \BibitemOpen
  \bibfield  {author} {\bibinfo {author} {\bibfnamefont {I.~U.}\ \bibnamefont
  {Heilmann}}, \bibinfo {author} {\bibfnamefont {J.~K.}\ \bibnamefont {Kjems}},
  \bibinfo {author} {\bibfnamefont {Y.}~\bibnamefont {Endoh}}, \bibinfo
  {author} {\bibfnamefont {G.~F.}\ \bibnamefont {Reiter}}, \bibinfo {author}
  {\bibfnamefont {G.}~\bibnamefont {Shirane}}, \ and\ \bibinfo {author}
  {\bibfnamefont {R.~J.}\ \bibnamefont {Birgeneau}},\ }\bibfield  {title}
  {\enquote {\bibinfo {title} {One- and two-magnon excitations in a
  one-dimensional antiferromagnet in a magnetic field},}\ }\href {\doibase
  10.1103/PhysRevB.24.3939} {\bibfield  {journal} {\bibinfo  {journal} {Phys.
  Rev. B}\ }\textbf {\bibinfo {volume} {24}},\ \bibinfo {pages} {3939--3953}
  (\bibinfo {year} {1981})}\BibitemShut {NoStop}%
\bibitem [{\citenamefont {Osano}\ \emph {et~al.}(1982)\citenamefont {Osano},
  \citenamefont {Shiba},\ and\ \citenamefont {Endoh}}]{Osano1982}%
  \BibitemOpen
  \bibfield  {author} {\bibinfo {author} {\bibfnamefont {K.}~\bibnamefont
  {Osano}}, \bibinfo {author} {\bibfnamefont {H.}~\bibnamefont {Shiba}}, \ and\
  \bibinfo {author} {\bibfnamefont {Y.}~\bibnamefont {Endoh}},\ }\bibfield
  {title} {\enquote {\bibinfo {title} {{Theory of One-Magnon and Two-Magnon
  Spectra of TMMC in a Magnetic Field}},}\ }\href {\doibase 10.1143/PTP.67.995}
  {\bibfield  {journal} {\bibinfo  {journal} {Progress of Theoretical Physics}\
  }\textbf {\bibinfo {volume} {67}},\ \bibinfo {pages} {995--1014} (\bibinfo
  {year} {1982})},\ \Eprint
  {http://arxiv.org/abs/https://academic.oup.com/ptp/article-pdf/67/4/995/5216089/67-4-995.pdf}
  {https://academic.oup.com/ptp/article-pdf/67/4/995/5216089/67-4-995.pdf}
  \BibitemShut {NoStop}%
\bibitem [{\citenamefont {Endoh}\ \emph {et~al.}(1984)\citenamefont {Endoh},
  \citenamefont {Ajiro}, \citenamefont {Shiba},\ and\ \citenamefont
  {Yoshizawa}}]{Endoh1984}%
  \BibitemOpen
  \bibfield  {author} {\bibinfo {author} {\bibfnamefont {Y.}~\bibnamefont
  {Endoh}}, \bibinfo {author} {\bibfnamefont {Y.}~\bibnamefont {Ajiro}},
  \bibinfo {author} {\bibfnamefont {H.}~\bibnamefont {Shiba}}, \ and\ \bibinfo
  {author} {\bibfnamefont {H.}~\bibnamefont {Yoshizawa}},\ }\bibfield  {title}
  {\enquote {\bibinfo {title} {Resonant coupling between one- and two-magnon
  excitations in tetramethylamine manganese trichloride {(TMMC)}},}\ }\href
  {\doibase 10.1103/PhysRevB.30.4074} {\bibfield  {journal} {\bibinfo
  {journal} {Phys. Rev. B}\ }\textbf {\bibinfo {volume} {30}},\ \bibinfo
  {pages} {4074--4076} (\bibinfo {year} {1984})}\BibitemShut {NoStop}%
\bibitem [{\citenamefont {Inami}\ \emph {et~al.}(1997)\citenamefont {Inami},
  \citenamefont {Kakurai},\ and\ \citenamefont {Tanaka}}]{Inami1997}%
  \BibitemOpen
  \bibfield  {author} {\bibinfo {author} {\bibfnamefont {T.}~\bibnamefont
  {Inami}}, \bibinfo {author} {\bibfnamefont {K.}~\bibnamefont {Kakurai}}, \
  and\ \bibinfo {author} {\bibfnamefont {H.}~\bibnamefont {Tanaka}},\
  }\bibfield  {title} {\enquote {\bibinfo {title} {Observation of non-linear
  effects in a quasi-one-dimensional antiferromagnet: magnetic excitations
  in},}\ }\href {\doibase 10.1088/0953-8984/9/6/019} {\bibfield  {journal}
  {\bibinfo  {journal} {Journal of Physics: Condensed Matter}\ }\textbf
  {\bibinfo {volume} {9}},\ \bibinfo {pages} {1357--1368} (\bibinfo {year}
  {1997})}\BibitemShut {NoStop}%
\bibitem [{\citenamefont {Bai}\ \emph {et~al.}(2021{\natexlab{a}})\citenamefont
  {Bai}, \citenamefont {Zhang}, \citenamefont {Dun}, \citenamefont {Zhang},
  \citenamefont {Huang}, \citenamefont {Zhou}, \citenamefont {Stone},
  \citenamefont {Kolesnikov}, \citenamefont {Ye}, \citenamefont {Batista},\
  and\ \citenamefont {Mourigal}}]{Bai2021}%
  \BibitemOpen
  \bibfield  {author} {\bibinfo {author} {\bibfnamefont {X.}~\bibnamefont
  {Bai}}, \bibinfo {author} {\bibfnamefont {S.-S.}\ \bibnamefont {Zhang}},
  \bibinfo {author} {\bibfnamefont {Z.}~\bibnamefont {Dun}}, \bibinfo {author}
  {\bibfnamefont {H.}~\bibnamefont {Zhang}}, \bibinfo {author} {\bibfnamefont
  {Q.}~\bibnamefont {Huang}}, \bibinfo {author} {\bibfnamefont
  {H.}~\bibnamefont {Zhou}}, \bibinfo {author} {\bibfnamefont {M.~B.}\
  \bibnamefont {Stone}}, \bibinfo {author} {\bibfnamefont {A.~I.}\ \bibnamefont
  {Kolesnikov}}, \bibinfo {author} {\bibfnamefont {F.}~\bibnamefont {Ye}},
  \bibinfo {author} {\bibfnamefont {C.~D.}\ \bibnamefont {Batista}}, \ and\
  \bibinfo {author} {\bibfnamefont {M.}~\bibnamefont {Mourigal}},\ }\bibfield
  {title} {\enquote {\bibinfo {title} {Hybridized quadrupolar excitations in
  the spin-anisotropic frustrated magnet {FeI}$_2$},}\ }\href {\doibase
  10.1038/s41567-020-01110-1} {\bibfield  {journal} {\bibinfo  {journal}
  {Nature Physics}\ } (\bibinfo {year} {2021}{\natexlab{a}}),\
  10.1038/s41567-020-01110-1}\BibitemShut {NoStop}%
\bibitem [{\citenamefont {Legros}\ \emph {et~al.}(2021)\citenamefont {Legros},
  \citenamefont {Zhang}, \citenamefont {Bai}, \citenamefont {Zhang},
  \citenamefont {Dun}, \citenamefont {Phelan}, \citenamefont {Batista},
  \citenamefont {Mourigal},\ and\ \citenamefont {Armitage}}]{Legros2020}%
  \BibitemOpen
  \bibfield  {author} {\bibinfo {author} {\bibfnamefont {A.}~\bibnamefont
  {Legros}}, \bibinfo {author} {\bibfnamefont {S.-S.}\ \bibnamefont {Zhang}},
  \bibinfo {author} {\bibfnamefont {X.}~\bibnamefont {Bai}}, \bibinfo {author}
  {\bibfnamefont {H.}~\bibnamefont {Zhang}}, \bibinfo {author} {\bibfnamefont
  {Z.}~\bibnamefont {Dun}}, \bibinfo {author} {\bibfnamefont {W.~A.}\
  \bibnamefont {Phelan}}, \bibinfo {author} {\bibfnamefont {C.~D.}\
  \bibnamefont {Batista}}, \bibinfo {author} {\bibfnamefont {M.}~\bibnamefont
  {Mourigal}}, \ and\ \bibinfo {author} {\bibfnamefont {N.~P.}\ \bibnamefont
  {Armitage}},\ }\bibfield  {title} {\enquote {\bibinfo {title} {Observation of
  4- and 6-magnon bound states in the spin-anisotropic frustrated
  antiferromagnet {FeI}$_{2}$},}\ }\href {\doibase
  10.1103/PhysRevLett.127.267201} {\bibfield  {journal} {\bibinfo  {journal}
  {Phys. Rev. Lett.}\ }\textbf {\bibinfo {volume} {127}},\ \bibinfo {pages}
  {267201} (\bibinfo {year} {2021})}\BibitemShut {NoStop}%
\bibitem [{\citenamefont {Bai}\ \emph {et~al.}(2021{\natexlab{b}})\citenamefont
  {Bai}, \citenamefont {Zhang}, \citenamefont {Zhang}, \citenamefont {Dun},
  \citenamefont {Phelan}, \citenamefont {Garlea}, \citenamefont {Mourigal},\
  and\ \citenamefont {Batista}}]{Bai2021decay}%
  \BibitemOpen
  \bibfield  {author} {\bibinfo {author} {\bibfnamefont {X.}~\bibnamefont
  {Bai}}, \bibinfo {author} {\bibfnamefont {S.-S.}\ \bibnamefont {Zhang}},
  \bibinfo {author} {\bibfnamefont {H.}~\bibnamefont {Zhang}}, \bibinfo
  {author} {\bibfnamefont {Z.}~\bibnamefont {Dun}}, \bibinfo {author}
  {\bibfnamefont {W.~A.}\ \bibnamefont {Phelan}}, \bibinfo {author}
  {\bibfnamefont {V.~O.}\ \bibnamefont {Garlea}}, \bibinfo {author}
  {\bibfnamefont {M.}~\bibnamefont {Mourigal}}, \ and\ \bibinfo {author}
  {\bibfnamefont {C.~D.}\ \bibnamefont {Batista}},\ }\href@noop {} {\enquote
  {\bibinfo {title} {Instabilities of heavy magnons in an anisotropic
  magnet},}\ } (\bibinfo {year} {2021}{\natexlab{b}}),\ \Eprint
  {http://arxiv.org/abs/arXiv:2107.05694} {arXiv:2107.05694} \BibitemShut
  {NoStop}%
\bibitem [{\citenamefont {Lee}\ \emph {et~al.}(2015)\citenamefont {Lee},
  \citenamefont {Han},\ and\ \citenamefont {Lee}}]{Lee2015paramagnet}%
  \BibitemOpen
  \bibfield  {author} {\bibinfo {author} {\bibfnamefont {H.}~\bibnamefont
  {Lee}}, \bibinfo {author} {\bibfnamefont {J.~H.}\ \bibnamefont {Han}}, \ and\
  \bibinfo {author} {\bibfnamefont {P.~A.}\ \bibnamefont {Lee}},\ }\bibfield
  {title} {\enquote {\bibinfo {title} {Thermal hall effect of spins in a
  paramagnet},}\ }\href {\doibase 10.1103/PhysRevB.91.125413} {\bibfield
  {journal} {\bibinfo  {journal} {Phys. Rev. B}\ }\textbf {\bibinfo {volume}
  {91}},\ \bibinfo {pages} {125413} (\bibinfo {year} {2015})}\BibitemShut
  {NoStop}%
\bibitem [{\citenamefont {Kim}\ \emph {et~al.}(2016)\citenamefont {Kim},
  \citenamefont {Ochoa}, \citenamefont {Zarzuela},\ and\ \citenamefont
  {Tserkovnyak}}]{Kim2016Kane}%
  \BibitemOpen
  \bibfield  {author} {\bibinfo {author} {\bibfnamefont {S.~K.}\ \bibnamefont
  {Kim}}, \bibinfo {author} {\bibfnamefont {H.}~\bibnamefont {Ochoa}}, \bibinfo
  {author} {\bibfnamefont {R.}~\bibnamefont {Zarzuela}}, \ and\ \bibinfo
  {author} {\bibfnamefont {Y.}~\bibnamefont {Tserkovnyak}},\ }\bibfield
  {title} {\enquote {\bibinfo {title} {Realization of the
  {H}aldane-{K}ane-{M}ele model in a system of localized spins},}\ }\href
  {\doibase 10.1103/PhysRevLett.117.227201} {\bibfield  {journal} {\bibinfo
  {journal} {Phys. Rev. Lett.}\ }\textbf {\bibinfo {volume} {117}},\ \bibinfo
  {pages} {227201} (\bibinfo {year} {2016})}\BibitemShut {NoStop}%
\bibitem [{\citenamefont {Sonnenschein}\ and\ \citenamefont
  {Reuther}(2017)}]{Sonnenschein2017}%
  \BibitemOpen
  \bibfield  {author} {\bibinfo {author} {\bibfnamefont {J.}~\bibnamefont
  {Sonnenschein}}\ and\ \bibinfo {author} {\bibfnamefont {J.}~\bibnamefont
  {Reuther}},\ }\bibfield  {title} {\enquote {\bibinfo {title} {Topological
  spinon bands and vison excitations in spin-orbit coupled quantum spin
  liquids},}\ }\href {\doibase 10.1103/PhysRevB.96.235113} {\bibfield
  {journal} {\bibinfo  {journal} {Phys. Rev. B}\ }\textbf {\bibinfo {volume}
  {96}},\ \bibinfo {pages} {235113} (\bibinfo {year} {2017})}\BibitemShut
  {NoStop}%
\bibitem [{\citenamefont {Joshi}\ and\ \citenamefont
  {Schnyder}(2019)}]{Joshi2019}%
  \BibitemOpen
  \bibfield  {author} {\bibinfo {author} {\bibfnamefont {D.~G.}\ \bibnamefont
  {Joshi}}\ and\ \bibinfo {author} {\bibfnamefont {A.~P.}\ \bibnamefont
  {Schnyder}},\ }\bibfield  {title} {\enquote {\bibinfo {title} {{Z}$_{2}$
  topological quantum paramagnet on a honeycomb bilayer},}\ }\href {\doibase
  10.1103/PhysRevB.100.020407} {\bibfield  {journal} {\bibinfo  {journal}
  {Phys. Rev. B}\ }\textbf {\bibinfo {volume} {100}},\ \bibinfo {pages}
  {020407} (\bibinfo {year} {2019})}\BibitemShut {NoStop}%
\bibitem [{\citenamefont {Romh{\'{a}}nyi}\ \emph {et~al.}(2015)\citenamefont
  {Romh{\'{a}}nyi}, \citenamefont {Penc},\ and\ \citenamefont
  {Ganesh}}]{Romhanyi2015}%
  \BibitemOpen
  \bibfield  {author} {\bibinfo {author} {\bibfnamefont {J.}~\bibnamefont
  {Romh{\'{a}}nyi}}, \bibinfo {author} {\bibfnamefont {K.}~\bibnamefont
  {Penc}}, \ and\ \bibinfo {author} {\bibfnamefont {R.}~\bibnamefont
  {Ganesh}},\ }\bibfield  {title} {\enquote {\bibinfo {title} {Hall effect of
  triplons in a dimerized quantum magnet},}\ }\href {\doibase
  10.1038/ncomms7805} {\bibfield  {journal} {\bibinfo  {journal} {Nature
  Communications}\ }\textbf {\bibinfo {volume} {6}} (\bibinfo {year} {2015}),\
  10.1038/ncomms7805}\BibitemShut {NoStop}%
\bibitem [{\citenamefont {McClarty}\ \emph {et~al.}(2017)\citenamefont
  {McClarty}, \citenamefont {Kr{\"u}ger}, \citenamefont {Guidi}, \citenamefont
  {Parker}, \citenamefont {Refson}, \citenamefont {Parker}, \citenamefont
  {Prabhakaran},\ and\ \citenamefont {Coldea}}]{mcclarty2017topological}%
  \BibitemOpen
  \bibfield  {author} {\bibinfo {author} {\bibfnamefont {P.~A.}\ \bibnamefont
  {McClarty}}, \bibinfo {author} {\bibfnamefont {F.}~\bibnamefont
  {Kr{\"u}ger}}, \bibinfo {author} {\bibfnamefont {T.}~\bibnamefont {Guidi}},
  \bibinfo {author} {\bibfnamefont {S.}~\bibnamefont {Parker}}, \bibinfo
  {author} {\bibfnamefont {K.}~\bibnamefont {Refson}}, \bibinfo {author}
  {\bibfnamefont {A.}~\bibnamefont {Parker}}, \bibinfo {author} {\bibfnamefont
  {D.}~\bibnamefont {Prabhakaran}}, \ and\ \bibinfo {author} {\bibfnamefont
  {R.}~\bibnamefont {Coldea}},\ }\bibfield  {title} {\enquote {\bibinfo {title}
  {Topological triplon modes and bound states in a {S}hastry--{S}utherland
  magnet},}\ }\href@noop {} {\bibfield  {journal} {\bibinfo  {journal} {Nature
  Physics}\ }\textbf {\bibinfo {volume} {13}},\ \bibinfo {pages} {736--741}
  (\bibinfo {year} {2017})}\BibitemShut {NoStop}%
\bibitem [{\citenamefont {Anisimov}\ \emph {et~al.}(2019)\citenamefont
  {Anisimov}, \citenamefont {Aust}, \citenamefont {Khaliullin},\ and\
  \citenamefont {Daghofer}}]{Anisimov2019}%
  \BibitemOpen
  \bibfield  {author} {\bibinfo {author} {\bibfnamefont {P.~S.}\ \bibnamefont
  {Anisimov}}, \bibinfo {author} {\bibfnamefont {F.}~\bibnamefont {Aust}},
  \bibinfo {author} {\bibfnamefont {G.}~\bibnamefont {Khaliullin}}, \ and\
  \bibinfo {author} {\bibfnamefont {M.}~\bibnamefont {Daghofer}},\ }\bibfield
  {title} {\enquote {\bibinfo {title} {Nontrivial triplon topology and triplon
  liquid in kitaev-heisenberg-type excitonic magnets},}\ }\href {\doibase
  10.1103/PhysRevLett.122.177201} {\bibfield  {journal} {\bibinfo  {journal}
  {Phys. Rev. Lett.}\ }\textbf {\bibinfo {volume} {122}},\ \bibinfo {pages}
  {177201} (\bibinfo {year} {2019})}\BibitemShut {NoStop}%
\bibitem [{\citenamefont {Song}\ \emph {et~al.}(2020)\citenamefont {Song},
  \citenamefont {He}, \citenamefont {Vishwanath},\ and\ \citenamefont
  {Wang}}]{Song2020spinon}%
  \BibitemOpen
  \bibfield  {author} {\bibinfo {author} {\bibfnamefont {X.-Y.}\ \bibnamefont
  {Song}}, \bibinfo {author} {\bibfnamefont {Y.-C.}\ \bibnamefont {He}},
  \bibinfo {author} {\bibfnamefont {A.}~\bibnamefont {Vishwanath}}, \ and\
  \bibinfo {author} {\bibfnamefont {C.}~\bibnamefont {Wang}},\ }\bibfield
  {title} {\enquote {\bibinfo {title} {From spinon band topology to the
  symmetry quantum numbers of monopoles in {D}irac spin liquids},}\ }\href
  {\doibase 10.1103/PhysRevX.10.011033} {\bibfield  {journal} {\bibinfo
  {journal} {Phys. Rev. X}\ }\textbf {\bibinfo {volume} {10}},\ \bibinfo
  {pages} {011033} (\bibinfo {year} {2020})}\BibitemShut {NoStop}%
\bibitem [{\citenamefont {Bhowmick}\ and\ \citenamefont
  {Sengupta}(2021)}]{Bhowmick2021Weyltriplon}%
  \BibitemOpen
  \bibfield  {author} {\bibinfo {author} {\bibfnamefont {D.}~\bibnamefont
  {Bhowmick}}\ and\ \bibinfo {author} {\bibfnamefont {P.}~\bibnamefont
  {Sengupta}},\ }\bibfield  {title} {\enquote {\bibinfo {title} {Weyl triplons
  in {SrCu}$_2$({BO}$_{3}$)$_{2}$},}\ }\href {\doibase
  10.1103/PhysRevB.104.085121} {\bibfield  {journal} {\bibinfo  {journal}
  {Phys. Rev. B}\ }\textbf {\bibinfo {volume} {104}},\ \bibinfo {pages}
  {085121} (\bibinfo {year} {2021})}\BibitemShut {NoStop}%
\bibitem [{\citenamefont {Haldar}\ \emph {et~al.}(2021)\citenamefont {Haldar},
  \citenamefont {Massarelli},\ and\ \citenamefont {Paramekanti}}]{Haldar2021}%
  \BibitemOpen
  \bibfield  {author} {\bibinfo {author} {\bibfnamefont {A.}~\bibnamefont
  {Haldar}}, \bibinfo {author} {\bibfnamefont {G.}~\bibnamefont {Massarelli}},
  \ and\ \bibinfo {author} {\bibfnamefont {A.}~\bibnamefont {Paramekanti}},\
  }\bibfield  {title} {\enquote {\bibinfo {title} {Higher-order topology and
  corner triplon excitations in two-dimensional quantum spin-dimer models},}\
  }\href {\doibase 10.1103/PhysRevB.104.184403} {\bibfield  {journal} {\bibinfo
   {journal} {Phys. Rev. B}\ }\textbf {\bibinfo {volume} {104}},\ \bibinfo
  {pages} {184403} (\bibinfo {year} {2021})}\BibitemShut {NoStop}%
\bibitem [{\citenamefont {Meier}\ and\ \citenamefont {Loss}(2003)}]{Meier2003}%
  \BibitemOpen
  \bibfield  {author} {\bibinfo {author} {\bibfnamefont {F.}~\bibnamefont
  {Meier}}\ and\ \bibinfo {author} {\bibfnamefont {D.}~\bibnamefont {Loss}},\
  }\bibfield  {title} {\enquote {\bibinfo {title} {Magnetization transport and
  quantized spin conductance},}\ }\href {\doibase
  10.1103/PhysRevLett.90.167204} {\bibfield  {journal} {\bibinfo  {journal}
  {Phys. Rev. Lett.}\ }\textbf {\bibinfo {volume} {90}},\ \bibinfo {pages}
  {167204} (\bibinfo {year} {2003})}\BibitemShut {NoStop}%
\bibitem [{\citenamefont {Katsura}\ \emph {et~al.}(2010)\citenamefont
  {Katsura}, \citenamefont {Nagaosa},\ and\ \citenamefont {Lee}}]{Katsura2010}%
  \BibitemOpen
  \bibfield  {author} {\bibinfo {author} {\bibfnamefont {H.}~\bibnamefont
  {Katsura}}, \bibinfo {author} {\bibfnamefont {N.}~\bibnamefont {Nagaosa}}, \
  and\ \bibinfo {author} {\bibfnamefont {P.~A.}\ \bibnamefont {Lee}},\
  }\bibfield  {title} {\enquote {\bibinfo {title} {Theory of the thermal {H}all
  effect in quantum magnets},}\ }\href {\doibase
  10.1103/PhysRevLett.104.066403} {\bibfield  {journal} {\bibinfo  {journal}
  {Phys. Rev. Lett.}\ }\textbf {\bibinfo {volume} {104}},\ \bibinfo {pages}
  {066403} (\bibinfo {year} {2010})}\BibitemShut {NoStop}%
\bibitem [{\citenamefont {van Hoogdalem}\ \emph {et~al.}(2013)\citenamefont
  {van Hoogdalem}, \citenamefont {Tserkovnyak},\ and\ \citenamefont
  {Loss}}]{Hoogdalem2013}%
  \BibitemOpen
  \bibfield  {author} {\bibinfo {author} {\bibfnamefont {K.~A.}\ \bibnamefont
  {van Hoogdalem}}, \bibinfo {author} {\bibfnamefont {Y.}~\bibnamefont
  {Tserkovnyak}}, \ and\ \bibinfo {author} {\bibfnamefont {D.}~\bibnamefont
  {Loss}},\ }\bibfield  {title} {\enquote {\bibinfo {title} {Magnetic
  texture-induced thermal {H}all effects},}\ }\href {\doibase
  10.1103/PhysRevB.87.024402} {\bibfield  {journal} {\bibinfo  {journal} {Phys.
  Rev. B}\ }\textbf {\bibinfo {volume} {87}},\ \bibinfo {pages} {024402}
  (\bibinfo {year} {2013})}\BibitemShut {NoStop}%
\bibitem [{\citenamefont {Zhang}\ \emph {et~al.}(2013)\citenamefont {Zhang},
  \citenamefont {Ren}, \citenamefont {Wang},\ and\ \citenamefont
  {Li}}]{Zhang2013}%
  \BibitemOpen
  \bibfield  {author} {\bibinfo {author} {\bibfnamefont {L.}~\bibnamefont
  {Zhang}}, \bibinfo {author} {\bibfnamefont {J.}~\bibnamefont {Ren}}, \bibinfo
  {author} {\bibfnamefont {J.-S.}\ \bibnamefont {Wang}}, \ and\ \bibinfo
  {author} {\bibfnamefont {B.}~\bibnamefont {Li}},\ }\bibfield  {title}
  {\enquote {\bibinfo {title} {Topological magnon insulator in insulating
  ferromagnet},}\ }\href {\doibase 10.1103/PhysRevB.87.144101} {\bibfield
  {journal} {\bibinfo  {journal} {Phys. Rev. B}\ }\textbf {\bibinfo {volume}
  {87}},\ \bibinfo {pages} {144101} (\bibinfo {year} {2013})}\BibitemShut
  {NoStop}%
\bibitem [{\citenamefont {Shindou}\ \emph
  {et~al.}(2013{\natexlab{b}})\citenamefont {Shindou}, \citenamefont {Ohe},
  \citenamefont {Matsumoto}, \citenamefont {Murakami},\ and\ \citenamefont
  {Saitoh}}]{Shindou13b}%
  \BibitemOpen
  \bibfield  {author} {\bibinfo {author} {\bibfnamefont {R.}~\bibnamefont
  {Shindou}}, \bibinfo {author} {\bibfnamefont {J.-i.}\ \bibnamefont {Ohe}},
  \bibinfo {author} {\bibfnamefont {R.}~\bibnamefont {Matsumoto}}, \bibinfo
  {author} {\bibfnamefont {S.}~\bibnamefont {Murakami}}, \ and\ \bibinfo
  {author} {\bibfnamefont {E.}~\bibnamefont {Saitoh}},\ }\bibfield  {title}
  {\enquote {\bibinfo {title} {Chiral spin-wave edge modes in dipolar magnetic
  thin films},}\ }\href {\doibase 10.1103/PhysRevB.87.174402} {\bibfield
  {journal} {\bibinfo  {journal} {Phys. Rev. B}\ }\textbf {\bibinfo {volume}
  {87}},\ \bibinfo {pages} {174402} (\bibinfo {year}
  {2013}{\natexlab{b}})}\BibitemShut {NoStop}%
\bibitem [{\citenamefont {Mook}\ \emph
  {et~al.}(2014{\natexlab{a}})\citenamefont {Mook}, \citenamefont {Henk},\ and\
  \citenamefont {Mertig}}]{Mook2014edge}%
  \BibitemOpen
  \bibfield  {author} {\bibinfo {author} {\bibfnamefont {A.}~\bibnamefont
  {Mook}}, \bibinfo {author} {\bibfnamefont {J.}~\bibnamefont {Henk}}, \ and\
  \bibinfo {author} {\bibfnamefont {I.}~\bibnamefont {Mertig}},\ }\bibfield
  {title} {\enquote {\bibinfo {title} {Edge states in topological magnon
  insulators},}\ }\href {\doibase 10.1103/PhysRevB.90.024412} {\bibfield
  {journal} {\bibinfo  {journal} {Phys. Rev. B}\ }\textbf {\bibinfo {volume}
  {90}},\ \bibinfo {pages} {024412} (\bibinfo {year}
  {2014}{\natexlab{a}})}\BibitemShut {NoStop}%
\bibitem [{\citenamefont {Shindou}\ and\ \citenamefont
  {Ohe}(2014)}]{Shindou14}%
  \BibitemOpen
  \bibfield  {author} {\bibinfo {author} {\bibfnamefont {R.}~\bibnamefont
  {Shindou}}\ and\ \bibinfo {author} {\bibfnamefont {J.-i.}\ \bibnamefont
  {Ohe}},\ }\bibfield  {title} {\enquote {\bibinfo {title} {Magnetostatic wave
  analog of integer quantum {H}all state in patterned magnetic films},}\ }\href
  {\doibase 10.1103/PhysRevB.89.054412} {\bibfield  {journal} {\bibinfo
  {journal} {Phys. Rev. B}\ }\textbf {\bibinfo {volume} {89}},\ \bibinfo
  {pages} {054412} (\bibinfo {year} {2014})}\BibitemShut {NoStop}%
\bibitem [{\citenamefont {Mook}\ \emph
  {et~al.}(2015{\natexlab{b}})\citenamefont {Mook}, \citenamefont {Henk},\ and\
  \citenamefont {Mertig}}]{Mook2015interfaces}%
  \BibitemOpen
  \bibfield  {author} {\bibinfo {author} {\bibfnamefont {A.}~\bibnamefont
  {Mook}}, \bibinfo {author} {\bibfnamefont {J.}~\bibnamefont {Henk}}, \ and\
  \bibinfo {author} {\bibfnamefont {I.}~\bibnamefont {Mertig}},\ }\bibfield
  {title} {\enquote {\bibinfo {title} {Topologically nontrivial magnons at an
  interface of two kagome ferromagnets},}\ }\href {\doibase
  10.1103/PhysRevB.91.224411} {\bibfield  {journal} {\bibinfo  {journal} {Phys.
  Rev. B}\ }\textbf {\bibinfo {volume} {91}},\ \bibinfo {pages} {224411}
  (\bibinfo {year} {2015}{\natexlab{b}})}\BibitemShut {NoStop}%
\bibitem [{\citenamefont {Owerre}(2016)}]{Owerre2016a}%
  \BibitemOpen
  \bibfield  {author} {\bibinfo {author} {\bibfnamefont {S.~A.}\ \bibnamefont
  {Owerre}},\ }\bibfield  {title} {\enquote {\bibinfo {title} {A first
  theoretical realization of honeycomb topological magnon insulator},}\ }\href
  {\doibase 10.1088/0953-8984/28/38/386001} {\bibfield  {journal} {\bibinfo
  {journal} {J. Phys.: Condens. Matter}\ }\textbf {\bibinfo {volume} {28}},\
  \bibinfo {pages} {386001} (\bibinfo {year} {2016})}\BibitemShut {NoStop}%
\bibitem [{\citenamefont {Mook}\ \emph
  {et~al.}(2016{\natexlab{a}})\citenamefont {Mook}, \citenamefont {Henk},\ and\
  \citenamefont {Mertig}}]{Mook2016Weyl}%
  \BibitemOpen
  \bibfield  {author} {\bibinfo {author} {\bibfnamefont {A.}~\bibnamefont
  {Mook}}, \bibinfo {author} {\bibfnamefont {J.}~\bibnamefont {Henk}}, \ and\
  \bibinfo {author} {\bibfnamefont {I.}~\bibnamefont {Mertig}},\ }\bibfield
  {title} {\enquote {\bibinfo {title} {Tunable magnon {W}eyl points in
  ferromagnetic pyrochlores},}\ }\href {\doibase
  10.1103/PhysRevLett.117.157204} {\bibfield  {journal} {\bibinfo  {journal}
  {Phys. Rev. Lett.}\ }\textbf {\bibinfo {volume} {117}},\ \bibinfo {pages}
  {157204} (\bibinfo {year} {2016}{\natexlab{a}})}\BibitemShut {NoStop}%
\bibitem [{\citenamefont {Xu}\ \emph {et~al.}(2016)\citenamefont {Xu},
  \citenamefont {Ohtsuki},\ and\ \citenamefont {Shindou}}]{Xu2016}%
  \BibitemOpen
  \bibfield  {author} {\bibinfo {author} {\bibfnamefont {B.}~\bibnamefont
  {Xu}}, \bibinfo {author} {\bibfnamefont {T.}~\bibnamefont {Ohtsuki}}, \ and\
  \bibinfo {author} {\bibfnamefont {R.}~\bibnamefont {Shindou}},\ }\bibfield
  {title} {\enquote {\bibinfo {title} {Integer quantum magnon hall
  plateau-plateau transition in a spin-ice model},}\ }\href {\doibase
  10.1103/PhysRevB.94.220403} {\bibfield  {journal} {\bibinfo  {journal} {Phys.
  Rev. B}\ }\textbf {\bibinfo {volume} {94}},\ \bibinfo {pages} {220403}
  (\bibinfo {year} {2016})}\BibitemShut {NoStop}%
\bibitem [{\citenamefont {Nakata}\ \emph
  {et~al.}(2017{\natexlab{a}})\citenamefont {Nakata}, \citenamefont
  {Klinovaja},\ and\ \citenamefont {Loss}}]{Nakata2017}%
  \BibitemOpen
  \bibfield  {author} {\bibinfo {author} {\bibfnamefont {K.}~\bibnamefont
  {Nakata}}, \bibinfo {author} {\bibfnamefont {J.}~\bibnamefont {Klinovaja}}, \
  and\ \bibinfo {author} {\bibfnamefont {D.}~\bibnamefont {Loss}},\ }\bibfield
  {title} {\enquote {\bibinfo {title} {Magnonic quantum {H}all effect and
  {W}iedemann-{F}ranz law},}\ }\href {\doibase 10.1103/PhysRevB.95.125429}
  {\bibfield  {journal} {\bibinfo  {journal} {Phys. Rev. B}\ }\textbf {\bibinfo
  {volume} {95}},\ \bibinfo {pages} {125429} (\bibinfo {year}
  {2017}{\natexlab{a}})}\BibitemShut {NoStop}%
\bibitem [{\citenamefont {Nakata}\ \emph
  {et~al.}(2017{\natexlab{b}})\citenamefont {Nakata}, \citenamefont {Kim},
  \citenamefont {Klinovaja},\ and\ \citenamefont {Loss}}]{Nakata2017AFM}%
  \BibitemOpen
  \bibfield  {author} {\bibinfo {author} {\bibfnamefont {K.}~\bibnamefont
  {Nakata}}, \bibinfo {author} {\bibfnamefont {S.~K.}\ \bibnamefont {Kim}},
  \bibinfo {author} {\bibfnamefont {J.}~\bibnamefont {Klinovaja}}, \ and\
  \bibinfo {author} {\bibfnamefont {D.}~\bibnamefont {Loss}},\ }\bibfield
  {title} {\enquote {\bibinfo {title} {Magnonic topological insulators in
  antiferromagnets},}\ }\href {\doibase 10.1103/PhysRevB.96.224414} {\bibfield
  {journal} {\bibinfo  {journal} {Phys. Rev. B}\ }\textbf {\bibinfo {volume}
  {96}},\ \bibinfo {pages} {224414} (\bibinfo {year}
  {2017}{\natexlab{b}})}\BibitemShut {NoStop}%
\bibitem [{\citenamefont {Mook}\ \emph
  {et~al.}(2017{\natexlab{a}})\citenamefont {Mook}, \citenamefont {Henk},\ and\
  \citenamefont {Mertig}}]{Mook2017nodal}%
  \BibitemOpen
  \bibfield  {author} {\bibinfo {author} {\bibfnamefont {A.}~\bibnamefont
  {Mook}}, \bibinfo {author} {\bibfnamefont {J.}~\bibnamefont {Henk}}, \ and\
  \bibinfo {author} {\bibfnamefont {I.}~\bibnamefont {Mertig}},\ }\bibfield
  {title} {\enquote {\bibinfo {title} {Magnon nodal-line semimetals and
  drumhead surface states in anisotropic pyrochlore ferromagnets},}\ }\href
  {\doibase 10.1103/PhysRevB.95.014418} {\bibfield  {journal} {\bibinfo
  {journal} {Phys. Rev. B}\ }\textbf {\bibinfo {volume} {95}},\ \bibinfo
  {pages} {014418} (\bibinfo {year} {2017}{\natexlab{a}})}\BibitemShut
  {NoStop}%
\bibitem [{\citenamefont {Li}\ and\ \citenamefont
  {Kovalev}(2018)}]{Li2018chiral}%
  \BibitemOpen
  \bibfield  {author} {\bibinfo {author} {\bibfnamefont {B.}~\bibnamefont
  {Li}}\ and\ \bibinfo {author} {\bibfnamefont {A.~A.}\ \bibnamefont
  {Kovalev}},\ }\bibfield  {title} {\enquote {\bibinfo {title} {Chiral
  topological insulator of magnons},}\ }\href {\doibase
  10.1103/PhysRevB.97.174413} {\bibfield  {journal} {\bibinfo  {journal} {Phys.
  Rev. B}\ }\textbf {\bibinfo {volume} {97}},\ \bibinfo {pages} {174413}
  (\bibinfo {year} {2018})}\BibitemShut {NoStop}%
\bibitem [{\citenamefont {Mook}\ \emph {et~al.}(2018)\citenamefont {Mook},
  \citenamefont {G\"obel}, \citenamefont {Henk},\ and\ \citenamefont
  {Mertig}}]{Mook2018duality}%
  \BibitemOpen
  \bibfield  {author} {\bibinfo {author} {\bibfnamefont {A.}~\bibnamefont
  {Mook}}, \bibinfo {author} {\bibfnamefont {B.}~\bibnamefont {G\"obel}},
  \bibinfo {author} {\bibfnamefont {J.}~\bibnamefont {Henk}}, \ and\ \bibinfo
  {author} {\bibfnamefont {I.}~\bibnamefont {Mertig}},\ }\bibfield  {title}
  {\enquote {\bibinfo {title} {Taking an electron-magnon duality shortcut from
  electron to magnon transport},}\ }\href {\doibase 10.1103/PhysRevB.97.140401}
  {\bibfield  {journal} {\bibinfo  {journal} {Phys. Rev. B}\ }\textbf {\bibinfo
  {volume} {97}},\ \bibinfo {pages} {140401} (\bibinfo {year}
  {2018})}\BibitemShut {NoStop}%
\bibitem [{\citenamefont {D\'{\i}az}\ \emph {et~al.}(2019)\citenamefont
  {D\'{\i}az}, \citenamefont {Klinovaja},\ and\ \citenamefont
  {Loss}}]{Diaz2019AFM}%
  \BibitemOpen
  \bibfield  {author} {\bibinfo {author} {\bibfnamefont {S.~A.}\ \bibnamefont
  {D\'{\i}az}}, \bibinfo {author} {\bibfnamefont {J.}~\bibnamefont
  {Klinovaja}}, \ and\ \bibinfo {author} {\bibfnamefont {D.}~\bibnamefont
  {Loss}},\ }\bibfield  {title} {\enquote {\bibinfo {title} {Topological
  magnons and edge states in antiferromagnetic skyrmion crystals},}\ }\href
  {\doibase 10.1103/PhysRevLett.122.187203} {\bibfield  {journal} {\bibinfo
  {journal} {Phys. Rev. Lett.}\ }\textbf {\bibinfo {volume} {122}},\ \bibinfo
  {pages} {187203} (\bibinfo {year} {2019})}\BibitemShut {NoStop}%
\bibitem [{\citenamefont {Kondo}\ \emph
  {et~al.}(2019{\natexlab{a}})\citenamefont {Kondo}, \citenamefont {Akagi},\
  and\ \citenamefont {Katsura}}]{Kondo20192d}%
  \BibitemOpen
  \bibfield  {author} {\bibinfo {author} {\bibfnamefont {H.}~\bibnamefont
  {Kondo}}, \bibinfo {author} {\bibfnamefont {Y.}~\bibnamefont {Akagi}}, \ and\
  \bibinfo {author} {\bibfnamefont {H.}~\bibnamefont {Katsura}},\ }\bibfield
  {title} {\enquote {\bibinfo {title} {${Z}_{2}$ topological invariant for
  magnon spin {H}all systems},}\ }\href {\doibase 10.1103/PhysRevB.99.041110}
  {\bibfield  {journal} {\bibinfo  {journal} {Phys. Rev. B}\ }\textbf {\bibinfo
  {volume} {99}},\ \bibinfo {pages} {041110} (\bibinfo {year}
  {2019}{\natexlab{a}})}\BibitemShut {NoStop}%
\bibitem [{\citenamefont {Kondo}\ \emph
  {et~al.}(2019{\natexlab{b}})\citenamefont {Kondo}, \citenamefont {Akagi},\
  and\ \citenamefont {Katsura}}]{Kondo2019}%
  \BibitemOpen
  \bibfield  {author} {\bibinfo {author} {\bibfnamefont {H.}~\bibnamefont
  {Kondo}}, \bibinfo {author} {\bibfnamefont {Y.}~\bibnamefont {Akagi}}, \ and\
  \bibinfo {author} {\bibfnamefont {H.}~\bibnamefont {Katsura}},\ }\bibfield
  {title} {\enquote {\bibinfo {title} {Three-dimensional topological magnon
  systems},}\ }\href {\doibase 10.1103/PhysRevB.100.144401} {\bibfield
  {journal} {\bibinfo  {journal} {Phys. Rev. B}\ }\textbf {\bibinfo {volume}
  {100}},\ \bibinfo {pages} {144401} (\bibinfo {year}
  {2019}{\natexlab{b}})}\BibitemShut {NoStop}%
\bibitem [{\citenamefont {Mook}\ \emph
  {et~al.}(2019{\natexlab{a}})\citenamefont {Mook}, \citenamefont {Henk},\ and\
  \citenamefont {Mertig}}]{Mook2019coplanar}%
  \BibitemOpen
  \bibfield  {author} {\bibinfo {author} {\bibfnamefont {A.}~\bibnamefont
  {Mook}}, \bibinfo {author} {\bibfnamefont {J.}~\bibnamefont {Henk}}, \ and\
  \bibinfo {author} {\bibfnamefont {I.}~\bibnamefont {Mertig}},\ }\bibfield
  {title} {\enquote {\bibinfo {title} {Thermal {H}all effect in noncollinear
  coplanar insulating antiferromagnets},}\ }\href {\doibase
  10.1103/PhysRevB.99.014427} {\bibfield  {journal} {\bibinfo  {journal} {Phys.
  Rev. B}\ }\textbf {\bibinfo {volume} {99}},\ \bibinfo {pages} {014427}
  (\bibinfo {year} {2019}{\natexlab{a}})}\BibitemShut {NoStop}%
\bibitem [{\citenamefont {Kim}\ \emph {et~al.}(2019)\citenamefont {Kim},
  \citenamefont {Nakata}, \citenamefont {Loss},\ and\ \citenamefont
  {Tserkovnyak}}]{Kim2019SkyrmionHallFerri}%
  \BibitemOpen
  \bibfield  {author} {\bibinfo {author} {\bibfnamefont {S.~K.}\ \bibnamefont
  {Kim}}, \bibinfo {author} {\bibfnamefont {K.}~\bibnamefont {Nakata}},
  \bibinfo {author} {\bibfnamefont {D.}~\bibnamefont {Loss}}, \ and\ \bibinfo
  {author} {\bibfnamefont {Y.}~\bibnamefont {Tserkovnyak}},\ }\bibfield
  {title} {\enquote {\bibinfo {title} {Tunable magnonic thermal {H}all effect
  in skyrmion crystal phases of ferrimagnets},}\ }\href {\doibase
  10.1103/PhysRevLett.122.057204} {\bibfield  {journal} {\bibinfo  {journal}
  {Phys. Rev. Lett.}\ }\textbf {\bibinfo {volume} {122}},\ \bibinfo {pages}
  {057204} (\bibinfo {year} {2019})}\BibitemShut {NoStop}%
\bibitem [{\citenamefont {Malki}\ and\ \citenamefont
  {Uhrig}(2019)}]{Malki2019}%
  \BibitemOpen
  \bibfield  {author} {\bibinfo {author} {\bibfnamefont {M.}~\bibnamefont
  {Malki}}\ and\ \bibinfo {author} {\bibfnamefont {G.~S.}\ \bibnamefont
  {Uhrig}},\ }\bibfield  {title} {\enquote {\bibinfo {title} {Topological
  magnon bands for magnonics},}\ }\href {\doibase 10.1103/PhysRevB.99.174412}
  {\bibfield  {journal} {\bibinfo  {journal} {Phys. Rev. B}\ }\textbf {\bibinfo
  {volume} {99}},\ \bibinfo {pages} {174412} (\bibinfo {year}
  {2019})}\BibitemShut {NoStop}%
\bibitem [{\citenamefont {D\'{\i}az}\ \emph {et~al.}(2020)\citenamefont
  {D\'{\i}az}, \citenamefont {Hirosawa}, \citenamefont {Klinovaja},\ and\
  \citenamefont {Loss}}]{Diaz2020FM}%
  \BibitemOpen
  \bibfield  {author} {\bibinfo {author} {\bibfnamefont {S.~A.}\ \bibnamefont
  {D\'{\i}az}}, \bibinfo {author} {\bibfnamefont {T.}~\bibnamefont {Hirosawa}},
  \bibinfo {author} {\bibfnamefont {J.}~\bibnamefont {Klinovaja}}, \ and\
  \bibinfo {author} {\bibfnamefont {D.}~\bibnamefont {Loss}},\ }\bibfield
  {title} {\enquote {\bibinfo {title} {Chiral magnonic edge states in
  ferromagnetic skyrmion crystals controlled by magnetic fields},}\ }\href
  {\doibase 10.1103/PhysRevResearch.2.013231} {\bibfield  {journal} {\bibinfo
  {journal} {Phys. Rev. Research}\ }\textbf {\bibinfo {volume} {2}},\ \bibinfo
  {pages} {013231} (\bibinfo {year} {2020})}\BibitemShut {NoStop}%
\bibitem [{\citenamefont {Kondo}\ \emph {et~al.}(2020)\citenamefont {Kondo},
  \citenamefont {Akagi},\ and\ \citenamefont {Katsura}}]{Kondo2020NH}%
  \BibitemOpen
  \bibfield  {author} {\bibinfo {author} {\bibfnamefont {H.}~\bibnamefont
  {Kondo}}, \bibinfo {author} {\bibfnamefont {Y.}~\bibnamefont {Akagi}}, \ and\
  \bibinfo {author} {\bibfnamefont {H.}~\bibnamefont {Katsura}},\ }\bibfield
  {title} {\enquote {\bibinfo {title} {{Non-Hermiticity and topological
  invariants of magnon Bogoliubov–de Gennes systems}},}\ }\href {\doibase
  10.1093/ptep/ptaa151} {\bibfield  {journal} {\bibinfo  {journal} {Progress of
  Theoretical and Experimental Physics}\ }\textbf {\bibinfo {volume} {2020}}
  (\bibinfo {year} {2020}),\ 10.1093/ptep/ptaa151},\ \bibinfo {note} {12A104},\
  \Eprint
  {http://arxiv.org/abs/https://academic.oup.com/ptep/article-pdf/2020/12/12A104/35937396/ptaa151.pdf}
  {https://academic.oup.com/ptep/article-pdf/2020/12/12A104/35937396/ptaa151.pdf}
  \BibitemShut {NoStop}%
\bibitem [{\citenamefont {Hirosawa}\ \emph {et~al.}(2020)\citenamefont
  {Hirosawa}, \citenamefont {D\'{\i}az}, \citenamefont {Klinovaja},\ and\
  \citenamefont {Loss}}]{Hirosawa2020}%
  \BibitemOpen
  \bibfield  {author} {\bibinfo {author} {\bibfnamefont {T.}~\bibnamefont
  {Hirosawa}}, \bibinfo {author} {\bibfnamefont {S.~A.}\ \bibnamefont
  {D\'{\i}az}}, \bibinfo {author} {\bibfnamefont {J.}~\bibnamefont
  {Klinovaja}}, \ and\ \bibinfo {author} {\bibfnamefont {D.}~\bibnamefont
  {Loss}},\ }\bibfield  {title} {\enquote {\bibinfo {title} {Magnonic
  quadrupole topological insulator in antiskyrmion crystals},}\ }\href
  {\doibase 10.1103/PhysRevLett.125.207204} {\bibfield  {journal} {\bibinfo
  {journal} {Phys. Rev. Lett.}\ }\textbf {\bibinfo {volume} {125}},\ \bibinfo
  {pages} {207204} (\bibinfo {year} {2020})}\BibitemShut {NoStop}%
\bibitem [{\citenamefont {Corticelli}\ \emph
  {et~al.}(2022{\natexlab{a}})\citenamefont {Corticelli}, \citenamefont
  {Moessner},\ and\ \citenamefont {McClarty}}]{Corticelli2022}%
  \BibitemOpen
  \bibfield  {author} {\bibinfo {author} {\bibfnamefont {A.}~\bibnamefont
  {Corticelli}}, \bibinfo {author} {\bibfnamefont {R.}~\bibnamefont
  {Moessner}}, \ and\ \bibinfo {author} {\bibfnamefont {P.~A.}\ \bibnamefont
  {McClarty}},\ }\bibfield  {title} {\enquote {\bibinfo {title} {Spin-space
  groups and magnon band topology},}\ }\href {\doibase
  10.1103/PhysRevB.105.064430} {\bibfield  {journal} {\bibinfo  {journal}
  {Phys. Rev. B}\ }\textbf {\bibinfo {volume} {105}},\ \bibinfo {pages}
  {064430} (\bibinfo {year} {2022}{\natexlab{a}})}\BibitemShut {NoStop}%
\bibitem [{\citenamefont {Lovesey}\ and\ \citenamefont
  {Springer}(1977)}]{Lovesey1977}%
  \BibitemOpen
  \bibinfo {editor} {\bibfnamefont {S.~W.}\ \bibnamefont {Lovesey}}\ and\
  \bibinfo {editor} {\bibfnamefont {T.}~\bibnamefont {Springer}},\ eds.,\ \href
  {\doibase 10.1007/978-3-642-81113-5} {\emph {\bibinfo {title} {Dynamics of
  Solids and Liquids by Neutron Scattering}}}\ (\bibinfo  {publisher} {Springer
  Berlin Heidelberg},\ \bibinfo {year} {1977})\BibitemShut {NoStop}%
\bibitem [{\citenamefont {Nag}\ \emph {et~al.}(2021)\citenamefont {Nag},
  \citenamefont {Nocera}, \citenamefont {Agrestini}, \citenamefont
  {Garcia-Fernandez}, \citenamefont {Walters}, \citenamefont {Cheong},
  \citenamefont {Johnston},\ and\ \citenamefont {Zhou}}]{Nag2021}%
  \BibitemOpen
  \bibfield  {author} {\bibinfo {author} {\bibfnamefont {A.}~\bibnamefont
  {Nag}}, \bibinfo {author} {\bibfnamefont {A.}~\bibnamefont {Nocera}},
  \bibinfo {author} {\bibfnamefont {S.}~\bibnamefont {Agrestini}}, \bibinfo
  {author} {\bibfnamefont {M.}~\bibnamefont {Garcia-Fernandez}}, \bibinfo
  {author} {\bibfnamefont {A.~C.}\ \bibnamefont {Walters}}, \bibinfo {author}
  {\bibfnamefont {S.-W.}\ \bibnamefont {Cheong}}, \bibinfo {author}
  {\bibfnamefont {S.}~\bibnamefont {Johnston}}, \ and\ \bibinfo {author}
  {\bibfnamefont {K.-J.}\ \bibnamefont {Zhou}},\ }\href@noop {} {\enquote
  {\bibinfo {title} {Quadrupolar magnetic excitations in an isotropic spin-1
  antiferromagnet},}\ } (\bibinfo {year} {2021}),\ \Eprint
  {http://arxiv.org/abs/arXiv:2111.03625} {arXiv:2111.03625} \BibitemShut
  {NoStop}%
\bibitem [{\citenamefont {Kecke}\ \emph {et~al.}(2007)\citenamefont {Kecke},
  \citenamefont {Momoi},\ and\ \citenamefont {Furusaki}}]{Kecke2007}%
  \BibitemOpen
  \bibfield  {author} {\bibinfo {author} {\bibfnamefont {L.}~\bibnamefont
  {Kecke}}, \bibinfo {author} {\bibfnamefont {T.}~\bibnamefont {Momoi}}, \ and\
  \bibinfo {author} {\bibfnamefont {A.}~\bibnamefont {Furusaki}},\ }\bibfield
  {title} {\enquote {\bibinfo {title} {Multimagnon bound states in the
  frustrated ferromagnetic one-dimensional chain},}\ }\href {\doibase
  10.1103/PhysRevB.76.060407} {\bibfield  {journal} {\bibinfo  {journal} {Phys.
  Rev. B}\ }\textbf {\bibinfo {volume} {76}},\ \bibinfo {pages} {060407}
  (\bibinfo {year} {2007})}\BibitemShut {NoStop}%
\bibitem [{\citenamefont {Rastelli}(2011)}]{Rastelli2011}%
  \BibitemOpen
  \bibfield  {author} {\bibinfo {author} {\bibfnamefont {E.}~\bibnamefont
  {Rastelli}},\ }\href {\doibase 10.1142/8189} {\emph {\bibinfo {title}
  {Statistical Mechanics of Magnetic Excitations}}}\ (\bibinfo  {publisher}
  {{WORLD} {SCIENTIFIC}},\ \bibinfo {year} {2011})\BibitemShut {NoStop}%
\bibitem [{\citenamefont {Chiu-Tsao}\ \emph {et~al.}(1975)\citenamefont
  {Chiu-Tsao}, \citenamefont {Levy},\ and\ \citenamefont
  {Paulson}}]{ChiuTsao1975}%
  \BibitemOpen
  \bibfield  {author} {\bibinfo {author} {\bibfnamefont {S.~T.}\ \bibnamefont
  {Chiu-Tsao}}, \bibinfo {author} {\bibfnamefont {P.~M.}\ \bibnamefont {Levy}},
  \ and\ \bibinfo {author} {\bibfnamefont {C.}~\bibnamefont {Paulson}},\
  }\bibfield  {title} {\enquote {\bibinfo {title} {Elementary excitations of
  high-degree pair interactions: The two-spin---deviation spectra for a spin-1
  ferromagnet},}\ }\href {\doibase 10.1103/PhysRevB.12.1819} {\bibfield
  {journal} {\bibinfo  {journal} {Phys. Rev. B}\ }\textbf {\bibinfo {volume}
  {12}},\ \bibinfo {pages} {1819--1838} (\bibinfo {year} {1975})}\BibitemShut
  {NoStop}%
\bibitem [{\citenamefont {Chiu-Tsao}\ and\ \citenamefont
  {Levy}(1976)}]{ChiuTsao1976}%
  \BibitemOpen
  \bibfield  {author} {\bibinfo {author} {\bibfnamefont {S.-T.}\ \bibnamefont
  {Chiu-Tsao}}\ and\ \bibinfo {author} {\bibfnamefont {P.~M.}\ \bibnamefont
  {Levy}},\ }\bibfield  {title} {\enquote {\bibinfo {title} {Excitation spectra
  for spin-3/2 systems with high-degree pair interactions},}\ }\href {\doibase
  10.1103/PhysRevB.13.3046} {\bibfield  {journal} {\bibinfo  {journal} {Phys.
  Rev. B}\ }\textbf {\bibinfo {volume} {13}},\ \bibinfo {pages} {3046--3055}
  (\bibinfo {year} {1976})}\BibitemShut {NoStop}%
\bibitem [{\citenamefont {Momoi}\ \emph {et~al.}(2006)\citenamefont {Momoi},
  \citenamefont {Sindzingre},\ and\ \citenamefont {Shannon}}]{Momoi2006}%
  \BibitemOpen
  \bibfield  {author} {\bibinfo {author} {\bibfnamefont {T.}~\bibnamefont
  {Momoi}}, \bibinfo {author} {\bibfnamefont {P.}~\bibnamefont {Sindzingre}}, \
  and\ \bibinfo {author} {\bibfnamefont {N.}~\bibnamefont {Shannon}},\
  }\bibfield  {title} {\enquote {\bibinfo {title} {Octupolar order in the
  multiple spin exchange model on a triangular lattice},}\ }\href {\doibase
  10.1103/PhysRevLett.97.257204} {\bibfield  {journal} {\bibinfo  {journal}
  {Phys. Rev. Lett.}\ }\textbf {\bibinfo {volume} {97}},\ \bibinfo {pages}
  {257204} (\bibinfo {year} {2006})}\BibitemShut {NoStop}%
\bibitem [{\citenamefont {Sudan}\ \emph {et~al.}(2009)\citenamefont {Sudan},
  \citenamefont {L\"uscher},\ and\ \citenamefont {L\"auchli}}]{Sudan2009}%
  \BibitemOpen
  \bibfield  {author} {\bibinfo {author} {\bibfnamefont {J.}~\bibnamefont
  {Sudan}}, \bibinfo {author} {\bibfnamefont {A.}~\bibnamefont {L\"uscher}}, \
  and\ \bibinfo {author} {\bibfnamefont {A.~M.}\ \bibnamefont {L\"auchli}},\
  }\bibfield  {title} {\enquote {\bibinfo {title} {Emergent multipolar spin
  correlations in a fluctuating spiral: The frustrated ferromagnetic
  spin-$\frac{1}{2}$ heisenberg chain in a magnetic field},}\ }\href {\doibase
  10.1103/PhysRevB.80.140402} {\bibfield  {journal} {\bibinfo  {journal} {Phys.
  Rev. B}\ }\textbf {\bibinfo {volume} {80}},\ \bibinfo {pages} {140402}
  (\bibinfo {year} {2009})}\BibitemShut {NoStop}%
\bibitem [{\citenamefont {T\'oth}\ \emph {et~al.}(2012)\citenamefont {T\'oth},
  \citenamefont {L\"auchli}, \citenamefont {Mila},\ and\ \citenamefont
  {Penc}}]{Toth2012}%
  \BibitemOpen
  \bibfield  {author} {\bibinfo {author} {\bibfnamefont {T.~A.}\ \bibnamefont
  {T\'oth}}, \bibinfo {author} {\bibfnamefont {A.~M.}\ \bibnamefont
  {L\"auchli}}, \bibinfo {author} {\bibfnamefont {F.}~\bibnamefont {Mila}}, \
  and\ \bibinfo {author} {\bibfnamefont {K.}~\bibnamefont {Penc}},\ }\bibfield
  {title} {\enquote {\bibinfo {title} {Competition between two- and
  three-sublattice ordering for $s=1$ spins on the square lattice},}\ }\href
  {\doibase 10.1103/PhysRevB.85.140403} {\bibfield  {journal} {\bibinfo
  {journal} {Phys. Rev. B}\ }\textbf {\bibinfo {volume} {85}},\ \bibinfo
  {pages} {140403} (\bibinfo {year} {2012})}\BibitemShut {NoStop}%
\bibitem [{\citenamefont {Shannon}\ \emph {et~al.}(2006)\citenamefont
  {Shannon}, \citenamefont {Momoi},\ and\ \citenamefont
  {Sindzingre}}]{Shannon2006}%
  \BibitemOpen
  \bibfield  {author} {\bibinfo {author} {\bibfnamefont {N.}~\bibnamefont
  {Shannon}}, \bibinfo {author} {\bibfnamefont {T.}~\bibnamefont {Momoi}}, \
  and\ \bibinfo {author} {\bibfnamefont {P.}~\bibnamefont {Sindzingre}},\
  }\bibfield  {title} {\enquote {\bibinfo {title} {Nematic order in square
  lattice frustrated ferromagnets},}\ }\href {\doibase
  10.1103/PhysRevLett.96.027213} {\bibfield  {journal} {\bibinfo  {journal}
  {Phys. Rev. Lett.}\ }\textbf {\bibinfo {volume} {96}},\ \bibinfo {pages}
  {027213} (\bibinfo {year} {2006})}\BibitemShut {NoStop}%
\bibitem [{\citenamefont {Holstein}\ and\ \citenamefont
  {Primakoff}(1940)}]{Holstein1940}%
  \BibitemOpen
  \bibfield  {author} {\bibinfo {author} {\bibfnamefont {T.}~\bibnamefont
  {Holstein}}\ and\ \bibinfo {author} {\bibfnamefont {H.}~\bibnamefont
  {Primakoff}},\ }\bibfield  {title} {\enquote {\bibinfo {title} {Field
  dependence of the intrinsic domain magnetization of a ferromagnet},}\ }\href
  {\doibase 10.1103/PhysRev.58.1098} {\bibfield  {journal} {\bibinfo  {journal}
  {Phys. Rev.}\ }\textbf {\bibinfo {volume} {58}},\ \bibinfo {pages}
  {1098--1113} (\bibinfo {year} {1940})}\BibitemShut {NoStop}%
\bibitem [{\citenamefont {Matsubara}\ and\ \citenamefont
  {Matsuda}(1956)}]{Matsubara1956}%
  \BibitemOpen
  \bibfield  {author} {\bibinfo {author} {\bibfnamefont {T.}~\bibnamefont
  {Matsubara}}\ and\ \bibinfo {author} {\bibfnamefont {H.}~\bibnamefont
  {Matsuda}},\ }\bibfield  {title} {\enquote {\bibinfo {title} {A lattice model
  of liquid helium, i},}\ }\href {\doibase 10.1143/ptp.16.569} {\bibfield
  {journal} {\bibinfo  {journal} {Progress of Theoretical Physics}\ }\textbf
  {\bibinfo {volume} {16}},\ \bibinfo {pages} {569--582} (\bibinfo {year}
  {1956})}\BibitemShut {NoStop}%
\bibitem [{\citenamefont {Mattis}(2006)}]{Mattis2006}%
  \BibitemOpen
  \bibfield  {author} {\bibinfo {author} {\bibfnamefont {D.~C.}\ \bibnamefont
  {Mattis}},\ }\href@noop {} {\emph {\bibinfo {title} {The Theory of Magnetism
  Made Simple - An Introduction to Physical Concepts and to Some Useful
  Mathematical Methods}}}\ (\bibinfo  {publisher} {World Scientific},\ \bibinfo
  {address} {Singapur},\ \bibinfo {year} {2006})\BibitemShut {NoStop}%
\bibitem [{\citenamefont {Kamra}\ \emph {et~al.}(2020)\citenamefont {Kamra},
  \citenamefont {Belzig},\ and\ \citenamefont {Brataas}}]{Kamra2020Niche}%
  \BibitemOpen
  \bibfield  {author} {\bibinfo {author} {\bibfnamefont {A.}~\bibnamefont
  {Kamra}}, \bibinfo {author} {\bibfnamefont {W.}~\bibnamefont {Belzig}}, \
  and\ \bibinfo {author} {\bibfnamefont {A.}~\bibnamefont {Brataas}},\
  }\bibfield  {title} {\enquote {\bibinfo {title} {Magnon-squeezing as a niche
  of quantum magnonics},}\ }\href {\doibase 10.1063/5.0021099} {\bibfield
  {journal} {\bibinfo  {journal} {Applied Physics Letters}\ }\textbf {\bibinfo
  {volume} {117}},\ \bibinfo {pages} {090501} (\bibinfo {year} {2020})},\
  \Eprint {http://arxiv.org/abs/https://doi.org/10.1063/5.0021099}
  {https://doi.org/10.1063/5.0021099} \BibitemShut {NoStop}%
\bibitem [{\citenamefont {Dzyaloshinsky}(1958)}]{Dzyaloshinsky58}%
  \BibitemOpen
  \bibfield  {author} {\bibinfo {author} {\bibfnamefont {I.}~\bibnamefont
  {Dzyaloshinsky}},\ }\bibfield  {title} {\enquote {\bibinfo {title} {A
  thermodynamic theory of {\textquotedblleft}weak{\textquotedblright}
  ferromagnetism of antiferromagnetics},}\ }\href {\doibase
  10.1016/0022-3697(58)90076-3} {\bibfield  {journal} {\bibinfo  {journal}
  {Journal of Physics and Chemistry of Solids}\ }\textbf {\bibinfo {volume}
  {4}},\ \bibinfo {pages} {241--255} (\bibinfo {year} {1958})}\BibitemShut
  {NoStop}%
\bibitem [{\citenamefont {Moriya}(1960)}]{Moriya60}%
  \BibitemOpen
  \bibfield  {author} {\bibinfo {author} {\bibfnamefont {T.}~\bibnamefont
  {Moriya}},\ }\bibfield  {title} {\enquote {\bibinfo {title} {Anisotropic
  superexchange interaction and weak ferromagnetism},}\ }\href {\doibase
  10.1103/PhysRev.120.91} {\bibfield  {journal} {\bibinfo  {journal} {Phys.
  Rev.}\ }\textbf {\bibinfo {volume} {120}},\ \bibinfo {pages} {91--98}
  (\bibinfo {year} {1960})}\BibitemShut {NoStop}%
\bibitem [{\citenamefont {Katsura}\ \emph {et~al.}(2005)\citenamefont
  {Katsura}, \citenamefont {Nagaosa},\ and\ \citenamefont
  {Balatsky}}]{Katsura2005}%
  \BibitemOpen
  \bibfield  {author} {\bibinfo {author} {\bibfnamefont {H.}~\bibnamefont
  {Katsura}}, \bibinfo {author} {\bibfnamefont {N.}~\bibnamefont {Nagaosa}}, \
  and\ \bibinfo {author} {\bibfnamefont {A.~V.}\ \bibnamefont {Balatsky}},\
  }\bibfield  {title} {\enquote {\bibinfo {title} {Spin current and
  magnetoelectric effect in noncollinear magnets},}\ }\href {\doibase
  10.1103/PhysRevLett.95.057205} {\bibfield  {journal} {\bibinfo  {journal}
  {Phys. Rev. Lett.}\ }\textbf {\bibinfo {volume} {95}},\ \bibinfo {pages}
  {057205} (\bibinfo {year} {2005})}\BibitemShut {NoStop}%
\bibitem [{\citenamefont {Fert}\ and\ \citenamefont {Levy}(1980)}]{Fert1980}%
  \BibitemOpen
  \bibfield  {author} {\bibinfo {author} {\bibfnamefont {A.}~\bibnamefont
  {Fert}}\ and\ \bibinfo {author} {\bibfnamefont {P.~M.}\ \bibnamefont
  {Levy}},\ }\bibfield  {title} {\enquote {\bibinfo {title} {Role of
  anisotropic exchange interactions in determining the properties of
  spin-glasses},}\ }\href {\doibase 10.1103/PhysRevLett.44.1538} {\bibfield
  {journal} {\bibinfo  {journal} {Phys. Rev. Lett.}\ }\textbf {\bibinfo
  {volume} {44}},\ \bibinfo {pages} {1538--1541} (\bibinfo {year}
  {1980})}\BibitemShut {NoStop}%
\bibitem [{\citenamefont {Levy}\ and\ \citenamefont {Fert}(1981)}]{Levy1981}%
  \BibitemOpen
  \bibfield  {author} {\bibinfo {author} {\bibfnamefont {P.~M.}\ \bibnamefont
  {Levy}}\ and\ \bibinfo {author} {\bibfnamefont {A.}~\bibnamefont {Fert}},\
  }\bibfield  {title} {\enquote {\bibinfo {title} {Anisotropy induced by
  nonmagnetic impurities in {CuMn} spin-glass alloys},}\ }\href {\doibase
  10.1103/PhysRevB.23.4667} {\bibfield  {journal} {\bibinfo  {journal} {Phys.
  Rev. B}\ }\textbf {\bibinfo {volume} {23}},\ \bibinfo {pages} {4667--4690}
  (\bibinfo {year} {1981})}\BibitemShut {NoStop}%
\bibitem [{\citenamefont {Zakeri}\ \emph {et~al.}(2010)\citenamefont {Zakeri},
  \citenamefont {Zhang}, \citenamefont {Prokop}, \citenamefont {Chuang},
  \citenamefont {Sakr}, \citenamefont {Tang},\ and\ \citenamefont
  {Kirschner}}]{Zakeri2010}%
  \BibitemOpen
  \bibfield  {author} {\bibinfo {author} {\bibfnamefont {K.}~\bibnamefont
  {Zakeri}}, \bibinfo {author} {\bibfnamefont {Y.}~\bibnamefont {Zhang}},
  \bibinfo {author} {\bibfnamefont {J.}~\bibnamefont {Prokop}}, \bibinfo
  {author} {\bibfnamefont {T.-H.}\ \bibnamefont {Chuang}}, \bibinfo {author}
  {\bibfnamefont {N.}~\bibnamefont {Sakr}}, \bibinfo {author} {\bibfnamefont
  {W.~X.}\ \bibnamefont {Tang}}, \ and\ \bibinfo {author} {\bibfnamefont
  {J.}~\bibnamefont {Kirschner}},\ }\bibfield  {title} {\enquote {\bibinfo
  {title} {Asymmetric spin-wave dispersion on {F}e(110): {D}irect evidence of
  the {D}zyaloshinskii-{M}oriya interaction},}\ }\href {\doibase
  10.1103/PhysRevLett.104.137203} {\bibfield  {journal} {\bibinfo  {journal}
  {Phys. Rev. Lett.}\ }\textbf {\bibinfo {volume} {104}},\ \bibinfo {pages}
  {137203} (\bibinfo {year} {2010})}\BibitemShut {NoStop}%
\bibitem [{\citenamefont {Wang}\ \emph {et~al.}(2020)\citenamefont {Wang},
  \citenamefont {Chen}, \citenamefont {Liu}, \citenamefont {Zhang},
  \citenamefont {Baumgaertl}, \citenamefont {Guo}, \citenamefont {Li},
  \citenamefont {Liu}, \citenamefont {Che}, \citenamefont {Tu}, \citenamefont
  {Liu}, \citenamefont {Gao}, \citenamefont {Han}, \citenamefont {Yu},
  \citenamefont {Wu}, \citenamefont {Grundler},\ and\ \citenamefont
  {Yu}}]{Wang2020YIGDMI}%
  \BibitemOpen
  \bibfield  {author} {\bibinfo {author} {\bibfnamefont {H.}~\bibnamefont
  {Wang}}, \bibinfo {author} {\bibfnamefont {J.}~\bibnamefont {Chen}}, \bibinfo
  {author} {\bibfnamefont {T.}~\bibnamefont {Liu}}, \bibinfo {author}
  {\bibfnamefont {J.}~\bibnamefont {Zhang}}, \bibinfo {author} {\bibfnamefont
  {K.}~\bibnamefont {Baumgaertl}}, \bibinfo {author} {\bibfnamefont
  {C.}~\bibnamefont {Guo}}, \bibinfo {author} {\bibfnamefont {Y.}~\bibnamefont
  {Li}}, \bibinfo {author} {\bibfnamefont {C.}~\bibnamefont {Liu}}, \bibinfo
  {author} {\bibfnamefont {P.}~\bibnamefont {Che}}, \bibinfo {author}
  {\bibfnamefont {S.}~\bibnamefont {Tu}}, \bibinfo {author} {\bibfnamefont
  {S.}~\bibnamefont {Liu}}, \bibinfo {author} {\bibfnamefont {P.}~\bibnamefont
  {Gao}}, \bibinfo {author} {\bibfnamefont {X.}~\bibnamefont {Han}}, \bibinfo
  {author} {\bibfnamefont {D.}~\bibnamefont {Yu}}, \bibinfo {author}
  {\bibfnamefont {M.}~\bibnamefont {Wu}}, \bibinfo {author} {\bibfnamefont
  {D.}~\bibnamefont {Grundler}}, \ and\ \bibinfo {author} {\bibfnamefont
  {H.}~\bibnamefont {Yu}},\ }\bibfield  {title} {\enquote {\bibinfo {title}
  {Chiral spin-wave velocities induced by all-garnet interfacial
  {D}zyaloshinskii-{M}oriya interaction in ultrathin yttrium iron garnet
  films},}\ }\href {\doibase 10.1103/PhysRevLett.124.027203} {\bibfield
  {journal} {\bibinfo  {journal} {Phys. Rev. Lett.}\ }\textbf {\bibinfo
  {volume} {124}},\ \bibinfo {pages} {027203} (\bibinfo {year}
  {2020})}\BibitemShut {NoStop}%
\bibitem [{\citenamefont {Wilson}\ \emph {et~al.}(2013)\citenamefont {Wilson},
  \citenamefont {Karhu}, \citenamefont {Lake}, \citenamefont {Quigley},
  \citenamefont {Meynell}, \citenamefont {Bogdanov}, \citenamefont {Fritzsche},
  \citenamefont {R\"o\ss{}ler},\ and\ \citenamefont {Monchesky}}]{Wilson2013}%
  \BibitemOpen
  \bibfield  {author} {\bibinfo {author} {\bibfnamefont {M.~N.}\ \bibnamefont
  {Wilson}}, \bibinfo {author} {\bibfnamefont {E.~A.}\ \bibnamefont {Karhu}},
  \bibinfo {author} {\bibfnamefont {D.~P.}\ \bibnamefont {Lake}}, \bibinfo
  {author} {\bibfnamefont {A.~S.}\ \bibnamefont {Quigley}}, \bibinfo {author}
  {\bibfnamefont {S.}~\bibnamefont {Meynell}}, \bibinfo {author} {\bibfnamefont
  {A.~N.}\ \bibnamefont {Bogdanov}}, \bibinfo {author} {\bibfnamefont
  {H.}~\bibnamefont {Fritzsche}}, \bibinfo {author} {\bibfnamefont {U.~K.}\
  \bibnamefont {R\"o\ss{}ler}}, \ and\ \bibinfo {author} {\bibfnamefont
  {T.~L.}\ \bibnamefont {Monchesky}},\ }\bibfield  {title} {\enquote {\bibinfo
  {title} {Discrete helicoidal states in chiral magnetic thin films},}\ }\href
  {\doibase 10.1103/PhysRevB.88.214420} {\bibfield  {journal} {\bibinfo
  {journal} {Phys. Rev. B}\ }\textbf {\bibinfo {volume} {88}},\ \bibinfo
  {pages} {214420} (\bibinfo {year} {2013})}\BibitemShut {NoStop}%
\bibitem [{\citenamefont {Meynell}\ \emph {et~al.}(2014)\citenamefont
  {Meynell}, \citenamefont {Wilson}, \citenamefont {Fritzsche}, \citenamefont
  {Bogdanov},\ and\ \citenamefont {Monchesky}}]{Meynell2014}%
  \BibitemOpen
  \bibfield  {author} {\bibinfo {author} {\bibfnamefont {S.~A.}\ \bibnamefont
  {Meynell}}, \bibinfo {author} {\bibfnamefont {M.~N.}\ \bibnamefont {Wilson}},
  \bibinfo {author} {\bibfnamefont {H.}~\bibnamefont {Fritzsche}}, \bibinfo
  {author} {\bibfnamefont {A.~N.}\ \bibnamefont {Bogdanov}}, \ and\ \bibinfo
  {author} {\bibfnamefont {T.~L.}\ \bibnamefont {Monchesky}},\ }\bibfield
  {title} {\enquote {\bibinfo {title} {Surface twist instabilities and skyrmion
  states in chiral ferromagnets},}\ }\href {\doibase
  10.1103/PhysRevB.90.014406} {\bibfield  {journal} {\bibinfo  {journal} {Phys.
  Rev. B}\ }\textbf {\bibinfo {volume} {90}},\ \bibinfo {pages} {014406}
  (\bibinfo {year} {2014})}\BibitemShut {NoStop}%
\bibitem [{\citenamefont {Reklis}(1974)}]{Reklis1974}%
  \BibitemOpen
  \bibfield  {author} {\bibinfo {author} {\bibfnamefont {R.~P.}\ \bibnamefont
  {Reklis}},\ }\bibfield  {title} {\enquote {\bibinfo {title} {Numerical
  calculation of two-spin-wave bound states in some two-dimensional heisenberg
  ferromagnets},}\ }\href {\doibase 10.1103/PhysRevB.9.4939} {\bibfield
  {journal} {\bibinfo  {journal} {Phys. Rev. B}\ }\textbf {\bibinfo {volume}
  {9}},\ \bibinfo {pages} {4939--4944} (\bibinfo {year} {1974})}\BibitemShut
  {NoStop}%
\bibitem [{\citenamefont {Kohmoto}(1985)}]{Kohmoto1985}%
  \BibitemOpen
  \bibfield  {author} {\bibinfo {author} {\bibfnamefont {M.}~\bibnamefont
  {Kohmoto}},\ }\bibfield  {title} {\enquote {\bibinfo {title} {Topological
  invariant and the quantization of the hall conductance},}\ }\href {\doibase
  https://doi.org/10.1016/0003-4916(85)90148-4} {\bibfield  {journal} {\bibinfo
   {journal} {Annals of Physics}\ }\textbf {\bibinfo {volume} {160}},\ \bibinfo
  {pages} {343--354} (\bibinfo {year} {1985})}\BibitemShut {NoStop}%
\bibitem [{\citenamefont {Fukui}\ \emph {et~al.}(2005)\citenamefont {Fukui},
  \citenamefont {Hatsugai},\ and\ \citenamefont {Suzuki}}]{Fukui2005}%
  \BibitemOpen
  \bibfield  {author} {\bibinfo {author} {\bibfnamefont {T.}~\bibnamefont
  {Fukui}}, \bibinfo {author} {\bibfnamefont {Y.}~\bibnamefont {Hatsugai}}, \
  and\ \bibinfo {author} {\bibfnamefont {H.}~\bibnamefont {Suzuki}},\
  }\bibfield  {title} {\enquote {\bibinfo {title} {Chern numbers in discretized
  {B}rillouin zone: Efficient method of computing (spin) {H}all
  conductances},}\ }\href {\doibase 10.1143/JPSJ.74.1674} {\bibfield  {journal}
  {\bibinfo  {journal} {Journal of the Physical Society of Japan}\ }\textbf
  {\bibinfo {volume} {74}},\ \bibinfo {pages} {1674--1677} (\bibinfo {year}
  {2005})},\ \Eprint
  {http://arxiv.org/abs/https://doi.org/10.1143/JPSJ.74.1674}
  {https://doi.org/10.1143/JPSJ.74.1674} \BibitemShut {NoStop}%
\bibitem [{\citenamefont
  {Hatsugai}(1993{\natexlab{a}})}]{hatsugaiChernNumberEdge1993}%
  \BibitemOpen
  \bibfield  {author} {\bibinfo {author} {\bibfnamefont {Y.}~\bibnamefont
  {Hatsugai}},\ }\bibfield  {title} {\enquote {\bibinfo {title} {Chern number
  and edge states in the integer quantum {{Hall}} effect},}\ }\href {\doibase
  10.1103/PhysRevLett.71.3697} {\bibfield  {journal} {\bibinfo  {journal}
  {Phys. Rev. Lett.}\ }\textbf {\bibinfo {volume} {71}},\ \bibinfo {pages}
  {3697--3700} (\bibinfo {year} {1993}{\natexlab{a}})}\BibitemShut {NoStop}%
\bibitem [{\citenamefont
  {Hatsugai}(1993{\natexlab{b}})}]{hatsugaiEdgeStatesInteger1993}%
  \BibitemOpen
  \bibfield  {author} {\bibinfo {author} {\bibfnamefont {Y.}~\bibnamefont
  {Hatsugai}},\ }\bibfield  {title} {\enquote {\bibinfo {title} {Edge states in
  the integer quantum {{Hall}} effect and the {{Riemann}} surface of the
  {{Bloch}} function},}\ }\href {\doibase 10.1103/PhysRevB.48.11851} {\bibfield
   {journal} {\bibinfo  {journal} {Phys. Rev. B}\ }\textbf {\bibinfo {volume}
  {48}},\ \bibinfo {pages} {11851--11862} (\bibinfo {year}
  {1993}{\natexlab{b}})}\BibitemShut {NoStop}%
\bibitem [{\citenamefont {Sun}\ and\ \citenamefont {Fradkin}(2008)}]{Sun2008}%
  \BibitemOpen
  \bibfield  {author} {\bibinfo {author} {\bibfnamefont {K.}~\bibnamefont
  {Sun}}\ and\ \bibinfo {author} {\bibfnamefont {E.}~\bibnamefont {Fradkin}},\
  }\bibfield  {title} {\enquote {\bibinfo {title} {Time-reversal symmetry
  breaking and spontaneous anomalous {H}all effect in {F}ermi fluids},}\ }\href
  {\doibase 10.1103/PhysRevB.78.245122} {\bibfield  {journal} {\bibinfo
  {journal} {Phys. Rev. B}\ }\textbf {\bibinfo {volume} {78}},\ \bibinfo
  {pages} {245122} (\bibinfo {year} {2008})}\BibitemShut {NoStop}%
\bibitem [{\citenamefont {Chong}\ \emph {et~al.}(2008)\citenamefont {Chong},
  \citenamefont {Wen},\ and\ \citenamefont {Solja\ifmmode \check{c}\else
  \v{c}\fi{}i\ifmmode~\acute{c}\else \'{c}\fi{}}}]{Chong2008}%
  \BibitemOpen
  \bibfield  {author} {\bibinfo {author} {\bibfnamefont {Y.~D.}\ \bibnamefont
  {Chong}}, \bibinfo {author} {\bibfnamefont {X.-G.}\ \bibnamefont {Wen}}, \
  and\ \bibinfo {author} {\bibfnamefont {M.}~\bibnamefont {Solja\ifmmode
  \check{c}\else \v{c}\fi{}i\ifmmode~\acute{c}\else \'{c}\fi{}}},\ }\bibfield
  {title} {\enquote {\bibinfo {title} {Effective theory of quadratic
  degeneracies},}\ }\href {\doibase 10.1103/PhysRevB.77.235125} {\bibfield
  {journal} {\bibinfo  {journal} {Phys. Rev. B}\ }\textbf {\bibinfo {volume}
  {77}},\ \bibinfo {pages} {235125} (\bibinfo {year} {2008})}\BibitemShut
  {NoStop}%
\bibitem [{\citenamefont {Sun}\ \emph {et~al.}(2009)\citenamefont {Sun},
  \citenamefont {Yao}, \citenamefont {Fradkin},\ and\ \citenamefont
  {Kivelson}}]{Sun2009}%
  \BibitemOpen
  \bibfield  {author} {\bibinfo {author} {\bibfnamefont {K.}~\bibnamefont
  {Sun}}, \bibinfo {author} {\bibfnamefont {H.}~\bibnamefont {Yao}}, \bibinfo
  {author} {\bibfnamefont {E.}~\bibnamefont {Fradkin}}, \ and\ \bibinfo
  {author} {\bibfnamefont {S.~A.}\ \bibnamefont {Kivelson}},\ }\bibfield
  {title} {\enquote {\bibinfo {title} {Topological insulators and nematic
  phases from spontaneous symmetry breaking in 2d {F}ermi systems with a
  quadratic band crossing},}\ }\href {\doibase 10.1103/PhysRevLett.103.046811}
  {\bibfield  {journal} {\bibinfo  {journal} {Phys. Rev. Lett.}\ }\textbf
  {\bibinfo {volume} {103}},\ \bibinfo {pages} {046811} (\bibinfo {year}
  {2009})}\BibitemShut {NoStop}%
\bibitem [{\citenamefont {Sticlet}\ \emph {et~al.}(2012)\citenamefont
  {Sticlet}, \citenamefont {Pi\'echon}, \citenamefont {Fuchs}, \citenamefont
  {Kalugin},\ and\ \citenamefont {Simon}}]{Sticlet2012}%
  \BibitemOpen
  \bibfield  {author} {\bibinfo {author} {\bibfnamefont {D.}~\bibnamefont
  {Sticlet}}, \bibinfo {author} {\bibfnamefont {F.}~\bibnamefont {Pi\'echon}},
  \bibinfo {author} {\bibfnamefont {J.-N.}\ \bibnamefont {Fuchs}}, \bibinfo
  {author} {\bibfnamefont {P.}~\bibnamefont {Kalugin}}, \ and\ \bibinfo
  {author} {\bibfnamefont {P.}~\bibnamefont {Simon}},\ }\bibfield  {title}
  {\enquote {\bibinfo {title} {Geometrical engineering of a two-band {C}hern
  insulator in two dimensions with arbitrary topological index},}\ }\href
  {\doibase 10.1103/PhysRevB.85.165456} {\bibfield  {journal} {\bibinfo
  {journal} {Phys. Rev. B}\ }\textbf {\bibinfo {volume} {85}},\ \bibinfo
  {pages} {165456} (\bibinfo {year} {2012})}\BibitemShut {NoStop}%
\bibitem [{\citenamefont {Silberglitt}\ and\ \citenamefont
  {Torrance}(1970)}]{Silberglitt1970}%
  \BibitemOpen
  \bibfield  {author} {\bibinfo {author} {\bibfnamefont {R.}~\bibnamefont
  {Silberglitt}}\ and\ \bibinfo {author} {\bibfnamefont {J.~B.}\ \bibnamefont
  {Torrance}},\ }\bibfield  {title} {\enquote {\bibinfo {title} {Effect of
  single-ion anisotropy on two-spin-wave bound state in a {H}eisenberg
  ferromagnet},}\ }\href {\doibase 10.1103/PhysRevB.2.772} {\bibfield
  {journal} {\bibinfo  {journal} {Phys. Rev. B}\ }\textbf {\bibinfo {volume}
  {2}},\ \bibinfo {pages} {772--778} (\bibinfo {year} {1970})}\BibitemShut
  {NoStop}%
\bibitem [{\citenamefont {Tonegawa}(1970)}]{Tonegawa1970}%
  \BibitemOpen
  \bibfield  {author} {\bibinfo {author} {\bibfnamefont {T.}~\bibnamefont
  {Tonegawa}},\ }\bibfield  {title} {\enquote {\bibinfo {title} {{Two-Magnon
  Bound States in the {H}eisenberg Ferromagnet with Anisotropic Exchange and
  Uniaxial Anisotropy Energies}},}\ }\href {\doibase 10.1143/PTPS.46.61}
  {\bibfield  {journal} {\bibinfo  {journal} {Progress of Theoretical Physics
  Supplement}\ }\textbf {\bibinfo {volume} {46}},\ \bibinfo {pages} {61--83}
  (\bibinfo {year} {1970})},\ \Eprint
  {http://arxiv.org/abs/https://academic.oup.com/ptps/article-pdf/doi/10.1143/PTPS.46.61/5304188/46-61.pdf}
  {https://academic.oup.com/ptps/article-pdf/doi/10.1143/PTPS.46.61/5304188/46-61.pdf}
  \BibitemShut {NoStop}%
\bibitem [{\citenamefont {Qi}\ \emph {et~al.}(2006)\citenamefont {Qi},
  \citenamefont {Wu},\ and\ \citenamefont {Zhang}}]{Qi2006}%
  \BibitemOpen
  \bibfield  {author} {\bibinfo {author} {\bibfnamefont {X.-L.}\ \bibnamefont
  {Qi}}, \bibinfo {author} {\bibfnamefont {Y.-S.}\ \bibnamefont {Wu}}, \ and\
  \bibinfo {author} {\bibfnamefont {S.-C.}\ \bibnamefont {Zhang}},\ }\bibfield
  {title} {\enquote {\bibinfo {title} {Topological quantization of the spin
  {H}all effect in two-dimensional paramagnetic semiconductors},}\ }\href
  {\doibase 10.1103/PhysRevB.74.085308} {\bibfield  {journal} {\bibinfo
  {journal} {Phys. Rev. B}\ }\textbf {\bibinfo {volume} {74}},\ \bibinfo
  {pages} {085308} (\bibinfo {year} {2006})}\BibitemShut {NoStop}%
\bibitem [{\citenamefont {Asb{\'{o}}th}\ \emph {et~al.}(2016)\citenamefont
  {Asb{\'{o}}th}, \citenamefont {Oroszl{\'{a}}ny},\ and\ \citenamefont
  {P{\'{a}}lyi}}]{Asboth2016}%
  \BibitemOpen
  \bibfield  {author} {\bibinfo {author} {\bibfnamefont {J.~K.}\ \bibnamefont
  {Asb{\'{o}}th}}, \bibinfo {author} {\bibfnamefont {L.}~\bibnamefont
  {Oroszl{\'{a}}ny}}, \ and\ \bibinfo {author} {\bibfnamefont {A.}~\bibnamefont
  {P{\'{a}}lyi}},\ }\href {\doibase 10.1007/978-3-319-25607-8} {\emph {\bibinfo
  {title} {A Short Course on Topological Insulators}}}\ (\bibinfo  {publisher}
  {Springer International Publishing},\ \bibinfo {year} {2016})\BibitemShut
  {NoStop}%
\bibitem [{\citenamefont {Li}\ \emph {et~al.}(2015)\citenamefont {Li},
  \citenamefont {Chen}, \citenamefont {Tong}, \citenamefont {Pi}, \citenamefont
  {Liu}, \citenamefont {Yang}, \citenamefont {Wang},\ and\ \citenamefont
  {Zhang}}]{Li2015Triangular}%
  \BibitemOpen
  \bibfield  {author} {\bibinfo {author} {\bibfnamefont {Y.}~\bibnamefont
  {Li}}, \bibinfo {author} {\bibfnamefont {G.}~\bibnamefont {Chen}}, \bibinfo
  {author} {\bibfnamefont {W.}~\bibnamefont {Tong}}, \bibinfo {author}
  {\bibfnamefont {L.}~\bibnamefont {Pi}}, \bibinfo {author} {\bibfnamefont
  {J.}~\bibnamefont {Liu}}, \bibinfo {author} {\bibfnamefont {Z.}~\bibnamefont
  {Yang}}, \bibinfo {author} {\bibfnamefont {X.}~\bibnamefont {Wang}}, \ and\
  \bibinfo {author} {\bibfnamefont {Q.}~\bibnamefont {Zhang}},\ }\bibfield
  {title} {\enquote {\bibinfo {title} {Rare-earth triangular lattice spin
  liquid: A single-crystal study of {YbMgGaO}$_{4}$},}\ }\href {\doibase
  10.1103/PhysRevLett.115.167203} {\bibfield  {journal} {\bibinfo  {journal}
  {Phys. Rev. Lett.}\ }\textbf {\bibinfo {volume} {115}},\ \bibinfo {pages}
  {167203} (\bibinfo {year} {2015})}\BibitemShut {NoStop}%
\bibitem [{\citenamefont {Li}\ \emph {et~al.}(2016{\natexlab{a}})\citenamefont
  {Li}, \citenamefont {Wang},\ and\ \citenamefont {Chen}}]{Li2016Triangular}%
  \BibitemOpen
  \bibfield  {author} {\bibinfo {author} {\bibfnamefont {Y.-D.}\ \bibnamefont
  {Li}}, \bibinfo {author} {\bibfnamefont {X.}~\bibnamefont {Wang}}, \ and\
  \bibinfo {author} {\bibfnamefont {G.}~\bibnamefont {Chen}},\ }\bibfield
  {title} {\enquote {\bibinfo {title} {Anisotropic spin model of strong
  spin-orbit-coupled triangular antiferromagnets},}\ }\href {\doibase
  10.1103/PhysRevB.94.035107} {\bibfield  {journal} {\bibinfo  {journal} {Phys.
  Rev. B}\ }\textbf {\bibinfo {volume} {94}},\ \bibinfo {pages} {035107}
  (\bibinfo {year} {2016}{\natexlab{a}})}\BibitemShut {NoStop}%
\bibitem [{\citenamefont {Maksimov}\ \emph {et~al.}(2019)\citenamefont
  {Maksimov}, \citenamefont {Zhu}, \citenamefont {White},\ and\ \citenamefont
  {Chernyshev}}]{Maksimov2019}%
  \BibitemOpen
  \bibfield  {author} {\bibinfo {author} {\bibfnamefont {P.~A.}\ \bibnamefont
  {Maksimov}}, \bibinfo {author} {\bibfnamefont {Z.}~\bibnamefont {Zhu}},
  \bibinfo {author} {\bibfnamefont {S.~R.}\ \bibnamefont {White}}, \ and\
  \bibinfo {author} {\bibfnamefont {A.~L.}\ \bibnamefont {Chernyshev}},\
  }\bibfield  {title} {\enquote {\bibinfo {title} {Anisotropic-exchange magnets
  on a triangular lattice: Spin waves, accidental degeneracies, and dual spin
  liquids},}\ }\href {\doibase 10.1103/PhysRevX.9.021017} {\bibfield  {journal}
  {\bibinfo  {journal} {Phys. Rev. X}\ }\textbf {\bibinfo {volume} {9}},\
  \bibinfo {pages} {021017} (\bibinfo {year} {2019})}\BibitemShut {NoStop}%
\bibitem [{\citenamefont {Wada}\ \emph {et~al.}(1975)\citenamefont {Wada},
  \citenamefont {Ishikawa},\ and\ \citenamefont {Oguchi}}]{Wada1975}%
  \BibitemOpen
  \bibfield  {author} {\bibinfo {author} {\bibfnamefont {K.}~\bibnamefont
  {Wada}}, \bibinfo {author} {\bibfnamefont {T.}~\bibnamefont {Ishikawa}}, \
  and\ \bibinfo {author} {\bibfnamefont {T.}~\bibnamefont {Oguchi}},\
  }\bibfield  {title} {\enquote {\bibinfo {title} {{Two-Magnon Bound States in
  the Triangular and Honeycomb Heisenberg Ferromagnets}},}\ }\href {\doibase
  10.1143/PTP.54.1589} {\bibfield  {journal} {\bibinfo  {journal} {Progress of
  Theoretical Physics}\ }\textbf {\bibinfo {volume} {54}},\ \bibinfo {pages}
  {1589--1598} (\bibinfo {year} {1975})},\ \Eprint
  {http://arxiv.org/abs/https://academic.oup.com/ptp/article-pdf/54/6/1589/5225388/54-6-1589.pdf}
  {https://academic.oup.com/ptp/article-pdf/54/6/1589/5225388/54-6-1589.pdf}
  \BibitemShut {NoStop}%
\bibitem [{\citenamefont {Ohgushi}\ \emph {et~al.}(2000)\citenamefont
  {Ohgushi}, \citenamefont {Murakami},\ and\ \citenamefont
  {Nagaosa}}]{Ohgushi2000}%
  \BibitemOpen
  \bibfield  {author} {\bibinfo {author} {\bibfnamefont {K.}~\bibnamefont
  {Ohgushi}}, \bibinfo {author} {\bibfnamefont {S.}~\bibnamefont {Murakami}}, \
  and\ \bibinfo {author} {\bibfnamefont {N.}~\bibnamefont {Nagaosa}},\
  }\bibfield  {title} {\enquote {\bibinfo {title} {Spin anisotropy and quantum
  hall effect in the kagom\'e lattice: Chiral spin state based on a
  ferromagnet},}\ }\href {\doibase 10.1103/PhysRevB.62.R6065} {\bibfield
  {journal} {\bibinfo  {journal} {Phys. Rev. B}\ }\textbf {\bibinfo {volume}
  {62}},\ \bibinfo {pages} {R6065--R6068} (\bibinfo {year} {2000})}\BibitemShut
  {NoStop}%
\bibitem [{\citenamefont {Mook}\ \emph
  {et~al.}(2014{\natexlab{b}})\citenamefont {Mook}, \citenamefont {Henk},\ and\
  \citenamefont {Mertig}}]{Mook2014}%
  \BibitemOpen
  \bibfield  {author} {\bibinfo {author} {\bibfnamefont {A.}~\bibnamefont
  {Mook}}, \bibinfo {author} {\bibfnamefont {J.}~\bibnamefont {Henk}}, \ and\
  \bibinfo {author} {\bibfnamefont {I.}~\bibnamefont {Mertig}},\ }\bibfield
  {title} {\enquote {\bibinfo {title} {Magnon {H}all effect and topology in
  kagome lattices: {A} theoretical investigation},}\ }\href {\doibase
  10.1103/PhysRevB.89.134409} {\bibfield  {journal} {\bibinfo  {journal} {Phys.
  Rev. B}\ }\textbf {\bibinfo {volume} {89}},\ \bibinfo {pages} {134409}
  (\bibinfo {year} {2014}{\natexlab{b}})}\BibitemShut {NoStop}%
\bibitem [{\citenamefont {Romh\'{a}nyi}(2019)}]{Romhanyi2019}%
  \BibitemOpen
  \bibfield  {author} {\bibinfo {author} {\bibfnamefont {J.}~\bibnamefont
  {Romh\'{a}nyi}},\ }\bibfield  {title} {\enquote {\bibinfo {title} {Multipolar
  edge states in the anisotropic kagome antiferromagnet},}\ }\href {\doibase
  10.1103/PhysRevB.99.014408} {\bibfield  {journal} {\bibinfo  {journal} {Phys.
  Rev. B}\ }\textbf {\bibinfo {volume} {99}},\ \bibinfo {pages} {014408}
  (\bibinfo {year} {2019})}\BibitemShut {NoStop}%
\bibitem [{\citenamefont {Hell}\ \emph {et~al.}(2013)\citenamefont {Hell},
  \citenamefont {Das},\ and\ \citenamefont {Wegewijs}}]{Hell2013}%
  \BibitemOpen
  \bibfield  {author} {\bibinfo {author} {\bibfnamefont {M.}~\bibnamefont
  {Hell}}, \bibinfo {author} {\bibfnamefont {S.}~\bibnamefont {Das}}, \ and\
  \bibinfo {author} {\bibfnamefont {M.~R.}\ \bibnamefont {Wegewijs}},\
  }\bibfield  {title} {\enquote {\bibinfo {title} {Transport of spin anisotropy
  without spin currents},}\ }\href {\doibase 10.1103/PhysRevB.88.115435}
  {\bibfield  {journal} {\bibinfo  {journal} {Phys. Rev. B}\ }\textbf {\bibinfo
  {volume} {88}},\ \bibinfo {pages} {115435} (\bibinfo {year}
  {2013})}\BibitemShut {NoStop}%
\bibitem [{\citenamefont {Misiorny}\ \emph {et~al.}(2013)\citenamefont
  {Misiorny}, \citenamefont {Hell},\ and\ \citenamefont
  {Wegewijs}}]{misiorny2013}%
  \BibitemOpen
  \bibfield  {author} {\bibinfo {author} {\bibfnamefont {M.}~\bibnamefont
  {Misiorny}}, \bibinfo {author} {\bibfnamefont {M.}~\bibnamefont {Hell}}, \
  and\ \bibinfo {author} {\bibfnamefont {M.~R.}\ \bibnamefont {Wegewijs}},\
  }\bibfield  {title} {\enquote {\bibinfo {title} {Spintronic magnetic
  anisotropy},}\ }\href@noop {} {\bibfield  {journal} {\bibinfo  {journal}
  {Nature Physics}\ }\textbf {\bibinfo {volume} {9}},\ \bibinfo {pages}
  {801--805} (\bibinfo {year} {2013})}\BibitemShut {NoStop}%
\bibitem [{\citenamefont {Gao}\ \emph {et~al.}(2018)\citenamefont {Gao},
  \citenamefont {Vanderbilt},\ and\ \citenamefont {Xiao}}]{Gao2018}%
  \BibitemOpen
  \bibfield  {author} {\bibinfo {author} {\bibfnamefont {Y.}~\bibnamefont
  {Gao}}, \bibinfo {author} {\bibfnamefont {D.}~\bibnamefont {Vanderbilt}}, \
  and\ \bibinfo {author} {\bibfnamefont {D.}~\bibnamefont {Xiao}},\ }\bibfield
  {title} {\enquote {\bibinfo {title} {Microscopic theory of spin toroidization
  in periodic crystals},}\ }\href {\doibase 10.1103/PhysRevB.97.134423}
  {\bibfield  {journal} {\bibinfo  {journal} {Phys. Rev. B}\ }\textbf {\bibinfo
  {volume} {97}},\ \bibinfo {pages} {134423} (\bibinfo {year}
  {2018})}\BibitemShut {NoStop}%
\bibitem [{\citenamefont {Shitade}\ \emph {et~al.}(2018)\citenamefont
  {Shitade}, \citenamefont {Watanabe},\ and\ \citenamefont
  {Yanase}}]{Shitade2018}%
  \BibitemOpen
  \bibfield  {author} {\bibinfo {author} {\bibfnamefont {A.}~\bibnamefont
  {Shitade}}, \bibinfo {author} {\bibfnamefont {H.}~\bibnamefont {Watanabe}}, \
  and\ \bibinfo {author} {\bibfnamefont {Y.}~\bibnamefont {Yanase}},\
  }\bibfield  {title} {\enquote {\bibinfo {title} {Theory of orbital magnetic
  quadrupole moment and magnetoelectric susceptibility},}\ }\href {\doibase
  10.1103/PhysRevB.98.020407} {\bibfield  {journal} {\bibinfo  {journal} {Phys.
  Rev. B}\ }\textbf {\bibinfo {volume} {98}},\ \bibinfo {pages} {020407}
  (\bibinfo {year} {2018})}\BibitemShut {NoStop}%
\bibitem [{\citenamefont {Onose}\ \emph {et~al.}(2010)\citenamefont {Onose},
  \citenamefont {Ideue}, \citenamefont {Katsura}, \citenamefont {Shiomi},
  \citenamefont {Nagaosa},\ and\ \citenamefont {Tokura}}]{Onose2010}%
  \BibitemOpen
  \bibfield  {author} {\bibinfo {author} {\bibfnamefont {Y.}~\bibnamefont
  {Onose}}, \bibinfo {author} {\bibfnamefont {T.}~\bibnamefont {Ideue}},
  \bibinfo {author} {\bibfnamefont {H.}~\bibnamefont {Katsura}}, \bibinfo
  {author} {\bibfnamefont {Y.}~\bibnamefont {Shiomi}}, \bibinfo {author}
  {\bibfnamefont {N.}~\bibnamefont {Nagaosa}}, \ and\ \bibinfo {author}
  {\bibfnamefont {Y.}~\bibnamefont {Tokura}},\ }\bibfield  {title} {\enquote
  {\bibinfo {title} {Observation of the magnon {H}all effect},}\ }\href
  {\doibase 10.1126/science.1188260} {\bibfield  {journal} {\bibinfo  {journal}
  {Science}\ }\textbf {\bibinfo {volume} {329}},\ \bibinfo {pages} {297--299}
  (\bibinfo {year} {2010})}\BibitemShut {NoStop}%
\bibitem [{\citenamefont {Ideue}\ \emph {et~al.}(2012)\citenamefont {Ideue},
  \citenamefont {Onose}, \citenamefont {Katsura}, \citenamefont {Shiomi},
  \citenamefont {Ishiwata}, \citenamefont {Nagaosa},\ and\ \citenamefont
  {Tokura}}]{Ideue2012}%
  \BibitemOpen
  \bibfield  {author} {\bibinfo {author} {\bibfnamefont {T.}~\bibnamefont
  {Ideue}}, \bibinfo {author} {\bibfnamefont {Y.}~\bibnamefont {Onose}},
  \bibinfo {author} {\bibfnamefont {H.}~\bibnamefont {Katsura}}, \bibinfo
  {author} {\bibfnamefont {Y.}~\bibnamefont {Shiomi}}, \bibinfo {author}
  {\bibfnamefont {S.}~\bibnamefont {Ishiwata}}, \bibinfo {author}
  {\bibfnamefont {N.}~\bibnamefont {Nagaosa}}, \ and\ \bibinfo {author}
  {\bibfnamefont {Y.}~\bibnamefont {Tokura}},\ }\bibfield  {title} {\enquote
  {\bibinfo {title} {Effect of lattice geometry on magnon {H}all effect in
  ferromagnetic insulators},}\ }\href {\doibase 10.1103/PhysRevB.85.134411}
  {\bibfield  {journal} {\bibinfo  {journal} {Phys. Rev. B}\ }\textbf {\bibinfo
  {volume} {85}},\ \bibinfo {pages} {134411} (\bibinfo {year}
  {2012})}\BibitemShut {NoStop}%
\bibitem [{\citenamefont {Li}\ \emph {et~al.}(2016{\natexlab{b}})\citenamefont
  {Li}, \citenamefont {Li}, \citenamefont {Kim}, \citenamefont {Balents},
  \citenamefont {Yu},\ and\ \citenamefont {Chen}}]{Li2016Weyl}%
  \BibitemOpen
  \bibfield  {author} {\bibinfo {author} {\bibfnamefont {F.-Y.}\ \bibnamefont
  {Li}}, \bibinfo {author} {\bibfnamefont {Y.-D.}\ \bibnamefont {Li}}, \bibinfo
  {author} {\bibfnamefont {Y.~B.}\ \bibnamefont {Kim}}, \bibinfo {author}
  {\bibfnamefont {L.}~\bibnamefont {Balents}}, \bibinfo {author} {\bibfnamefont
  {Y.}~\bibnamefont {Yu}}, \ and\ \bibinfo {author} {\bibfnamefont
  {G.}~\bibnamefont {Chen}},\ }\bibfield  {title} {\enquote {\bibinfo {title}
  {Weyl magnons in breathing pyrochlore antiferromagnets},}\ }\href {\doibase
  10.1038/ncomms12691} {\bibfield  {journal} {\bibinfo  {journal} {Nature
  Communications}\ }\textbf {\bibinfo {volume} {7}} (\bibinfo {year}
  {2016}{\natexlab{b}}),\ 10.1038/ncomms12691}\BibitemShut {NoStop}%
\bibitem [{\citenamefont {Neumann}\ \emph {et~al.}(2022)\citenamefont
  {Neumann}, \citenamefont {Mook}, \citenamefont {Henk},\ and\ \citenamefont
  {Mertig}}]{Neumann2022}%
  \BibitemOpen
  \bibfield  {author} {\bibinfo {author} {\bibfnamefont {R.~R.}\ \bibnamefont
  {Neumann}}, \bibinfo {author} {\bibfnamefont {A.}~\bibnamefont {Mook}},
  \bibinfo {author} {\bibfnamefont {J.}~\bibnamefont {Henk}}, \ and\ \bibinfo
  {author} {\bibfnamefont {I.}~\bibnamefont {Mertig}},\ }\bibfield  {title}
  {\enquote {\bibinfo {title} {Thermal {H}all effect of magnons in collinear
  antiferromagnetic insulators: {S}ignatures of magnetic and topological phase
  transitions},}\ }\href {\doibase 10.1103/PhysRevLett.128.117201} {\bibfield
  {journal} {\bibinfo  {journal} {Phys. Rev. Lett.}\ }\textbf {\bibinfo
  {volume} {128}},\ \bibinfo {pages} {117201} (\bibinfo {year}
  {2022})}\BibitemShut {NoStop}%
\bibitem [{\citenamefont {Chisnell}\ \emph {et~al.}(2015)\citenamefont
  {Chisnell}, \citenamefont {Helton}, \citenamefont {Freedman}, \citenamefont
  {Singh}, \citenamefont {Bewley}, \citenamefont {Nocera},\ and\ \citenamefont
  {Lee}}]{Chisnell2015}%
  \BibitemOpen
  \bibfield  {author} {\bibinfo {author} {\bibfnamefont {R.}~\bibnamefont
  {Chisnell}}, \bibinfo {author} {\bibfnamefont {J.~S.}\ \bibnamefont
  {Helton}}, \bibinfo {author} {\bibfnamefont {D.~E.}\ \bibnamefont
  {Freedman}}, \bibinfo {author} {\bibfnamefont {D.~K.}\ \bibnamefont {Singh}},
  \bibinfo {author} {\bibfnamefont {R.~I.}\ \bibnamefont {Bewley}}, \bibinfo
  {author} {\bibfnamefont {D.~G.}\ \bibnamefont {Nocera}}, \ and\ \bibinfo
  {author} {\bibfnamefont {Y.~S.}\ \bibnamefont {Lee}},\ }\bibfield  {title}
  {\enquote {\bibinfo {title} {Topological magnon bands in a kagome lattice
  ferromagnet},}\ }\href {\doibase 10.1103/PhysRevLett.115.147201} {\bibfield
  {journal} {\bibinfo  {journal} {Phys. Rev. Lett.}\ }\textbf {\bibinfo
  {volume} {115}},\ \bibinfo {pages} {147201} (\bibinfo {year}
  {2015})}\BibitemShut {NoStop}%
\bibitem [{\citenamefont {Chen}\ \emph {et~al.}(2018)\citenamefont {Chen},
  \citenamefont {Chung}, \citenamefont {Gao}, \citenamefont {Chen},
  \citenamefont {Stone}, \citenamefont {Kolesnikov}, \citenamefont {Huang},\
  and\ \citenamefont {Dai}}]{Chen2018CrI3}%
  \BibitemOpen
  \bibfield  {author} {\bibinfo {author} {\bibfnamefont {L.}~\bibnamefont
  {Chen}}, \bibinfo {author} {\bibfnamefont {J.-H.}\ \bibnamefont {Chung}},
  \bibinfo {author} {\bibfnamefont {B.}~\bibnamefont {Gao}}, \bibinfo {author}
  {\bibfnamefont {T.}~\bibnamefont {Chen}}, \bibinfo {author} {\bibfnamefont
  {M.~B.}\ \bibnamefont {Stone}}, \bibinfo {author} {\bibfnamefont {A.~I.}\
  \bibnamefont {Kolesnikov}}, \bibinfo {author} {\bibfnamefont
  {Q.}~\bibnamefont {Huang}}, \ and\ \bibinfo {author} {\bibfnamefont
  {P.}~\bibnamefont {Dai}},\ }\bibfield  {title} {\enquote {\bibinfo {title}
  {Topological spin excitations in honeycomb ferromagnet {CrI}$_{3}$},}\ }\href
  {\doibase 10.1103/PhysRevX.8.041028} {\bibfield  {journal} {\bibinfo
  {journal} {Phys. Rev. X}\ }\textbf {\bibinfo {volume} {8}},\ \bibinfo {pages}
  {041028} (\bibinfo {year} {2018})}\BibitemShut {NoStop}%
\bibitem [{\citenamefont {Cai}\ \emph {et~al.}(2021)\citenamefont {Cai},
  \citenamefont {Bao}, \citenamefont {Gu}, \citenamefont {Gao}, \citenamefont
  {Ma}, \citenamefont {Shangguan}, \citenamefont {Si}, \citenamefont {Dong},
  \citenamefont {Wang}, \citenamefont {Wu}, \citenamefont {Lin}, \citenamefont
  {Wang}, \citenamefont {Ran}, \citenamefont {Li}, \citenamefont {Adroja},
  \citenamefont {Xi}, \citenamefont {Yu}, \citenamefont {Wu}, \citenamefont
  {Li},\ and\ \citenamefont {Wen}}]{Cai2021CrBr3}%
  \BibitemOpen
  \bibfield  {author} {\bibinfo {author} {\bibfnamefont {Z.}~\bibnamefont
  {Cai}}, \bibinfo {author} {\bibfnamefont {S.}~\bibnamefont {Bao}}, \bibinfo
  {author} {\bibfnamefont {Z.-L.}\ \bibnamefont {Gu}}, \bibinfo {author}
  {\bibfnamefont {Y.-P.}\ \bibnamefont {Gao}}, \bibinfo {author} {\bibfnamefont
  {Z.}~\bibnamefont {Ma}}, \bibinfo {author} {\bibfnamefont {Y.}~\bibnamefont
  {Shangguan}}, \bibinfo {author} {\bibfnamefont {W.}~\bibnamefont {Si}},
  \bibinfo {author} {\bibfnamefont {Z.-Y.}\ \bibnamefont {Dong}}, \bibinfo
  {author} {\bibfnamefont {W.}~\bibnamefont {Wang}}, \bibinfo {author}
  {\bibfnamefont {Y.}~\bibnamefont {Wu}}, \bibinfo {author} {\bibfnamefont
  {D.}~\bibnamefont {Lin}}, \bibinfo {author} {\bibfnamefont {J.}~\bibnamefont
  {Wang}}, \bibinfo {author} {\bibfnamefont {K.}~\bibnamefont {Ran}}, \bibinfo
  {author} {\bibfnamefont {S.}~\bibnamefont {Li}}, \bibinfo {author}
  {\bibfnamefont {D.}~\bibnamefont {Adroja}}, \bibinfo {author} {\bibfnamefont
  {X.}~\bibnamefont {Xi}}, \bibinfo {author} {\bibfnamefont {S.-L.}\
  \bibnamefont {Yu}}, \bibinfo {author} {\bibfnamefont {X.}~\bibnamefont {Wu}},
  \bibinfo {author} {\bibfnamefont {J.-X.}\ \bibnamefont {Li}}, \ and\ \bibinfo
  {author} {\bibfnamefont {J.}~\bibnamefont {Wen}},\ }\bibfield  {title}
  {\enquote {\bibinfo {title} {Topological magnon insulator spin excitations in
  the two-dimensional ferromagnet {CrBr}$_{3}$},}\ }\href {\doibase
  10.1103/PhysRevB.104.L020402} {\bibfield  {journal} {\bibinfo  {journal}
  {Phys. Rev. B}\ }\textbf {\bibinfo {volume} {104}},\ \bibinfo {pages}
  {L020402} (\bibinfo {year} {2021})}\BibitemShut {NoStop}%
\bibitem [{\citenamefont {Zhu}\ \emph {et~al.}(2021)\citenamefont {Zhu},
  \citenamefont {Zhang}, \citenamefont {Wang}, \citenamefont {Dos~Santos},
  \citenamefont {Song}, \citenamefont {Mueller}, \citenamefont {Schmalzl},
  \citenamefont {Schmidt}, \citenamefont {Ivanov}, \citenamefont {Park} \emph
  {et~al.}}]{zhu2021topological}%
  \BibitemOpen
  \bibfield  {author} {\bibinfo {author} {\bibfnamefont {F.}~\bibnamefont
  {Zhu}}, \bibinfo {author} {\bibfnamefont {L.}~\bibnamefont {Zhang}}, \bibinfo
  {author} {\bibfnamefont {X.}~\bibnamefont {Wang}}, \bibinfo {author}
  {\bibfnamefont {F.~J.}\ \bibnamefont {Dos~Santos}}, \bibinfo {author}
  {\bibfnamefont {J.}~\bibnamefont {Song}}, \bibinfo {author} {\bibfnamefont
  {T.}~\bibnamefont {Mueller}}, \bibinfo {author} {\bibfnamefont
  {K.}~\bibnamefont {Schmalzl}}, \bibinfo {author} {\bibfnamefont {W.~F.}\
  \bibnamefont {Schmidt}}, \bibinfo {author} {\bibfnamefont {A.}~\bibnamefont
  {Ivanov}}, \bibinfo {author} {\bibfnamefont {J.~T.}\ \bibnamefont {Park}},
  \emph {et~al.},\ }\bibfield  {title} {\enquote {\bibinfo {title} {Topological
  magnon insulators in two-dimensional van der waals ferromagnets {CrSiTe}$_3$
  and {CrGeTe}$_3$: Toward intrinsic gap-tunability},}\ }\href@noop {}
  {\bibfield  {journal} {\bibinfo  {journal} {Science advances}\ }\textbf
  {\bibinfo {volume} {7}},\ \bibinfo {pages} {eabi7532} (\bibinfo {year}
  {2021})}\BibitemShut {NoStop}%
\bibitem [{\citenamefont {Bao}\ \emph {et~al.}(2018)\citenamefont {Bao},
  \citenamefont {Wang}, \citenamefont {Wang}, \citenamefont {Cai},
  \citenamefont {Li}, \citenamefont {Ma}, \citenamefont {Wang}, \citenamefont
  {Ran}, \citenamefont {Dong}, \citenamefont {Abernathy}, \citenamefont {Yu},
  \citenamefont {Wan}, \citenamefont {Li},\ and\ \citenamefont
  {Wen}}]{Bao2018}%
  \BibitemOpen
  \bibfield  {author} {\bibinfo {author} {\bibfnamefont {S.}~\bibnamefont
  {Bao}}, \bibinfo {author} {\bibfnamefont {J.}~\bibnamefont {Wang}}, \bibinfo
  {author} {\bibfnamefont {W.}~\bibnamefont {Wang}}, \bibinfo {author}
  {\bibfnamefont {Z.}~\bibnamefont {Cai}}, \bibinfo {author} {\bibfnamefont
  {S.}~\bibnamefont {Li}}, \bibinfo {author} {\bibfnamefont {Z.}~\bibnamefont
  {Ma}}, \bibinfo {author} {\bibfnamefont {D.}~\bibnamefont {Wang}}, \bibinfo
  {author} {\bibfnamefont {K.}~\bibnamefont {Ran}}, \bibinfo {author}
  {\bibfnamefont {Z.-Y.}\ \bibnamefont {Dong}}, \bibinfo {author}
  {\bibfnamefont {D.~L.}\ \bibnamefont {Abernathy}}, \bibinfo {author}
  {\bibfnamefont {S.-L.}\ \bibnamefont {Yu}}, \bibinfo {author} {\bibfnamefont
  {X.}~\bibnamefont {Wan}}, \bibinfo {author} {\bibfnamefont {J.-X.}\
  \bibnamefont {Li}}, \ and\ \bibinfo {author} {\bibfnamefont {J.}~\bibnamefont
  {Wen}},\ }\bibfield  {title} {\enquote {\bibinfo {title} {Discovery of
  coexisting dirac and triply degenerate magnons in a three-dimensional
  antiferromagnet},}\ }\href {\doibase 10.1038/s41467-018-05054-2} {\bibfield
  {journal} {\bibinfo  {journal} {Nature Communications}\ }\textbf {\bibinfo
  {volume} {9}} (\bibinfo {year} {2018}),\
  10.1038/s41467-018-05054-2}\BibitemShut {NoStop}%
\bibitem [{\citenamefont {Yao}\ \emph {et~al.}(2018)\citenamefont {Yao},
  \citenamefont {Li}, \citenamefont {Wang}, \citenamefont {Xue}, \citenamefont
  {Dan}, \citenamefont {Iida}, \citenamefont {Kamazawa}, \citenamefont {Li},
  \citenamefont {Fang},\ and\ \citenamefont {Li}}]{yao2018topological}%
  \BibitemOpen
  \bibfield  {author} {\bibinfo {author} {\bibfnamefont {W.}~\bibnamefont
  {Yao}}, \bibinfo {author} {\bibfnamefont {C.}~\bibnamefont {Li}}, \bibinfo
  {author} {\bibfnamefont {L.}~\bibnamefont {Wang}}, \bibinfo {author}
  {\bibfnamefont {S.}~\bibnamefont {Xue}}, \bibinfo {author} {\bibfnamefont
  {Y.}~\bibnamefont {Dan}}, \bibinfo {author} {\bibfnamefont {K.}~\bibnamefont
  {Iida}}, \bibinfo {author} {\bibfnamefont {K.}~\bibnamefont {Kamazawa}},
  \bibinfo {author} {\bibfnamefont {K.}~\bibnamefont {Li}}, \bibinfo {author}
  {\bibfnamefont {C.}~\bibnamefont {Fang}}, \ and\ \bibinfo {author}
  {\bibfnamefont {Y.}~\bibnamefont {Li}},\ }\bibfield  {title} {\enquote
  {\bibinfo {title} {Topological spin excitations in a three-dimensional
  antiferromagnet},}\ }\href@noop {} {\bibfield  {journal} {\bibinfo  {journal}
  {Nature Physics}\ }\textbf {\bibinfo {volume} {14}},\ \bibinfo {pages}
  {1011--1015} (\bibinfo {year} {2018})}\BibitemShut {NoStop}%
\bibitem [{\citenamefont {Wang}\ \emph {et~al.}(2019)\citenamefont {Wang},
  \citenamefont {Bo}, \citenamefont {Tang},\ and\ \citenamefont
  {Wan}}]{Wang2019Diracmagnon}%
  \BibitemOpen
  \bibfield  {author} {\bibinfo {author} {\bibfnamefont {D.}~\bibnamefont
  {Wang}}, \bibinfo {author} {\bibfnamefont {X.}~\bibnamefont {Bo}}, \bibinfo
  {author} {\bibfnamefont {F.}~\bibnamefont {Tang}}, \ and\ \bibinfo {author}
  {\bibfnamefont {X.}~\bibnamefont {Wan}},\ }\bibfield  {title} {\enquote
  {\bibinfo {title} {Calculated magnetic exchange interactions in the dirac
  magnon material {Cu}$_{3}${TeO}$_{6}$},}\ }\href {\doibase
  10.1103/PhysRevB.99.035160} {\bibfield  {journal} {\bibinfo  {journal} {Phys.
  Rev. B}\ }\textbf {\bibinfo {volume} {99}},\ \bibinfo {pages} {035160}
  (\bibinfo {year} {2019})}\BibitemShut {NoStop}%
\bibitem [{\citenamefont {Elliot}\ \emph {et~al.}(2021)\citenamefont {Elliot},
  \citenamefont {McClarty}, \citenamefont {Prabhakaran}, \citenamefont
  {Johnson}, \citenamefont {Walker}, \citenamefont {Manuel},\ and\
  \citenamefont {Coldea}}]{elliot2021order}%
  \BibitemOpen
  \bibfield  {author} {\bibinfo {author} {\bibfnamefont {M.}~\bibnamefont
  {Elliot}}, \bibinfo {author} {\bibfnamefont {P.~A.}\ \bibnamefont
  {McClarty}}, \bibinfo {author} {\bibfnamefont {D.}~\bibnamefont
  {Prabhakaran}}, \bibinfo {author} {\bibfnamefont {R.}~\bibnamefont
  {Johnson}}, \bibinfo {author} {\bibfnamefont {H.}~\bibnamefont {Walker}},
  \bibinfo {author} {\bibfnamefont {P.}~\bibnamefont {Manuel}}, \ and\ \bibinfo
  {author} {\bibfnamefont {R.}~\bibnamefont {Coldea}},\ }\bibfield  {title}
  {\enquote {\bibinfo {title} {Order-by-disorder from bond-dependent exchange
  and intensity signature of nodal quasiparticles in a honeycomb cobaltate},}\
  }\href@noop {} {\bibfield  {journal} {\bibinfo  {journal} {Nature
  Communications}\ }\textbf {\bibinfo {volume} {12}},\ \bibinfo {pages} {1--7}
  (\bibinfo {year} {2021})}\BibitemShut {NoStop}%
\bibitem [{\citenamefont {Malz}\ \emph {et~al.}(2019)\citenamefont {Malz},
  \citenamefont {Knolle},\ and\ \citenamefont {Nunnenkamp}}]{Malz2019}%
  \BibitemOpen
  \bibfield  {author} {\bibinfo {author} {\bibfnamefont {D.}~\bibnamefont
  {Malz}}, \bibinfo {author} {\bibfnamefont {J.}~\bibnamefont {Knolle}}, \ and\
  \bibinfo {author} {\bibfnamefont {A.}~\bibnamefont {Nunnenkamp}},\ }\bibfield
   {title} {\enquote {\bibinfo {title} {Topological magnon amplification},}\
  }\href {\doibase 10.1038/s41467-019-11914-2} {\bibfield  {journal} {\bibinfo
  {journal} {Nature Communications}\ }\textbf {\bibinfo {volume} {10}}
  (\bibinfo {year} {2019}),\ 10.1038/s41467-019-11914-2}\BibitemShut {NoStop}%
\bibitem [{\citenamefont {Perreault}\ \emph {et~al.}(2016)\citenamefont
  {Perreault}, \citenamefont {Knolle}, \citenamefont {Perkins},\ and\
  \citenamefont {Burnell}}]{Perreault2016}%
  \BibitemOpen
  \bibfield  {author} {\bibinfo {author} {\bibfnamefont {B.}~\bibnamefont
  {Perreault}}, \bibinfo {author} {\bibfnamefont {J.}~\bibnamefont {Knolle}},
  \bibinfo {author} {\bibfnamefont {N.~B.}\ \bibnamefont {Perkins}}, \ and\
  \bibinfo {author} {\bibfnamefont {F.~J.}\ \bibnamefont {Burnell}},\
  }\bibfield  {title} {\enquote {\bibinfo {title} {Raman scattering in
  correlated thin films as a probe of chargeless surface states},}\ }\href
  {\doibase 10.1103/PhysRevB.94.060408} {\bibfield  {journal} {\bibinfo
  {journal} {Phys. Rev. B}\ }\textbf {\bibinfo {volume} {94}},\ \bibinfo
  {pages} {060408} (\bibinfo {year} {2016})}\BibitemShut {NoStop}%
\bibitem [{\citenamefont {Feldmeier}\ \emph {et~al.}(2020)\citenamefont
  {Feldmeier}, \citenamefont {Natori}, \citenamefont {Knap},\ and\
  \citenamefont {Knolle}}]{Feldmeier2020}%
  \BibitemOpen
  \bibfield  {author} {\bibinfo {author} {\bibfnamefont {J.}~\bibnamefont
  {Feldmeier}}, \bibinfo {author} {\bibfnamefont {W.}~\bibnamefont {Natori}},
  \bibinfo {author} {\bibfnamefont {M.}~\bibnamefont {Knap}}, \ and\ \bibinfo
  {author} {\bibfnamefont {J.}~\bibnamefont {Knolle}},\ }\bibfield  {title}
  {\enquote {\bibinfo {title} {Local probes for charge-neutral edge states in
  two-dimensional quantum magnets},}\ }\href {\doibase
  10.1103/PhysRevB.102.134423} {\bibfield  {journal} {\bibinfo  {journal}
  {Phys. Rev. B}\ }\textbf {\bibinfo {volume} {102}},\ \bibinfo {pages}
  {134423} (\bibinfo {year} {2020})}\BibitemShut {NoStop}%
\bibitem [{\citenamefont {Rustagi}\ \emph {et~al.}(2020)\citenamefont
  {Rustagi}, \citenamefont {Bertelli}, \citenamefont {van~der Sar},\ and\
  \citenamefont {Upadhyaya}}]{Rustagi2020}%
  \BibitemOpen
  \bibfield  {author} {\bibinfo {author} {\bibfnamefont {A.}~\bibnamefont
  {Rustagi}}, \bibinfo {author} {\bibfnamefont {I.}~\bibnamefont {Bertelli}},
  \bibinfo {author} {\bibfnamefont {T.}~\bibnamefont {van~der Sar}}, \ and\
  \bibinfo {author} {\bibfnamefont {P.}~\bibnamefont {Upadhyaya}},\ }\bibfield
  {title} {\enquote {\bibinfo {title} {Sensing chiral magnetic noise via
  quantum impurity relaxometry},}\ }\href {\doibase
  10.1103/PhysRevB.102.220403} {\bibfield  {journal} {\bibinfo  {journal}
  {Phys. Rev. B}\ }\textbf {\bibinfo {volume} {102}},\ \bibinfo {pages}
  {220403} (\bibinfo {year} {2020})}\BibitemShut {NoStop}%
\bibitem [{\citenamefont {dos Santos}\ \emph {et~al.}(2018)\citenamefont {dos
  Santos}, \citenamefont {dos Santos~Dias}, \citenamefont {Guimar\~aes},
  \citenamefont {Bouaziz},\ and\ \citenamefont {Lounis}}]{DosSantos2018}%
  \BibitemOpen
  \bibfield  {author} {\bibinfo {author} {\bibfnamefont {F.~J.}\ \bibnamefont
  {dos Santos}}, \bibinfo {author} {\bibfnamefont {M.}~\bibnamefont {dos
  Santos~Dias}}, \bibinfo {author} {\bibfnamefont {F.~S.~M.}\ \bibnamefont
  {Guimar\~aes}}, \bibinfo {author} {\bibfnamefont {J.}~\bibnamefont
  {Bouaziz}}, \ and\ \bibinfo {author} {\bibfnamefont {S.}~\bibnamefont
  {Lounis}},\ }\bibfield  {title} {\enquote {\bibinfo {title} {Spin-resolved
  inelastic electron scattering by spin waves in noncollinear magnets},}\
  }\href {\doibase 10.1103/PhysRevB.97.024431} {\bibfield  {journal} {\bibinfo
  {journal} {Phys. Rev. B}\ }\textbf {\bibinfo {volume} {97}},\ \bibinfo
  {pages} {024431} (\bibinfo {year} {2018})}\BibitemShut {NoStop}%
\bibitem [{\citenamefont {Fert}\ \emph {et~al.}(1978)\citenamefont {Fert},
  \citenamefont {Bertrand}, \citenamefont {Leotin}, \citenamefont {Ousset},
  \citenamefont {Magari{\~{n}}o},\ and\ \citenamefont {Tuchendler}}]{Fert1978}%
  \BibitemOpen
  \bibfield  {author} {\bibinfo {author} {\bibfnamefont {A.}~\bibnamefont
  {Fert}}, \bibinfo {author} {\bibfnamefont {D.}~\bibnamefont {Bertrand}},
  \bibinfo {author} {\bibfnamefont {J.}~\bibnamefont {Leotin}}, \bibinfo
  {author} {\bibfnamefont {J.}~\bibnamefont {Ousset}}, \bibinfo {author}
  {\bibfnamefont {J.}~\bibnamefont {Magari{\~{n}}o}}, \ and\ \bibinfo {author}
  {\bibfnamefont {J.}~\bibnamefont {Tuchendler}},\ }\bibfield  {title}
  {\enquote {\bibinfo {title} {Excitation of two spin deviations by far
  infrared absorption in {FeI}$_2$},}\ }\href {\doibase
  10.1016/0038-1098(78)90721-4} {\bibfield  {journal} {\bibinfo  {journal}
  {Solid State Communications}\ }\textbf {\bibinfo {volume} {26}},\ \bibinfo
  {pages} {693--696} (\bibinfo {year} {1978})}\BibitemShut {NoStop}%
\bibitem [{\citenamefont {Petitgrand}\ \emph {et~al.}(1980)\citenamefont
  {Petitgrand}, \citenamefont {Brun},\ and\ \citenamefont
  {Meyer}}]{Petitgrand1980}%
  \BibitemOpen
  \bibfield  {author} {\bibinfo {author} {\bibfnamefont {D.}~\bibnamefont
  {Petitgrand}}, \bibinfo {author} {\bibfnamefont {A.}~\bibnamefont {Brun}}, \
  and\ \bibinfo {author} {\bibfnamefont {P.}~\bibnamefont {Meyer}},\ }\bibfield
   {title} {\enquote {\bibinfo {title} {Magnetic field dependence of spin waves
  and two magnon bound states in {FeI}$_2$},}\ }\href {\doibase
  10.1016/0304-8853(80)91097-5} {\bibfield  {journal} {\bibinfo  {journal}
  {Journal of Magnetism and Magnetic Materials}\ }\textbf {\bibinfo {volume}
  {15-18}},\ \bibinfo {pages} {381--382} (\bibinfo {year} {1980})}\BibitemShut
  {NoStop}%
\bibitem [{\citenamefont {Petitgrand}\ \emph {et~al.}(1979)\citenamefont
  {Petitgrand}, \citenamefont {Hennion},\ and\ \citenamefont
  {Escribe}}]{Petitgrand1979}%
  \BibitemOpen
  \bibfield  {author} {\bibinfo {author} {\bibfnamefont {D.}~\bibnamefont
  {Petitgrand}}, \bibinfo {author} {\bibfnamefont {B.}~\bibnamefont {Hennion}},
  \ and\ \bibinfo {author} {\bibfnamefont {C.}~\bibnamefont {Escribe}},\
  }\bibfield  {title} {\enquote {\bibinfo {title} {Neutron inelastic scattering
  from magnetic excitations of {FeI}$_2$},}\ }\href
  {http://inis.iaea.org/search/search.aspx?orig_q=RN:12586539} {\bibfield
  {journal} {\bibinfo  {journal} {Journal of Magnetism and Magnetic Materials}\
  ,\ \bibinfo {pages} {275--276}} (\bibinfo {year} {1979})}\BibitemShut
  {NoStop}%
\bibitem [{\citenamefont {Katsumata}\ \emph {et~al.}(2000)\citenamefont
  {Katsumata}, \citenamefont {Yamaguchi}, \citenamefont {Hagiwara},
  \citenamefont {Tokunaga}, \citenamefont {Mikeska}, \citenamefont {Goy},\ and\
  \citenamefont {Gross}}]{Katsumata2000}%
  \BibitemOpen
  \bibfield  {author} {\bibinfo {author} {\bibfnamefont {K.}~\bibnamefont
  {Katsumata}}, \bibinfo {author} {\bibfnamefont {H.}~\bibnamefont
  {Yamaguchi}}, \bibinfo {author} {\bibfnamefont {M.}~\bibnamefont {Hagiwara}},
  \bibinfo {author} {\bibfnamefont {M.}~\bibnamefont {Tokunaga}}, \bibinfo
  {author} {\bibfnamefont {H.-J.}\ \bibnamefont {Mikeska}}, \bibinfo {author}
  {\bibfnamefont {P.}~\bibnamefont {Goy}}, \ and\ \bibinfo {author}
  {\bibfnamefont {M.}~\bibnamefont {Gross}},\ }\bibfield  {title} {\enquote
  {\bibinfo {title} {Single-ion magnon bound states in an antiferromagnet with
  strong uniaxial anisotropy},}\ }\href {\doibase 10.1103/PhysRevB.61.11632}
  {\bibfield  {journal} {\bibinfo  {journal} {Phys. Rev. B}\ }\textbf {\bibinfo
  {volume} {61}},\ \bibinfo {pages} {11632--11636} (\bibinfo {year}
  {2000})}\BibitemShut {NoStop}%
\bibitem [{\citenamefont {Park}\ and\ \citenamefont {Yang}(2020)}]{Park2020}%
  \BibitemOpen
  \bibfield  {author} {\bibinfo {author} {\bibfnamefont {S.}~\bibnamefont
  {Park}}\ and\ \bibinfo {author} {\bibfnamefont {B.-J.}\ \bibnamefont
  {Yang}},\ }\bibfield  {title} {\enquote {\bibinfo {title} {Thermal {H}all
  effect from a two-dimensional {S}chwinger boson gas with {R}ashba spin-orbit
  interaction: {A}pplication to ferromagnets with in-plane
  {D}zyaloshinskii-{M}oriya interaction},}\ }\href {\doibase
  10.1103/PhysRevB.102.214421} {\bibfield  {journal} {\bibinfo  {journal}
  {Phys. Rev. B}\ }\textbf {\bibinfo {volume} {102}},\ \bibinfo {pages}
  {214421} (\bibinfo {year} {2020})}\BibitemShut {NoStop}%
\bibitem [{\citenamefont {Carnahan}\ \emph {et~al.}(2021)\citenamefont
  {Carnahan}, \citenamefont {Zhang},\ and\ \citenamefont
  {Xiao}}]{Carnahan2021}%
  \BibitemOpen
  \bibfield  {author} {\bibinfo {author} {\bibfnamefont {C.}~\bibnamefont
  {Carnahan}}, \bibinfo {author} {\bibfnamefont {Y.}~\bibnamefont {Zhang}}, \
  and\ \bibinfo {author} {\bibfnamefont {D.}~\bibnamefont {Xiao}},\ }\bibfield
  {title} {\enquote {\bibinfo {title} {Thermal {H}all effect of chiral spin
  fluctuations},}\ }\href {\doibase 10.1103/PhysRevB.103.224419} {\bibfield
  {journal} {\bibinfo  {journal} {Phys. Rev. B}\ }\textbf {\bibinfo {volume}
  {103}},\ \bibinfo {pages} {224419} (\bibinfo {year} {2021})}\BibitemShut
  {NoStop}%
\bibitem [{\citenamefont {Mook}\ \emph
  {et~al.}(2016{\natexlab{b}})\citenamefont {Mook}, \citenamefont {Henk},\ and\
  \citenamefont {Mertig}}]{Mook2016spindynamics}%
  \BibitemOpen
  \bibfield  {author} {\bibinfo {author} {\bibfnamefont {A.}~\bibnamefont
  {Mook}}, \bibinfo {author} {\bibfnamefont {J.}~\bibnamefont {Henk}}, \ and\
  \bibinfo {author} {\bibfnamefont {I.}~\bibnamefont {Mertig}},\ }\bibfield
  {title} {\enquote {\bibinfo {title} {Spin dynamics simulations of topological
  magnon insulators: From transverse current correlation functions to the
  family of magnon {H}all effects},}\ }\href {\doibase
  10.1103/PhysRevB.94.174444} {\bibfield  {journal} {\bibinfo  {journal} {Phys.
  Rev. B}\ }\textbf {\bibinfo {volume} {94}},\ \bibinfo {pages} {174444}
  (\bibinfo {year} {2016}{\natexlab{b}})}\BibitemShut {NoStop}%
\bibitem [{\citenamefont {Mook}\ \emph
  {et~al.}(2017{\natexlab{b}})\citenamefont {Mook}, \citenamefont {G\"obel},
  \citenamefont {Henk},\ and\ \citenamefont {Mertig}}]{Mook2017noncollinear}%
  \BibitemOpen
  \bibfield  {author} {\bibinfo {author} {\bibfnamefont {A.}~\bibnamefont
  {Mook}}, \bibinfo {author} {\bibfnamefont {B.}~\bibnamefont {G\"obel}},
  \bibinfo {author} {\bibfnamefont {J.}~\bibnamefont {Henk}}, \ and\ \bibinfo
  {author} {\bibfnamefont {I.}~\bibnamefont {Mertig}},\ }\bibfield  {title}
  {\enquote {\bibinfo {title} {Magnon transport in noncollinear spin textures:
  Anisotropies and topological magnon {H}all effects},}\ }\href {\doibase
  10.1103/PhysRevB.95.020401} {\bibfield  {journal} {\bibinfo  {journal} {Phys.
  Rev. B}\ }\textbf {\bibinfo {volume} {95}},\ \bibinfo {pages} {020401}
  (\bibinfo {year} {2017}{\natexlab{b}})}\BibitemShut {NoStop}%
\bibitem [{\citenamefont {Matsumoto}\ and\ \citenamefont
  {Murakami}(2011{\natexlab{a}})}]{Matsumoto2011a}%
  \BibitemOpen
  \bibfield  {author} {\bibinfo {author} {\bibfnamefont {R.}~\bibnamefont
  {Matsumoto}}\ and\ \bibinfo {author} {\bibfnamefont {S.}~\bibnamefont
  {Murakami}},\ }\bibfield  {title} {\enquote {\bibinfo {title} {Theoretical
  prediction of a rotating magnon wave packet in ferromagnets},}\ }\href
  {\doibase 10.1103/PhysRevLett.106.197202} {\bibfield  {journal} {\bibinfo
  {journal} {Phys. Rev. Lett.}\ }\textbf {\bibinfo {volume} {106}},\ \bibinfo
  {pages} {197202} (\bibinfo {year} {2011}{\natexlab{a}})}\BibitemShut
  {NoStop}%
\bibitem [{\citenamefont {Matsumoto}\ and\ \citenamefont
  {Murakami}(2011{\natexlab{b}})}]{Matsumoto2011b}%
  \BibitemOpen
  \bibfield  {author} {\bibinfo {author} {\bibfnamefont {R.}~\bibnamefont
  {Matsumoto}}\ and\ \bibinfo {author} {\bibfnamefont {S.}~\bibnamefont
  {Murakami}},\ }\bibfield  {title} {\enquote {\bibinfo {title} {Rotational
  motion of magnons and the thermal {H}all effect},}\ }\href {\doibase
  10.1103/PhysRevB.84.184406} {\bibfield  {journal} {\bibinfo  {journal} {Phys.
  Rev. B}\ }\textbf {\bibinfo {volume} {84}},\ \bibinfo {pages} {184406}
  (\bibinfo {year} {2011}{\natexlab{b}})}\BibitemShut {NoStop}%
\bibitem [{\citenamefont {Matsumoto}\ \emph {et~al.}(2014)\citenamefont
  {Matsumoto}, \citenamefont {Shindou},\ and\ \citenamefont
  {Murakami}}]{Matsumoto2014}%
  \BibitemOpen
  \bibfield  {author} {\bibinfo {author} {\bibfnamefont {R.}~\bibnamefont
  {Matsumoto}}, \bibinfo {author} {\bibfnamefont {R.}~\bibnamefont {Shindou}},
  \ and\ \bibinfo {author} {\bibfnamefont {S.}~\bibnamefont {Murakami}},\
  }\bibfield  {title} {\enquote {\bibinfo {title} {Thermal {H}all effect of
  magnons in magnets with dipolar interaction},}\ }\href {\doibase
  10.1103/PhysRevB.89.054420} {\bibfield  {journal} {\bibinfo  {journal} {Phys.
  Rev. B}\ }\textbf {\bibinfo {volume} {89}},\ \bibinfo {pages} {054420}
  (\bibinfo {year} {2014})}\BibitemShut {NoStop}%
\bibitem [{\citenamefont {Liang}\ \emph {et~al.}(2020)\citenamefont {Liang},
  \citenamefont {Wang}, \citenamefont {Du}, \citenamefont {Hallal},
  \citenamefont {Garcia}, \citenamefont {Chshiev}, \citenamefont {Fert},\ and\
  \citenamefont {Yang}}]{Liang2020}%
  \BibitemOpen
  \bibfield  {author} {\bibinfo {author} {\bibfnamefont {J.}~\bibnamefont
  {Liang}}, \bibinfo {author} {\bibfnamefont {W.}~\bibnamefont {Wang}},
  \bibinfo {author} {\bibfnamefont {H.}~\bibnamefont {Du}}, \bibinfo {author}
  {\bibfnamefont {A.}~\bibnamefont {Hallal}}, \bibinfo {author} {\bibfnamefont
  {K.}~\bibnamefont {Garcia}}, \bibinfo {author} {\bibfnamefont
  {M.}~\bibnamefont {Chshiev}}, \bibinfo {author} {\bibfnamefont
  {A.}~\bibnamefont {Fert}}, \ and\ \bibinfo {author} {\bibfnamefont
  {H.}~\bibnamefont {Yang}},\ }\bibfield  {title} {\enquote {\bibinfo {title}
  {Very large {D}zyaloshinskii-{M}oriya interaction in two-dimensional {J}anus
  manganese dichalcogenides and its application to realize skyrmion states},}\
  }\href {\doibase 10.1103/PhysRevB.101.184401} {\bibfield  {journal} {\bibinfo
   {journal} {Phys. Rev. B}\ }\textbf {\bibinfo {volume} {101}},\ \bibinfo
  {pages} {184401} (\bibinfo {year} {2020})}\BibitemShut {NoStop}%
\bibitem [{\citenamefont {Cui}\ \emph {et~al.}(2020)\citenamefont {Cui},
  \citenamefont {Liang}, \citenamefont {Shao}, \citenamefont {Cui},\ and\
  \citenamefont {Yang}}]{Cui2020}%
  \BibitemOpen
  \bibfield  {author} {\bibinfo {author} {\bibfnamefont {Q.}~\bibnamefont
  {Cui}}, \bibinfo {author} {\bibfnamefont {J.}~\bibnamefont {Liang}}, \bibinfo
  {author} {\bibfnamefont {Z.}~\bibnamefont {Shao}}, \bibinfo {author}
  {\bibfnamefont {P.}~\bibnamefont {Cui}}, \ and\ \bibinfo {author}
  {\bibfnamefont {H.}~\bibnamefont {Yang}},\ }\bibfield  {title} {\enquote
  {\bibinfo {title} {Strain-tunable ferromagnetism and chiral spin textures in
  two-dimensional {J}anus chromium dichalcogenides},}\ }\href {\doibase
  10.1103/PhysRevB.102.094425} {\bibfield  {journal} {\bibinfo  {journal}
  {Phys. Rev. B}\ }\textbf {\bibinfo {volume} {102}},\ \bibinfo {pages}
  {094425} (\bibinfo {year} {2020})}\BibitemShut {NoStop}%
\bibitem [{\citenamefont {Yuan}\ \emph {et~al.}(2020)\citenamefont {Yuan},
  \citenamefont {Yang}, \citenamefont {Cai}, \citenamefont {Wu}, \citenamefont
  {Chen}, \citenamefont {Yan},\ and\ \citenamefont {Shen}}]{Yuan2020Janus}%
  \BibitemOpen
  \bibfield  {author} {\bibinfo {author} {\bibfnamefont {J.}~\bibnamefont
  {Yuan}}, \bibinfo {author} {\bibfnamefont {Y.}~\bibnamefont {Yang}}, \bibinfo
  {author} {\bibfnamefont {Y.}~\bibnamefont {Cai}}, \bibinfo {author}
  {\bibfnamefont {Y.}~\bibnamefont {Wu}}, \bibinfo {author} {\bibfnamefont
  {Y.}~\bibnamefont {Chen}}, \bibinfo {author} {\bibfnamefont {X.}~\bibnamefont
  {Yan}}, \ and\ \bibinfo {author} {\bibfnamefont {L.}~\bibnamefont {Shen}},\
  }\bibfield  {title} {\enquote {\bibinfo {title} {Intrinsic skyrmions in
  monolayer janus magnets},}\ }\href {\doibase 10.1103/PhysRevB.101.094420}
  {\bibfield  {journal} {\bibinfo  {journal} {Phys. Rev. B}\ }\textbf {\bibinfo
  {volume} {101}},\ \bibinfo {pages} {094420} (\bibinfo {year}
  {2020})}\BibitemShut {NoStop}%
\bibitem [{\citenamefont {Zhang}\ \emph {et~al.}(2020)\citenamefont {Zhang},
  \citenamefont {Xu}, \citenamefont {Chen}, \citenamefont {Nahas},
  \citenamefont {Prokhorenko},\ and\ \citenamefont
  {Bellaiche}}]{Zhang2020Janus}%
  \BibitemOpen
  \bibfield  {author} {\bibinfo {author} {\bibfnamefont {Y.}~\bibnamefont
  {Zhang}}, \bibinfo {author} {\bibfnamefont {C.}~\bibnamefont {Xu}}, \bibinfo
  {author} {\bibfnamefont {P.}~\bibnamefont {Chen}}, \bibinfo {author}
  {\bibfnamefont {Y.}~\bibnamefont {Nahas}}, \bibinfo {author} {\bibfnamefont
  {S.}~\bibnamefont {Prokhorenko}}, \ and\ \bibinfo {author} {\bibfnamefont
  {L.}~\bibnamefont {Bellaiche}},\ }\bibfield  {title} {\enquote {\bibinfo
  {title} {Emergence of skyrmionium in a two-dimensional {CrGe(Se,Te)}$_{3}$
  janus monolayer},}\ }\href {\doibase 10.1103/PhysRevB.102.241107} {\bibfield
  {journal} {\bibinfo  {journal} {Phys. Rev. B}\ }\textbf {\bibinfo {volume}
  {102}},\ \bibinfo {pages} {241107} (\bibinfo {year} {2020})}\BibitemShut
  {NoStop}%
\bibitem [{\citenamefont {Shen}\ \emph {et~al.}(2021)\citenamefont {Shen},
  \citenamefont {Song}, \citenamefont {Xue}, \citenamefont {Wu}, \citenamefont
  {Wang},\ and\ \citenamefont {Zhong}}]{Shen2021Janus}%
  \BibitemOpen
  \bibfield  {author} {\bibinfo {author} {\bibfnamefont {Z.}~\bibnamefont
  {Shen}}, \bibinfo {author} {\bibfnamefont {C.}~\bibnamefont {Song}}, \bibinfo
  {author} {\bibfnamefont {Y.}~\bibnamefont {Xue}}, \bibinfo {author}
  {\bibfnamefont {Z.}~\bibnamefont {Wu}}, \bibinfo {author} {\bibfnamefont
  {J.}~\bibnamefont {Wang}}, \ and\ \bibinfo {author} {\bibfnamefont
  {Z.}~\bibnamefont {Zhong}},\ }\href@noop {} {\enquote {\bibinfo {title}
  {Strain tunable skyrmions and strong {D}zyaloshinskii-{M}oriya interaction in
  two-dimensional {J}anus {Cr(X,Y)}$_{3}$ trihalides monolayers},}\ } (\bibinfo
  {year} {2021}),\ \Eprint {http://arxiv.org/abs/arXiv:2109.00723}
  {arXiv:2109.00723} \BibitemShut {NoStop}%
\bibitem [{\citenamefont {Chen}\ \emph {et~al.}(2020)\citenamefont {Chen},
  \citenamefont {Mascaraque}, \citenamefont {Jia}, \citenamefont {Zimmermann},
  \citenamefont {Robertson}, \citenamefont {Conte}, \citenamefont {Hoffmann},
  \citenamefont {Barrio}, \citenamefont {Ding}, \citenamefont {Wiesendanger},
  \citenamefont {Michel}, \citenamefont {Blügel}, \citenamefont {Schmid},\
  and\ \citenamefont {Liu}}]{Chen2020chemi}%
  \BibitemOpen
  \bibfield  {author} {\bibinfo {author} {\bibfnamefont {G.}~\bibnamefont
  {Chen}}, \bibinfo {author} {\bibfnamefont {A.}~\bibnamefont {Mascaraque}},
  \bibinfo {author} {\bibfnamefont {H.}~\bibnamefont {Jia}}, \bibinfo {author}
  {\bibfnamefont {B.}~\bibnamefont {Zimmermann}}, \bibinfo {author}
  {\bibfnamefont {M.}~\bibnamefont {Robertson}}, \bibinfo {author}
  {\bibfnamefont {R.~L.}\ \bibnamefont {Conte}}, \bibinfo {author}
  {\bibfnamefont {M.}~\bibnamefont {Hoffmann}}, \bibinfo {author}
  {\bibfnamefont {M.~A.~G.}\ \bibnamefont {Barrio}}, \bibinfo {author}
  {\bibfnamefont {H.}~\bibnamefont {Ding}}, \bibinfo {author} {\bibfnamefont
  {R.}~\bibnamefont {Wiesendanger}}, \bibinfo {author} {\bibfnamefont {E.~G.}\
  \bibnamefont {Michel}}, \bibinfo {author} {\bibfnamefont {S.}~\bibnamefont
  {Blügel}}, \bibinfo {author} {\bibfnamefont {A.~K.}\ \bibnamefont {Schmid}},
  \ and\ \bibinfo {author} {\bibfnamefont {K.}~\bibnamefont {Liu}},\ }\bibfield
   {title} {\enquote {\bibinfo {title} {Large {D}zyaloshinskii-{M}oriya
  interaction induced by chemisorbed oxygen on a ferromagnet surface},}\ }\href
  {\doibase 10.1126/sciadv.aba4924} {\bibfield  {journal} {\bibinfo  {journal}
  {Science Advances}\ }\textbf {\bibinfo {volume} {6}},\ \bibinfo {pages}
  {eaba4924} (\bibinfo {year} {2020})}\BibitemShut {NoStop}%
\bibitem [{\citenamefont {Koch}\ \emph {et~al.}(2003)\citenamefont {Koch},
  \citenamefont {Waldmann}, \citenamefont {M\"uller}, \citenamefont {Reimann},\
  and\ \citenamefont {Saalfrank}}]{Koch2003}%
  \BibitemOpen
  \bibfield  {author} {\bibinfo {author} {\bibfnamefont {R.}~\bibnamefont
  {Koch}}, \bibinfo {author} {\bibfnamefont {O.}~\bibnamefont {Waldmann}},
  \bibinfo {author} {\bibfnamefont {P.}~\bibnamefont {M\"uller}}, \bibinfo
  {author} {\bibfnamefont {U.}~\bibnamefont {Reimann}}, \ and\ \bibinfo
  {author} {\bibfnamefont {R.~W.}\ \bibnamefont {Saalfrank}},\ }\bibfield
  {title} {\enquote {\bibinfo {title} {Ferromagnetic coupling and magnetic
  anisotropy in molecular {Ni(II)} squares},}\ }\href {\doibase
  10.1103/PhysRevB.67.094407} {\bibfield  {journal} {\bibinfo  {journal} {Phys.
  Rev. B}\ }\textbf {\bibinfo {volume} {67}},\ \bibinfo {pages} {094407}
  (\bibinfo {year} {2003})}\BibitemShut {NoStop}%
\bibitem [{\citenamefont {McGuire}(2017)}]{McGuire2017}%
  \BibitemOpen
  \bibfield  {author} {\bibinfo {author} {\bibfnamefont {M.}~\bibnamefont
  {McGuire}},\ }\bibfield  {title} {\enquote {\bibinfo {title} {Crystal and
  magnetic structures in layered, transition metal dihalides and trihalides},}\
  }\href {\doibase 10.3390/cryst7050121} {\bibfield  {journal} {\bibinfo
  {journal} {Crystals}\ }\textbf {\bibinfo {volume} {7}},\ \bibinfo {pages}
  {121} (\bibinfo {year} {2017})}\BibitemShut {NoStop}%
\bibitem [{\citenamefont {Gelard}\ \emph {et~al.}(1974)\citenamefont {Gelard},
  \citenamefont {Fert}, \citenamefont {Meriel},\ and\ \citenamefont
  {Allain}}]{Gelard1974}%
  \BibitemOpen
  \bibfield  {author} {\bibinfo {author} {\bibfnamefont {J.}~\bibnamefont
  {Gelard}}, \bibinfo {author} {\bibfnamefont {A.}~\bibnamefont {Fert}},
  \bibinfo {author} {\bibfnamefont {P.}~\bibnamefont {Meriel}}, \ and\ \bibinfo
  {author} {\bibfnamefont {Y.}~\bibnamefont {Allain}},\ }\bibfield  {title}
  {\enquote {\bibinfo {title} {Magnetic structure of {FeI}$_2$ by neutron
  diffraction experiments},}\ }\href {\doibase 10.1016/0038-1098(74)90213-0}
  {\bibfield  {journal} {\bibinfo  {journal} {Solid State Communications}\
  }\textbf {\bibinfo {volume} {14}},\ \bibinfo {pages} {187--189} (\bibinfo
  {year} {1974})}\BibitemShut {NoStop}%
\bibitem [{\citenamefont {Gallego}\ \emph {et~al.}(2019)\citenamefont
  {Gallego}, \citenamefont {Etxebarria}, \citenamefont {Elcoro}, \citenamefont
  {Tasci},\ and\ \citenamefont {Perez-Mato}}]{Gallego2019}%
  \BibitemOpen
  \bibfield  {author} {\bibinfo {author} {\bibfnamefont {S.~V.}\ \bibnamefont
  {Gallego}}, \bibinfo {author} {\bibfnamefont {J.}~\bibnamefont {Etxebarria}},
  \bibinfo {author} {\bibfnamefont {L.}~\bibnamefont {Elcoro}}, \bibinfo
  {author} {\bibfnamefont {E.~S.}\ \bibnamefont {Tasci}}, \ and\ \bibinfo
  {author} {\bibfnamefont {J.~M.}\ \bibnamefont {Perez-Mato}},\ }\bibfield
  {title} {\enquote {\bibinfo {title} {Automatic calculation of
  symmetry-adapted tensors in magnetic and non-magnetic materials: a new tool
  of the {B}ilbao {C}rystallographic {S}erver},}\ }\href {\doibase
  10.1107/s2053273319001748} {\bibfield  {journal} {\bibinfo  {journal} {Acta
  Crystallographica Section A Foundations and Advances}\ }\textbf {\bibinfo
  {volume} {75}},\ \bibinfo {pages} {438--447} (\bibinfo {year}
  {2019})}\BibitemShut {NoStop}%
\bibitem [{\citenamefont {Giamarchi}\ \emph {et~al.}(2008)\citenamefont
  {Giamarchi}, \citenamefont {R{\"u}egg},\ and\ \citenamefont
  {Tchernyshyov}}]{giamarchi2008bose}%
  \BibitemOpen
  \bibfield  {author} {\bibinfo {author} {\bibfnamefont {T.}~\bibnamefont
  {Giamarchi}}, \bibinfo {author} {\bibfnamefont {C.}~\bibnamefont
  {R{\"u}egg}}, \ and\ \bibinfo {author} {\bibfnamefont {O.}~\bibnamefont
  {Tchernyshyov}},\ }\bibfield  {title} {\enquote {\bibinfo {title}
  {Bose--{E}instein condensation in magnetic insulators},}\ }\href@noop {}
  {\bibfield  {journal} {\bibinfo  {journal} {Nature Physics}\ }\textbf
  {\bibinfo {volume} {4}},\ \bibinfo {pages} {198--204} (\bibinfo {year}
  {2008})}\BibitemShut {NoStop}%
\bibitem [{\citenamefont {Zapf}\ \emph {et~al.}(2014)\citenamefont {Zapf},
  \citenamefont {Jaime},\ and\ \citenamefont {Batista}}]{Zapf2014}%
  \BibitemOpen
  \bibfield  {author} {\bibinfo {author} {\bibfnamefont {V.}~\bibnamefont
  {Zapf}}, \bibinfo {author} {\bibfnamefont {M.}~\bibnamefont {Jaime}}, \ and\
  \bibinfo {author} {\bibfnamefont {C.~D.}\ \bibnamefont {Batista}},\
  }\bibfield  {title} {\enquote {\bibinfo {title} {Bose-{E}instein condensation
  in quantum magnets},}\ }\href {\doibase 10.1103/RevModPhys.86.563} {\bibfield
   {journal} {\bibinfo  {journal} {Rev. Mod. Phys.}\ }\textbf {\bibinfo
  {volume} {86}},\ \bibinfo {pages} {563--614} (\bibinfo {year}
  {2014})}\BibitemShut {NoStop}%
\bibitem [{\citenamefont {Nishida}\ \emph {et~al.}(2013)\citenamefont
  {Nishida}, \citenamefont {Kato},\ and\ \citenamefont
  {Batista}}]{Nishida2013}%
  \BibitemOpen
  \bibfield  {author} {\bibinfo {author} {\bibfnamefont {Y.}~\bibnamefont
  {Nishida}}, \bibinfo {author} {\bibfnamefont {Y.}~\bibnamefont {Kato}}, \
  and\ \bibinfo {author} {\bibfnamefont {C.~D.}\ \bibnamefont {Batista}},\
  }\bibfield  {title} {\enquote {\bibinfo {title} {Efimov effect in quantum
  magnets},}\ }\href {\doibase 10.1038/nphys2523} {\bibfield  {journal}
  {\bibinfo  {journal} {Nature Physics}\ }\textbf {\bibinfo {volume} {9}},\
  \bibinfo {pages} {93--97} (\bibinfo {year} {2013})}\BibitemShut {NoStop}%
\bibitem [{\citenamefont {Corticelli}\ \emph
  {et~al.}(2022{\natexlab{b}})\citenamefont {Corticelli}, \citenamefont
  {Moessner},\ and\ \citenamefont {McClarty}}]{Corticelli2022b}%
  \BibitemOpen
  \bibfield  {author} {\bibinfo {author} {\bibfnamefont {A.}~\bibnamefont
  {Corticelli}}, \bibinfo {author} {\bibfnamefont {R.}~\bibnamefont
  {Moessner}}, \ and\ \bibinfo {author} {\bibfnamefont {P.~A.}\ \bibnamefont
  {McClarty}},\ }\href {\doibase 10.48550/ARXIV.2203.06678} {\enquote {\bibinfo
  {title} {Identifying, and constructing, complex magnon band topology},}\ }
  (\bibinfo {year} {2022}{\natexlab{b}})\BibitemShut {NoStop}%
\bibitem [{\citenamefont {Zyuzin}\ and\ \citenamefont
  {Kovalev}(2016)}]{Zyuzin2016SNE}%
  \BibitemOpen
  \bibfield  {author} {\bibinfo {author} {\bibfnamefont {V.~A.}\ \bibnamefont
  {Zyuzin}}\ and\ \bibinfo {author} {\bibfnamefont {A.~A.}\ \bibnamefont
  {Kovalev}},\ }\bibfield  {title} {\enquote {\bibinfo {title} {Magnon spin
  {N}ernst effect in antiferromagnets},}\ }\href {\doibase
  10.1103/PhysRevLett.117.217203} {\bibfield  {journal} {\bibinfo  {journal}
  {Phys. Rev. Lett.}\ }\textbf {\bibinfo {volume} {117}},\ \bibinfo {pages}
  {217203} (\bibinfo {year} {2016})}\BibitemShut {NoStop}%
\bibitem [{\citenamefont {Cheng}\ \emph {et~al.}(2016)\citenamefont {Cheng},
  \citenamefont {Okamoto},\ and\ \citenamefont {Xiao}}]{Cheng2016SNE}%
  \BibitemOpen
  \bibfield  {author} {\bibinfo {author} {\bibfnamefont {R.}~\bibnamefont
  {Cheng}}, \bibinfo {author} {\bibfnamefont {S.}~\bibnamefont {Okamoto}}, \
  and\ \bibinfo {author} {\bibfnamefont {D.}~\bibnamefont {Xiao}},\ }\bibfield
  {title} {\enquote {\bibinfo {title} {Spin {N}ernst effect of magnons in
  collinear antiferromagnets},}\ }\href {\doibase
  10.1103/PhysRevLett.117.217202} {\bibfield  {journal} {\bibinfo  {journal}
  {Phys. Rev. Lett.}\ }\textbf {\bibinfo {volume} {117}},\ \bibinfo {pages}
  {217202} (\bibinfo {year} {2016})}\BibitemShut {NoStop}%
\bibitem [{\citenamefont {Zhang}\ \emph {et~al.}(2018)\citenamefont {Zhang},
  \citenamefont {Okamoto},\ and\ \citenamefont {Xiao}}]{Zhang2018}%
  \BibitemOpen
  \bibfield  {author} {\bibinfo {author} {\bibfnamefont {Y.}~\bibnamefont
  {Zhang}}, \bibinfo {author} {\bibfnamefont {S.}~\bibnamefont {Okamoto}}, \
  and\ \bibinfo {author} {\bibfnamefont {D.}~\bibnamefont {Xiao}},\ }\bibfield
  {title} {\enquote {\bibinfo {title} {Spin-nernst effect in the paramagnetic
  regime of an antiferromagnetic insulator},}\ }\href {\doibase
  10.1103/PhysRevB.98.035424} {\bibfield  {journal} {\bibinfo  {journal} {Phys.
  Rev. B}\ }\textbf {\bibinfo {volume} {98}},\ \bibinfo {pages} {035424}
  (\bibinfo {year} {2018})}\BibitemShut {NoStop}%
\bibitem [{\citenamefont {Zyuzin}\ and\ \citenamefont
  {Kovalev}(2018)}]{Zyuzin2018}%
  \BibitemOpen
  \bibfield  {author} {\bibinfo {author} {\bibfnamefont {V.~A.}\ \bibnamefont
  {Zyuzin}}\ and\ \bibinfo {author} {\bibfnamefont {A.~A.}\ \bibnamefont
  {Kovalev}},\ }\bibfield  {title} {\enquote {\bibinfo {title} {Spin {H}all and
  {N}ernst effects of {W}eyl magnons},}\ }\href {\doibase
  10.1103/PhysRevB.97.174407} {\bibfield  {journal} {\bibinfo  {journal} {Phys.
  Rev. B}\ }\textbf {\bibinfo {volume} {97}},\ \bibinfo {pages} {174407}
  (\bibinfo {year} {2018})}\BibitemShut {NoStop}%
\bibitem [{\citenamefont {Mook}\ \emph
  {et~al.}(2019{\natexlab{b}})\citenamefont {Mook}, \citenamefont {Neumann},
  \citenamefont {Henk},\ and\ \citenamefont {Mertig}}]{Mook2019SSE}%
  \BibitemOpen
  \bibfield  {author} {\bibinfo {author} {\bibfnamefont {A.}~\bibnamefont
  {Mook}}, \bibinfo {author} {\bibfnamefont {R.~R.}\ \bibnamefont {Neumann}},
  \bibinfo {author} {\bibfnamefont {J.}~\bibnamefont {Henk}}, \ and\ \bibinfo
  {author} {\bibfnamefont {I.}~\bibnamefont {Mertig}},\ }\bibfield  {title}
  {\enquote {\bibinfo {title} {Spin {S}eebeck and spin {N}ernst effects of
  magnons in noncollinear antiferromagnetic insulators},}\ }\href {\doibase
  10.1103/PhysRevB.100.100401} {\bibfield  {journal} {\bibinfo  {journal}
  {Phys. Rev. B}\ }\textbf {\bibinfo {volume} {100}},\ \bibinfo {pages}
  {100401} (\bibinfo {year} {2019}{\natexlab{b}})}\BibitemShut {NoStop}%
\bibitem [{\citenamefont {Li}\ \emph {et~al.}(2020{\natexlab{a}})\citenamefont
  {Li}, \citenamefont {Sandhoefner},\ and\ \citenamefont
  {Kovalev}}]{Li2020SNE}%
  \BibitemOpen
  \bibfield  {author} {\bibinfo {author} {\bibfnamefont {B.}~\bibnamefont
  {Li}}, \bibinfo {author} {\bibfnamefont {S.}~\bibnamefont {Sandhoefner}}, \
  and\ \bibinfo {author} {\bibfnamefont {A.~A.}\ \bibnamefont {Kovalev}},\
  }\bibfield  {title} {\enquote {\bibinfo {title} {Intrinsic spin {N}ernst
  effect of magnons in a noncollinear antiferromagnet},}\ }\href {\doibase
  10.1103/PhysRevResearch.2.013079} {\bibfield  {journal} {\bibinfo  {journal}
  {Phys. Rev. Research}\ }\textbf {\bibinfo {volume} {2}},\ \bibinfo {pages}
  {013079} (\bibinfo {year} {2020}{\natexlab{a}})}\BibitemShut {NoStop}%
\bibitem [{\citenamefont {Mook}\ \emph
  {et~al.}(2020{\natexlab{b}})\citenamefont {Mook}, \citenamefont {Neumann},
  \citenamefont {Johansson}, \citenamefont {Henk},\ and\ \citenamefont
  {Mertig}}]{Mook2020MSHE}%
  \BibitemOpen
  \bibfield  {author} {\bibinfo {author} {\bibfnamefont {A.}~\bibnamefont
  {Mook}}, \bibinfo {author} {\bibfnamefont {R.~R.}\ \bibnamefont {Neumann}},
  \bibinfo {author} {\bibfnamefont {A.}~\bibnamefont {Johansson}}, \bibinfo
  {author} {\bibfnamefont {J.}~\bibnamefont {Henk}}, \ and\ \bibinfo {author}
  {\bibfnamefont {I.}~\bibnamefont {Mertig}},\ }\bibfield  {title} {\enquote
  {\bibinfo {title} {Origin of the magnetic spin {H}all effect: {S}pin current
  vorticity in the fermi sea},}\ }\href {\doibase
  10.1103/PhysRevResearch.2.023065} {\bibfield  {journal} {\bibinfo  {journal}
  {Phys. Rev. Research}\ }\textbf {\bibinfo {volume} {2}},\ \bibinfo {pages}
  {023065} (\bibinfo {year} {2020}{\natexlab{b}})}\BibitemShut {NoStop}%
\bibitem [{\citenamefont {Li}\ \emph {et~al.}(2020{\natexlab{b}})\citenamefont
  {Li}, \citenamefont {Mook}, \citenamefont {Raeliarijaona},\ and\
  \citenamefont {Kovalev}}]{Li2020Edelstein}%
  \BibitemOpen
  \bibfield  {author} {\bibinfo {author} {\bibfnamefont {B.}~\bibnamefont
  {Li}}, \bibinfo {author} {\bibfnamefont {A.}~\bibnamefont {Mook}}, \bibinfo
  {author} {\bibfnamefont {A.}~\bibnamefont {Raeliarijaona}}, \ and\ \bibinfo
  {author} {\bibfnamefont {A.~A.}\ \bibnamefont {Kovalev}},\ }\bibfield
  {title} {\enquote {\bibinfo {title} {Magnonic analog of the {E}delstein
  effect in antiferromagnetic insulators},}\ }\href {\doibase
  10.1103/PhysRevB.101.024427} {\bibfield  {journal} {\bibinfo  {journal}
  {Phys. Rev. B}\ }\textbf {\bibinfo {volume} {101}},\ \bibinfo {pages}
  {024427} (\bibinfo {year} {2020}{\natexlab{b}})}\BibitemShut {NoStop}%
\bibitem [{\citenamefont {Neumann}\ \emph {et~al.}(2020)\citenamefont
  {Neumann}, \citenamefont {Mook}, \citenamefont {Henk},\ and\ \citenamefont
  {Mertig}}]{Neumann2020}%
  \BibitemOpen
  \bibfield  {author} {\bibinfo {author} {\bibfnamefont {R.~R.}\ \bibnamefont
  {Neumann}}, \bibinfo {author} {\bibfnamefont {A.}~\bibnamefont {Mook}},
  \bibinfo {author} {\bibfnamefont {J.}~\bibnamefont {Henk}}, \ and\ \bibinfo
  {author} {\bibfnamefont {I.}~\bibnamefont {Mertig}},\ }\bibfield  {title}
  {\enquote {\bibinfo {title} {Orbital magnetic moment of magnons},}\ }\href
  {\doibase 10.1103/PhysRevLett.125.117209} {\bibfield  {journal} {\bibinfo
  {journal} {Phys. Rev. Lett.}\ }\textbf {\bibinfo {volume} {125}},\ \bibinfo
  {pages} {117209} (\bibinfo {year} {2020})}\BibitemShut {NoStop}%
\bibitem [{\citenamefont {Cohen}\ and\ \citenamefont
  {Ruvalds}(1969)}]{Cohen1969}%
  \BibitemOpen
  \bibfield  {author} {\bibinfo {author} {\bibfnamefont {M.~H.}\ \bibnamefont
  {Cohen}}\ and\ \bibinfo {author} {\bibfnamefont {J.}~\bibnamefont
  {Ruvalds}},\ }\bibfield  {title} {\enquote {\bibinfo {title} {Two-phonon
  bound states},}\ }\href {\doibase 10.1103/PhysRevLett.23.1378} {\bibfield
  {journal} {\bibinfo  {journal} {Phys. Rev. Lett.}\ }\textbf {\bibinfo
  {volume} {23}},\ \bibinfo {pages} {1378--1381} (\bibinfo {year}
  {1969})}\BibitemShut {NoStop}%
\bibitem [{\citenamefont {Ziman}(2001)}]{Ziman2001}%
  \BibitemOpen
  \bibfield  {author} {\bibinfo {author} {\bibfnamefont {J.}~\bibnamefont
  {Ziman}},\ }\href {\doibase 10.1093/acprof:oso/9780198507796.001.0001} {\emph
  {\bibinfo {title} {Electrons and Phonons}}}\ (\bibinfo  {publisher} {Oxford
  University Press},\ \bibinfo {year} {2001})\BibitemShut {NoStop}%
\bibitem [{\citenamefont {Cohen-Tannoudji}\ \emph {et~al.}(1998)\citenamefont
  {Cohen-Tannoudji}, \citenamefont {Dupont-Roc},\ and\ \citenamefont
  {Grynberg}}]{CohenTannoudji1998}%
  \BibitemOpen
  \bibfield  {author} {\bibinfo {author} {\bibfnamefont {C.}~\bibnamefont
  {Cohen-Tannoudji}}, \bibinfo {author} {\bibfnamefont {J.}~\bibnamefont
  {Dupont-Roc}}, \ and\ \bibinfo {author} {\bibfnamefont {G.}~\bibnamefont
  {Grynberg}},\ }\href {\doibase 10.1002/9783527617197} {\emph {\bibinfo
  {title} {Atom{\textemdash}Photon Interactions}}}\ (\bibinfo  {publisher}
  {Wiley},\ \bibinfo {year} {1998})\BibitemShut {NoStop}%
\bibitem [{\citenamefont {Derzhko}\ \emph {et~al.}(2006)\citenamefont
  {Derzhko}, \citenamefont {Verkholyak}, \citenamefont {Krokhmalskii},\ and\
  \citenamefont {B\"uttner}}]{Derzhko2006}%
  \BibitemOpen
  \bibfield  {author} {\bibinfo {author} {\bibfnamefont {O.}~\bibnamefont
  {Derzhko}}, \bibinfo {author} {\bibfnamefont {T.}~\bibnamefont {Verkholyak}},
  \bibinfo {author} {\bibfnamefont {T.}~\bibnamefont {Krokhmalskii}}, \ and\
  \bibinfo {author} {\bibfnamefont {H.}~\bibnamefont {B\"uttner}},\ }\bibfield
  {title} {\enquote {\bibinfo {title} {Dynamic probes of quantum spin chains
  with the {D}zyaloshinskii-{M}oriya interaction},}\ }\href {\doibase
  10.1103/PhysRevB.73.214407} {\bibfield  {journal} {\bibinfo  {journal} {Phys.
  Rev. B}\ }\textbf {\bibinfo {volume} {73}},\ \bibinfo {pages} {214407}
  (\bibinfo {year} {2006})}\BibitemShut {NoStop}%
\bibitem [{\citenamefont {Soltani}\ \emph {et~al.}(2019)\citenamefont
  {Soltani}, \citenamefont {{Khastehdel Fumani}},\ and\ \citenamefont
  {Mahdavifar}}]{Soltani2019}%
  \BibitemOpen
  \bibfield  {author} {\bibinfo {author} {\bibfnamefont {M.}~\bibnamefont
  {Soltani}}, \bibinfo {author} {\bibfnamefont {F.}~\bibnamefont {{Khastehdel
  Fumani}}}, \ and\ \bibinfo {author} {\bibfnamefont {S.}~\bibnamefont
  {Mahdavifar}},\ }\bibfield  {title} {\enquote {\bibinfo {title} {Ising in a
  transverse field with added transverse {D}zyaloshinskii-{M}oriya
  interaction},}\ }\href {\doibase https://doi.org/10.1016/j.jmmm.2018.12.019}
  {\bibfield  {journal} {\bibinfo  {journal} {Journal of Magnetism and Magnetic
  Materials}\ }\textbf {\bibinfo {volume} {476}},\ \bibinfo {pages} {580--588}
  (\bibinfo {year} {2019})}\BibitemShut {NoStop}%
\bibitem [{\citenamefont {Jordan}\ and\ \citenamefont
  {Wigner}(1928)}]{Jordan1928}%
  \BibitemOpen
  \bibfield  {author} {\bibinfo {author} {\bibfnamefont {P.}~\bibnamefont
  {Jordan}}\ and\ \bibinfo {author} {\bibfnamefont {E.}~\bibnamefont
  {Wigner}},\ }\bibfield  {title} {\enquote {\bibinfo {title} {{\"U}ber das
  {P}aulische {{\"A}}quivalenzverbot},}\ }\href {\doibase 10.1007/BF01331938}
  {\bibfield  {journal} {\bibinfo  {journal} {Zeitschrift f{\"u}r Physik}\
  }\textbf {\bibinfo {volume} {47}},\ \bibinfo {pages} {631--651} (\bibinfo
  {year} {1928})}\BibitemShut {NoStop}%
\bibitem [{\citenamefont {Kulkarni}\ \emph {et~al.}(1999)\citenamefont
  {Kulkarni}, \citenamefont {Schmidt},\ and\ \citenamefont
  {Tsui}}]{Kulkarni1999}%
  \BibitemOpen
  \bibfield  {author} {\bibinfo {author} {\bibfnamefont {D.}~\bibnamefont
  {Kulkarni}}, \bibinfo {author} {\bibfnamefont {D.}~\bibnamefont {Schmidt}}, \
  and\ \bibinfo {author} {\bibfnamefont {S.-K.}\ \bibnamefont {Tsui}},\
  }\bibfield  {title} {\enquote {\bibinfo {title} {Eigenvalues of tridiagonal
  pseudo-{T}oeplitz matrices},}\ }\href {\doibase
  10.1016/s0024-3795(99)00114-7} {\bibfield  {journal} {\bibinfo  {journal}
  {Linear Algebra and its Applications}\ }\textbf {\bibinfo {volume} {297}},\
  \bibinfo {pages} {63--80} (\bibinfo {year} {1999})}\BibitemShut {NoStop}%
\bibitem [{\citenamefont {Noschese}\ \emph {et~al.}(2012)\citenamefont
  {Noschese}, \citenamefont {Pasquini},\ and\ \citenamefont
  {Reichel}}]{Noschese2012}%
  \BibitemOpen
  \bibfield  {author} {\bibinfo {author} {\bibfnamefont {S.}~\bibnamefont
  {Noschese}}, \bibinfo {author} {\bibfnamefont {L.}~\bibnamefont {Pasquini}},
  \ and\ \bibinfo {author} {\bibfnamefont {L.}~\bibnamefont {Reichel}},\
  }\bibfield  {title} {\enquote {\bibinfo {title} {Tridiagonal {T}oeplitz
  matrices: properties and novel applications},}\ }\href {\doibase
  10.1002/nla.1811} {\bibfield  {journal} {\bibinfo  {journal} {Numerical
  Linear Algebra with Applications}\ }\textbf {\bibinfo {volume} {20}},\
  \bibinfo {pages} {302--326} (\bibinfo {year} {2012})}\BibitemShut {NoStop}%
\bibitem [{\citenamefont {Suzuki}\ \emph {et~al.}(2013)\citenamefont {Suzuki},
  \citenamefont {ichi Inoue},\ and\ \citenamefont {Chakrabarti}}]{Suzuki2013}%
  \BibitemOpen
  \bibfield  {author} {\bibinfo {author} {\bibfnamefont {S.}~\bibnamefont
  {Suzuki}}, \bibinfo {author} {\bibfnamefont {J.}~\bibnamefont {ichi Inoue}},
  \ and\ \bibinfo {author} {\bibfnamefont {B.~K.}\ \bibnamefont
  {Chakrabarti}},\ }\href {\doibase 10.1007/978-3-642-33039-1} {\emph {\bibinfo
  {title} {Quantum Ising Phases and Transitions in Transverse Ising Models}}}\
  (\bibinfo  {publisher} {Springer Berlin Heidelberg},\ \bibinfo {year}
  {2013})\BibitemShut {NoStop}%
\end{thebibliography}%

\end{document}